**DIPOLE-LOADED MONOPOLE OPTIMIZED USING VSO, v.3**

**R. A. Formato**

Consulting Engineer & Registered Patent Attorney
P.O. Box 1714
Harwich, MA 02645 USA
rf2@ieee.org

**Abstract —** A dipole-loaded monopole antenna is optimized for uniform hemispherical coverage using VSO, a new global search design and optimization algorithm. The antenna's performance is compared to genetic algorithm and hill-climber optimized loaded monopoles, and VSO is tested against two suites of benchmark functions and several other algorithms.

## 1. INTRODUCTION

This paper describes the design optimization of a dipole-loaded monopole using a new design and optimization (D&O) methodology, Very Simple Optimization (VSO). The prototype antenna was introduced by Altshuler [1], and its design subsequently refined using Genetic Algorithm (GA) optimization [2]. VSO is introduced as a new deterministic, elitist D&O methodology and tested against the dipole-loaded monopole problem and against two benchmark suites using several other algorithms. GA is fundamentally stochastic, and consequently produces a different antenna design every time it is run. Such uncertainty can be a significant impediment because there is no good way of knowing why any particular antenna design is different from others. It could be the result of changing the fitness function, or, what is more likely, due to GA's inherent randomness. The effects of different objective functions are best investigated using deterministic D&O algorithms like VSO because every run with the same setup returns the same results.

## 2. THE DIPOLE-LOADED MONOPOLE

The VSO-optimized loaded monopole is shown schematically in Figure 1. It comprises eight straight wire segments (the long wire parallel to the X-axis consists of two separate wires along the ±X axes). Additional geometry details may be found in [2], Figure 1. All wires are perfectly electrically conducting (PEC), and the antenna is fed at its base against an infinite PEC ground in the plane Z=0 (X-Y plane).

VSO is reminiscent of stochastic Hill Climbing (private communication, Dr. Robert C. Green II, EE/CS, Univ. of Toledo) and consequently merits comparison to that algorithm as well as to GA. Therefore the loaded monopole also was optimized using SAHC, a hill climbing variant described in §4. Perspective views of the VSO, GA, and SAHC-optimized antennas appear in Figures 2(a-c), respectively (axis length 0.2 meter, frequency 299.8 MHz). While the VSO antenna is quite different from the GA and SAHC designs, whose geometries are similar, the radiation patterns of all three monopoles are quite similar as discussed below.

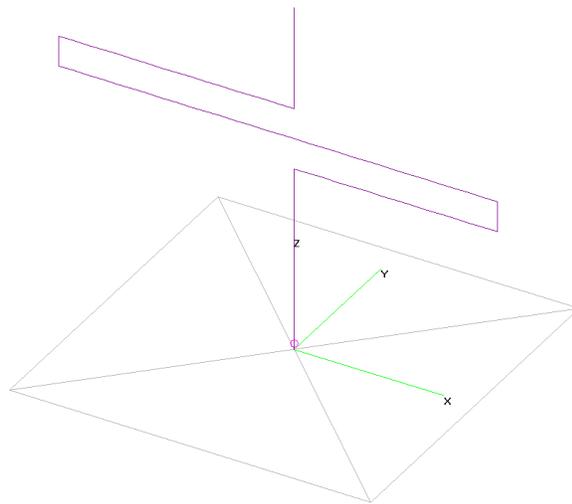

**Figure 1.** Geometry of VSO Dipole-Loaded Monopole.

The VSO/SAHC-optimized monpoles' performance was computed using the Numerical Electromagnetics Code version 4.2 double precision (NEC-4.2D) [3] as the optimizer's modeling engine. An earlier version of NEC was used in [2], and details of the GA setup can be found there. The design objective was a uniform hemispherical radiation pattern without regard to input impedance. The resulting NEC input files for the three optimized monopoles appear in Figures 3(a-c). The VSO/SAHC fitness function (to be maximized) was $f(G) = \dfrac{1}{\sum (G(\theta, \phi) - G_{avg})}$ where $G(\theta, \phi)$ is the power gain as a function of the angles $(\theta, \phi)$ in NEC's standard right-handed spherical polar coordinates. $G_{avg}$ is the average gain over all calculation angles. Gain was computed at 5° increments in $\theta$ and 45° in $\phi$. This fitness function minimizes the difference between the actual and average gains at each calculation point, which has the effect of smoothing the pattern.

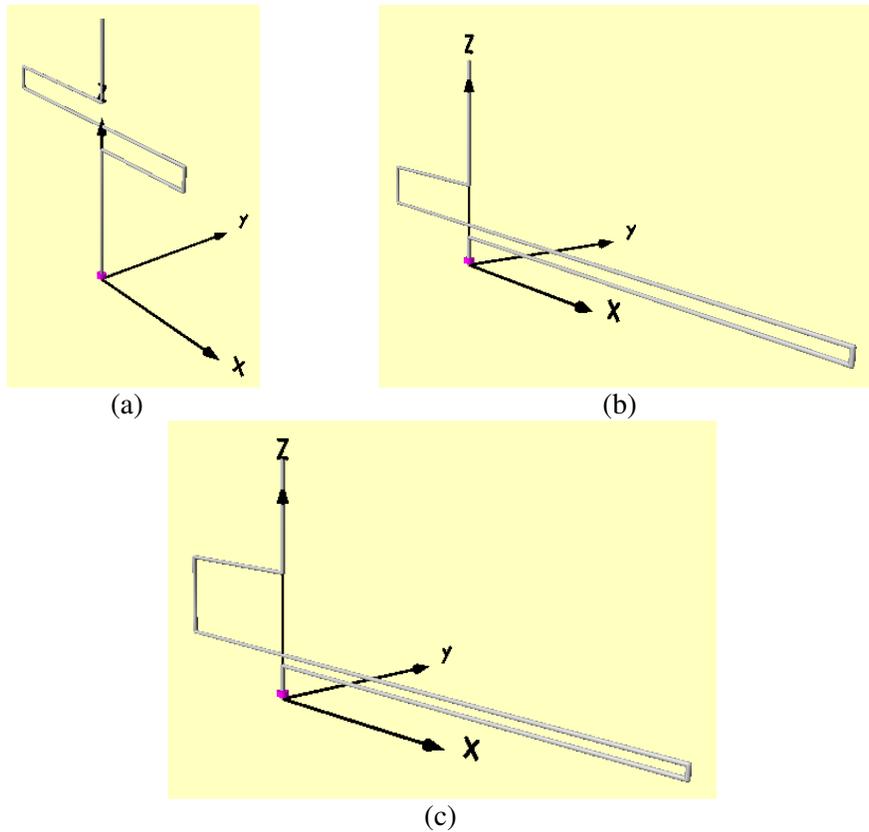

(a)

(b)

(c)

**Figure 2.** VSO, GA, and SAHC Optimized Dipole-Loaded Monopoles.

The antenna designer must specify a suitable objective (fitness) function, but doing so can be problematic [4]. What appears to be a perfectly reasonable function for achieving a desired balance between various antenna performance measures may, in fact, be quite poor. Because stochastic D&O algorithms return different results on every run, there is no good way to test the effects of changing the fitness function, for example, by changing coefficient values or by combining parameters differently. In addition, unlike benchmark testing that uses purely analytical formulations, antenna design usually requires a stand-alone numerical modeling engine, in this case NEC, that often imposes long computation times thus making it even more difficult to evaluate the quality of a specific objective function. VSO addresses these issues.

The VSO/SAHC loaded monopole designs are compared to the GA-optimized monopole by using NEC-4.2D to compute the GA antenna's performance with the geometry data in Figure 3(b) (taken from Table 1 in [2]). Figures 4(a-c), respectively, show the VSO, GA, and SAHC-optimized radiation patterns (total power gain) in an azimuth plane containing the maximum gain, while Figure 5 provides three-dimensional (3-D) perspective views. All three antennas meet the performance objective of providing very uniform coverage of the upper hemisphere. The VSO's gain ranges from a minimum of 2.22 dBi to a maximum of 3.58 dBi (1.36 dB spread), while GA's is between 2.12 and 4.89 dBi (2.77 dB spread), and the corresponding SAHC values are 2.73 to 3.46 dBi (0.73 dB spread). The VSO monopole's gain is 1.41 dB more uniform than the GA's, but less uniform than the SAHC's by 0.68 dB. VSO required far fewer function evaluations than SAHC (8,600 vs. 183,120) because it is deterministic. Even though it was not included as a design objective, NEC outputs antenna input impedance, the VSO, GA and SAHC values, respectively, being $38.4 + j159.4$ , $298.4 + j755.3$ , and $70.2 + j215$ $\Omega$.

## 3. VERY SIMPLE OPTIMIZATION (VSO)

VSO is a new deterministic, iterative, elitist D&O algorithm based on a very simple idea, and it appears to work well for a wide range of functions. This section describes VSO and some benchmark test results. Many global search and optimization metaheuristics are drawn from Nature, prime examples being GA, Particle Swarm Optimization (PSO), Ant Colony Optimization (ACO), Synthetic Annealing (SA), and Biogeography Based Optimization (BBO) as representative metaheuristics. These algorithms are inherently stochastic because they "explore" a decision space in a random manner, just as "survival of the fittest" randomly improves a species, or just as bees or ants randomly search for food, or as other natural processes seek minima, for example, energy level, or maxima, for example, a friendlier habitat. These approaches necessarily return different results on successive runs because some variables must be computed from a probability distribution and cannot be known in advance (true randomness is quite different from pseudo or quasirandomness; see, for example, [5]). VSO by contrast is entirely deterministic because it is purely geometrical in nature (although some measure of randomness can be included if the algorithm designer wishes).

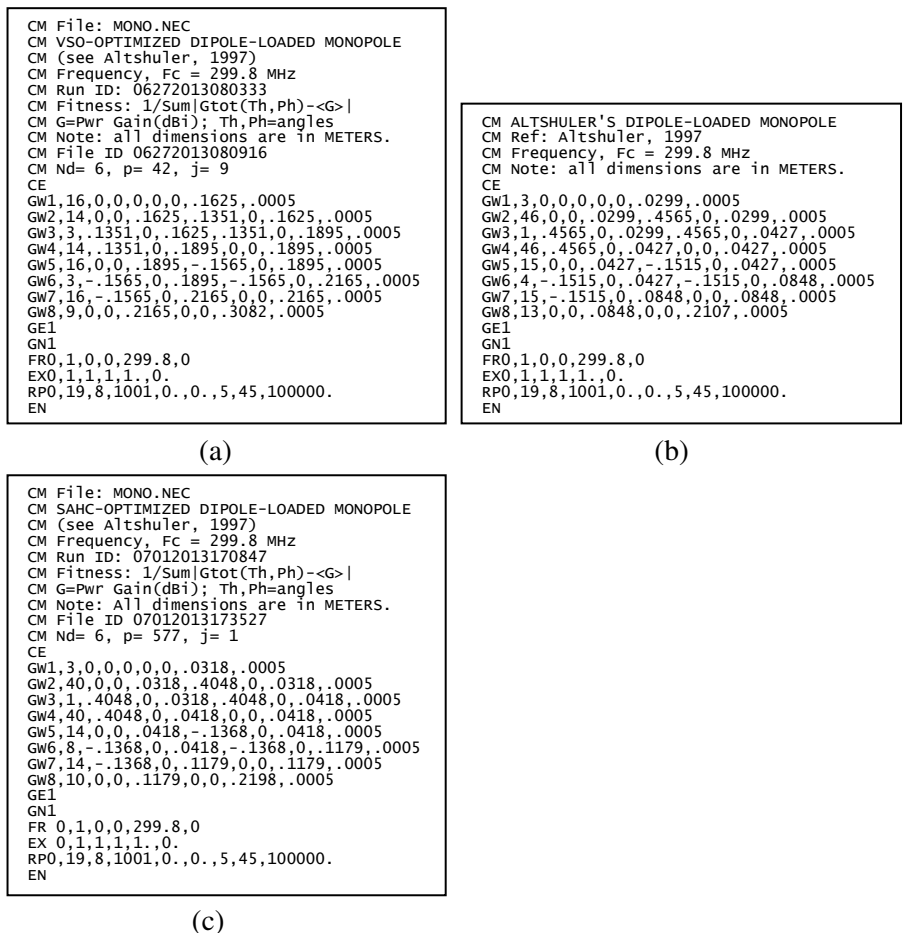

```
CM File: MONO.NEC
CM VSO-OPTIMIZED DIPOLE-LOADED MONOPOLE
CM (see Altshuler, 1997)
CM Frequency, Fc = 299.8 MHz
CM Fitness: 1/Sum|Gtot(Th,Ph)-<G>|
CM Run ID: 06272013080333
CM G=Pwr Gain(dBi); Th,Ph=angles
CM Note: all dimensions are in METERS.
CM File ID 06272013080916
CM Nd= 6, p= 42, j= 9
CE
GW1,16,0,0,0,0,0,.1625,.0005
GW2,14,0,0,.1625,.1351,0,.1625,.0005
GW3,3,.1351,0,.1625,.1351,0,.1895,.0005
GW4,14,.1351,0,.1895,0,0,.1895,.0005
GW5,16,0,0,.1895,-.1515,0,.1895,.0005
GW6,3,-.1565,0,.1895,-.1565,0,.2165,.0005
GW7,16,-.1565,0,.2165,0,0,.2165,.0005
GW8,9,0,0,.2165,0,0,.3082,.0005
GE1
GN1
FR0,1,0,0,299.8,0
EX0,1,1,1,1,.0.
RP0,19,8,1001,0.,0.,5,45,100000.
EN
```

(a)

```
CM ALTSHULER'S DIPOLE-LOADED MONOPOLE
CM Ref: Altshuler, 1997
CM Frequency = 299.8 MHz
CM Note: all dimensions are in METERS.
CE
GW1,3,0,0,0,0,0,.0299,.0005
GW2,46,0,0,.0299,.4565,0,.0299,.0005
GW3,1,.4565,0,.0299,.4565,0,.0427,.0005
GW4,46,.4565,0,.0427,0,0,.0427,.0005
GW5,15,0,0,.0427,-.1515,0,.0427,.0005
GW6,4,-.1515,0,.0427,-.1515,0,.0848,.0005
GW7,15,-.1515,0,.0848,0,0,.0848,.0005
GW8,13,0,0,.0848,0,0,.2107,.0005
GE1
GN1
FR0,1,0,0,299.8,0
EX0,1,1,1,1,.0.
RP0,19,8,1001,0.,0.,5,45,100000.
EN
```

(b)

```
CM File: MONO.NEC
CM SAHC-OPTIMIZED DIPOLE-LOADED MONOPOLE
CM (see Altshuler, 1997)
CM Frequency, Fc = 299.8 MHz
CM Run ID: 07012013170847
CM Fitness: 1/Sum|Gtot(Th,Ph)-<G>|
CM G=Pwr Gain(dBi); Th,Ph=angles
CM Note: All dimensions are in METERS.
CM File ID 07012013173527
CM Nd= 6, p= 577, j= 1
CE
GW1,3,0,0,0,0,0,.0318,.0005
GW2,40,0,0,.0318,.4048,0,.0318,.0005
GW3,1,.4048,0,.0318,.4048,0,.0418,.0005
GW4,40,.4048,0,.0418,0,0,.0418,.0005
GW5,14,0,0,.0418,-.1368,0,.0418,.0005
GW6,8,-.1368,0,.0418,-.1368,0,.1179,.0005
GW7,14,-.1368,0,.1179,0,0,.1179,.0005
GW8,10,0,0,.1179,0,0,.2198,.0005
GE1
GN1
FR 0,1,0,0,299.8,0
EX 0,1,1,1,1,.0.
RP0,19,8,1001,0.,0.,5,45,100000.
EN
```

(c)

**Figure 3.** Optimized Monopole VSO, GA and SAHC NEC Input Files.

The basic idea underlying VSO is that at every iteration each sample point in the decision space (DS) is moved along a straight line towards the point with the overall best fitness up to the previous iteration (which remains fixed). The best fitness can be a maximum or minimum (many algorithms perform minimization), but here VSO is used to maximize an objective function. The VSO algorithm is shown diagrammatically in Figure 6 (O is the coordinate system origin).

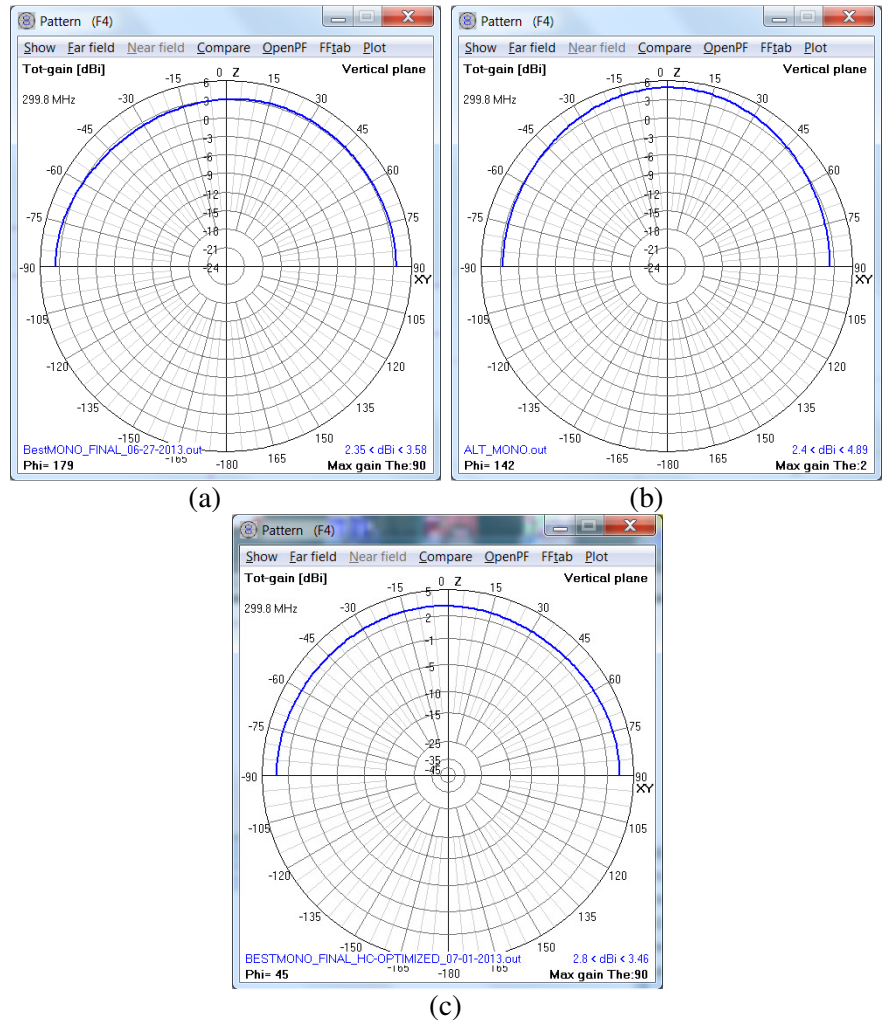

**Figure 4.** Patterns for VSO, GA, and SAHC-optimized Monopoles.

DS is defined by $x_i^{\min} \le x_i \le x_i^{\max}$, $1 \le i \le N_d$, where $i$ is the coordinate number and $N_d$ DS's dimensionality. At VSO's $j^{th}$ iteration the location of each point where the objective function $f(\vec{R})$ is evaluated is specified by a position vector $\vec{R}_j^p = \sum_{k=1}^{N_d} x_k^{p,j} \hat{e}_k$, where $p$ is the sample point number, $x_k^{p,j}$ its coordinates, and $\hat{e}_k$ the unit vector along the $x_k$ axis. The best overall best fitness is $F^* = f(\vec{R}^*)$ at the point $\vec{R}^* = \sum_{k=1}^{N_d} x_k^* \hat{e}_k$, where $f(\cdot)$ is the objective function being maximized. Sample point $p$'s position is updated according to $\vec{R}_j^p = \vec{R}_{j-1}^p + \rho(\vec{R}^* - \vec{R}_{j-1}^p)$ in which the parameter $0 \le \rho \le 1$ determines where along the line joining $\vec{R}_{j-1}^p$ and $\vec{R}^*$ the sample point is relocated.

When $\rho = 0$ the sample point remains in its original position (no movement), whereas when $\rho = 1$ it is moved to the location of the overall best fitness. When $\rho = 0.5$ the sample point is moved to the midpoint of the line joining $\vec{R}_{j-1}^p$ and $\vec{R}^*$. While this value was used for all VSO runs reported here, other fixed values may work better, or perhaps a variable value would. The question of exactly how $\rho$ should be specified has not been investigated further because $\rho = 0.5$ provides good results by bisecting at each step the distance between the sample point's starting location and the location of the best fitness.

Figure 7 shows VSO pseudocode. It is evident that VSO is quite simple; hence its name. Initialization begins with the specification of an Initial Sample Point Distribution (ISPD) against which the initial fitnesses are calculated to determine the best starting fitness. This distribution subsequently is updated at each step by moving each sample point towards the point with the best overall fitness. VSO thus is an elitist algorithm because the overall best fitness always is retained as the point towards which all other sample points are moved.

The ISPD can be deterministic, as it is here, or stochastic or hybrid in nature. As a general proposition, randomness increases an algorithm's ability to explore DS, which is why Nature-based stochastic metaheuristics are so popular. By comparison, VSO offers the "best of both worlds" because at any iteration full or partial randomness can be added if doing so improves performance. For the runs reported here a variant of the deterministic "probe line" distribution described in [6,7] was employed to create ISPD.

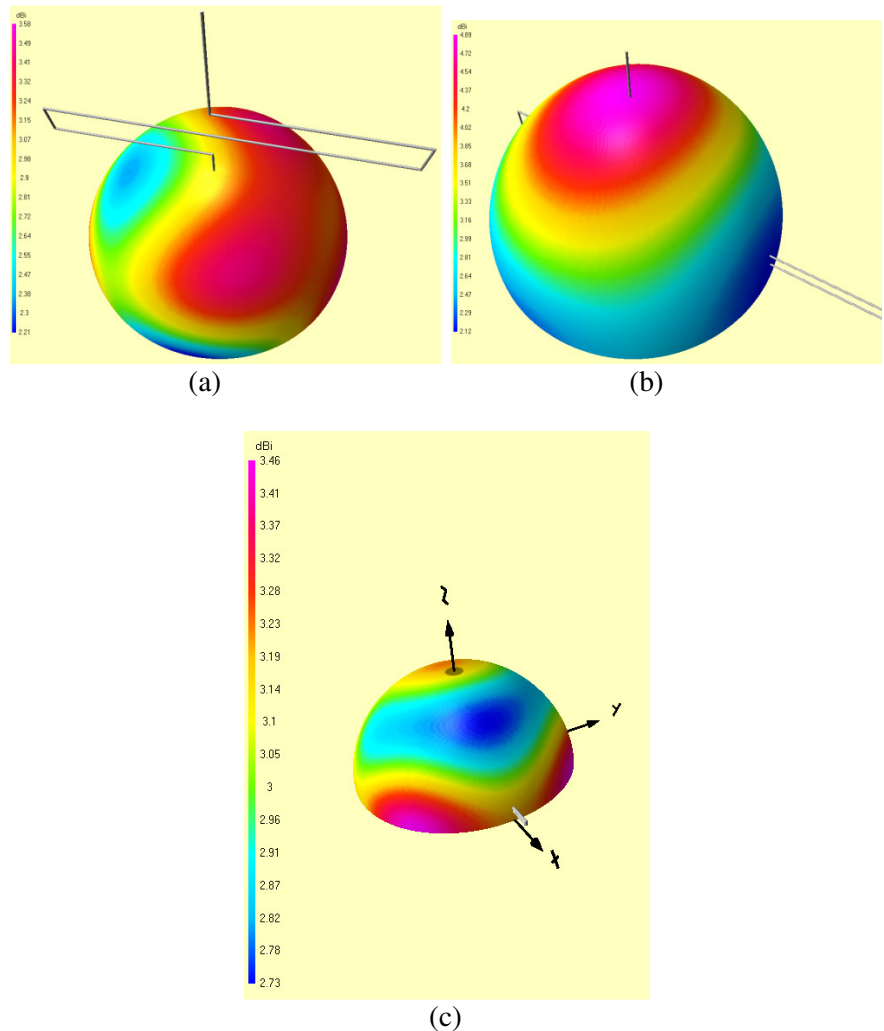

(a)                    (b)

(c)

**Figure 5.** 3-D Patterns for VSO, GA, and SAHC-optimized Monopoles.

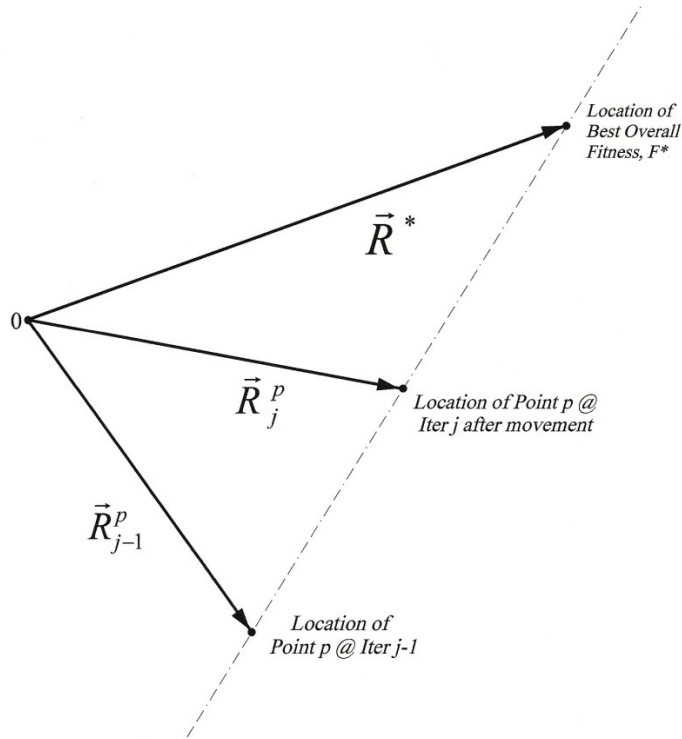

**Figure 6.** VSO's Simple Sample Point Relocation Scheme.

```
Algorithm VSO

I - Initialization:
    j ← 0
    (a)  ISPD
    (b)  Compute initial fitnesses
    (c)  Best fitness: F* = f(R⃗*)
II - Do Until (Termination Criterion)
    j ← j+1
    (a)  Reposition sample points:
         R⃗ⱼᵖ = R⃗ⱼ₋₁ᵖ + ρ(R⃗* − R⃗ⱼ₋₁ᵖ)
    (b)  Compute fitnesses using current
         sample point distribution.
    (c)  Update best fitness:
         If f(R⃗*) ≥ F* ∴ F* = f(R⃗*)
```

**Figure 7.** VSO Pseudocode.

VSO's ISPD is generated by placing sample points uniformly along lines parallel to the coordinate system axes that intersect at a point along the decision space principal diagonal. The intersection point is

$$\vec{D} = \vec{X}_{\min} + \gamma(\vec{X}_{\max} - \vec{X}_{\min}) \quad \text{where} \quad \vec{X}_{\min} = \sum_{i=1}^{N_d} x_i^{\min} \hat{e}_i \quad \text{and} \quad \vec{X}_{\max} = \sum_{i=1}^{N_d} x_i^{\max} \hat{e}_i$$

are the diagonal's endpoint vectors. Parameter $0 \le \gamma \le 1$ determines where along the diagonal the orthogonal sample point array is placed. For the runs reported here, ten $\gamma$ values were used uniformly spaced in the intervals $0.05 \le \gamma < 0.49$ and $0.51 \le \gamma < 0.95$. To avoid the possibility of a biased ISPD, $\gamma = 0.5$ is excluded intentionally so that no sample points are placed at the origin in a symmetrical DS, and an even number of sample points, in this case 14 per line, was used for the same reason. A typical three-dimensional ISPD using these parameter values is shown in Figure 8 (the oblique line is the principal diagonal). VSO then evolves ISPD iteration-by-iteration according to the simple repositioning scheme described above.

A VSO run ends when a user-specified termination criterion is met, often a predetermined number of iterations or saturation of the best returned fitness. The runs reported here terminated on the earlier of 15 iterations or fitness saturation within 0.001 over 4 iterations tested every third step, that is, when $j \, \text{MOD} \, 3 = 0$. In almost every case this early termination criterion was met in fewer than 15 iterations.

## 4. BECHMARK FUNCTIONS

Global search and optimization algorithms typically are tested against standard suites of benchmark functions. Two sets of benchmarks were used to test VSO. The six-function suite in [8] was selected because it provides a direct comparison to vibrational PSO (v-PSO), which is a new state-of-the-art D&O methodology. Table I shows the form of the test functions, DS, and the location and value of the known maximum (same notation as [8]). Note that the signs have been changed because VSO performs maximization whereas v-PSO performs minimization, and that the form of $f_1$ in [8] is missing the last two terms and DS is slightly smaller (see $f_{10}$ in [9]).

Because v-PSO is stochastic, in [8] it was tested statistically by making 100 runs and averaging the results. The algorithm was applied to each benchmark in 10, 20 and 30 dimensions using a total of 200,000 function evaluations per run (10,000 generations, 20 particle swarm), thus resulting in a total of 20,000,000 evaluations for each of the six benchmarks. Results for v-PSO are reported in Table IV in [8] and reproduced here in Table II, along with VSO's results for the same cases.

VSO out-performed v-PSO in every case, both in terms of the quality of its solution and by far in the number of function evaluations, $N_{eval}$. Compared to 20,000,000 evaluations per function for v-PSO, VSO required a worst case maximum of only 67,200 for the 30-dimensional Schwefel and returned much better results. VSO's determinism greatly reduces the number of function evaluations, but at the same time VSO may not explore DS as well as a stochastic algorithm. It is evident that how VSO's ISPD is configured is an important consideration, perhaps the most important single factor in setting up effective VSO runs, an issue that requires further investigation.

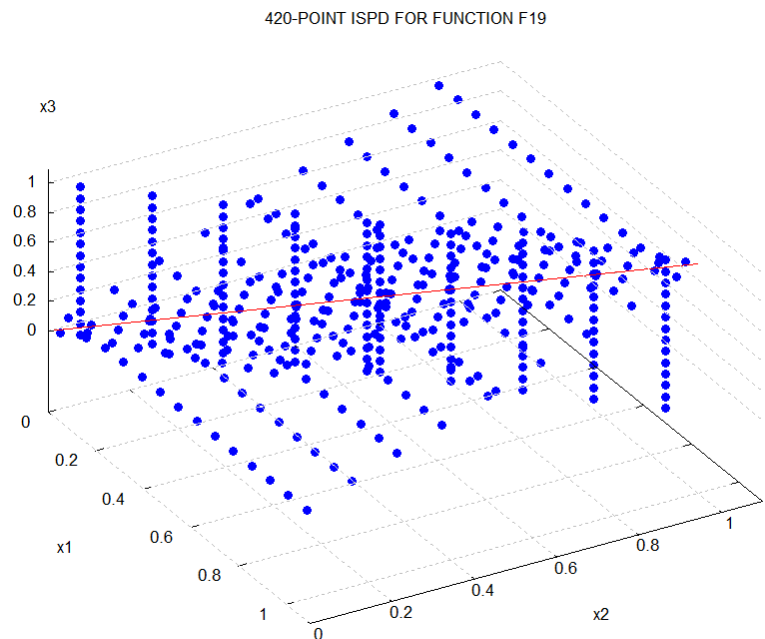

**Figure 8.** VSO Three Dimensional ISPD

In addition to testing against the v-PSO benchmarks, VSO also was tested against a recognized suite of twenty three benchmarks. The results are summarized in Table III. This suite's description appears in detail in [6,9,10] and consequently is not repeated here (note that there is some overlap with the v-PSO suite). In Table III, VSO is compared directly to Central Force Optimization (CFO) and to Group Search Optimizer (GSO), which is based on a metaphor of animal foraging [10], and indirectly to two other GA and PSO variants. CFO was selected because, like VSO, it is deterministic. GSO was selected because it has been extensively tested and was itself compared to two other algorithms, GA and PSO, as described in [10].

The test functions in Table III are numbered as they are in the GSO paper. $f_{max}$ is the known global maximum (note that the negative of each function is used because, unlike the other algorithms, CFO, like VSO, searches for maxima instead of minima). The $< \cdot >$ brackets denote mean value because all the algorithms discussed in [10] are stochastic, thus requiring a statistical assessment. The data in Table III for the other algorithms are reproduced from [10]. As noted, the high dimensionality results are averages over 1,000 runs, whereas the lower dimensionality data are 50 run averages. The GSO experimental setup is described in detail in [10]. By contrast, because both CFO and VSO are deterministic, their results are repeatable over runs with the same parameters so that only a single run is required. The same VSO setup described above was used for the 23-benchmark suite.

The results in Table III speak for themselves. VSO returned the best or equal fitness on 12 of the benchmarks, a very robust performance in view of the relatively small number of function evaluations (maximum of 67,200 in only two cases). In every other case, VSO's best fitness was close to the actual maximum, except for benchmark $f_{14}$ where its returned fitness was poor compared to the known maximum. It is likely that VSO would perform considerably better against $f_{14}$ using a different ISPD, which again highlights ISPD's importance.

A version of Steepest Ascent Hill Climbing with Replacement (SAHC) [11] was implemented using essentially the same setup parameters as VSO. The number of sample points was equalized by increasing the number per dimension from 14 to 140 because VSO utilized 10 gamma values in its ISPD, while gamma is not a parameter in SAHC.

Its ISPD was computed as $\vec{R}_0^p = \sum_{k=1}^{N_d} [x_k^{min} + r_0 (x_k^{max} - x_k^{min})]\hat{e}_k, \ p = 1,...,N_p,$

where $0.05 \leq r_0 < 0.95$ is a uniformly distributed random variable (RV). At each step the sample points' locations were tweaked according to $\vec{R}_j^{\,p} = \sum_{k=1}^{N_d} (x_k^{p,j-1} + r_j\, L_{diag})\hat{e}_k$ where $-0.1 \leq r_j < 0.1$ is a uniformly distributed RV and $L_{diag} = \sqrt{\sum_{k=1}^{N_d} (x_j^{\max} - x_j^{\min})^2}$ the length of DS's principal diagonal. The same VSO early termination criterion was applied. A total of 1,000 independent SAHC runs was made, each with a new random ISPD, and the best fitness over all runs returned.

Results comparing VSO and SAHC using the GSO benchmark suite appear in Table IV. VSO returned the best fitness against all but five benchmarks, and those were all low-dimensionality functions. Perhaps the most important data in Table IV are the number of function evaluations. VSO never required more than 67,200 calculations, whereas SAHC required no fewer than 1,120,840, which is a staggering difference. On the 30-D functions the difference is even greater, with SAHC requiring at least 16,850,400 function evaluations yielding solutions that generally were quite poor compared to VSO. It is evident that VSO markedly outperforms SAHC for real-world problems like antenna design where the number of modeling engine runs can be a significant impediment in formulating an effective objective function.

## 5. CONCLUSION

This paper applies VSO to the optimization of a dipole-loaded monopole antenna with very good results. It describes VSO as a new, easily implemented deterministic, iterative D&O algorithm, and reports test data using two benchmark suites that compare it to other stochastic algorithms, again with very good results. VSO holds promise as an effective D&O methodology, especially for problems requiring the formulation of a suitable objective function. VSO merits further study, in particular with respect to how ISPD should be specified. The Appendix contains a complete source code listing for the VSO implementation used here, and an electronic listing is available upon request to the author (rf2@ieee.org).

## TABLE I.  v-PSO Benchmark Functions

| $f$ | Function | $f(x)$ | DS | $x^*$ | $f(x^*)$ |
|---|---|---|---|---|---|
| $f_1$ | Ackley | $20\exp\left(-0.2\sqrt{\frac{1}{N_d}\sum_{i=1}^{N_d}x_i^2}\right)+$ $\exp\left(\frac{1}{N_d}\sum_{i=1}^{N_d}\cos(2\pi x_i)\right)-20-e$ | $[-30,30]^{N_d}$ | $[0]^{N_d}$ | $0$ |
| $f_2$ | Cosine Mixture | $-\sum_{i=1}^{N_d}x_i^2+0.1\sum_{i=1}^{N_d}\cos(5\pi x_i)$ | $[-1,1]^{N_d}$ | $[0]^{N_d}$ | $0.1N_d$ |
| $f_3$ | Exponential | $\exp(-0.5\sum_{i=1}^{N_d}x_i^2)$ | $[-1,1]^{N_d}$ | $[0]^{N_d}$ | $1$ |
| $f_4$ | Griewank | $-\frac{1}{4000}\sum_{i=1}^{N_d}(x_i-100)^2+$ $\prod_{i=1}^{N_d}\cos\left(\frac{x_i-100}{\sqrt{i}}\right)-1$ | $[-600,600]^{N_d}$ | $[0]^{N_d}$ | |
| $f_5$ | Rastrigin | $-\sum_{i=1}^{N_d}[x_i^2-10\cos(2\pi x_i)+10]$ | $[-5.12,5.12]^{N_d}$ | $[0]^{N_d}$ | $0$ |
| $f_6$ | Schwefel | $-418.9829\,N_d+\sum_{i=1}^{N_d}[x_i\sin(\sqrt{|x_i|})]$ | $[-500,500]^{N_d}$ | $[420.9687]^{N_d}$ | $0$ |

## TABLE II.  VSO Results for v-PSO Benchmarks

| $f$ | $N_d$ | v-PSO* | VSO | Neval |
|---|---|---|---|---|
| $f_1$ | 10 | −1.84e-15±2.9e-16 | 4.77e-18 | 9,800 |
| | 20 | −2.84e-15±1.5e-16 | 4.77e-18 | 19,600 |
| | 30 | −4.93e-15±3.4e-16 | 4.77e-18 | 29,400 |
| $f_2$ | 10 | 1±0 | 1 | 9,800 |
| | 20 | 2±0 | 2 | 19,600 |
| | 30 | 3±0 | 3 | 29,400 |
| $f_3$ | 10 | 1±0 | 1 | 9,800 |
| | 20 | 1±3e-18 | 1 | 19,600 |
| | 30 | 1±1e-17 | 1 | 29,400 |
| $f_4$ | 10 | −0.020±0.006 | 0 | 9,800 |
| | 20 | −0.0026±0.002 | 0 | 19,600 |
| | 30 | −8.8568e-4±0.001 | 0 | 29,400 |
| $f_5$ | 10 | 0±0 | 0 | 9,800 |
| | 20 | 0±0 | 0 | 19,600 |
| | 30 | −5.6843e-16±1e-15 | 0 | 29,400 |
| $f_6$ | 10 | −620.8131±50.4 | −1.305e-4 | 18,200 |
| | 20 | −1.3384e+3±68.5 | −2.551e-4 | 44,800 |
| | 30 | −2.1395e+3±103.3 | −3.827e-4 | 67,200 |

\* average best fitness; data reproduced from Table IV in [8]

TABLE III
Comparative Results for GSO 23-Function Benchmark Suite

| $f$ | $N_d$ | $f_{max}^{(1)}$ | <Best Fitness>/Other Algorithm | CFO | | VSO | |
|---|---|---|---|---|---|---|---|
| | | | | Best Fitness | $N_{eval}$ | Best Fitness | $N_{eval}$ |
| Unimodal Functions (*other algorithms:* average of 1000 runs) | | | | | | | |
| $f_1$ | 30 | 0 | **-3.6927x10$^{-37}$** / PSO | -2.6592x10$^{-2}$ | 108,660 | -1.2037x10$^{-35}$ | 29,400 |
| $f_2$ | 30 | 0 | **-2.9168x10$^{-24}$** / PSO | -4x10$^{-8}$ | 161,640 | -4.3368x10$^{-19}$ | 29,400 |
| $f_3$ | 30 | 0 | -1.1979x10$^{-3}$ / PSO | -6x10$^{-8}$ | 239,340 | **0** | 29,400 |
| $f_4$ | 30 | 0 | -0.1078 / GSO | -4.2x10$^{-7}$ | 59,160 | **-3.4694x10$^{-18}$** | 29,400 |
| $f_5$ | 30 | 0 | -37.3582 / PSO | **-2.17187x10$^{-2}$** | 164,160 | -1.3154x10$^{-5}$ | 54,600 |
| $f_6$ | 30 | 0 | -1.6000x10$^{-2}$ / GSO | **0** | 73,620 | **0** | 29,600 |
| $f_7$ | 30 | 0 | -9.9024x10$^{-3}$ / PSO | -3.55996x10$^{-3}$ | 66,660 | **-7.7487x10$^{-4}$** | 29,400 |
| Multimodal Functions, Many Local Maxima (*other algorithms:* average of 1000 runs) | | | | | | | |
| $f_8$ | 30 | 12,569.5 | **12,569.4882** / GSO | 12,569.4852 | 69,720 | 12,569.4866 | 67,200 |
| $f_9$ | 30 | 0 | -0.6509 / GA | -3.52x10$^{-6}$ | 117,120 | **0** | 29,400 |
| $f_{10}$ | 30 | 0 | -2.6548x10$^{-5}$ / GSO | -1.5x10$^{-7}$ | 111,660 | **4.7705x10$^{-18}$** | 29,400 |
| $f_{11}$ | 30 | 0 | **-3.0792x10$^{-2}$** / GSO | -2.00124 | 160,680 | -8.2269x10$^{-2}$ | 67,200 |
| $f_{12}$ | 30 | 0 | **-2.7648x10$^{-11}$** / GSO | -0.105859 | 68,220 | -5.0111x10$^{-6}$ | 29,400 |
| $f_{13}$ | 30 | 0 | -4.6948x10$^{-5}$ / GSO | -6.5966x10$^{-2}$ | 103,320 | **-3.2007x10$^{-6}$** | 42,000 |
| Multimodal Functions, Few Local Maxima (*other algorithms:* average of 50 runs) | | | | | | | |
| $f_{14}$ | 2 | -1 | **-0.9980** / GSO | -1.005284 | 12,824 | -6.9034 | 2,800 |
| $f_{15}$ | 4 | -0.0003075 | **-3.7713x10$^{-4}$** / GSO | -2.41631x10$^{-3}$ | 19,920 | -1.6333x10$^{-3}$ | 3,920 |
| $f_{16}$ | 2 | 1.0316285 | **1.031628** / GSO | 1.031607 | 9,256 | 1.0316239 | 2,800 |
| $f_{17}$ | 2 | -0.398 | -0.3979 / GSO | **-0.398** | 8,824 | -0.3979 | 2,800 |
| $f_{18}$ | 2 | -3 | **-3** / GSO | **-3** | 15,784 | **-3** | 2,800 |
| $f_{19}$ | 3 | 3.86 | **3.8628** / GSO | 3.86157 | 12,612 | 3.774 | 4,200 |
| $f_{20}$ | 6 | 3.32 | 3.2697 / GSO | **3.31976** | 52,128 | 3.0333 | 10,920 |
| $f_{21}$ | 4 | 10 | 7.5439 / PSO | 10.1466 | 25,376 | **10.1532** | 7,280 |
| $f_{22}$ | 4 | 10 | 8.3553 / PSO | 10.4028 | 29,168 | **10.4029** | 7,280 |
| $f_{23}$ | 4 | 10 | 8.9439 / PSO | 10.5362 | 24,784 | **10.5364** | 7,280 |

**(1)** **Negative of the functions in [10] computed by CFO/VSO because they search for maxima instead of minima.**

TABLE IV
Comparative Results SAHC & VSO

| $f$ | $N_d$ | $f_{max}^{(1)}$ | SAHC | | VSO | |
|---|---|---|---|---|---|---|
| | | | Best Fitness (over 1,000 runs) | $N_{eval}$ | Best Fitness | $N_{eval}$ |
| Unimodal Functions | | | | | | |
| $f_1$ | 30 | 0 | -3.1296x10$^4$ | 16,850,400 | **-1.2037x10$^{-35}$** | 29,400 |
| $f_2$ | 30 | 0 | -118.1927 | 16,863,000 | **-4.3368x10$^{-19}$** | 29,400 |
| $f_3$ | 30 | 0 | -2.9379x10$^4$ | 16,900,800 | **0** | 29,400 |
| $f_4$ | 30 | 0 | -56.9567 | 16,888,200 | **-3.4694x10$^{-18}$** | 29,400 |
| $f_5$ | 30 | 0 | -1.9033x10$^{14}$ | 16,938,600 | **-1.3154x10$^{-5}$** | 54,600 |
| $f_6$ | 30 | 0 | -2.6567x10$^4$ | 16,888,200 | **0** | 29,600 |
| $f_7$ | 30 | 0 | -23.7855 | 16,900,800 | **-7.7487x10$^{-4}$** | 29,400 |
| Multimodal Functions, Many Local Maxima | | | | | | |
| $f_8$ | 30 | 12,569.5 | 5,462.47 | 16,913,400 | **12,569.4866** | 67,200 |
| $f_9$ | 30 | 0 | -4,340.43 | 16,850,400 | **0** | 29,400 |
| $f_{10}$ | 30 | 0 | -18.8522 | 16,850,400 | **4.7705x10$^{-18}$** | 29,400 |
| $f_{11}$ | 30 | 0 | -263.789 | 16,913,400 | **-8.2269x10$^{-2}$** | 67,200 |
| $f_{12}$ | 30 | 0 | -5.0287x10$^7$ | 16,913,400 | **-5.0111x10$^{-6}$** | 29,400 |
| $f_{13}$ | 30 | 0 | -1.2672x10$^8$ | 16,888,200 | **-3.2007x10$^{-4}$** | 42,000 |
| Multimodal Functions, Few Local Maxima | | | | | | |
| $f_{14}$ | 2 | -1 | **-0.998** | 1,126,720 | -6.9034 | 2,800 |
| $f_{15}$ | 4 | -0.0003075 | **-9.3807x10$^{-4}$** | 2,246,720 | -1.6333x10$^{-3}$ | 3,920 |
| $f_{16}$ | 2 | 1.0316285 | 1.0316071 | 1,125,880 | **1.0316239** | 2,800 |
| $f_{17}$ | 2 | -0.398 | **-0.397986** | 1,120,840 | -0.3979 | 2,800 |
| $f_{18}$ | 2 | -3 | -3.00144 | 1,125,880 | **-3** | 2,800 |
| $f_{19}$ | 3 | 3.86 | **3.8625** | 1,690,080 | 3.774 | 4,200 |
| $f_{20}$ | 6 | 3.32 | **3.2756** | 3,372,600 | 3.0333 | 10,920 |
| $f_{21}$ | 4 | 10 | 8.9088 | 2,248,400 | **10.1532** | 7,280 |
| $f_{22}$ | 4 | 10 | 9.2653 | 2,250,080 | **10.4029** | 7,280 |
| $f_{23}$ | 4 | 10 | 9.5548 | 2,245,040 | **10.5364** | 7,280 |

# APPENDIX - VSO SOURCE CODE LISTING

```
'Program 'VSO_07-02-2013.BAS' compiled with
'Power Basic/Windows Compiler 10.04.0108.
'www.PowerBasic.com

'VERY SIMPLE OPTIMIZATION (VSO)'
'==============================

'(C) Richard A. Formato 2013
'All rights reserved worldwide.

'This software is FREEWARE that may be freely copied and
'distributed as long as NO FEE or CHARGE of any kind is
'requested or paid as long as this copyright notice
'is included.

'PROGRAM USED TO GENERATE DATA FOR PAPER
'"A DIPOLE-LOADED MONOPOLE OPTIMIZED USING VSO," VERSION 2, 02 July 2013.

'IMPORTANT NOTE
'==============
'THIS IS A MODIFIED VERSION OF 'CFO_04-07-2010(WinIBenchmarks)_VER2.bas.'

'IT CONTAINS SOME CFO-SPECIFIC ROUTINES THAT ARE NOT USED BUT HAVE NOT
'BEEN REMOVED.  VARIABLE NAMES ARE TAKEN FROM THE CFO CODE AND ADAPTED
'TO VSO AS MUCH AS POSSIBLE.

'LAST MOD 07-02-2013 ~0620 HRS EDT

'================================================================

#COMPILE EXE

#DIM ALL

%USEMACROS = 1

#INCLUDE "win32api.inc"

DEFEXT A-Z

'---------- EQUATES ----------
%IDC_FRAME1          = 101
%IDC_FRAME2          = 102
%IDC_Function_Number1 = 121
%IDC_Function_Number2 = 122
%IDC_Function_Number3 = 123
%IDC_Function_Number4 = 124
%IDC_Function_Number5 = 125
%IDC_Function_Number6 = 126
%IDC_Function_Number7 = 127
%IDC_Function_Number8 = 128
%IDC_Function_Number9 = 129
%IDC_Function_Number10 = 130
%IDC_Function_Number11 = 131
%IDC_Function_Number12 = 132
%IDC_Function_Number13 = 133
%IDC_Function_Number14 = 134
%IDC_Function_Number15 = 135
%IDC_Function_Number16 = 136
%IDC_Function_Number17 = 137
%IDC_Function_Number18 = 138
%IDC_Function_Number19 = 139
%IDC_Function_Number20 = 140
%IDC_Function_Number21 = 141
%IDC_Function_Number22 = 142
%IDC_Function_Number23 = 143
%IDC_Function_Number24 = 144
%IDC_Function_Number25 = 145
%IDC_Function_Number26 = 146
%IDC_Function_Number27 = 147
%IDC_Function_Number28 = 148
%IDC_Function_Number29 = 149
%IDC_Function_Number30 = 150
%IDC_Function_Number31 = 151
%IDC_Function_Number32 = 152
%IDC_Function_Number33 = 153
%IDC_Function_Number34 = 154
%IDC_Function_Number35 = 155
%IDC_Function_Number36 = 156
%IDC_Function_Number37 = 157
%IDC_Function_Number38 = 158
%IDC_Function_Number39 = 159
%IDC_Function_Number40 = 160
%IDC_Function_Number41 = 161
%IDC_Function_Number42 = 162
%IDC_Function_Number43 = 163
%IDC_Function_Number44 = 164
%IDC_Function_Number45 = 165
%IDC_Function_Number46 = 166
%IDC_Function_Number47 = 167
%IDC_Function_Number48 = 168
%IDC_Function_Number49 = 169
%IDC_Function_Number50 = 170

'-------------------------- GLOBAL CONSTANTS & SYMBOLS --------------------------

GLOBAL XIoffset() AS EXT 'offset array for Rosenbrock F6 function

GLOBAL XiMin(), XiMax(), DiagLength, StartingXiMin(), StartingXiMax() AS EXT 'decision space boundaries, length of diagonal

GLOBAL Aijc() AS EXT 'array for Shekel's Foxholes function

GLOBAL EulerConst, PI, PI2, PI4, TwoPI, FourPI, e, Root2, FivePI AS EXT 'mathematical constants

GLOBAL Alphabet$, Digits$, RunID$  'upper/lower case alphabet, digits 0-9 & Run ID

GLOBAL Quote$, SpecialCharacters$ 'quotation mark & special symbols

GLOBAL MuO, EpsO, c, etaO AS EXT    'E&M constants

GLOBAL Rad2Deg, Deg2Rad, Feet2Meters, Meters2Feet, Inches2Meters, Meters2Inches AS EXT 'conversion factors

GLOBAL Miles2Meters, Meters2Miles, NautMi2Meters, Meters2NautMi AS EXT               'conversion factors

GLOBAL ScreenWidth&, ScreenHeight& 'screen width & height

GLOBAL xOffset&, yOffset&       'offsets for probe plot windows

GLOBAL FunctionNumber&&

GLOBAL AddNoiseToPBM2$

GLOBAL VSOversion$, ISPDtype$ 'VSO Initial Sample Point Distribution (Deterministic or Random)

'-------------------------- TEST FUNCTION DECLARATIONS --------------------------

DECLARE FUNCTION F1(R(),Nd&&,pA&,j&&)       'F1 (n-D)

DECLARE FUNCTION F2(R(),Nd&&,pA&,j&&)       'F2(n-D)

DECLARE FUNCTION F3(R(),Nd&&,pA&,j&&)       'F3 (n-D)

DECLARE FUNCTION F4(R(),Nd&&,pA&,j&&)       'F4 (n-D)

DECLARE FUNCTION F5(R(),Nd&&,pA&,j&&)       'F5 (n-D)

DECLARE FUNCTION F6(R(),Nd&&,pA&,j&&)       'F6 (n-D)

DECLARE FUNCTION F7(R(),Nd&&,pA&,j&&)       'F7 (n-D)

DECLARE FUNCTION F8(R(),Nd&&,pA&,j&&)       'F8 (n-D)
```

```
DECLARE FUNCTION F9(R(),Nd&&,p&&,j&&)              'F9 (n-D)
DECLARE FUNCTION F10(R(),Nd&&,p&&,j&&)             'F10 (n-D)
DECLARE FUNCTION F11(R(),Nd&&,p&&,j&&)             'F11 (n-D)
DECLARE FUNCTION F12(R(),Nd&&,p&&,j&&)             'F12 (n-D)
DECLARE FUNCTION u(xi,a,k,m)                       'Auxiliary function for F12 & F13
DECLARE FUNCTION F13(R(),Nd&&,p&&,j&&)             'F13 (n-D)
DECLARE FUNCTION F14(R(),Nd&&,p&&,j&&)             'F14 (n-D)
DECLARE FUNCTION F15(R(),Nd&&,p&&,j&&)             'F15 (n-D)
DECLARE FUNCTION F16(R(),Nd&&,p&&,j&&)             'F16 (n-D)
DECLARE FUNCTION F17(R(),Nd&&,p&&,j&&)             'F17 (n-D)
DECLARE FUNCTION F18(R(),Nd&&,p&&,j&&)             'F18 (n-D)
DECLARE FUNCTION F19(R(),Nd&&,p&&,j&&)             'F19 (n-D)
DECLARE FUNCTION F20(R(),Nd&&,p&&,j&&)             'F20 (n-D)
DECLARE FUNCTION F21(R(),Nd&&,p&&,j&&)             'F21 (n-D)
DECLARE FUNCTION F22(R(),Nd&&,p&&,j&&)             'F22 (n-D)
DECLARE FUNCTION F23(R(),Nd&&,p&&,j&&)             'F23 (n-D)
DECLARE FUNCTION F24(R(),Nd&&,p&&,j&&)             'F24 (n-D)
DECLARE FUNCTION F25(R(),Nd&&,p&&,j&&)             'F25 (n-D)
DECLARE FUNCTION F26(R(),Nd&&,p&&,j&&)             'F26 (n-D)
DECLARE FUNCTION F27(R(),Nd&&,p&&,j&&)             'F27 (n-D)
DECLARE FUNCTION COSINEMIX(R(),Nd&&,p&&,j&&)       'Cosine Mixture (n-D)
DECLARE FUNCTION EXPONENTIAL(R(),Nd&&,p&&,j&&)     'Exponential (n-D)
DECLARE FUNCTION RASTRIGIN(R(),Nd&&,p&&,j&&)       'Rastrigin (n-D)
DECLARE FUNCTION ACKLEY(R(),Nd&&,p&&,j&&)          'Ackley (n-D)
DECLARE FUNCTION Parrott4(R(),Nd&&,p&&,j&&)        'Parrott F4 (1-D)
DECLARE FUNCTION SGO(R(),Nd&&,p&&,j&&)             'SGO Function (2-D)
DECLARE FUNCTION GoldsteinPrice(R(),Nd&&,p&&,j&&)  'Goldstein-Price Function (2-D)
DECLARE FUNCTION StepFunction(R(),Nd&&,p&&,j&&)    'Step Function (n-D)
DECLARE FUNCTION Schwefel226(R(),Nd&&,p&&,j&&)     'Schwefel Prob. 2.26 (n-D)
DECLARE FUNCTION Colville(R(),Nd&&,p&&,j&&)        'Colville Function (4-D)
DECLARE FUNCTION Griewank(R(),Nd&&,p&&,j&&)        'Griewank (n-D)
DECLARE FUNCTION Himmelblau(R(),Nd&&,p&&,j&&)      'Himmelblau (2-D)
DECLARE FUNCTION Rosenbrock(R(),Nd&&,p&&,j&&)      'Rosenbrock (n-D)
DECLARE FUNCTION Sphere(R(),Nd&&,p&&,j&&)          'Sphere (n-D)
DECLARE FUNCTION HimmelblauNLO(R(),Nd&&,p&&,j&&)   'Himmelblau NLO (5-D)
DECLARE FUNCTION Tripod(R(),Nd&&,p&&,j&&)          'Tripod (2-D)
DECLARE FUNCTION Sign(X)                           'Auxiliary function for Tripod
DECLARE FUNCTION RosenbrockF6(R(),Nd&&,p&&,j&&)    'Rosenbrock F6 (1D-D)
DECLARE FUNCTION CompressionSpring(R(),Nd&&,p&&,j&&)'Compression Spring (3-D)
DECLARE FUNCTION GearTrain(R(),Nd&&,p&&,j&&)       'Gear Train (4-D)
DECLARE FUNCTION PBM_1(R(),Nd&&,p&&,j&&)           'PBM Benchmark #1
DECLARE FUNCTION PBM_2(R(),Nd&&,p&&,j&&)           'PBM Benchmark #2
DECLARE FUNCTION PBM_3(R(),Nd&&,p&&,j&&)           'PBM Benchmark #3
DECLARE FUNCTION PBM_4(R(),Nd&&,p&&,j&&)           'PBM Benchmark #4
DECLARE FUNCTION PBM_5(R(),Nd&&,p&&,j&&)           'PBM Benchmark #5
DECLARE FUNCTION LO_MONO(R(),Nd&&,p&&,j&&)         'DIPOLE-LOADED MONOPOLE (Altshuler, 1997)

'--------------------------------- SUB DECLARATIONS ---------------------------------

DECLARE SUB CheckCosineMix(Nd&&)

DECLARE SUB VSOrunStats(FunctionName$,Neval&&,j&&,Fstar,Nd&&,Rstar(),StopTime,StartTime)

DECLARE SUB SAHCrunStats(FunctionName$,Neval&&,BestFstar,BestRstar(),NumIndependentSAHCruns&&,Nd&&,StopTime,StartTime)

DECLARE SUB PrintGWcard(Nd&&,WireNum&&,xEnd1,yEnd1,zEnd1,xEnd2,yEnd2,a_wvln)

DECLARE SUB ShrinkDecisionSpace(ShrinkFactor,BestFitnessCoords())

DECLARE SUB CopyBestNCinputFile(NECFile$)

DECLARE SUB GetNECdata(NECoutputFile$,NumFreqs&&,NumRadPattAngles&&,Zo,FrequencyMHZ(),RadEfficiencyPCT(),MaxGainDBI(),...
                       MinGainDBI(),RinOhms(),XinOhms(),VSWR(),ForwardGainDBI(),RearGainDBI(),FileStatus$,FileID$)

DECLARE SUB SampleObjectiveSpace(FunctionName$,Nd&&,Neval&&,Lumpiness)

DECLARE SUB CopyBestMatrices(Np&&,Nd&&,Niter&&,R(),M(),Rbest(),Mbest())

DECLARE SUB CheckNECFiles(NECFileError$)

DECLARE SUB GetTestFunctionNumber(FunctionName$)

DECLARE SUB FillArrayaij

DECLARE SUB Plot3DbestProbePaths(NumPaths&&,M(),R(),Np&&,Nd&&,LastIteration&&,FunctionName$)

DECLARE SUB Plot2DbestProbePaths(NumPaths&&,M(),R(),Np&&,Nd&&,LastIteration&&,FunctionName$)

DECLARE SUB Plot2DindividualProbePaths(NumPaths&&,M(),R(),Np&&,Nd&&,LastIteration&&,FunctionName$)

DECLARE SUB Show2DsamplePoints(R(),Np&&,Nd&&,Niter&&,j&&,FunctionName$)

DECLARE SUB Show3DsamplePoints(R(),Np&&,Nd&&,Niter&&,j&&,FunctionName$)

DECLARE SUB StatusWindow(FunctionName$,StatusWindowHandle???)

DECLARE SUB PlotResults(FunctionName$,Nd&&,BestFitnessOverall,BestNpNd&&,BestGamma,Neval&&,Rbest(),Mbest(),BestSamplePointNumberOverall&&,...
                        BestIterationOverall&&,LastIterationBestRun&&,Alpha,Beta)

DECLARE SUB
DisplayRunParameters(FunctionName$,Nd&&,Np&&,Niter&&,G,DeltaT,Alpha,Beta,Frep,R(),A(),M(),PlaceInitialPoints$,InitialAcceleration$,RepositionFactor$,RunVSO$,...
ShrinkDS$,CheckForEarlyTermination$)

DECLARE SUB GetBestFitness(M(),Np&&,StepNumber&&,BestFitness,BestProbeNumber&&,BestIteration&&)

DECLARE SUB
TabulateIDprobeCoordinates(MaxIDsamplePointsPlotted&&,Nd&&,Np&&,LastIteration&&,G,DeltaT,Alpha,Beta,Frep,R(),M(),PlaceInitialPoints$,InitialAcceleration$,Rep
ositionFactor$,FunctionName$,Gamma)

DECLARE SUB
GetPlotAnnotation(PlotAnnotation$,Nd&&,Np&&,Niter&&,G,DeltaT,Alpha,Beta,Frep,M(),PlaceInitialPoints$,InitialAcceleration$,RepositionFactor$,FunctionName$,Gam
ma)
```

```
DECLARE SUB
ChangeRunParameters(NumPointsPerDimension&&,Np&&,Niter&&,G,Alpha,Beta,DeltaT,Frep,PlaceInitialPoints$,InitialAcceleration$,RepositionFactor$,FunctionNam
e$)

DECLARE SUB CLEANUP

DECLARE SUB
PlotID probePositions(MaxIDsamplePointsPlotted&&,Np&&,Niter&&,G,DeltaT,Alpha,Beta,Frep,R(),M(),PlaceInitialPoints$,InitialAcceleration$,Repositio
nFactor$,FunctionName$,Gamma)

DECLARE SUB DisplayMatrix(Np&&,Niter&&,M())

DECLARE SUB DisplayNbestMatrix(Np&&,Niter&&,Mbest())

DECLARE SUB DisplayMatrixThisIteration(Np&&,j&&,M())

DECLARE SUB DisplayAmatrix(Np&&,Niter&&,A())

DECLARE SUB DisplayAmatrixThisIteration(Np&&,j&&,A())

DECLARE SUB DisplayRmatrix(Np&&,Niter&&,R())

DECLARE SUB DisplayRmatrixThisIteration(Np&&,Niter&&,j&&,R(),Gamma)

DECLARE SUB DisplayXiMinMax(Ni&&,XiMin(),XiMax())

DECLARE SUB DisplayRunParameters2(FunctionName$,Np&&,Niter&&,G,DeltaT,Alpha,Beta,Frep,PlaceInitialPoints$,InitialAcceleration$,RepositionFactor$

DECLARE SUB
PlotBestProbeVsIteration(Ni&&,Np&&,LastIteration&&,G,DeltaT,Alpha,Beta,Frep,M(),PlaceInitialPoints$,InitialAcceleration$,RepositionFactor$,FunctionName$,Gamm
a)

DECLARE SUB
PlotBestFitnessEvolution(Ni&&,Np&&,LastIteration&&,G,DeltaT,Alpha,Beta,Frep,M(),PlaceInitialPoints$,InitialAcceleration$,RepositionFactor$,FunctionName$,Gamm
a)

DECLARE SUB
PlotAverageDistance(Ni&&,Np&&,LastIteration&&,G,DeltaT,Alpha,Beta,Frep,M(),PlaceInitialPoints$,InitialAcceleration$,RepositionFactor$,FunctionName$,R(),DiagL
ength,Gamma)

DECLARE SUB Plot2Dfunction(FunctionName$)

DECLARE SUB Plot1Dfunction(FunctionName$)

DECLARE SUB GetFunctionRunParameters(FunctionName$,Ni&&,DiagLength)

DECLARE SUB InitialProbeDistribution(Np&&,Ni&&,Niter&&,R(),PlaceInitialPoints$,Gamma)

DECLARE SUB RetrieveErrantPoints(Np&&,Ni&&,j&&,R(),Frep)

DECLARE SUB RetrieveErrantsamplePoints2(Np&&,Ni&&,j&&,R(),A(),Frep)

DECLARE SUB IPD(Np&&,Ni&&,Niter&&,R(),Gamma)

DECLARE SUB ESPO(NumPointsPerDimension&&,NumGammas&&,Np&&,j&&)

DECLARE SUB ResetDecisionSpaceBoundaries(Ni&&)

DECLARE SUB ThreeDplot(PlotFilename&,PlotTitle$,Annotation$,xCoord$,yCoord$,zCoord$,_
                          xaxisLabel$,Yaxisabel$,ZaxisLabel$,ZHin$,ZMax$,GnuPlotEXE$,A$)

DECLARE SUB ThreeDplot2(PlotFileName&,PlotTitle$,Annotation$,xCoord$,yCoord$,zCoord$,xaxisLabel$,_
                          YaxisLabel$,ZaxisLabel$,ZaxisLabel$,ZMin$,zMax$,GnuPlotEXE$,A$,xStart$,xStop$,yStart$,ystop$)

DECLARE SUB TwoDplot(PlotFileName&,PlotTitle$,xCoord$,yCoord$,YaxisLabel$,_
                       LogXaxis$,LogYaxis$,xMin$,xMax$,yMin$,yMax$,xTics$,yTics$,GnuPlotEXE$,LineType$,Annotation$)

DECLARE SUB TwoDplot2Curves(PlotFilename&,PlotFileName2$,PlotTitle$,Annotation$,xCoord$,yCoord$,xaxisLabel$,YaxisLabel$,_
                             LogXaxis$,LogYaxis$,xMin$,xMax$,yMin$,yMax$,xTics$,yTics$,GnuPlotEXE$,LineSize)

DECLARE SUB TwoDplot3Curves(NumCurves&&,PlotFileName1$,PlotFileName2$,PlotFileName3$,PlotTitle$,Annotation$,xCoord$,yCoord$,_
                             xaxisLabel$,YaxisLabel$,LogXaxis$,LogYaxis$,xMin$,xMax$,yMin$,yMax$,xTics$,yTics$,GnuPlotEXE$)

DECLARE SUB CreateGnuPlotINIfile(PlotWindowULC_X&&,PlotWindowULC_Y&&,PlotWindowWidth&&,PlotWindowHeight&&)

DECLARE SUB Delay(NumSecs)

DECLARE SUB MathematicalConstants

DECLARE SUB AlphabetAndDigits

DECLARE SUB SpecialSymbols

DECLARE SUB EMconstants

DECLARE SUB ConversionFactors

DECLARE SUB ShowConstants

DECLARE SUB ComplexMultiply(ReA,ImA,ReB,ImB,ReC,ImC)

DECLARE SUB ComplexDivide(ReA,ImA,ReB,ImB,ReC,ImC)

DECLARE SUB VSO(FunctionName$,Ni&&)

DECLARE SUB SAHC(FunctionName$,Ni&&)

'---------- FUNCTION DECLARATIONS ----------

DECLARE FUNCTION StandingWaveRatio(Zo,ReZ,ImZ)

DECLARE FUNCTION Int2STR$(N&&)

DECLARE FUNCTION FP2STR$(x)

DECLARE FUNCTION FP2STR2$(x!)

DECLARE FUNCTION ProbeWeight2(Nd&&,Np&&,R(),M(),p&&,j&&)

DECLARE FUNCTION SlopeRatio(M(),Np&&,StepNumber&&) 'computes a "weighting factor" based on probe's position (greater weight if closer to decision space boundary)

DECLARE CALLBACK FUNCTION DlgProc

DECLARE FUNCTION HasFITNESSsaturated$(NavgSteps&,j&&,Np&&,Nd&&,M(),R(),DiagLength)

DECLARE FUNCTION HasbAvGsaturated$(NavgSteps&,j&&,Np&&,Nd&&,M(),R(),DiagLength)

DECLARE FUNCTION OscillationInDavg$(j&&,Np&&,Nd&&,M(),R(),DiagLength)

DECLARE FUNCTION DavgThisStep(j&&,Np&&,Nd&&,M(),R(),DiagLength)

DECLARE FUNCTION NoSpaces$(X,NumDigits&&)

DECLARE FUNCTION FormatFP$(X,Ndigits&&)

DECLARE FUNCTION FormatInteger$(N&&)

DECLARE FUNCTION TerminateNowForSaturation$(j&&,Nd&&,Np&&,Niter&&,G,DeltaT,Alpha,Beta,R(),A(),M())

DECLARE FUNCTION MagVector(V(),Nd&&)

DECLARE FUNCTION UniformDeviate(u&&)

DECLARE FUNCTION RandomNum(a,b)

DECLARE FUNCTION GaussianDeviate(Mu,Sigma)

DECLARE FUNCTION UnitStep(X)

DECLARE FUNCTION Fibonacci&&(N&&)

DECLARE FUNCTION ObjectiveFunction(R(),Nd&&,p&&,j&&,FunctionName$)
```

```
DECLARE FUNCTION UnitStep(X)

'==========================================================================================
'----- MAIN PROGRAM -----

FUNCTION PBMAIN () AS LONG

    LOCAL N&&, Np&, Niter&&

    LOCAL FunctionName$  'name of objective function

    LOCAL N&&, i&&, YN&

    LOCAL A$, RunVSO$, NECfileError$, ALG$
'   ------------------------------ Global Constants -----------------------------

    REDIM Aij(1 TO 2, 1 TO 25) 'Global array for Shekel's Foxholes function

    CALL FillArrayAij

    CALL MathematicalConstants 'NOTE: Calling order is important!!

    CALL AlphabetAndDigits

    CALL SpecialSymbols

    CALL EMconstants

    CALL ConversionFactors ': CALL ShowConstants 'to verify constants have been set
'   ------------------------------ General Setup -----------------------------

    RunID$      = REMOVE$(DATE$+TIME$,ANY Alphabet$+" -:/")

    VSOversion$ = "VSO Ver. 07-02-2013"

    RANDOMIZE TIMER  'seed random number generator with program start time

    DESKTOP GET SIZE TO ScreenWidth&, ScreenHeight&  'get screen size (global variables)

    IF DIR$("wgnuplot.exe") = "" THEN
        MSGBOX("WARNING!  'wgnuplot.exe' not found.  Run terminated.") : EXIT FUNCTION
    END IF
'   ---------- VSO RUN PARAMETERS ----------
    CALL GetTestFunctionNumber(FunctionName$)
    CALL GetFunctionRunParameters(FunctionName$,N&&,DiagLength)
    MSGBOX("Function is "+REMOVE$(STR$(N&&),ANY " ")+"-dim "+FunctionName$+$CRLF+"Diag = "+STR$(DiagLength))
'   ---------- PLOT 1D and 2D FUNCTIONS ON-SCREEN FOR VISUALIZATION ---------
    IF N&& = 2 AND INSTR(FunctionName$,"PBM_") > 0 THEN
        CALL CheckNECfiles(NECfileError$)
        IF NECfileError$ = "YES" THEN
            EXIT FUNCTION
        ELSE
            MSGBOX("Begin computing plot of function "+FunctionName$+"?  May take a while - be patient...")
        END IF
    END IF

    SELECT CASE N&&
        CASE 1 : CALL Plot1Dfunction(FunctionName$)
        CASE 2 : CALL Plot2Dfunction(FunctionName$)
    END SELECT
'   ------------------- RUN VSO -------------------

    ALG$ = ""
    DO UNTIL ALG$ = "V" OR ALG$ = "H"
        ALG$ = UCASE$(INPUTBOX$("VSO ('v') or"+$CRLF+"Hill Climber ('H')?","SELECT ALGORITHM","V"))
    LOOP

    IF ALG$ = "V" THEN ALG$ = "VSO"
    IF ALG$ = "H" THEN ALG$ = "Hill Climber"

    YN& = MSGBOX("Run "+ALG$+" ON FUNCTION " + FunctionName$ + "?"+CHR$(13)+CHR$(13)+"Get some coffee & sit back...",%MB_YESNO,"CONFIRM RUN")
    IF YN& = %IDYES AND ALG$ = "VSO" THEN CALL VSO(FunctionName$,N&&)
    IF YN& = %IDYES AND ALG$ = "Hill Climber" THEN CALL SAHC(FunctionName$,N&&)

END FUNCTION 'PBMAIN()

'========================================= VSO SUBROUTINE ==========================================

SUB VSO(FunctionName$,N&&)

LOCAL p&&, i&&, j&& 'Standard Indices: Sample Point #, Coordinate #, Iteration #

LOCAL Neval&&, NumGammas&&, NumPointsPerDimension&&, Niter&&, Np&& 'Number of SAMPLE POINTS

LOCAL Fstar, Rstar(), Rho, R(), Fitness(), Fbest1, Fbest2 AS EXT

LOCAL StartTime, StopTime AS EXT

LOCAL A$, N&&

StartTime = TIMER

'-------------- Initial Parameter Values -----------------

REDIM Rstar(1 TO N&&, 0 TO Niter&&) 'Rstar() is coordinate matrix of best fitness
Niter&&                 = 15
NumGammas&&             = 20 'EVEN #
NumPointsPerDimension&& = 14 'EVEN # to avoid points at origin in DS's
Np&&                    = NumPointsPerDimension&&*N&&&*NumGammas&&

MSGBOX("INFO:  Np*Niter = "+STR$(Np&&*(Niter&&+1)))

REDIM R(1 TO Np&&, 1 TO N&&&, 0 TO Niter&&), Fitness(1 TO Np&&, 0 TO Niter&&) 're-initializes Position vector/Fitness matrices to zero

Fstar = -1E42I00 : Rho = 0.5## : Neval&& = 0

'STEP (A1) -------------------- Compute DETERMINISTIC ISPD at Iteration #0 -----------------------
        ISPDtype$ = "Deterministic" : CALL ISPD(NumPointsPerDimension&&,NumGammas&&,Np&&,N&&&,R(),0)
        IF N&& = 2 THEN CALL Show2DsamplePoints(R(),N&&&,Niter&&,0,FunctionName$)
        IF N&& = 3 THEN CALL Show3DsamplePoints(R(),Np&&,N&&&,0,FunctionName$)
        IF N&& = 2 OR N&& = 3 THEN MSGBOX("any key...")

'STEP (A2) -------------------- Compute Initial Fitness Matrix (Iteration #0) ---------------------
        FOR p&& = 1 TO Np&&
            Fitness(p&&,0)  = ObjectiveFunction(R(),N&&&,p&&,0,FunctionName$) : Neval&& = Neval&& + 1
            IF Fitness(p&&,0) >= Fstar THEN
                Fstar        = Fitness(p&&,0)
                FOR i&& = 1 TO N&&&
                    Rstar(i&&) = R(p&&,i&&,0)
                NEXT i&&
                Fbest1 = Fstar : Fbest2 = -1E42I00
            END IF
        NEXT p&&

'ITERATION LOOP (STARTING AT ITERATION #1)
'---------------------------------------------------------------
FOR j&& = 1 TO Niter&&

'STEP (B) ----------- Compute Sample Point Position Vectors for j-th Iteration --------
        FOR p&& = 1 TO Np&&
            FOR i&& = 1 TO N&&&: R(p&&,i&&,j&&) = R(p&&,i&&,j&&-1) + Rho*(Rstar(i&&)-R(p&&,i&&,j&&-1)) : NEXT i&&
        NEXT p&&

'STEP (C) -------------- Compute Fitness Matrix for Current Sample Point Distribution -----------
        FOR p&& = 1 TO Np&&
            Fitness(p&&,j&&)  = ObjectiveFunction(R(),N&&&,p&&,j&&,FunctionName$) : Neval&& = Neval&& + 1
            Fstar             = Fitness(p&&,j&&)
            FOR i&& = 1 TO N&&& : Rstar(i&&) = R(p&&,i&&,j&&) : NEXT i&&
            IF FunctionName$ = "LD_MONO" THEN CALL CopyBestNECinputFile("MONO.NEC")
```

```
                END IF
            NEXT p&&

            IF j&& MOD 3 = 0 THEN 'check for early termination
                Fbest1 = Fbest2 : Fbest2 = Fstar
                IF Fbest2-Fbest1 <= 0.001## THEN EXIT FOR
            END IF
    NEXT j&& 'END ITERATION LOOP

    StopTime = TIMER

    CALL VSOrunStats(FunctionName$,Neval&&,j&&,Fstar,N&&,Rstar(),StopTime,StartTime)

END SUB 'VSO()

'--------------

SUB VSOrunStats(FunctionName$,Neval&&,j&&,Fstar,N&&,Rstar(),StopTime,StartTime)

LOCAL A$, i&&, N&&

    A$ = "VSO results for FUNCTION "+FunctionName$+SCRLF+SCRLF+"neval ="+STR$(Neval&&)+SCRLF+SCRLF+"# iter = "+STR$(j&&)+_
         SCRLF+SCRLF+CHR$"F*= "+STR$(Fstar)+" @"+SCRLF
    FOR i&& = 1 TO N&& : A$ = A$+"          x("+STR$(i&&))+")="+STR$(Rstar(i&&))+SCRLF : NEXT i&&
    A$ = A$+SCRLF+"Runtime = "+STR$(ROUND(StopTime-StartTime,4))+" secs   ["+_
                             REMOVE$(STR$(ROUND((StopTime-StartTime)/60##,1)),ANY" ")+" min]"

    IF FunctionName$ = "LD_MONO" THEN
        N&& = FREEFILE
        OPEN "BestMONO.NEC" FOR APPEND AS #N&&
            PRINT #N&&,"" : PRINT #N&&, A$
        CLOSE #N&&
    END IF

    MSGBOX(A$)

END SUB 'VSOrunStats()

'================================  HILL-CLIMBER SUBROUTINE  =================================

SUB SAHC(FunctionName$,N&&)

'Ref: Sean Luke, 2013, Essentials of Metaheuristics, LuLu, second edition
'http://cs.gmu.edu/~sean/book/metaheuristics/ (no charge)

LOCAL p&&, i&&, j&& 'Standard Indices: Sample Point #, Coordinate #, Iteration #

LOCAL Neval&&, NumPointsPerDimension&&, Niter&&, Np&& 'Number of SAMPLE POINTS

LOCAL R(), Fitness(), Fbest1, Fbest2, BestFstar, BestFstar() AS EXT

LOCAL StartTime, StopTime AS EXT

LOCAL A$, k&&, SAHCrunNumber&&, NumIndependentSAHCruns&&

    StartTime = TIMER

    '-------------- Initial Parameter Values ----------------
    BestFstar            = -1E4200
    Niter&&              = 15
    NumPointsPerDimension&& = 140 'EVEN # to avoid points at origin in OS's
    Np&&                 = NumPointsPerDimension&&*N&&

    MSGBOX("INFO: Np*Niter = "+STR$(Np&&*(Niter&&+1))+" per SAHC run.")

    REDIM BestFstar(1 TO N&&)

    Neval&& = 0

    NumIndependentSAHCruns&& = 1000 : IF FunctionName$ = "LD_MONO" THEN NumIndependentSAHCruns&& = 50 'to avoid excessive runtime

    FOR SAHCrunNumber&& = 1 TO NumIndependentSAHCruns&&

    REDIM R(1 TO Np&&, 0 TO Niter&&), Fitness(1 TO Np&&, 0 TO Niter&&) 're-initializes Position Vector/Fitness matrices to zero

    'STEP (A1) -------------- Compute RANDOM ISPD at Iteration #0 ----------------------
        ISPDtype$ = "Random"
        FOR p&& = 1 TO Np&& 'randomize sample points
            FOR i&& = 1 TO N&&
                R(p&&,i&&,0) = XiMin(i&&) + RandomNum(0.05&&,0.95&&)*(XiMax(i&&)-XiMin(i&&))
            NEXT i&&
        NEXT p&&

    'STEP (A2) --------------- Compute Initial Fitness Matrix (Iteration #0) ----------------------
        FOR p&& = 1 TO Np&&
            Fitness(p&&,0)  = ObjectiveFunction(R(),p&&,0,FunctionName$) : Neval&& = Neval&& + 1
            IF Fitness(p&&,0) > BestFstar THEN
                BestFstar    = Fitness(p&&,0)
                FOR i&& = 1 TO N&& : BestFstar(i&&) = R(p&&,i&&,0) : NEXT i&&
                Fbest1       = BestFstar : Fbest2 = -1E4200
            END IF
        NEXT p&&

    'ITERATION LOOP (STARTING AT ITERATION #1)

    FOR j&& = 1 TO Niter&&

    'STEP (B) ----------- Compute Sample Point Position Vectors for j-th Iteration ---------
            FOR p&& = 1 TO Np&& 'tweak each SP with a location small change (max 10% of OS diagonal length)
                FOR i&& = 1 TO N&&
                    R(p&&,i&&,j&&) = MIN(MAX(R(p&&,i&&,j&&-1) + RandomNum(-0.1##,0.1##)*DiagLength,XiMin(i&&)),XiMax(i&&)) 'stay inside OS
                NEXT i&&
            NEXT p&&

    'STEP (C) -------------- Compute Fitness Matrix for Current Sample Point Distribution ----------------
            FOR p&& = 1 TO Np&&
                Fitness(p&&,j&&)  = ObjectiveFunction(R(),p&&,j&&,FunctionName$) : Neval&& = Neval&& + 1
                IF Fitness(p&&,j&&) >= BestFstar THEN
                    BestFstar
                    FOR i&& = 1 TO N&& : BestFstar(i&&) = R(p&&,i&&,j&&) : NEXT i&&
                    IF FunctionName$ = "LD_MONO" THEN CALL CopyBestNECtinputfile("MONO.NEC")
                END IF
            NEXT p&&

            IF j&& MOD 3 = 0 THEN 'check for early termination
                Fbest1 = Fbest2 : Fbest2 = BestFstar
                IF Fbest2-Fbest1 <= 0.001## THEN EXIT FOR
            END IF
    NEXT j&& 'END ITERATION LOOP

    NEXT SAHCrunNumber&& 'END HC RUN LOOP

    StopTime = TIMER

    CALL SAHCrunStats(FunctionName$,Neval&&,BestFstar,BestFstar(),NumIndependentSAHCruns&&,N&&,StopTime,StartTime)

END SUB 'SAHC()

'--------------------

SUB SAHCrunStats(FunctionName$,Neval&&,BestFstar,BestFstar(),NumIndependentSAHCruns&&,N&&,StopTime,StartTime)

LOCAL A$, i&&, N&&

    A$ = "Hill Climber results for FUNCTION "+FunctionName$+SCRLF+SCRLF+_
         "neval ="+STR$(Neval&&)+SCRLF+SCRLF+_
         "best F*= "+STR$(BestFstar)+" over "+int2STR$(NumIndependentSAHCruns&&)+" runs @"+SCRLF+SCRLF
    FOR i&& = 1 TO N&& : A$ = A$+"          x("+STR$(i&&))+")="+STR$(BestFstar(i&&))+SCRLF : NEXT i&&
    A$ = A$+SCRLF+"Runtime = "+STR$(ROUND(StopTime-StartTime,4))+" secs   ["+_
                             REMOVE$(STR$(ROUND((StopTime-StartTime)/60##,1)),ANY" ")+" min]"

    IF FunctionName$ = "LD_MONO" THEN
        N&& = FREEFILE
        OPEN "BestMONO.NEC" FOR APPEND AS #N&&
            PRINT #N&&,"" : PRINT #N&&, A$
        CLOSE #N&&
```

```
        END IF
    MSGBOX(A$)

END SUB 'SAHCrunStats()

'---------------------------
SUB ShrinkDecisionSpace(ShrinkFactor,%d&&,BestFitnessCoords())
LOCAL i&&
    FOR i&& = 1 TO NumDimensions&&
        XiWin(i&&) = MIN(MAX(BestFitnessCoords(i&&) - (1##-ShrinkFactor)*(BestFitnessCoords(i&&)-XiMin(i&&)),XiMin(i&&)),XiMax(i&&))
        XiMax(i&&) = MAX(MIN(BestFitnessCoords(i&&) + (1##-ShrinkFactor)*(XiMax(i&&)-BestFitnessCoords(i&&)),XiMax(i&&)),XiMin(i&&))
    NEXT i&&
END SUB

'-------
SUB ISPD(NumPointsPerDimension&&,NumGammas&&,%p&&,%d&&,R(),j&&) 'Initial Sample Point Distribution (ISPD) on "Point Lines"
                                                                'parallel to coordinate axes using internal gamma values.

LOCAL DeltaX1, DelXi, DelX2, Di, Gamma, GammaMin1, GammaMax1, GammaMin2, GammaMax2, DelGamma1, DelGamma2 AS EXT

LOCAL p&&, i&&, k&&, m&&, NumXipoints&&, NumX2point&&, x1pointnum&&, x2pointnum&&, GammaNum&&, StartingPointOffset&&

LOCAL A$, B$, C$

    IF NumGammas&& MOD 2 <> 0 THEN
        MSGBOX("WARNING! # gammas must be an even number.  Execution terminated.") : END
    END IF

    GammaMin1 = 0.05## : GammaMax1 = 0.49## : DelGamma1 = (GammaMax1-GammaMin1)/(NumGammas&&\2-1)
    GammaMin2 = 0.51## : GammaMax2 = 0.95## : DelGamma2 = (GammaMax2-GammaMin2)/(NumGammas&&\2-1)

    StartingPointOffset&& = (GammaNum&&-1)*NumPointsPerDimension&&*%d&&

    FOR GammaNum&& = 1 TO NumGammas&&
        IF GammaNum&& <= NumGammas&&\2 THEN
            Gamma = GammaMin1 + (GammaNum&&-1)*DelGamma1
        ELSE
            Gamma = GammaMin2 + (GammaNum&&-NumGammas&&\2-1)*DelGamma2
        END IF

        StartingPointOffset&& = (GammaNum&&-1)*NumPointsPerDimension&&*%d&&

            FOR i&& = 1 TO %d&&
                FOR k&& = 1 TO NumGammas&&
                    R(i&&+StartingPointOffset&&,i&&,j&&) = XiWin(i&&) + Gamma*(XiMax(i&&)-XiWin(i&&))
                NEXT i&&

            FOR i&& = 1 TO %d&& 'place sample points probe line line-by-line (i&& is dimension [coordinate] number)
                DeltaX = (XiMax(i&&)-XiWin(i&&))/(NumPointsPerDimension&&-1)
                FOR k&& = 1 TO NumPointsPerDimension&&
                    p&& = k&& + NumPointsPerDimension&&*(i&&-1) 'point # without offset
                    R(p&&+StartingPointOffset&&,i&&,j&&) = XiWin(i&&) + (k&&-1)*DeltaX
                NEXT k&&
            NEXT i&&
    NEXT GammaNum&&

END SUB 'ISPD()

'-------------
FUNCTION Int2STR%(m&&) : Int2STR% = REMOVE$(STR$(m&&),ANY" ") : END FUNCTION
FUNCTION FP2STR$(x) : FP2STR$ = REMOVE$(STR$(x),ANY" ") : END FUNCTION
FUNCTION FP2STR2$(x!) : FP2STR2$ = REMOVE$(STR$(x!),ANY" ") : END FUNCTION

'============================================================================================================================
SUB SampleObjectiveSpace(FunctionName%,%d&&,%eval&&,Luminess)

LOCAL X1(), X2(), FofX1, FofX2, DeltaX1, DeltaX, MaxSlope, MinSlope, DeltaSlope, DiagFraction, Slopes(), NormalizedSlope() AS EXT

LOCAL F(), Fmin, Fmax, DeltaF, VarFunctionValue, VarNormalizedSlope AS EXT

LOCAL AvgFunctionValue, SigFunctionValue, SigmaNormalizedSlope AS EXT

LOCAL NormalizedFunctionValue(), FunctionValueHistogram(), NormalizedFunctionValueHistogram(), SlopeHistogram(), NormalizedSlopeHistogram() AS EXT

LOCAL BucketBottom, BucketTop AS EXT

LOCAL NumSamplePoints&&, SP&&, i&&, k&&

LOCAL AverageFunctionValue$, SigmaFunctionValue$, AverageNormalizedSlope$, SigmaNormalizedSlope$, LuminessIndex%, A$

LOCAL NumHistogramBuckets&&, Bucket%&&, MaxBucketPoints&&

LOCAL FuncValHistAVG, FuncValHistSIG, FuncValHistVAR, SlopeValHistVAR, SlopeValHistSIG, SlopeValHistSIG AS EXT

LOCAL FuncValHistSIGMA$, FuncValHistAVERAGE$, SlopeValHistAVERAGE$, SlopeValHistSIGMA$

LOCAL InterDecileNormFUNCT2DreleFreq, InterDecileNormSLOPEreleFreq, TenPctPoint, NinetyPctPoint AS EXT

LOCAL Ten2NinetynormFUNCreleFreq$, Ten2NinetynormSLOPEreleFreq$

'------------------------------ SETUP ------------------------------

NumSamplePoints&& = MAX(100*%d&&,500)

REDIM Slopes(1 TO NumSamplePoints&&), NormalizedSlope(1 TO NumSamplePoints&&)

REDIM F(1 TO NumSamplePoints&&), NormalizedFunctionValue(1 TO NumSamplePoints&&)

REDIM X1(1 TO NumSamplePoints&&, 1 TO %d&&, 0 TO 0), X2(1 TO NumSamplePoints&&, 1 TO %d&&, 0 TO 0)

Diagfraction = 0.05## : NumHistogramBuckets&& = 100^1000

'  ---------------- Randomly Sample Objective Space -----------------
    FOR SP&& = 1 TO NumSamplePoints&& 'create random sample points X1
        FOR i&& = 1 TO %d&& : X1(SP&&,i&&,0) = XiWin(i&&)+RandomNum(D##,1##)*(XiMax(i&&)-XiWin(i&&)) : NEXT i&&
    NEXT SP&&

    DeltaX1 = Diagfraction*DiagLength
    FOR SP&& = 1 TO NumSamplePoints&& 'offset SP to create new set X2
        FOR i&& = 1 TO %d&&
            X2(SP&&,i&&,0) = X2(SP&&,i&&,0) = Sgn(RandomNum(-1##,1##))*DeltaX1 'randomly add/subtract DeltaX from SP coordinates
            X2(SP&&,i&&,0) = MIN(MAX(X2(SP&&,i&&,0),XiWin(i&&)),XiMax(i&&)) 'keep coordinate value inside DS
        NEXT i&&
    NEXT SP&&

'  ----------- Get Min/Max Function Values/Slopes at Each SP -----------------------
    Fmax = -1E420O    : Fmin  = -Fmax        'get min/max function values
    MaxSlope = -1E420O : MinSlope = -MaxSlope 'get min/max slopes
    FOR SP&& = 1 TO NumSamplePoints&&        'compute fitnesses at point SP&& and "average" slope at each sample point
        DeltaX = D## : FOR i&& = 1 TO %d&& : DeltaX = DeltaX + (X1(SP&&,i&&,0)-X2(SP&&,i&&,0))^2 : NEXT i&& : DeltaX = SQR(DeltaX)
        FofX1 = ObjectiveFunction(X1(),%d&&,SP&&,0,FunctionName%)
        FofX2 = ObjectiveFunction(X2(),%d&&,SP&&,0,FunctionName%)
        F(SP&&)      = FofX1
        IF FofX1     > Fmax THEN Fmax = FofX1
        IF FofX1     < Fmin THEN Fmin = FofX1
        Slopes(SP&&) = ABS(FofX1-FofX2)/DeltaX
        IF Slopes(SP&&) > MaxSlope THEN MaxSlope = Slopes(SP&&)
        IF Slopes(SP&&) < MinSlope THEN MinSlope = Slopes(SP&&)
    NEXT SP&&
    DeltaF     = Fmax - Fmin        'range of function values
    DeltaSlope = MaxSlope-MinSlope 'range of slopes

'  ------------- Normalized Function Values at Each SP -----------------------
    FOR SP&& = 1 TO NumSamplePoints&&
        NormalizedFunctionValue(SP&&) = (F(SP&&)-Fmin)/DeltaF 'note: y=(x-a)/(b-a) maps interval [a,b] into [0,1]
    NEXT SP&&

    AvgFunctionValue        = D## : FOR SP&& = 1 TO NumSamplePoints&& : AvgFunctionValue = AvgFunctionValue + NormalizedFunctionValue(SP&&) : NEXT SP&& 'AVG
Function value
    AvgFunctionValue        = AvgFunctionValue/NumSamplePoints&&
    AverageFunctionValue$   = "0"+REMOVE$(STR$(ROUND(AvgFunctionValue,3)),ANY" ")
    IF AvgFunctionValue     = D## THEN AverageFunctionValue$ = "0.0"
```

```
           VarFunctionValue     = 0##  : FOR SP&& = 1 TO NumSamplePoints&& : VarFunctionValue = VarFunctionValue + (NormalizedFunctionValue(SP&&)-AvgFunctionValue)^2
           : NEXT SP&& 'VAR Function Value
           VarFunctionValue     = VarFunctionValue/NumSamplePoints&&
           SigFunctionValue     = SQR(VarFunctionValue) 'standard deviation
           SigmaFunctionValue$  = "0"+REMOVE$(STR$(ROUND(SigFunctionValue,3)),ANY" ")
           IF SigFunctionValue = 0## THEN SigmaFunctionValue$ = "0.0"

'----------------------------------------- Normalized Function Value Histogram -----------------------------------------
           REDIM FunctionValueHistogram(1 TO NumHistogramBuckets&&)
           FOR SP&& = 1 TO NumSamplePoints&&
                BucketNum&& = 1 TO NumHistogramBuckets&&-1
                     BucketBottom = (BucketNum&&-1)/NumHistogramBuckets&& : BucketTop = BucketNum&&/NumHistogramBuckets&&
                     IF NormalizedFunctionValue(SP&&) >= BucketBottom AND NormalizedFunctionValue(SP&&) < BucketTop THEN _
                          FunctionValueHistogram(BucketNum&&) = FunctionValueHistogram(BucketNum&&) + 1
                NEXT BucketNum&&
                BucketBottom = (NumHistogramBuckets&&-1)/NumHistogramBuckets&& : BucketTop = 1##
                IF NormalizedFunctionValue(SP&&) >= BucketBottom AND NormalizedFunctionValue(SP&&) <= BucketTop THEN _
                     FunctionValueHistogram(NumHistogramBuckets&&) = FunctionValueHistogram(NumHistogramBuckets&&) + 1
           NEXT SP&&
           MaxBucketPoints&& = 0
           FOR BucketNum&& = 1 TO NumHistogramBuckets&& 'get max # points in a bucket for normalization
                IF FunctionValueHistogram(BucketNum&&) >= MaxBucketPoints&& THEN MaxBucketPoints&& = FunctionValueHistogram(BucketNum&&)
           NEXT BucketNum&&
           FOR BucketNum&& = 1 TO NumHistogramBuckets&& 'normalize FUNCTION VALUE histogram
                NormalizedFunctionValueHistogram(BucketNum&&) = FunctionValueHistogram(BucketNum&&)/MaxBucketPoints&&
                IF NormalizedFunctionValueHistogram(BucketNum&&-1) <= 0.1## AND NormalizedFunctionValueHistogram(BucketNum&&-1)/2## _
                     THEN InterDeclNormFUNCTIONrelIfreq = (NormalizedFunctionValueHistogram(BucketNum&&)-NormalizedFunctionValueHistogram(BucketNum&&))/2##
           NEXT BucketNum&&
           InterDeclNormFUNCTIONrelIfreq    = ABS(NinetyctPoint-TenPctPoint)
           Ten2NinetyNormFUNCrelIfreq$  = "0"+REMOVE$(STR$(ROUND(InterDeclNormFUNCTIONrelIfreq,3)),ANY" ")
           IF InterDeclNormFUNCTIONrelIfreq = 0## THEN Ten2NinetyNormFUNCrelIfreq$ = "0.0"

           FuncValHistAVG           = 0##
           FOR BucketNum&& = 1 TO NumHistogramBuckets&& 'AVG of normalized FUNCTION VALUE histogram
                FuncValHistAVG = FuncValHistAVG + NormalizedFunctionValueHistogram(BucketNum&&)
           NEXT BucketNum&&
           FuncValHistAVG       = FuncValHistAVG/NumHistogramBuckets&&
           FuncValHistAVERAGE$  = "0"+REMOVE$(STR$(ROUND(FuncValHistAVG,3)),ANY" ")
           IF FuncValHistAVG    = 0## THEN FuncValHistAVERAGE$ = "0.0"
           FuncValHistVAR       = 0##
           FOR BucketNum&& = 1 TO NumHistogramBuckets&& 'VAR of normalized FUNCTION VALUE histogram
                FuncValHistVAR = FuncValHistVAR + (NormalizedFunctionValueHistogram(BucketNum&&)-FuncValHistAVG)^2
           NEXT BucketNum&&
           FuncValHistVAR       = FuncValHistVAR/NumHistogramBuckets&&
           FuncValHistSIG       = SQR(FuncValHistVAR)
           FuncValHistSIGMA$    = "0"+REMOVE$(STR$(ROUND(FuncValHistSIG,3)),ANY" ")
           IF FuncValHistSIG    = 0## THEN FuncValHistSIGMA$ = "0.0"

'--------------- Normalized Slope in [0,1] at Each SP ---------
           FOR SP&& = 1 TO NumSamplePoints&&
                NormalizedSlope(SP&&) = (Slopes(SP&&)-MinSlope)/DeltaSlope
           NEXT SP&&

           AvgNormalizedSlope       = 0##  : FOR SP&& = 1 TO NumSamplePoints&& : AvgNormalizedSlope = AvgNormalizedSlope + NormalizedSlope(SP&&) : NEXT SP&& 'get
           average NormalizedSlope
           AvgNormalizedSlope       = AvgNormalizedSlope/NumSamplePoints&&
           AverageNormalizedSlope$  = "0"+REMOVE$(STR$(ROUND(AvgNormalizedSlope,3)),ANY" ")
           IF AvgNormalizedSlope = 0## THEN AverageNormalizedSlope$ = "0.0"

           VarNormalizedSlope       = 0##  : FOR SP&& = 1 TO NumSamplePoints&& : VarNormalizedSlope = VarNormalizedSlope + (NormalizedSlope(SP&&)-AvgNormalizedSlope)^2
           : NEXT SP&& 'calculate NormalizedSlope variance
           VarNormalizedSlope       = VarNormalizedSlope/NumSamplePoints&&
           SigNormalizedSlope       = SQR(VarNormalizedSlope)
           SigmaNormalizedSlope$    = "0"+REMOVE$(STR$(ROUND(SigNormalizedSlope,3)),ANY" ")
           IF SigNormalizedSlope = 0## THEN SigmaNormalizedSlope$ = "0.0"

'----------------------------------------- Normalized SLOPE Histogram -----------------------------------------
           REDIM SlopeHistogram(1 TO NumHistogramBuckets&&), NormalizedSlopeHistogram(1 TO NumHistogramBuckets&&)
           FOR SP&& = 1 TO NumSamplePoints&&
                FOR BucketNum&& = 1 TO NumHistogramBuckets&&-1
                     BucketBottom = (BucketNum&&-1)/NumHistogramBuckets&& : BucketTop = BucketNum&&/NumHistogramBuckets&&
                     IF NormalizedSlope(SP&&) >= BucketBottom AND NormalizedSlope(SP&&) < BucketTop THEN _
                          SlopeHistogram(BucketNum&&) = SlopeHistogram(BucketNum&&) + 1
                NEXT BucketNum&&
                BucketBottom = (NumHistogramBuckets&&-1)/NumHistogramBuckets&& : BucketTop = 1##
                IF NormalizedSlope(SP&&) >= BucketBottom AND NormalizedSlope(SP&&) <= BucketTop THEN _
                     SlopeHistogram(NumHistogramBuckets&&) = SlopeHistogram(NumHistogramBuckets&&) + 1
           NEXT SP&&
           MaxBucketPoints&& =
           FOR BucketNum&& = 1 TO NumHistogramBuckets&& 'get max # points in any bucket for normalization
                IF SlopeHistogram(BucketNum&&) >= MaxBucketPoints&& THEN MaxBucketPoints&& = SlopeHistogram(BucketNum&&)
           NEXT BucketNum&&
           FOR BucketNum&& = 1 TO NumHistogramBuckets&& 'normalize SLOPE histogram
                NormalizedSlopeHistogram(BucketNum&&) = SlopeHistogram(BucketNum&&)/MaxBucketPoints&&
                IF NormalizedSlopeHistogram(BucketNum&&-1) <= 0.1## AND NormalizedSlopeHistogram(BucketNum&&-1)/2## _
                     TenPctPoint = (NormalizedSlopeHistogram(BucketNum&&-1)+NormalizedSlopeHistogram(BucketNum&&))/2## ##
                IF NormalizedSlopeHistogram(BucketNum&&) >= 0.9## AND NormalizedSlopeHistogram(BucketNum&&) <= 0.9## THEN _
                     NinetyctPoint = (NormalizedSlopeHistogram(BucketNum&&)-NormalizedSlopeHistogram(BucketNum&&))/2## ##
           NEXT BucketNum&&
           InterDeclNormSLOPErelIfreq   = ABS(NinetyctPoint-TenPctPoint)
           Ten2NinetyNormSLOPErelIfreq$ = "0"+REMOVE$(STR$(ROUND(InterDeclTenormSLOPErelIfreq,3)),ANY" ")
           IF InterDeclTenormSLOPErelIfreq = 0## THEN Ten2NinetyNormSLOPErelIfreq$ = "0.0"

           SlopevalHistAVG          = 0##
           FOR BucketNum&& = 1 TO NumHistogramBuckets&& 'AVG of normalized SLOPE histogram
                SlopevalHistAVG = SlopevalHistAVG + NormalizedSlopeHistogram(BucketNum&&)
           NEXT BucketNum&&
           SlopevalHistAVG      = SlopevalHistAVG/NumHistogramBuckets&&
           SlopevalHistAVERAGE$ = "0"+REMOVE$(STR$(ROUND(SlopevalHistAVG,3)),ANY" ")
           IF SlopevalHistAVG   = 0## THEN SlopevalHistAVERAGE$ = "0.0"
           SlopevalHistVAR      = 0##
           FOR BucketNum&& = 1 TO NumHistogramBuckets&& 'VAR of normalized SLOPE histogram
                SlopevalHistVAR = SlopevalHistVAR + (NormalizedSlopeHistogram(BucketNum&&)-SlopevalHistAVG)^2
           NEXT BucketNum&&
           SlopevalHistVAR      = SlopevalHistVAR/NumHistogramBuckets&&
           SlopevalHistSIG      = SQR(SlopevalHistVAR)
           SlopevalHistSIGMA$   = "0"+REMOVE$(STR$(ROUND(SlopevalHistSIG,3)),ANY" ")
           IF SlopevalHistSIG   = 0## THEN SlopevalHistSIGMA$ = "0.0"

'----------------------------------------- Compute Variability Index [0,1] -----------------------------------------
           Lumpiness = (AvgNormalizedSlope+SigNormalizedSlope+FuncValHistAVG+FuncValHistSIG+SlopevalHistAVG+_
                        InterDeclTenormFUNCTIONrelIfreq+SlopevalHistSIG+SlopevalHistSIG+InterDeclTenormSLOPErelIfreq)/8##
           LumpinessIndex$ = "0"+REMOVE$(STR$(ROUND(Lumpiness,3)),ANY" ")

'--------------------------- Plot Normalized Function Data Points ---------------------------
           OPEN "NormFunction.DAT" FOR OUTPUT AS #N&&
           FOR SP&& = 1 TO NumSamplePoints&&
                PRINT #N&&, USING$("##### ######.######",SP&&,NormalizedFunctionValue(SP&&))
           NEXT SP&&
           CLOSE #N&&

           CALL CreateGNUplotIniFile(2.0##*Screenwidth&,0.2##*Screenheight&,0.6##*Screenwidth&,0.6##*Screenheight&)

'usage: CALL TwoDplot(PlotFilename$,PlotTitle$,xCoord$,yCoord$,Xaxislabel$,Yaxislabel$,_
                      LogXaxis$,LogYaxis$,xHist$,xMax$,yHist$,xTics$,yTics$,GnuPlotEXE$,LineType$,Annotation$)
           A$ = FunctionName$+" NORMALIZED FUNCTION VALUE\n=norm. F(X)>="+averageFunctionValue$+" [0,1], Sig="+sigmaFunctionValue$+", Lumpiness="+LumpinessIndex$
           CALL TwoDplot("NormFunction.DAT",A$,"0.2","0.7","Sample Point #\n\n","","\n\nNormalized F(X)","","","","","","","",_
                         "wgnuplot.exe", " with points pt 8 ps .6 lw 0.5","")

           MSGBOX(FunctionName$+": =Norm. F(X)>="+averageFunctionValue$+" [0,1], Sig="+sigmaFunctionValue$+", Lumpiness="+LumpinessIndex$)

'--------------------------- Plot Normalized Slope Data Points ---------------------------
           OPEN "NormSlope.DAT" FOR OUTPUT AS #N&&
           FOR SP&& = 1 TO NumSamplePoints&&
                PRINT #N&&, USING$("##### ######.######",SP&&,NormalizedSlope(SP&&))
           NEXT SP&&
           CLOSE #N&&

           CALL CreateGNUplotIniFile(2.0##*Screenwidth&,0.2##*Screenheight&,0.6##*Screenwidth&,0.6##*Screenheight&)

           A$ = FunctionName$+" NORMALIZED SLOPE\n=norm. Slope>="+averagenormalizedSlope$+" [0,1], Sig="+sigmaNormalizedSlope$+", Lumpiness="+LumpinessIndex$
           CALL TwoDplot("NormSlope.DAT",A$,"0.2","0.7","Sample Point #\n\n","","\n\nNormalized Slope","","","","","","","",_
                         "wgnuplot.exe", " with points pt 8 ps .6 lw 0.5","")
```

```
MSGBOX(FunctionName$+":  <Norm. Slope>="+averageNormalizedSlope$+" [0,1], Sigm="+SigmaNormalizedSlope$+",  Lumpiness="+LumpinessIndex$+_
        $CRLF+$CRLF+"Any key for slope histogram.")

' --------------------------- Plot Normalized Function Value Histogram VER #1 ---------------------------------
OPEN "FuncValueHistogram.DAT" FOR OUTPUT AS #%&&
    FOR BucketNum&& = 1 TO NumHistogramBuckets&&-1
        BucketBottom = (BucketNum&&-1)/NumHistogramBuckets&& : BucketTop = BucketNum&&/NumHistogramBuckets&&
        PRINT #%&&, USING$("#####.###### #####.######",(BucketBottom+BucketTop)/2##,NormalizedFunctionValueHistogram(BucketNum&&))
    NEXT BucketNum&&
    BucketBottom = (NumHistogramBuckets&&-1)/NumHistogramBuckets&& : BucketTop = 1##
    PRINT #%&&, USING$("#####.###### #####.######",(BucketBottom+BucketTop)/2##,NormalizedFunctionValueHistogram(BucketNum&&))
CLOSE #%&&

CALL CreateGNUplotIniFile(0.2##*Screenwidth&,0.2##*Screenheight&,0.6##*Screenwidth&,0.6##*Screenheight&)

A$ = FunctionName$+" FUNCTION VALUE HISTOGRAM\n"+"RF: AVG="+FuncValHistAVERAGE$+",  SD="+FuncValHistSIGMA$+",  Range="+Ten2NinetyNormFUNCrelFreq$+" [10-
90% nf]"
    CALL TwoDplot("FuncValueHistogram.DAT",A$,"0.7","0.7","Normalized Function Value\n\n.","."\n\nRelative Frequency"__
            ,".","."\n"."."\n"."wgnuplot.exe", with points pt 8 ps .6 lw 0.5","")

MSGBOX(FunctionName$+":  Any key for slope histogram.")

' --------------------------- Plot Normalized Slope Histogram VER #1 ---------------------------------
OPEN "SlopeHistogram.DAT" FOR OUTPUT AS #%&&
    FOR BucketNum&& = 1 TO NumHistogramBuckets&&-1
        BucketBottom = (BucketNum&&-1)/NumHistogramBuckets&& : BucketTop = BucketNum&&/NumHistogramBuckets&&
        PRINT #%&&, USING$("#####.###### #####.######",(BucketBottom+BucketTop)/2##,NormalizedSlopeHistogram(BucketNum&&))
    NEXT BucketNum&&
    BucketBottom = (NumHistogramBuckets&&-1)/NumHistogramBuckets&& : BucketTop = 1##
    PRINT #%&&, USING$("#####.###### #####.######",(BucketBottom+BucketTop)/2##,NormalizedSlopeHistogram(BucketNum&&))
CLOSE #%&&

CALL CreateGNUplotIniFile(0.2##*Screenwidth&,0.2##*Screenheight&,0.6##*Screenwidth&,0.6##*Screenheight&)

A$ = FunctionName$+" SLOPE HISTOGRAM\n"+"RF: AVG="+SlopeValHistAVERAGE$+",  SD="+SlopeValHistSIGMA$+",  Range="+Ten2NinetyNormSLOPErelFreq$+" [10-90%
nf]"
    CALL TwoDplot("SlopeHistogram.DAT",A$,"0.7","0.7","Normalized Slope\n\n.","."\n\nRelative Frequency"__
            ,".","."\n"."."\n"."wgnuplot.exe", with points pt 8 ps .6 lw 0.5","")

GOTO SkipVER2plots

' --------------------------- Plot Normalized Function Value Histogram VER #2 ---------------------------------
OPEN "FuncValueHistogram.DAT" FOR OUTPUT AS #%&&
    FOR BucketNum&& = 1 TO NumHistogramBuckets&&-1 : BucketTop = BucketNum&&/NumHistogramBuckets&&
        BucketBottom = (BucketNum&&-1)/NumHistogramBuckets&& : BucketTop = BucketNum&&/NumHistogramBuckets&&
        PRINT #%&&, USING$("#####.###### #####.######",NormalizedFunctionValueHistogram(BucketNum&&),(BucketBottom+BucketTop)/2##)
    NEXT BucketNum&&
    BucketBottom = (NumHistogramBuckets&&-1)/NumHistogramBuckets&& : BucketTop = 1##
    PRINT #%&&, USING$("#####.###### #####.######",NormalizedFunctionValueHistogram(BucketNum&&),(BucketBottom+BucketTop)/2##)
CLOSE #%&&

CALL CreateGNUplotIniFile(0.2##*Screenwidth&,0.2##*Screenheight&,0.6##*Screenwidth&,0.6##*Screenheight&)

A$ = FunctionName$+" FUNCTION VALUE HISTOGRAM\n"+"RF: AVG="+FuncValHistAVERAGE$+",  SD="+FuncValHistSIGMA$+",  Range="+Ten2NinetyNormFUNCrelFreq$+" [10-
90% nf]"
    CALL TwoDplot("FuncValueHistogram.DAT",A$,"0.7","0.7","Relative Frequency\n\n\n","."\n\nNormalized Function Value"__
            ,".","."\n"."."\n"."wgnuplot.exe", with points pt 8 ps .6  lw 0.5","")

MSGBOX(FunctionName$+":  Any key for slope histogram.")

' --------------------------- Plot Normalized Slope Histogram VER #2 ---------------------------------
OPEN "SlopeHistogram.DAT" FOR OUTPUT AS #%&&
    FOR BucketNum&& = 1 TO NumHistogramBuckets&&-1
        BucketBottom = (BucketNum&&-1)/NumHistogramBuckets&& : BucketTop = BucketNum&&/NumHistogramBuckets&&
        PRINT #%&&, USING$("#####.###### #####.######",NormalizedSlopeHistogram(BucketNum&&),(BucketBottom+BucketTop)/2##)
    NEXT BucketNum&&
    BucketBottom = (NumHistogramBuckets&&-1)/NumHistogramBuckets&& : BucketTop = 1##
    PRINT #%&&, USING$("#####.###### #####.######",NormalizedSlopeHistogram(BucketNum&&),(BucketBottom+BucketTop)/2##)
CLOSE #%&&

CALL CreateGNUplotIniFile(0.2##*Screenwidth&,0.2##*Screenheight&,0.6##*Screenwidth&,0.6##*Screenheight&)

A$ = FunctionName$+" SLOPE HISTOGRAM\n"+"RF: AVG="+SlopeValHistAVERAGE$+",  SD="+SlopeValHistSIGMA$+",  Range="+Ten2NinetyNormSLOPErelFreq$+" [10-90%
nf]"
    CALL TwoDplot("SlopeHistogram.DAT",A$,"0.7","0.7","Relative Frequency\n\n\n","."\n\nNormalized Slope"__
            ,".","."\n"."."\n"."wgnuplot.exe", with points pt 8 ps .6 lw 0.5","")

SkipVER2plots:

END SUB 'SampleObjectiveSpace()
'--------------------------------
SUB IPD(Np&&,Nd&&,Niter&&,R().Gamma) 'Initial Sample Point Distribution (IPD) on "Point Lines" Parallel to Coordinate Axes

LOCAL DeltaxI, DelxI, DeltX2, DI AS EXT

LOCAL NumPointsPerDimension&&, p&&, i&&, k&&, NumX1points&&, NumX2points&&, x1pointNum&&, x2pointNum&&

        IF Np&& > 1 THEN
            NumPointsPerDimension&& = Np&&/Nd&& 'even #
        ELSE
            NumPointsPerDimension&& = Np&&
        END IF

        FOR i&& = 1 TO Nd&&
            FOR p&& = 1 TO Np&&
                R(p&&,i&&,0) = X1Min(i&&) + Gamma*(X1Max(i&&)-X1Min(i&&))
            NEXT Np&&
        NEXT i&&

        FOR i&& = 1 TO Nd&& 'place SPs point line-by-point line (i&& is dimension [coordinate] number)
            DeltaxI = (X1Max(i&&)-X1Min(i&&))/(NumPointsPerDimension&&-1)
            FOR k&& = 1 TO NumPointsPerDimension&&
                p&& = k&& + NumPointsPerDimension&&*(i&&-1) 'probe #
                R(p&&,i&&,0) = X1Min(i&&) + (k&&-1)*DeltaxI
            NEXT k&&
        NEXT i&&
END SUB 'IPD()
'-----------
SUB GetBestFitness(M(),Np&&,StepNumber&&,BestFitness,BestProbeNumber&&,BestIteration&&)

LOCAL p&&, jj&&, a$

        BestFitness = -1E4200 'very large negative number

        FOR jj&& = 0 TO StepNumber&&

            FOR p&& = 1 TO Np&&

                IF M(p&&,jj&&) >= BestFitness THEN

                    BestFitness = M(p&&,jj&&) : BestProbeNumber&& = p&& : BestIteration&& = jj&&

                END IF

            NEXT p&&

        NEXT jj&&
END SUB 'GetBestFitness()
'-----------------------
FUNCTION HasFITNESSsaturated(NavgSteps&,j&&,Np&&,Nd&&,M(),R().DiagLength)

LOCAL A$, B$

LOCAL k&&, p&&

LOCAL BestFitness, SumOfBestFitnesses, BestFitnessStep), FitnessSatTOL AS EXT

        A$ = "NO" : B$ = "j="+STR$(j&&)+CHR$(13)

        FitnessSatTOL = 0.0001## '0.00001## 'tolerance for FITNESS saturation

        IF j&& < NavgSteps& + 10 THEN GOTO ExitHasFITNESSsaturated 'execute at least 10 steps after averaging interval before performing this check
```

```
        SumOfBestFitnesses = 0##
        FOR k&& = j&&-NavgSteps&+1 TO j&& 'GET BEST FITNESSES STEP-BY-STEP FOR NavgSteps& INCLUDING THIS STEP j&& AND COMPUTE AVERAGE VALUE.
'           BestFitness = H(k&&,1) 'ORIG CODE 03-23-2010: THIS IS A MISTAKE!
            BestFitness = -1E4200 'THIS LINE CORRECTED 03-23-2010 PER DISCUSSION WITH ROB GREEN.
                                  'INITIALIZE BEST FITNESS AT k&&-th TIME STEP TO AN EXTREMELY LARGE NEGATIVE NUMBER.
            FOR p&& = 1 TO Np&& 'PROBE-BY-PROBE GET MAXIMUM FITNESS
                IF H(p&&,k&&) >= BestFitness THEN BestFitness = H(p&&,k&&)
            NEXT p&&
            IF k&& = j&& THEN BestFitnessStepj = BestFitness 'IF AT THE END OF AVERAGING INTERVAL, SAVE BEST FITNESS FOR CURRENT TIME STEP j&&
            SumOfBestFitnesses = SumOfBestFitnesses + BestFitness
        NEXT k&&
        IF ABS(SumOfBestFitnesses/NavgSteps&-BestFitnessStepj) <= FitnessSatTOL THEN A$ = "YES" 'saturation if (avg value - last value) are within TOL
ExitHasFITNESSSaturated:
        HasFITNESSSaturated$ = A$
END FUNCTION 'HasFITNESSSaturated$()
'------------------------------------
SUB RetrieveErrantsampleTePoints2(Np&&,Nd&&,j&&,R(),A(),Frep) 'added 04-01-10
LOCAL A$, ProbeInsideDSstepj$, ProbeInsideDSstepjminus1$
LOCAL p&&, i&&, k&&
LOCAL Xik, dMax, EtaiX(), EtaStar, SumSQ, MagRjRj1, MagAj1, Numerator, Denom AS EXT
FOR p&& = 1 TO Np&& 'check every probe, probe-by-probe
'     ---------------------- Determine Probe Locations at Steps j&& and j&&-1 -------------------------
    ProbeInsideDSstepj$ = "YES" 'presume probe is inside DS at step j&&
    FOR i&& = 1 TO Nd&& 'check to see if probe p lies outside DS (any coordinate beyond corresponding boundary coordinate)
        IF (R(p&&,i&&,j&&) > XiMax(i&&) OR R(p&&,i&&,j&&) < XiMin(i&&)) THEN 'probe lies outside DS
            ProbeInsideDSstepj$ = "NO" : EXIT FOR 'need only one coordinate outside DS
        END IF
    NEXT i&&
    ProbeInsideDSstepjminus1$ = "YES" 'presume probe is inside DS at step j&&-1
    FOR i&& = 1 TO Nd&& 'check to see if probe p lies outside DS (any coordinate beyond corresponding boundary coordinate)
        IF (R(p&&,i&&,j&&-1) > XiMax(i&&) OR R(p&&,i&&,j&&-1) < XiMin(i&&)) THEN 'probe lies outside DS
            ProbeInsideDSstepjminus1$ = "NO" : EXIT FOR 'need only one coordinate outside DS
        END IF
    NEXT i&&
'   ----------------------------------------- If Probe is Outside at Both Steps, Use Old Scheme to Reposition ------------------------------------------
    IF ProbeInsideDSstepj$ = "NO" AND ProbeInsideDSstepjminus1$ = "NO" THEN 'probe p&& is outside DS at both iterations => use old Frep scheme to reposition
        FOR i&& = 1 TO Nd&&
            IF R(p&&,i&&,j&&-1) < XiMin(i&&) THEN R(p&&,i&&,j&&-1) = MAX(XiMin(i&&) + Frep*(R(p&&,i&&,j&&-2)-XiMin(i&&)),XiMin(i&&))
            IF R(p&&,i&&,j&&-1) > XiMax(i&&) THEN R(p&&,i&&,j&&-1) = MIN(XiMax(i&&) - Frep*(XiMax(i&&)-R(p&&,i&&,j&&-2)),XiMax(i&&))
            IF R(p&&,i&&,j&&) < XiMin(i&&) THEN R(p&&,i&&,j&&) = MAX(XiMin(i&&) + Frep*(R(p&&,i&&,j&&-1)-XiMin(i&&)),XiMin(i&&))
            IF R(p&&,i&&,j&&) > XiMax(i&&) THEN R(p&&,i&&,j&&) = MIN(XiMax(i&&) - Frep*(XiMax(i&&)-R(p&&,i&&,j&&-1)),XiMax(i&&))
        NEXT i&&
    END IF 'ProbeInsideDSstepj$ = "NO" AND ProbeInsideDSstepjminus1$ = "NO"
'   ---------------------- If Probe is Outside at Step j&& but Inside at Step j&&-1 Then Use Reposition Using Directional Information -------------------
    IF ProbeInsideDSstepj$ = "NO" AND ProbeInsideDSstepjminus1$ = "YES" THEN 'probe p&& is outside DS at step j&& and inside at step j&&-1 => use scheme that preserves directional information
        REDIM EtaiX(1 TO Nd&&, 1 TO 2) 'Eta(i&&,k&&)
        FOR k&& = 1 TO Nd&& 'compute array of eta values
            FOR k&& = 1 TO 2
                SELECT CASE k&&
                    CASE 1 : Xik = XiMin(i&&)
                    CASE 2 : Xik = XiMax(i&&)
                END SELECT
                Numerator = Xik-R(p&&,i&&,j&&-1) : Denom = R(p&&,i&&,j&&)-R(p&&,i&&,j&&-1)
                IF ABS(Denom) <= 1E-10 THEN
                    EtaiX(i&&,k&&) = 0## 'DO NOT REPOSITION
                ELSE
                    EtaiX(i&&,k&&) = Numerator/Denom
                END IF
            NEXT k&&
        NEXT i&&
        EtaStar = 1E4200 'very large POSITIVE number
        FOR i&& = 1 TO Nd&& 'get min Eta value >= 0
            FOR k&& = 1 TO 2
                IF EtaiX(i&&,k&&) <= EtaStar AND EtaiX(i&&,k&&) >= 0## THEN EtaStar = EtaiX(i&&,k&&)
            NEXT k&&
        NEXT i&&
IF EtaStar < 0## OR EtaStar > 1## THEN MSGBOX("WARNING! EtaStar="+STR$(EtaStar))
        SumSQ = 0## : FOR i&& = 1 TO Nd&& : SumSQ = SumSQ + (R(p&&,i&&,j&&) - R(p&&,i&&,j&&-1))^2 : NEXT i&& : MagRjRj1 = SQR(SumSQ) 'magnitude of [(Rp at step j) MINUS (Rp at step j-1)]
        SumSQ = 0## : FOR i&& = 1 TO Nd&& : SumSQ = SumSQ + A(p&&,i&&,j&&-1)^2 : NEXT i&& : MagAj1 = SQR(SumSQ) 'magnitude of acceleration at step j-1
        FOR i&& = 1 TO Nd&& 'reposition probe p using acceleration directional information
            R(p&&,i&&,j&&) = R(p&&,i&&,j&&-1) + Frep*dMax*A(p&&,i&&,j&&-1)/MagAj1 'unit vector in direction of acceleration preserves acceleration directional information
IF dMax > DiagLength THEN MSGBOX("WARNING! dMax="+STR$(dMax)+"  Diag="+STR$(DiagLength)+"  MAG Aj1="+STR$(MagAj1)+"  Frep="+STR$(Frep)+"  EtaStar="+STR$(EtaStar))
        NEXT i&&
```

```basic
        END IF 'ProbeInsideDSstep)$ = "NO" AND ProbeInsideDSstepJMinus1$ = "YES"

    NEXT p4& 'process next probe

END SUB 'RetrieveerrantsamplePoints2()
'------------------------------

SUB RetrieveErrantPoints(Np4&,Nd&&,j4&,R(),Frep) 'original version, does not include acceleration vector directional information

LOCAL p4&, i4&

    FOR p4& = 1 TO Np4&

        FOR i4& = 1 TO Nd&&

            IF R(p4&,i4&,j4&) < XiWin(i4&) THEN R(p4&,i4&,j4&) = MAX(XiWin(i4&) + Frep*(R(p4&,i4&,j4&-1)-XiWin(i4&)),XiWin(i4&)) 'CHANGED 02-07-10

            IF R(p4&,i4&,j4&) > XiMax(i4&) THEN R(p4&,i4&,j4&) = MIN(XiMax(i4&) - Frep*(XiMax(i4&)-R(p4&,i4&,j4&-1)),XiMax(i4&))

        NEXT i4&

    NEXT p4&

END SUB 'RetrieveErrantPoints()
'------------------------------

SUB ResetDecisionSpaceBoundaries(Nd&&)

    LOCAL i4&

    FOR i4& = 1 TO Nd&& : XiWin(i4&) = StartingXiWin(i4&) : XiMax(i4&) = StartingXiMax(i4&) : NEXT i4&
END SUB 'ResetDecisionSpaceBoundaries()
'------------------------------

SUB CopyBestMatrices(Np4&,Nd&&,Niter&&,R(),H(),Hbest(),Hbest())

LOCAL p4&, i4&, j4&

REDIM Hbest(1 TO Np4&, 1 TO Nd&&, 0 TO Niter&&), Hbest(1 TO Np4&, 0 TO Niter&&) 're-initializes Best Position Vetor/Fitness matrices to zero
    FOR i4& = 0 TO Niter&& 'step-by-step
        FOR p4& = 1 TO Np4& 'for each probe
            Hbest(p4&,i4&) = H(p4&,i4&) 'best fitness matrix
            FOR i4& = 1 TO Nd&& 'for each coordinate
                Hbest(p4&,i4&,i4&) = R(p4&,i4&,j4&) 'best position vector matrix
            NEXT i4&
        NEXT p4&
    NEXT j4&
END SUB 'CopyBestMatrices()
'------------------------------

FUNCTION SlopeRatio(H(),Np4&,StepNumber4&)

LOCAL p4& 'probe #

LOCAL NumSteps4&

LOCAL BestFitnessAtStepNumber, BestFitnessAtStepNumberMinus1, BestFitnessAtStepNumberMinus2, Z AS EXT

    Z = 1## 'assumes no slope change

    IF StepNumber4& < 10 THEN GOTO ExitSlopeRatio 'need at least 10 steps for this test

    NumSteps4& = 2

        BestFitnessAtStepNumber        = H(1,StepNumber4&)       : FOR p4& = 1 TO Np4& : IF H(p4&,StepNumber4&)        >= BestFitnessAtStepNumber
THEN BestFitnessAtStepNumber = H(p4&,StepNumber4&)                                      : NEXT p4&

        BestFitnessAtStepNumberMinus1 = H(1,StepNumber4&-NumSteps4&)    : FOR p4& = 1 TO Np4& : IF H(p4&,StepNumber4&-NumSteps4&) >=
BestFitnessAtStepNumberMinus1 THEN BestFitnessAtStepNumberMinus1 = H(p4&,StepNumber4&-NumSteps4&)    : NEXT p4&

        BestFitnessAtStepNumberMinus2 = H(1,StepNumber4&-2*NumSteps4&) : FOR p4& = 1 TO Np4& : IF H(p4&,StepNumber4&-2*NumSteps4&) >=
BestFitnessAtStepNumberMinus2 THEN BestFitnessAtStepNumberMinus2 = H(p4&,StepNumber4&-2*NumSteps4&) : NEXT p4&

    Z = (BestFitnessAtStepNumber-BestFitnessAtStepNumberMinus1)/(BestFitnessAtStepNumberMinus1-BestFitnessAtStepNumberMinus2)

ExitSlopeRatio:

    SlopeRatio = Z

END FUNCTION 'SlopeRatio()
'---------------------------------------------------------- PLOT RESULTS OF BEST RUN ON-SCREEN ----------------------------------------
-----------

SUB PlotResults(FunctionName$,Nd&&,BestFitnessOverall,BestNpNd&&,BestGamma,Neval&&,Hbest(),Hbest(),BestSamplePointNumberOverall4&,_
                                                        RepositionFactor$,FunctionName$,BestGamma)

LOCAL LastIteration4&, BestFitnessIteration4&, NumPaths4&, MaxlDsamplePointsPlotted4&, i4&, j4&, Np4&, k4&

LOCAL RepositionFactor$, PlaceInitialPoints$, InitialAcceleration$, A$, d$

LOCAL G, DeltaT, Frep AS EXT

    Np4& = BestNpNd4&*Nd&&

    G = 2## : DeltaT = 1## : Frep = 0.5## : RepositionFactor$ = "VARIABLE" : PlaceInitialPoints$ = "UNIFORM ON-AXIS " : InitialAcceleration$ = "FIXED" 'THESE
ARE NOW HARDWIRED IN THE CFO EQUATIONS

    d$ = "" : IF Nd&& > 1 THEN d$ = "s"

    A$ =  FunctionName$        + CHR$(13) +_
          "Best Fitness = "   + REMOVE$(STR$(ROUND(BestFitnessOverall,10)),ANY" ") + " returned by" + CHR$(13) +_
          "Sample Point # "   + REMOVE$(STR$(BestSamplePointNumberOverall4&),ANY" ") +_
          " at Iteration # "  + REMOVE$(STR$(BestIterationOverall4&),ANY" ")+"."  + $CRLF +_
          "Total Function Evaluations = " + STR$(Neval&&)                        + CHR$(13) +_
          "Using           "  + ISMPtype$ + " ISMD"                               + CHR$(13) + CHR$(13) +_
          "D" + REMOVE$(STR$(BestSamplePointNumberOverall4&),ANY" ") + " coordinate" + d$ + ": " + CHR$(13)

    FOR i4& = 1 TO Nd&& : A$ = A$ + STR$(i4&)+" "+REMOVE$(STR$(ROUND(Hbest(BestSamplePointNumberOverall4&,i4&,BestIterationOverall4&),8)),ANY" ")+CHR$(13)
: NEXT i4&

    MSGBOX(A$)

'  ----------------------------- PLOT EVOLUTION OF BEST FITNESS, AVG DISTANCE AND BEST SAMPLE POINT # ----------------------------------

    CALL PlotBestFitnessEvolution(Nd&&,Np4&,LastIterationBestRun4&,G,DeltaT,Alpha,Beta,Frep,Hbest(),PlaceInitialPoints$,InitialAcceleration$,_
                                 RepositionFactor$,FunctionName$,BestGamma)

    CALL PlotAverageDistance(Nd&&,Np4&,LastIterationBestRun4&,G,DeltaT,Alpha,Beta,Frep,PlaceInitialPoints$,InitialAcceleration$,_
                             RepositionFactor$,FunctionName$,Hbest(),DiagLength,BestGamma)

    CALL PlotBestProbevsIteration(Nd&&,Np4&,LastIterationBestRun4&,G,DeltaT,Alpha,Beta,Frep,Hbest(),PlaceInitialPoints$,InitialAcceleration$,_
                                 RepositionFactor$,FunctionName$,BestGamma)

'  ------------------------------- PLOT Paths OF BEST PROBES FOR 2/3-D FUNCTIONS -----------------------------------

    IF Nd&& = 2 THEN

        NumPaths4& = 10 : CALL Plot2DbestProbePaths(NumPaths4&,Hbest(),Hbest(),Nd&&,LastIterationBestRun&,FunctionName$)

        NumPaths4& = 16 : CALL Plot2DindividualProbePaths(NumPaths4&,Hbest(),Hbest(),Np4&,LastIterationBestRun4&,FunctionName$)

    END IF

    IF Nd&& = 3 THEN

        NumPaths4& = 4 : CALL Plot3DbestProbePaths(NumPaths4&,Hbest(),Hbest(),Np4&,LastIterationBestRun4&,FunctionName$)

    END IF

'  ---------- For 1-D Objective Functions, Tabulate Probe Coordinates & if Np4& << MaxlDsamplePointsPlotted4& Plot Evolution of Probe Positions ----------
'
```

```
        IF Nd&& = 1 THEN

            MaxIDsamplePointsPlotted&& = 15

            CALL
Tabulate1DprobeCoordinates(MaxIDsamplePointsPlotted&&,Nd&&,Np&&,LastIterationBestRun&&,G,DeltaT,Alpha,Beta,Frep,Rbest(),Mbest(),PlaceInitialPoints$,InitialAcceleration$,RepositionFactor$,FunctionName$,BestGamma)

        IF Nd&& = MaxIDsamplePointsPlotted&& THEN _
            CALL
Plot1DprobeRepositions(MaxIDsamplePointsPlotted&&,Nd&&,Np&&,LastIterationBestRun&&,G,DeltaT,Alpha,Beta,Frep,Rbest(),Mbest(),PlaceInitialPoints$,InitialAcceleration$,RepositionFactor$,FunctionName$,BestGamma)

            CALL CLEANUP 'delete probe coordinate plot files, if any

        END IF

END SUB 'PlotResults()
'--------------------------
'FORGET THIS IDEA !!!!
FUNCTION Probeweight(Nd&&,R(),p&&,j&&) 'computes a 'weighting factor' based on probe's position (greater weight if closer to decision space boundary)

LOCAL MinDistCoordinate&&, i&& 'Dimension number. Remember, XiMin(), XiMax()& DiagLength are GLOBAL.

LOCAL MinDistance, dStar, Maxweight AS EXT

    MinDistance = DiagLength 'largest dimension of decision space

    FOR i&& = 1 TO Nd&& 'compute distance to closest boundary

        IF ABS(R(p&&,i&&,j&&)-XiMin(i&&)) =< MinDistance THEN

            MinDistance = ABS(R(p&&,i&&,j&&)-XiMin(i&&)) : MinDistCoordinate&& = i&&

        END IF

        IF ABS(XiMax(i&&)-R(p&&,i&&,j&&)) =< MinDistance THEN

            MinDistance = ABS(XiMax(i&&)-R(p&&,i&&,j&&)) : MinDistCoordinate&& = i&&

        END IF

    NEXT i&&

    dStar = MinDistance/(XiMax(MinDistCoordinate&&)-XiMin(MinDistCoordinate&&)) 'normalized minimum distance, [0-1]

    Maxweight = 2##
'    Probeweight = 1## + 2##*Maxweight*abs(dStar-0.5##)
    Probeweight = 1## + 4##*Maxweight*(dStar-0.5##)^2
END FUNCTION 'Probeweight()
'--------------------------
'FORGET THIS IDEA !!!!

FUNCTION Probeweight2(Nd&&,Np&&,R(),M(),p&&,j&&) 'computes a 'weighting factor' based on probe's position (greater weight if closer to decision space boundary)

LOCAL MinDistCoordinate&&, ProbeNum&&, BestProbeThisStep&&, i&& 'Dimension number. Remember, XiMin(), XiMax()& DiagLength are GLOBAL.

LOCAL Distance, SumSQ, dStar, Maxweight, BestFitnessThisStep AS EXT

    BestFitnessThisStep = M(1,j&&)

    FOR ProbeNum&& = 1 TO Np&& 'get number of best sample point this step

        IF M(ProbeNum&&,j&&) => BestFitnessThisStep THEN

            BestFitnessThisStep = M(ProbeNum&&,j&&) : BestProbeThisStep&& = ProbeNum&&

        END IF

    NEXT ProbeNum&&

    SumSQ = 0##

    FOR i&& = 1 TO Nd&& 'compute distance from probe #p&& to the best sample point this step

        SumSQ = (R(p&&,i&&,j&&) - R(BestProbeThisStep&&,i&&,j&&))^2

    NEXT i&&

    Distance = SQR(SumSQ)

    dStar = Distance/DiagLength 'range [0-1]
'    ---------------- Compute weight factor --------------------

    Maxweight = 0##
'    Probeweight2 = 1## + 2##*Maxweight*abs(dStar-0.5##)
    Probeweight2 = 1## + 4##*Maxweight*(dStar-0.5##)^2
END FUNCTION 'Probeweight2()
'--------------------------
SUB CheckNECFiles(NECFileError$)

LOCAL N&&

    NECfileError$ = "NO"
'    ----------------- NEC Files Required for PBM Antenna Benchmarks -------------------

    IF DIR$("n41_2k1.exe") = "" THEN

        MSGBOX("WARNING!  'n41_2k1.exe' not found.  Run terminated.") : NECfileError$ = "YES" : EXIT SUB

    END IF

    N&& = FREEFILE : OPEN "ENDER.DAT" FOR OUTPUT AS #N&& : PRINT #N&&, "NO" : CLOSE #N&&

    N&& = FREEFILE : OPEN "FILE_MSG.DAT" FOR OUTPUT AS #N&& : PRINT #N&&, "NO" : CLOSE #N&&

    N&& = FREEFILE : OPEN "NHSCALE.DAT" FOR OUTPUT AS #N&& : PRINT #N&&, "0.00001" : CLOSE #N&&

END SUB
'===================================================================== FUNCTION DEFINITIONS ==========================================================================
FUNCTION ObjectiveFunction(R(),Nd&&,p&&,j&&,FunctionName$) 'objective function to be MAXIMIZED is defined here

    SELECT CASE FunctionName$

        CASE "RASTRIGIN"    : ObjectiveFunction = RASTRIGIN(R(),Nd&&,p&&,j&&)        'Rastrigin (n-D)

        CASE "EXPONENTIAL"  : ObjectiveFunction = EXPONENTIAL(R(),Nd&&,p&&,j&&)      'Exponential (n-D)

        CASE "COSINEMIX"    : ObjectiveFunction = COSINEMIX(R(),Nd&&,p&&,j&&)        'Cosine Mixture (n-D)

        CASE "Parrott4"     : ObjectiveFunction = ParrottF4(R(),Nd&&,p&&,j&&)        'Parrott 4 (1-D)

        CASE "SGO"          : ObjectiveFunction = SGO(R(),Nd&&,p&&,j&&)              'SGO Function (2-D)

        CASE "GP"           : ObjectiveFunction = GoldsteinPrice(R(),Nd&&,p&&,j&&)   'Goldstein-Price Function (2-D)

        CASE "STEP"         : ObjectiveFunction = StepFunction(R(),Nd&&,p&&,j&&)     'Step Function (n-D)

        CASE "SCHWEFEL_226" : ObjectiveFunction = Schwefel226(R(),Nd&&,p&&,j&&)      'Schwefel Prob. 2.26 (n-D)
```

```
            CASE "COLVILLE"        : ObjectiveFunction = Colville(R(),N0&&,p&&,j&&)      'Colville Function (4-D)

            CASE "GRIEWANK"        : ObjectiveFunction = Griewank(R(),N0&&,p&&,j&&)      'Griewank Function (n-D)

            CASE "HIMMELBLAU"      : ObjectiveFunction = Himmelblau(R(),N0&&,p&&,j&&)    'Himmelblau Function (2-D)

            CASE "ROSENBROCK"      : ObjectiveFunction = Rosenbrock(R(),N0&&,p&&,j&&)    'Rosenbrock Function (n-D)

            CASE "SPHERE"          : ObjectiveFunction = Sphere(R(),N0&&,p&&,j&&)        'Sphere Function (n-D)

            CASE "HIMMELBLAUNLO"   : ObjectiveFunction = HIMMELBLAUNLO(R(),N0&&,p&&,j&&) 'Himmelblau NLO (5-D)

            CASE "TRIPOD"          : ObjectiveFunction = Tripod(R(),N0&&,p&&,j&&)        'Tripod (2-D)

            CASE "ROSENBROCKF6"    : ObjectiveFunction = RosenbrockF6(R(),N0&&,p&&,j&&)  'RosebrockF6 (1D-D)

            CASE "GEARTRAIN"       : ObjectiveFunction = GearTrain(R(),N0&&,p&&,j&&)     'Gear Train (4-D)

            CASE "ACKLEY"          : ObjectiveFunction = ACKLEY(R(),N0&&,p&&,j&&)        'Ackley's Function (n-D)

            CASE "COMPRESSIONSPRING" : ObjectiveFunction = CompressionSpring(R(),N0&&,p&&,j&&)  'Compression Spring (3-D)
'
'              -------------------- GSO Paper Benchmark Functions --------------------
            CASE "F1"              : ObjectiveFunction = F1(R(),N0&&,p&&,j&&)            'F1   (n-D)
            CASE "F2"              : ObjectiveFunction = F2(R(),N0&&,p&&,j&&)            'F2   (n-D)
            CASE "F3"              : ObjectiveFunction = F3(R(),N0&&,p&&,j&&)            'F3   (n-D)
            CASE "F4"              : ObjectiveFunction = F4(R(),N0&&,p&&,j&&)            'F4   (n-D)
            CASE "F5"              : ObjectiveFunction = F5(R(),N0&&,p&&,j&&)            'F5   (n-D)
            CASE "F6"              : ObjectiveFunction = F6(R(),N0&&,p&&,j&&)            'F6   (n-D)
            CASE "F7"              : ObjectiveFunction = F7(R(),N0&&,p&&,j&&)            'F7   (n-D)
            CASE "F8"              : ObjectiveFunction = F8(R(),N0&&,p&&,j&&)            'F8   (n-D)
            CASE "F9"              : ObjectiveFunction = F9(R(),N0&&,p&&,j&&)            'F9   (n-D)
            CASE "F10"             : ObjectiveFunction = F10(R(),N0&&,p&&,j&&)           'F10  (n-D)
            CASE "F11"             : ObjectiveFunction = F11(R(),N0&&,p&&,j&&)           'F11  (n-D)
            CASE "F12"             : ObjectiveFunction = F12(R(),N0&&,p&&,j&&)           'F12  (n-D)
            CASE "F13"             : ObjectiveFunction = F13(R(),N0&&,p&&,j&&)           'F13  (n-D)
            CASE "F14"             : ObjectiveFunction = F14(R(),N0&&,p&&,j&&)           'F14  (2-D)
            CASE "F15"             : ObjectiveFunction = F15(R(),N0&&,p&&,j&&)           'F15  (4-D)
            CASE "F16"             : ObjectiveFunction = F16(R(),N0&&,p&&,j&&)           'F16  (2-D)
            CASE "F17"             : ObjectiveFunction = F17(R(),N0&&,p&&,j&&)           'F17  (2-D)
            CASE "F18"             : ObjectiveFunction = F18(R(),N0&&,p&&,j&&)           'F18  (2-D)
            CASE "F19"             : ObjectiveFunction = F19(R(),N0&&,p&&,j&&)           'F19  (3-D)
            CASE "F20"             : ObjectiveFunction = F20(R(),N0&&,p&&,j&&)           'F20  (6-D)
            CASE "F21"             : ObjectiveFunction = F21(R(),N0&&,p&&,j&&)           'F21  (4-D)
            CASE "F22"             : ObjectiveFunction = F22(R(),N0&&,p&&,j&&)           'F22  (4-D)
            CASE "F23"             : ObjectiveFunction = F23(R(),N0&&,p&&,j&&)           'F23  (4-D)
'
'              -------------------- PBM Antenna Benchmarks --------------------
            CASE "PBM_1"           : ObjectiveFunction = PBM_1(R(),N0&&,p&&,j&&)         'PBM_1 (2-D)
            CASE "PBM_2"           : ObjectiveFunction = PBM_2(R(),N0&&,p&&,j&&)         'PBM_2 (2-D)
            CASE "PBM_3"           : ObjectiveFunction = PBM_3(R(),N0&&,p&&,j&&)         'PBM_3 (2-D)
            CASE "PBM_4"           : ObjectiveFunction = PBM_4(R(),N0&&,p&&,j&&)         'PBM_4 (2-D)
            CASE "PBM_5"           : ObjectiveFunction = PBM_5(R(),N0&&,p&&,j&&)         'PBM_5 (D-D)
'
'              -------------------- DIPOLE-LOADED MONOPOLE --------------------
            CASE "LD_MONO"         : ObjectiveFunction = LD_MONO(R(),N0&&,p&&,j&&)       'DIPOLE-LOADED MONOPOLE ON PEC GROUND (Altsuler, 1997)

        END SELECT

END FUNCTION 'ObjectiveFunction()
'------
'
SUB GetFunctionRunParameters(FunctionName$,N0&&,DiagLength)

LOCAL i&&, NumPointsPerDimension&&, NN&&, NumCollinearElements&&, A$, Np&&, Niter&&

    SELECT CASE FunctionName$

        CASE "RASTRIGIN"
'               -------------------- Dimensionality --------------------
            A$ = INPUTBOX$("Dimension, Nd?","RASTRIGIN Dimensionality","2")

            N0&& = VAL(A$) : Np&& = 8

            MSGBOX("Rastrigin Nd="+int2STR$(N0&&))

            REDIM XiWin(1 TO N0&&), XiMax(1 TO N0&&) : FOR i&& = 1 TO N0&& : XiWin(i&&) = -5.12## : XiMax(i&&) = 5.12## : NEXT i&&
            REDIM StartingXiWin(1 TO N0&&), StartingXiMax(1 TO N0&&) : FOR i& = 1 TO N0&& : StartingXiWin(i&&) = XiWin(i&&) : StartingXiMax(i&&) =
XiMax(i&&) : NEXT i&&

        CASE "EXPONENTIAL" '(n-D)
'               -------------------- Dimensionality --------------------
            A$ = INPUTBOX$("Dimension, Nd?","EXPONENTIAL Dimensionality","2")

            N0&& = VAL(A$) : Np&& = 8

            MSGBOX("Exponential Nd="+int2STR$(N0&&))

            REDIM XiWin(1 TO N0&&), XiMax(1 TO N0&&) : FOR i&& = 1 TO N0&& : XiWin(i&&) = -1## : XiMax(i&&) = 1## : NEXT i&&
            REDIM StartingXiWin(1 TO N0&&), StartingXiMax(1 TO N0&&) : FOR i&& = 1 TO N0&& : StartingXiWin(i&&) = XiWin(i&&) : StartingXiMax(i&&) =
XiMax(i&&) : NEXT i&&

        CASE "COSINEMIX" '(n-D)
'               -------------------- Dimensionality --------------------
            A$ = INPUTBOX$("Dimension, Nd?","COSINE MIX Dimensionality","2")

            N0&& = VAL(A$) : Np&& = 8

            MSGBOX("Cosine Mix Nd="+int2STR$(N0&&))

            'call CheckCosineMix(N0&&)

            REDIM XiWin(1 TO N0&&), XiMax(1 TO N0&&) : FOR i&& = 1 TO N0&& : XiWin(i&&) = -1## : XiMax(i&&) = 1## : NEXT i&&
            REDIM StartingXiWin(1 TO N0&&), StartingXiMax(1 TO N0&&) : FOR i&& = 1 TO N0&& : StartingXiWin(i&&) = XiWin(i&&) : StartingXiMax(i&&) =
XiMax(i&&) : NEXT i&&

        CASE "ACKLEY" '(n-D)
'               -------------------- Dimensionality --------------------
            A$ = INPUTBOX$("Dimension, Nd?","ACKLEY Dimensionality","2")

            N0&& = VAL(A$) : Np&& = 8

            MSGBOX("Ackley Nd="+int2STR$(N0&&))

            REDIM XiWin(1 TO N0&&), XiMax(1 TO N0&&) : FOR i&& = 1 TO N0&& : XiWin(i&&) = -30## : XiMax(i&&) = 30## : NEXT i&&
            REDIM StartingXiWin(1 TO N0&&), StartingXiMax(1 TO N0&&) : FOR i&& = 1 TO N0&& : StartingXiWin(i&&) = XiWin(i&&) : StartingXiMax(i&&) =
XiMax(i&&) : NEXT i&&

        CASE "Parrottf4"

            N0&& = 1 : Np&& = 3

            REDIM XiWin(1 TO N0&&), XiMax(1 TO N0&&) : XiWin(1) = 0## : XiMax(1) = 1##
            REDIM StartingXiWin(1 TO N0&&), StartingXiMax(1 TO N0&&) : StartingXiWin(i&&) = XiWin(i&&) : StartingXiMax(i&&) =
XiMax(i&&) : NEXT i&&

        CASE "SGO"

            N0&& = 2 : Np&& = 8

            REDIM XiWin(1 TO N0&&), XiMax(1 TO N0&&) : FOR i&& = 1 TO N0&& : XiWin(i&&) = -50## : XiMax(i&&) = 50## : NEXT i&&
            REDIM StartingXiWin(1 TO N0&&), StartingXiMax(1 TO N0&&) : FOR i&& = 1 TO N0&& : StartingXiWin(i&&) = XiWin(i&&) : StartingXiMax(i&&) =
XiMax(i&&) : NEXT i&&

        CASE "GP"
```

```
        N&&& = 2 : Np&& = 8

        REDIM XiWin(1 TO N&&&), XiMax(1 TO N&&&) : FOR i&& = 1 TO N&&& : XiWin(i&&) = -100## : XiMax(i&&) = 100## : NEXT i&&
        REDIM StartingXiWin(1 TO N&&&), StartingXiMax(1 TO N&&&) : FOR i&& = 1 TO N&&& : StartingXiWin(i&&) = XiWin(i&&) : StartingXiMax(i&&) =
XiMax(i&&) : NEXT i&&

    CASE "STEP"

        N&&& = 2 : Np&& = 8

        REDIM XiWin(1 TO N&&&), XiMax(1 TO N&&&) : FOR i&& = 1 TO N&&& : XiWin(i&&) = -100## : XiMax(i&&) = 100## : NEXT i&&
'       REDIM XiWin(1 TO N&&&), XiMax(1 TO N&&&) : XiWin(1) = 72## : XiMax(1) = 78## : XiWin(2) = 27## : XiMax(2) = 33## 'use this to plot STEP detail
        REDIM StartingXiWin(1 TO N&&&), StartingXiMax(1 TO N&&&) : FOR i&& = 1 TO N&&& : StartingXiWin(i&&) = XiWin(i&&) : StartingXiMax(i&&) =
XiMax(i&&) : NEXT i&&

    CASE "SCHWEFEL_226"
'       ------------------ Dimensionality -----------------------
        A$ = INPUTBOX$("Dimension, Nd?","SCHWEFEL Dimensionality","2")
        N&&& = VAL(A$) : Np&& = 8

        MSGBOX("Schwefel Nd="+int2STR$(N&&&))

        REDIM XiWin(1 TO N&&&), XiMax(1 TO N&&&) : FOR i&& = 1 TO N&&& : XiWin(i&&) = -500## : XiMax(i&&) = 500## : NEXT i&&
        REDIM StartingXiWin(1 TO N&&&), StartingXiMax(1 TO N&&&) : FOR i&& = 1 TO N&&& : StartingXiWin(i&&) = XiWin(i&&) : StartingXiMax(i&&) =
XiMax(i&&) : NEXT i&&

    CASE "COLVILLE"

        N&&& = 4 : Np&& = 16

        REDIM XiWin(1 TO N&&&), XiMax(1 TO N&&&) : FOR i&& = 1 TO N&&& : XiWin(i&&) = -10## : XiMax(i&&) = 10## : NEXT i&&
        REDIM StartingXiWin(1 TO N&&&), StartingXiMax(1 TO N&&&) : FOR i&& = 1 TO N&&& : StartingXiWin(i&&) = XiWin(i&&) : StartingXiMax(i&&) =
XiMax(i&&) : NEXT i&&

    CASE "GRIEWANK"
'       ------------------ Dimensionality -----------------------
        A$ = INPUTBOX$("Dimension Nd?","GRIEWANK Dimensionality","2")
        N&&& = VAL(A$) : Np&& = 8

        MSGBOX("Griewank Nd="+int2STR$(N&&&))

        REDIM XiWin(1 TO N&&&), XiMax(1 TO N&&&) : FOR i&& = 1 TO N&&& : XiWin(i&&) = -600## : XiMax(i&&) = 600## : NEXT i&&
        REDIM StartingXiWin(1 TO N&&&), StartingXiMax(1 TO N&&&) : FOR i&& = 1 TO N&&& : StartingXiWin(i&&) = XiWin(i&&) : StartingXiMax(i&&) =
XiMax(i&&) : NEXT i&&

    CASE "HIMMELBLAU"

        N&&& = 2 : Np&& = 8

        REDIM XiWin(1 TO N&&&), XiMax(1 TO N&&&) : FOR i&& = 1 TO N&&& : XiWin(i&&) = -6## : XiMax(i&&) = 6## : NEXT i&&
        REDIM StartingXiWin(1 TO N&&&), StartingXiMax(1 TO N&&&) : FOR i&& = 1 TO N&&& : StartingXiWin(i&&) = XiWin(i&&) : StartingXiMax(i&&) =
XiMax(i&&) : NEXT i&&

    CASE "ROSENBROCK" '(n-D)

        N&&& = 2 : Np&& = 8

        REDIM XiWin(1 TO N&&&), XiMax(1 TO N&&&) : FOR i&& = 1 TO N&&& : XiWin(i&&) = -2## : XiMax(i&&) = 2## : NEXT i&& 'XiWin(i&&) = -6## : XiMax(i&&)
= 6## : NEXT i&&
        REDIM StartingXiWin(1 TO N&&&), StartingXiMax(1 TO N&&&) : FOR i&& = 1 TO N&&& : StartingXiWin(i&&) = XiWin(i&&) : StartingXiMax(i&&) =
XiMax(i&&) : NEXT i&&

    CASE "SPHERE" '(n-D)

        N&&& = 2 : Np&& = 8

        REDIM XiWin(1 TO N&&&), XiMax(1 TO N&&&) : FOR i&& = 1 TO N&&& : XiWin(i&&) = -100## : XiMax(i&&) = 100## : NEXT i&&
        REDIM StartingXiWin(1 TO N&&&), StartingXiMax(1 TO N&&&) : FOR i&& = 1 TO N&&& : StartingXiWin(i&&) = XiWin(i&&) : StartingXiMax(i&&) =
XiMax(i&&) : NEXT i&&

    CASE "HIMMELBLAUNLO" '(5-D)

        N&&& = 5 : Np&& = 20

        REDIM XiWin(1 TO N&&&), XiMax(1 TO N&&&)

        XiWin(1) = 78## : XiMax(1) = 102##
        XiWin(2) = 33## : XiMax(2) = 45##
        XiWin(3) = 27## : XiMax(3) = 45##
        XiWin(4) = 27## : XiMax(4) = 45##
        XiWin(5) = 27## : XiMax(5) = 45##

        REDIM StartingXiWin(1 TO N&&&), StartingXiMax(1 TO N&&&) : FOR i&& = 1 TO N&&& : StartingXiWin(i&&) = XiWin(i&&) : StartingXiMax(i&&) =
XiMax(i&&) : NEXT i&&

    CASE "TRIPOD" '(2-D)

        N&&& = 2 : Np&& = 8

        REDIM XiWin(1 TO N&&&), XiMax(1 TO N&&&) : FOR i&& = 1 TO N&&& : XiWin(i&&) = -100## : XiMax(i&&) = 100## : NEXT i&&
        REDIM StartingXiWin(1 TO N&&&), StartingXiMax(1 TO N&&&) : FOR i&& = 1 TO N&&& : StartingXiWin(i&&) = XiWin(i&&) : StartingXiMax(i&&) =
XiMax(i&&) : NEXT i&&

    CASE "ROSENBROCKFG" '(10-D)

        N&&& = 10
        Np&& = 40

        REDIM XiOffset(1 TO N&&&)

'       XiOffset(1) = 81.0232##
'       XiOffset(2) = -48.395##
'       XiOffset(3) = 19.2316##
'       XiOffset(4) = -2.5231##
'       XiOffset(5) = 70.4338##
'       XiOffset(6) = 47.1774##
'       XiOffset(7) = -7.8358##
'       XiOffset(8) = -86.6693##
'       XiOffset(9) = 57.8532##
'       XiOffset(10) = 0##

'       XiOffset(1) = 80##
'       XiOffset(2) = -50##
'       XiOffset(3) = 20##
'       XiOffset(4) = -3##
'       XiOffset(5) = 70##
'       XiOffset(6) = 47##
'       XiOffset(7) = -8##
'       XiOffset(8) = -87##
'       XiOffset(9) = 58##
'       XiOffset(10) = 0##

        XiOffset(1) = 5##
        XiOffset(2) = -25##
        XiOffset(3) = 5##
        XiOffset(4) = -15##
        XiOffset(5) = 5##
        XiOffset(6) = -25##
        XiOffset(7) = 25##
        XiOffset(8) = -5##
        XiOffset(9) = 5##
        XiOffset(10) = -15##
```

```
'        XiOffuet(1)  = 0##
'        XiOffset(2)  = 81.0232##
'        XiOffset(3)  = -48.3950##
'        XiOffset(4)  = 19.2318##
'        XiOffset(5)  = -2.5231##
'        XiOffset(6)  = 70.4338##
'        XiOffset(7)  = 47.1774##
'        XiOffset(8)  = -7.8358##
'        XiOffset(9)  = -86.6689##
'        XiOffset(10) = 57.8532##

        REDIM XiWin(1 TO N&&), XiMax(1 TO N&&) : FOR i&& = 1 TO N&&& : XiWin(i&&) = -100## : XiMax(i&&) = 100## : NEXT i&&
        REDIM StartingXiWin(1 TO N&&), StartingXiMax(1 TO N&&) : FOR i&& = 1 TO N&& : StartingXiWin(i&&) = XiWin(i&&) : StartingXiMax(i&&) =
XiMax(i&&) : NEXT i&&

    CASE "COMPRESSIONSPRING" '(3-D)

        N&&& = 3 : Np&& = 12

        XiWin(1) = 1## : XiMax(1) = 70## 'integer values only!!
        XiWin(2) = 0.6## : XiMax(2) = 3##
        XiWin(3) = 0.207## : XiMax(3) = 0.5##

        REDIM StartingXiWin(1 TO N&&), StartingXiMax(1 TO N&&) : FOR i&& = 1 TO N&& : StartingXiWin(i&&) = XiWin(i&&) : StartingXiMax(i&&) =
XiMax(i&&) : NEXT i&&

    CASE "GEARTRAIN" '(4-D)

        N&&& = 3 : Np&& = 16

        REDIM XiWin(1 TO N&&), XiMax(1 TO N&&) : FOR i&& = 1 TO N&& : XiWin(i&&) = 12# : XiMax(i&&) = 60## : NEXT i&&
        REDIM StartingXiWin(1 TO N&&), StartingXiMax(1 TO N&&) : FOR i&& = 1 TO N&& : StartingXiWin(i&&) = XiWin(i&&) : StartingXiMax(i&&) =
XiMax(i&&) : NEXT i&&

    CASE "F1" '(n-D)

        N&&& = 30 : Np&& = 60

        REDIM XiWin(1 TO N&&), XiMax(1 TO N&&) : FOR i&& = 1 TO N&& : XiWin(i&&) = -100## : XiMax(i&&) = 100## : NEXT i&&
        REDIM StartingXiWin(1 TO N&&), StartingXiMax(1 TO N&&) : FOR i&& = 1 TO N&& : StartingXiWin(i&&) = XiWin(i&&) : StartingXiMax(i&&) =
XiMax(i&&) : NEXT i&&

    CASE "F2" '(n-D)

        N&&& = 30 : Np&& = 60

        REDIM XiWin(1 TO N&&), XiMax(1 TO N&&) : FOR i&& = 1 TO N&& : XiWin(i&&) = -10## : XiMax(i&&) = 10## : NEXT i&&
        REDIM StartingXiWin(1 TO N&&), StartingXiMax(1 TO N&&) : FOR i&& = 1 TO N&& : StartingXiWin(i&&) = XiWin(i&&) : StartingXiMax(i&&) =
XiMax(i&&) : NEXT i&&

    CASE "F3" '(n-D)

        N&&& = 30 : Np&& = 60

        REDIM XiWin(1 TO N&&), XiMax(1 TO N&&) : FOR i&& = 1 TO N&& : XiWin(i&&) = -100## : XiMax(i&&) = 100## : NEXT i&&
        REDIM StartingXiWin(1 TO N&&), StartingXiMax(1 TO N&&) : FOR i&& = 1 TO N&& : StartingXiWin(i&&) = XiWin(i&&) : StartingXiMax(i&&) =
XiMax(i&&) : NEXT i&&

    CASE "F4" '(n-D)

        N&&& = 30 : Np&& = 60

        REDIM XiWin(1 TO N&&), XiMax(1 TO N&&) : FOR i&& = 1 TO N&& : XiWin(i&&) = -100## : XiMax(i&&) = 100## : NEXT i&&
        REDIM StartingXiWin(1 TO N&&), StartingXiMax(1 TO N&&) : FOR i&& = 1 TO N&& : StartingXiWin(i&&) = XiWin(i&&) : StartingXiMax(i&&) =
XiMax(i&&) : NEXT i&&

    CASE "F5" '(n-D)

        N&&& = 30 : Np&& = 60

        REDIM XiWin(1 TO N&&), XiMax(1 TO N&&) : FOR i&& = 1 TO N&& : XiWin(i&&) = -30## : XiMax(i&&) = 30## : NEXT i&&
        REDIM StartingXiWin(1 TO N&&), StartingXiMax(1 TO N&&) : FOR i&& = 1 TO N&& : StartingXiWin(i&&) = XiWin(i&&) : StartingXiMax(i&&) =
XiMax(i&&) : NEXT i&&

    CASE "F6" '(n-D) STEP

        N&&& = 30 : Np&& = 60

        REDIM XiWin(1 TO N&&), XiMax(1 TO N&&) : FOR i&& = 1 TO N&& : XiWin(i&&) = -100## : XiMax(i&&) = 100## : NEXT i&&
        REDIM StartingXiWin(1 TO N&&), StartingXiMax(1 TO N&&) : FOR i&& = 1 TO N&& : StartingXiWin(i&&) = XiWin(i&&) : StartingXiMax(i&&) =
XiMax(i&&) : NEXT i&&

    CASE "F7" '(n-D)

        N&&& = 30 : Np&& = 60

        REDIM XiWin(1 TO N&&), XiMax(1 TO N&&) : FOR i&& = 1 TO N&& : XiWin(i&&) = -1.28## : XiMax(i&&) = 1.28## : NEXT i&&
        REDIM StartingXiWin(1 TO N&&), StartingXiMax(1 TO N&&) : FOR i&& = 1 TO N&& : StartingXiWin(i&&) = XiWin(i&&) : StartingXiMax(i&&) =
XiMax(i&&) : NEXT i&&

    CASE "F8" '(n-D)

        N&&& = 30 : Np&& = 60

        REDIM XiWin(1 TO N&&), XiMax(1 TO N&&) : FOR i&& = 1 TO N&& : XiWin(i&&) = -500## : XiMax(i&&) = 500## : NEXT i&&
        REDIM StartingXiWin(1 TO N&&), StartingXiMax(1 TO N&&) : FOR i&& = 1 TO N&& : StartingXiWin(i&&) = XiWin(i&&) : StartingXiMax(i&&) =
XiMax(i&&) : NEXT i&&

    CASE "F9" '(n-D)

        N&&& = 30 : Np&& = 60

        REDIM XiWin(1 TO N&&), XiMax(1 TO N&&) : FOR i&& = 1 TO N&& : XiWin(i&&) = -5.12## : XiMax(i&&) = 5.12## : NEXT i&&
        REDIM StartingXiWin(1 TO N&&), StartingXiMax(1 TO N&&) : FOR i&& = 1 TO N&& : StartingXiWin(i&&) = XiWin(i&&) : StartingXiMax(i&&) =
XiMax(i&&) : NEXT i&&

    CASE "F10" '(n-D) Ackley's Function

        N&&& = 30 : Np&& = 60

        REDIM XiWin(1 TO N&&), XiMax(1 TO N&&) : FOR i&& = 1 TO N&& : XiWin(i&&) = -32## : XiMax(i&&) = 32## : NEXT i&&
        REDIM StartingXiWin(1 TO N&&), StartingXiMax(1 TO N&&) : FOR i&& = 1 TO N&& : StartingXiWin(i&&) = XiWin(i&&) : StartingXiMax(i&&) =
XiMax(i&&) : NEXT i&&

    CASE "F11" '(n-D)

        N&&& = 30 : Np&& = 60

        REDIM XiWin(1 TO N&&), XiMax(1 TO N&&) : FOR i&& = 1 TO N&& : XiWin(i&&) = -600## : XiMax(i&&) = 600## : NEXT i&&
        REDIM StartingXiWin(1 TO N&&), StartingXiMax(1 TO N&&) : FOR i&& = 1 TO N&& : StartingXiWin(i&&) = XiWin(i&&) : StartingXiMax(i&&) =
XiMax(i&&) : NEXT i&&

    CASE "F12" '(n-D) Penalized #1

        N&&& = 30 : Np&& = 60

        REDIM XiWin(1 TO N&&), XiMax(1 TO N&&) : FOR i&& = 1 TO N&& : XiWin(i&&) = -50## : XiMax(i&&) = 50## : NEXT i&&
'       REDIM XiWin(1 TO N&&), XiMax(1 TO N&&) : FOR i&& = 1 TO N&& : XiWin(i&&) = -5## : XiMax(i&&) = 5## 'use this interval for second
run to improve performance
        REDIM StartingXiWin(1 TO N&&), StartingXiMax(1 TO N&&) : FOR i&& = 1 TO N&& : StartingXiWin(i&&) = XiWin(i&&) : StartingXiMax(i&&) =
XiMax(i&&) : NEXT i&&
```

```
CASE "F13" '(n-D) Penalized #2

    Nd&& = 30 : Np&& = 60

    REDIM XiWin(1 TO Nd&&), XiMax(1 TO Nd&&) : FOR i&& = 1 TO Nd&& : XiWin(i&&) = -50## : XiMax(i&&) = 50## : NEXT i&&
    REDIM StartingXiWin(1 TO Nd&&), StartingXiMax(1 TO Nd&&) : FOR i&& = 1 TO Nd&& : StartingXiWin(i&&) = XiWin(i&&) : StartingXiMax(i&&) =
XiMax(i&&) : NEXT i&&

CASE "F14" '(2-D) Shekel's Foxholes

    Nd&& = 2 : Np&& = 8

    REDIM XiWin(1 TO Nd&&), XiMax(1 TO Nd&&) : FOR i&& = 1 TO Nd&& : XiWin(i&&) = -65.536## : XiMax(i&&) = 65.536## : NEXT i&&
    REDIM StartingXiWin(1 TO Nd&&), StartingXiMax(1 TO Nd&&) : FOR i&& = 1 TO Nd&& : StartingXiWin(i&&) = XiWin(i&&) : StartingXiMax(i&&) =
XiMax(i&&) : NEXT i&&

CASE "F15" '(4-D) Kowalik's Function

    Nd&& = 4
    Np&& = 16

    REDIM XiWin(1 TO Nd&&), XiMax(1 TO Nd&&) : FOR i&& = 1 TO Nd&& : XiWin(i&&) = -5## : XiMax(i&&) = 5## : NEXT i&&
    REDIM StartingXiWin(1 TO Nd&&), StartingXiMax(1 TO Nd&&) : FOR i&& = 1 TO Nd&& : StartingXiWin(i&&) = XiWin(i&&) : StartingXiMax(i&&) =
XiMax(i&&) : NEXT i&&

CASE "F16" '(2-D) Camel Back

    Nd&& = 2 : Np&& = 8

    REDIM XiWin(1 TO Nd&&), XiMax(1 TO Nd&&) : FOR i&& = 1 TO Nd&& : XiWin(i&&) = -5## : XiMax(i&&) = 5## : NEXT i&&
    REDIM StartingXiWin(1 TO Nd&&), StartingXiMax(1 TO Nd&&) : FOR i&& = 1 TO Nd&& : StartingXiWin(i&&) = XiWin(i&&) : StartingXiMax(i&&) =
XiMax(i&&) : NEXT i&&

CASE "F17" '(2-D) Branin

    Nd&& = 2 : Np&& = 8

    REDIM XiWin(1 TO Nd&&), XiMax(1 TO Nd&&) : XiWin(1) = -5## : XiMax(1) = 10## : XiWin(2) = 0## : XiMax(2) = 15##
    REDIM StartingXiWin(1 TO Nd&&), StartingXiMax(1 TO Nd&&) : FOR i&& = 1 TO Nd&& : StartingXiWin(i&&) = XiWin(i&&) : StartingXiMax(i&&) =
XiMax(i&&) : NEXT i&&

CASE "F18" '(2-D) Goldstein-Price

    Nd&& = 2 : Np&& = 8

    REDIM XiWin(1 TO Nd&&), XiMax(1 TO Nd&&) : XiWin(1) = -2## : XiMax(1) = 2## : XiWin(2) = -2## : XiMax(2) = 2##
    REDIM StartingXiWin(1 TO Nd&&), StartingXiMax(1 TO Nd&&) : FOR i&& = 1 TO Nd&& : StartingXiWin(i&&) = XiWin(i&&) : StartingXiMax(i&&) =
XiMax(i&&) : NEXT i&&

CASE "F19" '(3-D) Hartman's Family #1

    Nd&& = 3 : Np&& = 12

    REDIM XiWin(1 TO Nd&&), XiMax(1 TO Nd&&) : FOR i&& = 1 TO Nd&& : XiWin(i&&) = 0## : XiMax(i&&) = 1## : NEXT i&&
    REDIM StartingXiWin(1 TO Nd&&), StartingXiMax(1 TO Nd&&) : FOR i&& = 1 TO Nd&& : StartingXiWin(i&&) = XiWin(i&&) : StartingXiMax(i&&) =
XiMax(i&&) : NEXT i&&

CASE "F20" '(6-D) Hartman's Family #2

    Nd&& = 6 : Np&& = 24

    REDIM XiWin(1 TO Nd&&), XiMax(1 TO Nd&&) : FOR i&& = 1 TO Nd&& : XiWin(i&&) = 0## : XiMax(i&&) = 1## : NEXT i&&
    REDIM StartingXiWin(1 TO Nd&&), StartingXiMax(1 TO Nd&&) : FOR i&& = 1 TO Nd&& : StartingXiWin(i&&) = XiWin(i&&) : StartingXiMax(i&&) =
XiMax(i&&) : NEXT i&&

CASE "F21" '(4-D) Shekel's Family m=5

    Nd&& = 4 : Np&& = 16

    REDIM XiWin(1 TO Nd&&), XiMax(1 TO Nd&&) : FOR i&& = 1 TO Nd&& : XiWin(i&&) = 0## : XiMax(i&&) = 10## : NEXT i&&
    REDIM StartingXiWin(1 TO Nd&&), StartingXiMax(1 TO Nd&&) : FOR i&& = 1 TO Nd&& : StartingXiWin(i&&) = XiWin(i&&) : StartingXiMax(i&&) =
XiMax(i&&) : NEXT i&&

CASE "F22" '(4-D) Shekel's Family m=7

    Nd&& = 4 : Np&& = 16

    REDIM XiWin(1 TO Nd&&), XiMax(1 TO Nd&&) : FOR i&& = 1 TO Nd&& : XiWin(i&&) = 0## : XiMax(i&&) = 10## : NEXT i&&
    REDIM StartingXiWin(1 TO Nd&&), StartingXiMax(1 TO Nd&&) : FOR i&& = 1 TO Nd&& : StartingXiWin(i&&) = XiWin(i&&) : StartingXiMax(i&&) =
XiMax(i&&) : NEXT i&&

CASE "F23" '(4-D) Shekel's Family m=10

    Nd&& = 4 : Np&& = 16

    REDIM XiWin(1 TO Nd&&), XiMax(1 TO Nd&&) : FOR i&& = 1 TO Nd&& : XiWin(i&&) = 0## : XiMax(i&&) = 10## : NEXT i&&
    REDIM StartingXiWin(1 TO Nd&&), StartingXiMax(1 TO Nd&&) : FOR i&& = 1 TO Nd&& : StartingXiWin(i&&) = XiWin(i&&) : StartingXiMax(i&&) =
XiMax(i&&) : NEXT i&&

CASE "PBM_1" '2-D

    Nd&&                     = 2
    NumPointsPerDimension&&  = 2 '4 '20
    Np&&                     = NumPointsPerDimension&&*Nd&&

    Niter&&    = 100
'.        G       = 2##
'.        Alpha   = 2##
'.        Beta    = 1##
'.        DeltaT  = 1##
'.        Frep    = 0.5##

'.     PlaceInitialPoints$  = "UNIFORM ON-AXIS"
'.     InitialAcceleration$ = "ZERO"
'.     RepositionFactor$    = "VARIABLE" '"FIXED"

    Np&& = NumPointsPerDimension&&*Nd&&

    REDIM XiWin(1 TO Nd&&), XiMax(1 TO Nd&&)

    XiWin(1) = 0.5## : XiMax(1) = 3## 'dipole length, L, in wavelengths
    XiWin(2) = 0## : XiMax(2) = PI2 'polar angle, Theta, in Radians
    REDIM StartingXiWin(1 TO Nd&&), StartingXiMax(1 TO Nd&&) : FOR i&& = 1 TO Nd&& : StartingXiWin(i&&) = XiWin(i&&) : StartingXiMax(i&&) =
XiMax(i&&) : NEXT i&&

    Nd&& = FREEFILE : OPEN "INFILE.DAT" FOR OUTPUT AS #Nd&& : PRINT #Nd&&,"PBM1.NEC" : PRINT #Nd&&,"PBM1.OUT" : CLOSE #Nd&& 'NEC Input/Output Files

CASE "PBM_2" '2-D

    AddNoiseTOPBM2$ = "NO" '"YES" '"NO" '"YES"

    Nd&&                     = 2
    NumPointsPerDimension&&  = 4 '20
    Np&&                     = NumPointsPerDimension&&*Nd&&

    Niter&&    = 100
'.        G       = 2##
'.        Alpha   = 2##
'.        Beta    = 2##
'.        DeltaT  = 1##
'.        Frep    = 0.5##

'.     PlaceInitialPoints$  = "UNIFORM ON-AXIS"
'.     InitialAcceleration$ = "ZERO"
'.     RepositionFactor$    = "VARIABLE" '"FIXED"
```

```
        Np&& = NumPointsPerDimension&&^Nd&&

        REDIM XiWin(1 TO Nd&&), XiMax(1 TO Nd&&)

        XiWin(1) = 5## : XiMax(1) = 15## 'dipole separation, D, in wavelengths
        XiWin(2) = 0## : XiMax(2) = PI   'polar angle, Theta, in Radians
        REDIM StartingXiWin(1 TO Nd&&), StartingXiMax(1 TO Nd&&) : FOR i&& = 1 TO Nd&& : StartingXiWin(i&&) = XiWin(i&&) : StartingXiMax(i&&) =
XiMax(i&&) : NEXT i&&

        Nn&& = FREEFILE : OPEN "INFILE.DAT" FOR OUTPUT AS #Nn&& : PRINT #Nn&&,"PBM2.NEC" : PRINT #Nn&&,"PBM2.OUT" : CLOSE #Nn&&

    CASE "PBM_3" '2-D
        Nd&&                    = 2
        NumPointsPerDimension&& = 4 '20
        Np&&                    = NumPointsPerDimension&&^Nd&&

        Niter&&    = 100
'       G          = 2##
'       Alpha      = 2##
'       Beta       = 2##
'       DeltaT     = 1##
'       Frep       = 0.5##

'       PlaceInitialPoints$   = "UNIFORM ON-AXIS"
'       InitialAcceleration$  = "ZERO"
'       RepositionFactor$     = "VARIABLE" '"FIXED"

        Np&& = NumPointsPerDimension&&^Nd&&

        REDIM XiWin(1 TO Nd&&), XiMax(1 TO Nd&&)

        XiWin(1) = 0## : XiMax(1) = 4## 'Phase Parameter, Beta (0-4)
        XiWin(2) = 0## : XiMax(2) = PI  'polar angle, Theta, in Radians
        REDIM StartingXiWin(1 TO Nd&&), StartingXiMax(1 TO Nd&&) : FOR i&& = 1 TO Nd&& : StartingXiWin(i&&) = XiWin(i&&) : StartingXiMax(i&&) =
XiMax(i&&) : NEXT i&&

        Nn&& = FREEFILE : OPEN "INFILE.DAT" FOR OUTPUT AS #Nn&& : PRINT #Nn&&,"PBM3.NEC" : PRINT #Nn&&,"PBM3.OUT" : CLOSE #Nn&&

    CASE "PBM_4" '2-D
        Nd&&                    = 2
        NumPointsPerDimension&& = 4 '6 '2 '4 '20
        Np&&                    = NumPointsPerDimension&&^Nd&&

        Niter&&    = 100
'       G          = 2##
'       Alpha      = 2##
'       Beta       = 2##
'       DeltaT     = 1##
'       Frep       = 0.5##

'       PlaceInitialPoints$   = "UNIFORM ON-AXIS"
'       InitialAcceleration$  = "ZERO"
'       RepositionFactor$     = "VARIABLE" '"FIXED"

        Np&& = NumPointsPerDimension&&^Nd&&

        REDIM XiWin(1 TO Nd&&), XiMax(1 TO Nd&&)

        XiWin(1) = 0.5## : XiMax(1) = 1.5## 'ARM LENGTH (NOT Total Length), wavelengths (0.5-1.5)
        XiWin(2) = PI/18## : XiMax(2) = PI/2## 'Inner angle, Alpha, in Radians (PI/18-PI/2)
        REDIM StartingXiWin(1 TO Nd&&), StartingXiMax(1 TO Nd&&) : FOR i&& = 1 TO Nd&& : StartingXiWin(i&&) = XiWin(i&&) : StartingXiMax(i&&) =
XiMax(i&&) : NEXT i&&

        Nn&& = FREEFILE : OPEN "INFILE.DAT" FOR OUTPUT AS #Nn&& : PRINT #Nn&&,"PBM4.NEC" : PRINT #Nn&&,"PBM4.OUT" : CLOSE #Nn&&

    CASE "PBM_5"
        NumCollinearElements&& = 6 '30 'EVEN or ODD: 6,7,10,13,16,24 used by PBM

        Nd&&                    = NumCollinearElements&& - 1
        NumPointsPerDimension&& = 4 '20
        Np&&                    = NumPointsPerDimension&&^Nd&&

        Niter&&    = 100
'       G          = 2##
'       Alpha      = 2##
'       Beta       = 2##
'       DeltaT     = 1##
'       Frep       = 0.5##

'       PlaceInitialPoints$   = "UNIFORM ON-AXIS"
'       InitialAcceleration$  = "ZERO"
'       RepositionFactor$     = "VARIABLE" '"FIXED"

        Nd&& = NumCollinearElements&& - 1

        REDIM XiWin(1 TO Nd&&), XiMax(1 TO Nd&&) : FOR i&& = 1 TO Nd&& : XiWin(i&&) = 0.5## : XiMax(i&&) = 1.5## : NEXT i&&

        Nn&& = FREEFILE : OPEN "INFILE.DAT" FOR OUTPUT AS #Nn&& : PRINT #Nn&&,"PBM5.NEC" : PRINT #Nn&&,"PBM5.OUT" : CLOSE #Nn&&

    CASE "LD_MONO" 'DIPOLE-LOADED MONOPOLE

'       -------- Check for NEC Executable ---------

        IF DIR$("N42D_1k5segs_1k1D_1k1S.EXE") = "" THEN
            MSGBOX("WARNING! NEC4.1D executable NEC41D_4K_D53011.EXE not found!"+CHR$(13)+CHR$(13)+_
                   RUN TERMINATED.")+CHR$(13)+CHR$(13)
            EXIT SUB
        END IF

        Nd&& = 6

        REDIM XiWin(1 TO Nd&&), XiMax(1 TO Nd&&)
        XiWin(1) = 0.05## : XiMax(1) = 0.50## 'X1, wavelengths
        XiWin(2) = 0.03## : XiMax(2) = 0.35## 'Z1, wavelengths
        XiWin(3) = 0.03## : XiMax(3) = 0.35## 'Z1, wavelengths
        XiWin(4) = 0.01## : XiMax(4) = 0.10## 'Z2, wavelengths
        XiWin(5) = 0.03## : XiMax(5) = 0.10## 'Z3, wavelengths
        XiWin(6) = 0.05## : XiMax(6) = 0.10## 'Z4, wavelengths

'       XiWin(1) = 0.05## : XiMax(1) = 0.70## 'X1, wavelengths
'       XiWin(2) = 0.05## : XiMax(2) = 0.70## 'X2, wavelengths
'       XiWin(3) = 0.03## : XiMax(3) = 0.50## 'Z1, wavelengths
'       XiWin(4) = 0.03## : XiMax(4) = 0.30## 'Z2, wavelengths
'       XiWin(5) = 0.03## : XiMax(5) = 0.10## 'Z3, wavelengths
'       XiWin(6) = 0.01## : XiMax(6) = 0.70## 'Z4, wavelengths

        REDIM StartingXiWin(1 TO Nd&&), StartingXiMax(1 TO Nd&&)
        FOR i&& = 1 TO Nd&& : StartingXiWin(i&&) = XiWin(i&&) : StartingXiMax(i&&) = XiMax(i&&) : NEXT i&&

        Nn&& = FREEFILE : OPEN "INFILE.DAT" FOR OUTPUT AS #Nn&& : PRINT #Nn&&,"MONO.OUT" : CLOSE #Nn&&

'   =============================================================================================
'   NOTE - DON'T FORGET TO ADD NEW TEST FUNCTIONS TO FUNCTION ObjectiveFunction() ABOVE !!
'   =============================================================================================

    END SELECT

    IF Nd&& > 100 THEN Niter&& = MIN(Niter&&,200) 'to avoid array dimensioning problems

    DiagLength = 0## : FOR i&& = 1 TO Nd&& : DiagLength = DiagLength + (XiMax(i&&)-XiWin(i&&))^2 : NEXT i&& : DiagLength = SQR(DiagLength) 'compute length of
decision space principal diagonal

END SUB 'GetFunctionNumParameters()

'--------------------------------
```

```
FUNCTION RASTRIGIN(R(),Nd&&,p&&,j&&) '(n-D)

'MAXIMUM = ZERO @ [O]^Nd (n-D CASE).

'Reference:
'Pehlivanoglu, Y. V., "N New Particel Swarm Optimization Method Enhanced with a
'Periodic Mutation Strategy and Neural Networks," IEEE Trans. Evol. Computation,
'vol. 17, No. 3, June 2013, pp. 436-452.

    LOCAL Z, Xi AS EXT

    LOCAL i&&

    Z = 0##

    FOR i&& = 1 TO Nd&&

        Xi = R(p&&,i&&,j&&)

        Z = Z + Xi^2 - 10##*COS(TwoPi*Xi)

    NEXT i&&

    RASTRIGIN = -(Z+10##*Nd&&)

END FUNCTION 'RASTRIGIN

'---------------------

SUB CheckCosineMix(Nd&&)
LOCAL R() AS EXT
LOCAL i&&
    REDIM R(1 TO 1, 1 TO Nd&&, 0 TO 0)
    FOR i&& = 1 TO Nd&& : R(1,i&&,0) = 0## : NEXT i&&
    MSGBOX("Nd = "+STR$(Nd&&)+",  CosMix(0^Nd)="+STR$(COSINEMIX(R(),Nd&&,1,0)))
END SUB 'CheckCosineMix()

'---------------------

FUNCTION COSINEMIX(R(),Nd&&,p&&,j&&) '(n-D)

'MAXIMUM = 0.1*Nd @ [O]^Nd (n-D CASE).

'Reference:
'Pehlivanoglu, Y. V., "N New Particel Swarm Optimization Method Enhanced with a
'Periodic Mutation Strategy and Neural Networks," IEEE Trans. Evol. Computation,
'vol. 17, No. 3, June 2013, pp. 436-452.

    LOCAL Z, Xi, Sum1, Sum2 AS EXT

    LOCAL i&&

    Z = 0## : Sum1 = 0## : Sum2 = 0##

    FOR i&& = 1 TO Nd&&

        Xi = R(p&&,i&&,j&&)

        Sum1 = Sum1 + Xi^2

        Sum2 = Sum2 + COS(FivePi*Xi)

    NEXT i&&

    Z = Sum1 - 0.1##*Sum2

    COSINEMIX = -Z

END FUNCTION 'COSINEMIX

'---------------------

FUNCTION EXPONENTIAL(R(),Nd&&,p&&,j&&) '(n-D)

'MAXIMUM = 1 @ [O]^Nd (n-D CASE).

'Reference:
'Pehlivanoglu, Y. V., "N New Particel Swarm Optimization Method Enhanced with a
'Periodic Mutation Strategy and Neural Networks," IEEE Trans. Evol. Computation,
'vol. 17, No. 3, June 2013, pp. 436-452.

    LOCAL Z, Xi, Sum AS EXT

    LOCAL i&&

    Z = 0## : Sum = 0##

    FOR i&& = 1 TO Nd&&

        Xi = R(p&&,i&&,j&&)

        Sum = Sum + Xi^2

    NEXT i&&

    Z = -EXP(-0.5##*Sum)

    EXPONENTIAL = -Z

END FUNCTION 'EXPONENTIAL

'---------------------

FUNCTION ACKLEY(R(),Nd&&,p&&,j&&) '(n-D)

'MAXIMUM = ZERO (n-D CASE).

'References:
'Yao, X., Liu, Y., and Lin, G., "Evolutionary Programming Made Faster,"
'IEEE Trans. Evolutionary Computation, Vol. 3, No. 2, 82-102, Jul. 1999.

'Pehlivanoglu, Y. V., "N New Particel Swarm Optimization Method Enhanced with a
'Periodic Mutation Strategy and Neural Networks," IEEE Trans. Evol. Computation,
'vol. 17, No. 3, June 2013, pp. 436-452.

    LOCAL Z, Xi, Sum1, Sum2 AS EXT

    LOCAL i&&

    Z = 0## : Sum1 = 0## : Sum2 = 0##

    FOR i&& = 1 TO Nd&&

        Xi = R(p&&,i&&,j&&)

        Sum1 = Sum1 + Xi^2

        Sum2 = Sum2 + COS(TwoPi*Xi)

    NEXT i&&

    Z = -20##*EXP(-0.2##*SQR(Sum1/Nd&&)) - EXP(Sum2/Nd&&) + 20## + e

    ACKLEY = -Z

END FUNCTION 'ACKLEY

'---------------------

FUNCTION ParrotF4(R(),Nd&&,p&&,j&&) 'Parrott F4 (1-D)

'MAXIMUM = 1 AT ~0.0796875... WITH ZERO OFFSET (SEEMS TO WORK BEST WITH JUST 3 PROBES, BUT NOT ALLOWED IN THIS VERSION...)

'References:

'Beasley, D., D. R. Bull, and R. R. Martin, "A Sequential Niche Technique for Multimodal
'Function Optimization," Evol. Comp. (MIT Press), Vol. 1, No. 2, 1993, pp. 101-125
```

```
'(online at http://citeseer.ist.psu.edu/beasley93sequential.html).

'Parrott, D., and X. Li, "Locating and Tracking Multiple Dynamic Optima by a Particle Swarm
'Model using Speciation," IEEE Trans. Evol. Computation, vol. 10, no. 4, Aug. 2006, pp. 440-458.

LOCAL Z, x, offset AS EXT

    offset = O##

    x = R(p&&,1,j&&)

    Z = EXP(-2##*LOG(2##)*((x-0.08##-offset)/0.854##)^2)*(SIN(5##*PI*((x-offset)^0.75##-0.05##))^6  'WARNING! This is a NATURAL LOG, NOT Log10!!!

    ParrottF4 = Z

END FUNCTION 'ParrottF4()
'----------------------------
FUNCTION SGO(R(),Nd&&,p&&,j&&) 'SGO Function (2-D)

'MAXIMUM =130.8323226... @ ~(-2.8362075...,-2.8362075...) WITH ZERO OFFSET.

'Reference:
'Hsiao, Y., Chuang, C., Jiang, J., and Chien, C., "A Novel Optimization Algorithm: Space
'Gravitational Optimization," Proc. of 2005 IEEE International Conference on Systems, Man,
'and Cybernetics, 3, 2323-2328. (2005)

    LOCAL x1, x2, Z, t1, t2, SGOx1offset, SGOx2offset AS EXT

'   SGOx1offset = 40## : SGOx2offset = 10##

    x1 = R(p&&,1,j&&) - SGOx1offset : x2 = R(p&&,2,j&&) - SGOx2offset

    t1 = x1^4 - 16##*x1^2 + 0.5##*x1 : t2 = x2^4 - 16##*x2^2 + 0.5##*x2

    Z = -(t1 + t2)

    SGO = Z

END FUNCTION 'SGO()
'-----------------
FUNCTION GoldsteinPrice(R(),Nd&&,p&&,j&&) 'Goldstein-Price Function (2-D)

'MAXIMUM = -3 @ (0,-1) WITH ZERO OFFSET.

'Reference:
'Cui, Z., Zeng, J., and Sun, G. (2006) 'A Fast Particle Swarm Optimization,' Int'l. J.
'Innovative Computing, Information and Control, vol. 2, no. 6, December, pp. 1365-1380.

    LOCAL Z, x1, x2, offset1, offset2, t1, t2 AS EXT

    offset1 = O## : offset2 = O##

'   offset1 = 20## : offset2 = -10##

    x1 = R(p&&,1,j&&)-offset1 : x2 = R(p&&,2,j&&)-offset2

    t1 = 1##+(x1+x2+1##)^2*(19##-14##*x1+3##*x1^2-14##*x2+6##*x1*x2+3##*x2^2)

    t2 = 30##+(2##*x1-3##*x2)^2*(18##-32##*x1+12##*x1^2+48##*x2-36##*x1*x2+27##*x2^2)

    Z = t1*t2

    GoldsteinPrice = -Z

END FUNCTION 'GoldsteinPrice()
'----------
FUNCTION StepFunction(R(),Nd&&,p&&,j&&) 'Step Function (n-D)

'MAXIMUM VALUE = 0 @ [offset]n.

'Reference:
'Yao, X., Liu, Y., and Lin, G., "Evolutionary Programming Made Faster,"
'IEEE Trans. Evolutionary Computation, Vol. 3, No. 2, 82-102, Jul. 1999.

    LOCAL Offset, Z AS EXT

    LOCAL i&&

    Z = O## : Offset = O## '75.123## 'O##

    FOR i&& = 1 TO Nd&&

        IF Nd&& = 2 AND i&& = 1 THEN Offset = 75 '75##

        IF Nd&& = 2 AND i&& = 2 THEN Offset = 35 '30 '35##

        Z = Z + INT(CR(p&&,i&&,j&&)-Offset) + 0.5##)^2

    NEXT i&&

    StepFunction = -Z

END FUNCTION 'StepFunction()
'----------
FUNCTION Schwefel226(R(),Nd&&,p&&,j&&) 'Schwefel Problem 2.26 (n-D)

'MAXIMUM = 0 @ [420.9687]^Nd (Nd CASE).

'References:
'Yao, X., Liu, Y., and Lin, G., "Evolutionary Programming Made Faster,"
'IEEE Trans. Evolutionary Computation, Vol. 3, No. 2, 82-102, Jul. 1999.

'Pehlivanoglu, Y. V., "A New Particel Swarm Optimization Method Enhanced with a
'Periodic Mutation Strategy and Neural Networks," IEEE Trans. Evol. Computation,
'Vol. 17, No. 3, June 2013, pp. 436-452.

    LOCAL Z, Xi AS EXT

    Z = O##

    FOR i&& = 1 TO Nd&&

        Xi = R(p&&,i&&,j&&)

        Z = Z + Xi*SIN(SQR(ABS(Xi)))

    NEXT i&&

    Schwefel226 = Z - 418.9829##*Nd&&

END FUNCTION 'SCHWEFEL226()
'------------------------
FUNCTION Colville(R(),Nd&&,p&&,j&&) 'Colville Function (4-D)

'MAXIMUM = 0 @ (1,1,1,1) WITH ZERO OFFSET.

'Reference: Doo-Hyun, and Se-Young, O., "A New Mutation Rule for Evolutionary Programming Motivated
'from Backpropagation Learning," IEEE Trans. Evolutionary Computation, Vol. 4, No. 2, pp. 188-190,
'July 2000.
```

```
    LOCAL Z, x1, x2, x3, x4, offset AS EXT

    offset = 0## '7.123##

    x1 = R(p&&.1.j&&)-offset : x2 = R(p&&.2.j&&)-offset : x3 = R(p&&.3.j&&)-offset : x4 = R(p&&.4.j&&)-offset

    Z =  100##*(x2-x1^2)^2 + (1##-x1)^2  + _
              90##*(x4-x3^2)^2 + (1##-x3)^2  + _
              10.1##*((x2-1##)^2 + (x4-1##)^2) + _
              19.8##*(x2-1##)*(x4-1##)

    Colville = -Z

END FUNCTION 'Colville()
'-----------
FUNCTION Griewank(R(),Nd&&,p&&,j&&) 'Griewank (n-D)

'Max of zero at (0,...,0)

'References:
'Yao, X., Liu, Y., and Lin, G., "Evolutionary Programming Made Faster,"
'IEEE Trans. Evolutionary Computation, Vol. 3, No. 2, 82-102, Jul. 1999.

'Pehlivanoglu, Y. V., "A New Particle Swarm Optimization Method Enhanced with a
'Periodic Mutation Strategy and Neural Networks," IEEE Trans. Evol. Computation,
'Vol. 17, No. 3, June 2013, pp. 436-452.

    LOCAL Offset, Sum, Prod, Z, Xi AS EXT

    LOCAL i&&

    Sum = 0## : Prod = 1##

    Offset = 0## '75.123##

    FOR i&& = 1 TO Nd&&

        Xi = R(p&&.i&&.j&&) - Offset

        Sum = Sum + Xi^2

        Prod = Prod*COS(Xi/SQR(i&&))

    NEXT i&&

    Z = Sum/4000## - Prod + 1##

    Griewank = -Z

END FUNCTION 'Griewank()
'-----------
FUNCTION Himmelblau(R(),Nd&&,p&&,j&&) 'HimmelblAu (2-D)

    LOCAL Z, x1, x2, offset AS EXT

    offset = 0##

    x1 = R(p&&.1.j&&)-offset : x2 = R(p&&.2.j&&)-offset

    Z = 200## - (x1^2 + x2 -11##)^2 - (x1+x2^2-7##)^2

    Himmelblau = z

END FUNCTION 'Himmelblau()
'-----------
FUNCTION Rosenbrock(R(),Nd&&,p&&,j&&) 'Rosenbrock (n-D)

'MAXIMUM = 0 @ [1,...,1]^n (n-D CASE).
'Reference: Yao, X., Liu, Y., and Lin, G., "Evolutionary Programming Made Faster,"
'IEEE Trans. Evolutionary Computation, Vol. 3, No. 2, 82-102, Jul. 1999.

    LOCAL Z, Xi, Xi1 AS EXT

    LOCAL i&&

    Z = 0##

    FOR i&& = 1 TO Nd&&-1

        Xi = R(p&&.i&&.j&&) : Xi1 = R(p&&.i&&+1,j&&)

        Z = Z + 100##*(Xi1-Xi^2)^2 + (Xi-1##)^2

    NEXT i&&

    Rosenbrock = -Z

END FUNCTION 'ROSENBROCK()
'-----------
FUNCTION Sphere(R(),Nd&&,p&&,j&&) 'Sphere (n-D)

'MAXIMUM = 0 @ [0,...,0]^n (n-D CASE).
'Reference: Yao, X., Liu, Y., and Lin, G., "Evolutionary Programming Made Faster,"
'IEEE Trans. Evolutionary Computation, Vol. 3, No. 2, 82-102, Jul. 1999.

    LOCAL Z, Xi, Xi1 AS EXT

    LOCAL i&&

    Z = 0##

    FOR i&& = 1 TO Nd&&

        Xi = R(p&&.i&&,j&&)

        Z = Z + Xi^2

    NEXT i&&

    Sphere = -Z

END FUNCTION 'SPHERE()
'-----------
FUNCTION HimmelblauNLO(R(),Nd&&,p&&,j&&) 'Himmelblau non-linear optimization (5-D)

'MAXIMUM = 31025.5562644972 @ (78.0,33.0,27.0709971052,45.0,44.9692425501)
'Reference: "Constrained Optimization using CODEQ," Mahamed G.H. Omran & Ayed Salman,
'Chaos, Solitons and Fractals, 42(2009), 662-668

    LOCAL Z, x1, x2, x3, x4, x5, g1, g2, g3 AS EXT

    Z = 1E4200

    x1 = R(p&&.1.j&&) : x2 = R(p&&.2.j&&) : x3 = R(p&&.3.j&&) : x4 = R(p&&.4.j&&) : x5 = R(p&&.5.j&&)

    g1 = 85.334407## + 0.0056858##*x2*x5 + 0.00026##*x1*x4  - 0.0022053##*x3*x5

    g2 = 80.51249## + 0.0071317##*x2*x5 + 0.0029955##*x1*x2 + 0.0021813##*x3*x3

    g3 = 9.300961## + 0.0047026##*x3*x5 + 0.0012547##*x1*x3 + 0.0019085##*x3*x4
```

```
    IF g1## < 0 OR g1 > 92## OR g2 < 90## OR g2 > 110## OR g3 < 20## OR g3 > 25## THEN GOTO ExitHimmelblauNLO
    Z = 5.3578547##*x3^3 + 0.8356891##*x1*x5 + 37.29329##*x1 - 40792.141##

ExitHimmelblauNLO:
    HimmelblauNLO = -Z
END FUNCTION 'HimmelblauNLO()
'------------
FUNCTION Tripod(R(),Nd&&,p&&,j&&) 'Tripod (2-D)
'MAXIMUM = 0 at (0,-50)
'Reference: "Appendix: A mini-benchmark," Maurice Clerc
    LOCAL Z, x1, x2, s1, s2, t1, t2, t3 AS EXT
    x1 = R(p&&,1,j&&) : x2 = R(p&&,2,j&&)
    s1 = Sign(x1) : s2 = Sign(x2)
    t1 = (1##-s2)*(ABS(x1)+ABS(x2-50##))
    t2 = 0.5##*(1##+s2)*(1##-s1)*(1##+ABS(x1+50##)+ABS(x2-50##))
    t3 = (1##+s1)*(2##+ABS(x1-50##)+ABS(x2-50##))
    Z = 0.5##*(t1 + t2 + t3)
    Tripod = -Z
END FUNCTION 'Tripod()
'--------------------
FUNCTION Sign(x)
LOCAL Z AS EXT
    Z = 1## : IF X =< 0## THEN Z = -1##
    Sign = Z
END FUNCTION
'------------
FUNCTION RosenbrockF6(R(),Nd&&,p&&,j&&) 'Rosenbrock F6 (10-D)
'WARNING !!  03-19-10  THIS FUNCTION CONTAINS ERRORS.  SEE CLERC's EMAIL!
'MAXIMUM = 394 at (0,-50)
'Reference: "Appendix: A mini-benchmark," Maurice Clerc (NOTE: Uses his notation...)
    LOCAL Z, Xi, Xi1, Zi, Zi1, Sum AS EXT
    LOCAL i&&
    Sum = 0##
    FOR i&& = 2 TO Nd&&
        Xi = R(p&&,i&&,j&&) : Xi1 = R(p&&,i&&-1,j&&)
        Zi = Xi - Xioffset(i&&) + 1## : Zi1 = Xi1 - Xioffset(i&&-1) + 1##
        Sum = Sum + 100##*(Zi1^2 - Zi)^2  + (Zi1-1##)^2
    NEXT i&&
    Z = 390## + Sum
    RosenbrockF6 = -Z
END FUNCTION 'RosenbrockF6()
'--------------------
FUNCTION CompressionSpring(R(),Nd&&,p&&,j&&) 'Compression Spring (3-D)
'MAXIMUM = 394 at (0,-50)
'Reference: "Appendix: A mini-benchmark," Maurice Clerc (NOTE: Uses his notation...)
LOCAL Z, x1, x2, x3, g1, g2, g3, g4, g5, Cf, Fmax, S, Lf, Lmax, Sigp, SigPM, Fp, K, Sigw AS EXT
    Z = 1E42OO
    x1 = ROUND(R(p&&,1,j&&),0) : x2 = R(p&&,2,j&&)  : x3 = ROUND(R(p&&,3,j&&),3)
    Cf = 1## + 0.75##*x3/(x2-x3)+0.615##*x3/x2
    Fmax = 1000## : S = 189000## : Lmax = 14## : SigPM = 6## : Fp = 300## : Sigw = 1.25##
    K = 11.5##*1E6*x3^4/(8##*x1^2*x3^3)
    Lf = Fmax/K + 1.05##*(x1+2##)*x3
    SigP = Fp/K
    g1 = 8##*Cf*Fmax*x2/(Pi*x3^3) - S
    g2 = Lf - Lmax
    g3 = SigP - SigPM
    g4 = SigP - Fp/K  'WARNING!  03-19-10.  THIS IS SATISFIED EXACTLY (SEE CLERC's EMAIL - TYPO IN HIS BENCHMARKS!)
    g5 = Sigw - (Fmax-Fp)/K
    IF g1 > 0## OR g2 > 0## OR g3 > 0## OR g4 > 0## OR g5 > 0## THEN GOTO ExitCompressionSpring
    Z = Pi*x2^2*x3^2*(x1+1##)/4##
ExitCompressionSpring:
    CompressionSpring = -Z
END FUNCTION 'CompressionSpring
'--------------------
FUNCTION GearTrain(R(),Nd&&,p&&,j&&) 'GearTrain (4-D)
'MAXIMUM = 394 at (0,-50)
'Reference: "Appendix: A mini-benchmark," Maurice Clerc (NOTE: Uses his notation...)
LOCAL Z, x1, x2, x3, x4 AS EXT
    x1 = ROUND(R(p&&,1,j&&),0) : x2 = ROUND(R(p&&,2,j&&),0)
    x3 = ROUND(R(p&&,3,j&&),0) : x4 = ROUND(R(p&&,4,j&&),0)
    Z = (1##/6.931##-x1*x2/(x3*x4))^2
    GearTrain = -Z
END FUNCTION 'GearTrain
'--------------------
FUNCTION F1(R(),Nd&&,p&&,j&&) 'F1 (n-D)
'MAXIMUM = ZERO (n-D CASE).
```

```
'Reference:
    LOCAL Z, Xi AS EXT
    LOCAL i&&
    Z = 0##
    FOR i&& = 1 TO N0&&
        Xi = R(p&&,i&&,j&&)
        Z = Z + Xi^2
    NEXT i&&
    F1 = -Z
END FUNCTION 'F1
'-----------
FUNCTION F2(R(),N0&&,p&&,j&&) 'F2 (n-0 CASE).
'MAXIMUM = ZERO (n-0 CASE).
'Reference:
    LOCAL Sum, prod, Z, Xi AS EXT
    LOCAL i&&
    Z = 0## : Sum = 0## : Prod = 1##
    FOR i&& = 1 TO N0&&
        Xi = R(p&&,i&&,j&&)
        Sum = Sum+ ABS(Xi)
        Prod = Prod*ABS(Xi)
    NEXT i&&
    Z = Sum + Prod
    F2 = -Z
END FUNCTION 'F2
'-----------
FUNCTION F3(R(),N0&&,p&&,j&&) 'F3 (n-0)
'MAXIMUM = ZERO (n-0 CASE).
'Reference:
    LOCAL Z, Xk, Sum AS EXT
    LOCAL i&&, k&&
    Z = 0##
    FOR i&& = 1 TO N0&&
        Sum = 0##
        FOR k&& = 1 TO i&&
            Xk = R(p&&,k&&,j&&)
            Sum = Sum + Xk
        NEXT k&&
        Z = Z + Sum^2
    NEXT i&&
    F3 = -Z
END FUNCTION 'F3

'-----------
FUNCTION F4(R(),N0&&,p&&,j&&) 'F4 (n-0)
'MAXIMUM = ZERO (n-0 CASE).
'Reference:
    LOCAL Z, Xi, MaxXi AS EXT
    LOCAL i&&
    MaxXi = -1E+200
    FOR i&& = 1 TO N0&&
        Xi = R(p&&,i&&,j&&)
        IF ABS(Xi) >= MaxXi THEN MaxXi = ABS(Xi)
    NEXT i&&
    F4 = -MaxXi
END FUNCTION 'F4
'-----------
FUNCTION F5(R(),N0&&,p&&,j&&) 'F5 (n-0)
'MAXIMUM = ZERO (n-0 CASE).
'Reference:
    LOCAL Z, Xi, XiPlus1 AS EXT
    LOCAL i&&
    Z = 0##
    FOR i&& = 1 TO N0&&-1
        Xi      = R(p&&,i&&,j&&)
        XiPlus1 = R(p&&,i&&+1,j&&)
        Z = Z + (100##*(XiPlus1-Xi^2)^2+(Xi-1##))^2
    NEXT i&&
    F5 = -Z
END FUNCTION 'F5
'-----------
FUNCTION F6(R(),N0&&,p&&,j&&) 'F6 (n-0 STEP)
'MAXIMUM VALUE = 0 @ [offset]^n.

'Reference:
```

```
"Yao, X., Liu, Y., and Lin, G., "Evolutionary Programming Made Faster,"
'IEEE Trans. Evolutionary Computation, Vol. 3, No. 2, 82-102, Jul. 1999.

    LOCAL Z AS EXT
    LOCAL i&&
    Z = O##
    FOR i&& = 1 TO N&&
        Z = Z + INT(R(p&&,i&&,j&&) + 0.5##)^2
    NEXT i&&
    F6 = -Z
END FUNCTION 'F6
'----------
FUNCTION F7(R(),N&&&,p&&,j&&) 'F7
'MAXIMUM VALUE = 0 0 [offset]^n.
'Reference:
'Yao, X., Liu, Y., and Lin, G., "Evolutionary Programming Made Faster,"
'IEEE Trans. Evolutionary Computation, Vol. 3, No. 2, 82-102, Jul. 1999.

    LOCAL Z, Xi AS EXT
    LOCAL i&&
    Z = O##
    FOR i&& = 1 TO N&&&
        Xi = R(p&&,i&&,j&&)
        Z = Z + i&&*Xi^4
    NEXT i&&
    F7 = -Z - RandomNum(O##,1##)
END FUNCTION 'F7
'----------
FUNCTION F8(R(),N&&&,p&&,j&&) '(n-D) F8 [Schwefel Problem 2.26]
'MAXIMUM = 12,569.5 0 [420.8687]^30 (30-D CASE).
'Reference:
'Yao, X., Liu, Y., and Lin, G., "Evolutionary Programming Made Faster,"
'IEEE Trans. Evolutionary Computation, Vol. 3, No. 2, 82-102, Jul. 1999.

    LOCAL Z, Xi AS EXT
    Z = O##
    FOR i&& = 1 TO N&&&
        Xi = R(p&&,i&&,j&&)
        Z = Z - Xi*SIN(SQR(ABS(Xi)))
    NEXT i&&
    F8 = -Z
END FUNCTION 'F8
'----------
FUNCTION F9(R(),N&&&,p&&,j&&) '(n-D) F9 [Rastrigin]
'MAXIMUM = ZERO (n-D CASE).
'Reference:
'Yao, X., Liu, Y., and Lin, G., "Evolutionary Programming Made Faster,"
'IEEE Trans. Evolutionary Computation, Vol. 3, No. 2, 82-102, Jul. 1999.

    LOCAL Z, Xi AS EXT
    LOCAL i&&
    Z = O##
    FOR i&& = 1 TO N&&&
        Xi = R(p&&,i&&,j&&)
        Z = Z + (Xi^2 - 10##*COS(TwoPi*Xi) + 10##)^2
    NEXT i&&
    F9 = -Z
END FUNCTION 'F9
'----------
FUNCTION F10(R(),N&&&,p&&,j&&) '(n-D) F10 [Ackley's Function]
'MAXIMUM = ZERO (n-D CASE).
'Reference:
'Yao, X., Liu, Y., and Lin, G., "Evolutionary Programming Made Faster,"
'IEEE Trans. Evolutionary Computation, Vol. 3, No. 2, 82-102, Jul. 1999.

    LOCAL Z, Xi, Sum1, Sum2 AS EXT
    LOCAL i&&
    Z = O## : Sum1 = O## : Sum2 = O##
    FOR i&& = 1 TO N&&&
        Xi   = R(p&&,i&&,j&&)
        Sum1 = Sum1 + Xi^2
        Sum2 = Sum2 + COS(TwoPi*Xi)
    NEXT i&&
    Z = -20##*EXP(-0.2##*SQR(Sum1/N&&&)) - EXP(Sum2/N&&&) + 20## + e
    F10 = -Z
END FUNCTION 'F10
'----------
FUNCTION F11(R(),N&&&,p&&,j&&) '(n-D) F11
'MAXIMUM = ZERO (n-D CASE).
'Reference:
```

```
'Yao, X., Liu, Y., and Lin, G., "Evolutionary Programming Made Faster,"
'IEEE Trans. Evolutionary Computation, Vol. 3, No. 2, 82-102, Jul. 1999.

    LOCAL Z, Xi, Sum, Prod AS EXT

    LOCAL i&&

    Z = 0## : Sum = 0## : Prod = 1##

    FOR i&& = 1 TO n&&

        Xi    = R(p&&,i&&,j&&)

        Sum = Sum + (Xi-100##)^2

        Prod = Prod*COS((Xi-100##)/SQR(i&&))

    NEXT i&&

    Z = Sum/4000## - Prod + 1##

    F11 = -Z

END FUNCTION 'F11
'-----
FUNCTION u(Xi,a,k,m)

LOCAL Z AS EXT

    Z = 0##

    SELECT CASE Xi

        CASE > a   : Z = k*(Xi-a)^m

        CASE < -a  : Z = k*(-Xi-a)^m

    END SELECT

    u = Z

END FUNCTION
'----------
FUNCTION F12(K(),n&&,p&&,j&&) '(n=0) F12, Penalized #1
'Ref: Yao(1999).  Max=0 @ (-1,-1,...,-1), -50=<Xi=<50.

    LOCAL Offset, Sum1, Sum2, Z, X1, Y1, Xn, Yn, Xi, Yi, XiPlus1, YiPlus1 AS EXT

    LOCAL i&&, m&&, A$

    X1 = R(p&&,1,j&&)    : Y1 = 1## + (X1+1##)/4##

    Xn = R(p&&,N&&,j&&)  : Yn = 1## + (Xn-1##)/4##

    Sum1 = 0##

    FOR i&& = 1 TO N&&-1

        Xi       = R(p&&,i&&,j&&)    : Yi      = 1## + (Xi+1##)/4##

        XiPlus1 = R(p&&,i&&+1,j&&): YiPlus1 = 1## + (XiPlus1+1##)/4##

        Sum1 = Sum1 + (Yi-1##)^2*(1##+10##*(SIN(Pi*YiPlus1))^2)

    NEXT i&&

    Sum1 = Sum1 + 10##*(SIN(Pi*Y1))^2 + (Yn-1##)^2

    Sum1 = Pi*Sum1/N&&

    Sum2 = 0##

    FOR i&& = 1 TO N&&

        Xi = R(p&&,i&&,j&&)

        Sum2 = u(Xi,10##,100##,4##)

    NEXT i&&

    Z = Sum1 + Sum2

    F12 = -Z

END FUNCTION 'F12()
'-----------------
FUNCTION F13(K(),n&&,p&&,j&&) '(n=0) F13, Penalized #2
'Ref: Yao(1999).  Max=0 @ (1,1,...,1), -50=<Xi=<50.

    LOCAL Offset, Sum1, Sum2, Z, Xi, Xn, XiPlus1, X1 AS EXT

    LOCAL i&&, m&&, A$

    X1 = R(p&&,1,j&&) : Xn = R(p&&,N&&,j&&)

    Sum1 = 0##

    FOR i&& = 1 TO N&&-1

        Xi = R(p&&,i&&,j&&) : XiPlus1 = R(p&&,i&&+1,j&&)

        Sum1 = Sum1 + (Xi-1##)^2*(1##+(SIN(3##*Pi*XiPlus1))^2)

    NEXT i&&

    Sum1 = Sum1 + (SIN(Pi*3##*X1))^2 +(Xn-1##)^2*(1##+(SIN(TwoPi*Xn))^2)

    Sum2 = 0##

    FOR i&& = 1 TO N&&

        Xi = R(p&&,i&&,j&&)

        Sum2 = Sum2 + u(Xi,5##,100##,4##)

    NEXT i&&

    Z = Sum1/10## + Sum2

    F13 = -Z

END FUNCTION 'F13()
'-----------------
SUB FillArrayAij  'needed for function #14, Shekel's Foxholes

    Aij(1,1)=-32## : Aij(1,2)=-16## : Aij(1,3)=0## : Aij(1,4)=16## : Aij(1,5)=32##
    Aij(1,6)=-32## : Aij(1,7)=-16## : Aij(1,8)=0## : Aij(1,9)=16## : Aij(1,10)=32##
    Aij(1,11)=-32## : Aij(1,12)=-16## : Aij(1,13)=0## : Aij(1,14)=16## : Aij(1,15)=32##
    Aij(1,16)=-32## : Aij(1,17)=-16## : Aij(1,18)=0## : Aij(1,19)=16## : Aij(1,20)=32##
    Aij(1,21)=-32## : Aij(1,22)=-16## : Aij(1,23)=0## : Aij(1,24)=16## : Aij(1,25)=32##

    Aij(2,1)=-32## : Aij(2,2)=-32## : Aij(2,3)=-32## : Aij(2,4)=-32## : Aij(2,5)=-32##
    Aij(2,6)=-16## : Aij(2,7)=-16## : Aij(2,8)=-16## : Aij(2,9)=-16## : Aij(2,10)=-16##
    Aij(2,11)=0## : Aij(2,12)=0## : Aij(2,13)=0## : Aij(2,14)=0## : Aij(2,15)=0##
    Aij(2,16)=16## : Aij(2,17)=16## : Aij(2,18)=16## : Aij(2,19)=16## : Aij(2,20)=16##
    Aij(2,21)=32## : Aij(2,22)=32## : Aij(2,23)=32## : Aij(2,24)=32## : Aij(2,25)=32##
```

```
END SUB
'-----
FUNCTION F14(R(),Nd&&,p&&,j&&) 'F14 (2-D) Shekel's Foxholes (INVERTED...)
    LOCAL Sum1, Sum2, Z, Xi AS EXT
    LOCAL i&&, jj&&
    Sum1 = 0##
    FOR jj&& = 1 TO 25
        Sum2 = 0##
        FOR i&& = 1 TO 2
            Xi = R(p&&,i&&,j&&)
            Sum2 = Sum2 + (Xi-Aij(i&&,jj&&))^6
        NEXT i&&
        Sum1 = Sum1 + 1##/(jj&&+Sum2)
    NEXT j&&
    Z = 1##/(0.002##+Sum1)
    F14 = -Z
END FUNCTION 'F14
'----------
FUNCTION F16(R(),Nd&&,p&&,j&&) 'F16 (2-D) 6-Hump Camel-Back
    LOCAL x1, x2, Z AS EXT
    x1 = R(p&&,1,j&&) : x2 = R(p&&,2,j&&)
    Z = 4##*x1^2 - 2.1##*x1^4 + x1^6/3## + x1*x2 - 4*x2^2 + 4*x2^4
    F16 = -Z
END FUNCTION 'F16
'----------
FUNCTION F15(R(),Nd&&,p&&,j&&) 'F15 (4-D) Kowalik's Function
'Global maximum = -0.0003075 @ (0.1928,0.1908,0.1231,0.1358)
    LOCAL x1, x2, x3, x4, Num, Denom, Z, Aj(), Bj() AS EXT
    LOCAL jj&&
    REDIM Aj(1 TO 11), Bj(1 TO 11)
    Aj(1) = 0.1957##  : Bj(1)  = 1##/0.25##
    Aj(2) = 0.1947##  : Bj(2)  = 1##/0.50##
    Aj(3) = 0.1735##  : Bj(3)  = 1##/1.00##
    Aj(4) = 0.1600##  : Bj(4)  = 1##/2.00##
    Aj(5) = 0.0844##  : Bj(5)  = 1##/4.00##
    Aj(6) = 0.0627##  : Bj(6)  = 1##/6.00##
    Aj(7) = 0.0456##  : Bj(7)  = 1##/8.00##
    Aj(8) = 0.0342##  : Bj(8)  = 1##/10.0##
    Aj(9) = 0.0323##  : Bj(9)  = 1##/12.0##
    Aj(10) = 0.0235## : Bj(10) = 1##/14.0##
    Aj(11) = 0.0246## : Bj(11) = 1##/16.0##
    Z = 0##
    x1 = R(p&&,1,j&&) : x2 = R(p&&,2,j&&) : x3 = R(p&&,3,j&&) : x4 = R(p&&,4,j&&)
    FOR jj&& = 1 TO 11
        Num   = x1*(Bj(jj&&)^2+Bj(jj&&)*x2)
        Denom = Bj(jj&&)^2+Bj(jj&&)*x3+x4
        Z = Z + (Aj(jj&&)-Num/Denom)^2
    NEXT jj&&
    F15 = -Z
END FUNCTION 'F15
'----------
FUNCTION F17(R(),Nd&&,p&&,j&&) 'F17, (2-D) Branin
'Global maximum = -0.398 @ (-5) 1.142,12.275), (3.142,2.275), (9.425,2.425)
    LOCAL x1, x2, Z AS EXT
    x1 = R(p&&,1,j&&) : x2 = R(p&&,2,j&&)
    Z = (x2-5.1##*x1^2/(4##*Pi^2)+5##*x1/Pi-6##)^2 + 10##*(1##-1##/(8##*Pi))*COS(x1) + 10##
    F17 = -Z
END FUNCTION 'F17
'----------
FUNCTION F18(R(),Nd&&,p&&,j&&) 'Goldstein-Price 2-D Test Function
'Global maximum = -3 @ (0,-1)
    LOCAL Z, x1, x2, t1, t2 AS EXT
    x1 = R(p&&,1,j&&) : x2 = R(p&&,2,j&&)
    t1 = 1##+(x1+x2+1##)^2*(19##-14##*x1+3##*x1^2-14##*x2+6##*x1*x2+3##*x2^2)
    t2 = 30##+(2##*x1-3##*x2)^2*(18##-32##*x1+12##*x1^2+48##*x2-36##*x1*x2+27##*x2^2)
    Z  = t1*t2
    F18 = -Z
END FUNCTION 'F18()
'----------
FUNCTION F19(R(),Nd&&,p&&,j&&) 'F19 (3-D) Hartman's Family #1
'Global maximum = 3.86 @ (0.114,0.556,0.852)
    LOCAL Xi, Z, Sum, Aji(), Cj(), Pji() AS EXT
    LOCAL i&&, jj&&, m&&
    REDIM Aji(1 TO 4, 1 TO 3), Cj(1 TO 4), Pji(1 TO 4, 1 TO 3)
    Aji(1,1) = 3.0##  : Aji(1,2) = 10## : Aji(1,3) = 30## : Cj(1) = 1.0##
    Aji(2,1) = 0.1##  : Aji(2,2) = 10## : Aji(2,3) = 35## : Cj(2) = 1.2##
    Aji(3,1) = 3.0##  : Aji(3,2) = 10## : Aji(3,3) = 30## : Cj(3) = 3.0##
    Aji(4,1) = 0.1##  : Aji(4,2) = 10## : Aji(4,3) = 35## : Cj(4) = 3.2##

    Pji(1,1) = 0.36890## : Pji(1,2) = 0.1170## : Pji(1,3) = 0.2673##
    Pji(2,1) = 0.46990## : Pji(2,2) = 0.4387## : Pji(2,3) = 0.7470##
    Pji(3,1) = 0.10910## : Pji(3,2) = 0.8732## : Pji(3,3) = 0.5547##
    Pji(4,1) = 0.03815## : Pji(4,2) = 0.5743## : Pji(4,3) = 0.8828##
```

```
        Z = 0##
        FOR j&& = 1 TO 4
            Sum = 0##
            FOR i&& = 1 TO 3
                Xi = R(p&&,i&&,j&&)
                Sum = Sum + Aji(j&&,i&&)*(Xi-Pji(j&&,i&&))^2
            NEXT i&&
            Z = Z + Cj(j&&)*EXP(-Sum)
        NEXT j&&
        F19 = Z
END FUNCTION 'F19
'-----------
FUNCTION F20(R(),N&&,p&&,j&&) 'F20 (6-D) Hartman's Family #2
'Global maximum = 3.32 @ (0.201,0.150,0.477,0.275,0.311,0.657)
        LOCAL Xi, Z, Sum, Aji(), Cj(), Pji() AS EXT
        LOCAL i&&, j&&, m&&
        REDIM Aji(1 TO 4, 1 TO 6), Cj(1 TO 4), Pji(1 TO 4, 1 TO 6)
            Aji(1,1) = 10.0## : Aji(1,2) = 3.0## : Aji(1,3) = 17.0## : Cj(1) = 1.0##
            Aji(2,1) = 0.05## : Aji(2,2) = 10.0## : Aji(2,3) = 17.0## : Cj(2) = 1.2##
            Aji(3,1) = 3.00## : Aji(3,2) = 3.50## : Aji(3,3) = 1.70## : Cj(3) = 3.0##
            Aji(4,1) = 17.0## : Aji(4,2) = 8.00## : Aji(4,3) = 0.05## : Cj(4) = 3.2##

            Aji(1,4) = 3.5## : Aji(1,5) = 1.7## : Aji(1,6) = 8##
            Aji(2,4) = 0.1## : Aji(2,5) = 8## : Aji(2,6) = 14##
            Aji(3,4) = 10## : Aji(3,5) = 17## : Aji(3,6) = 8##
            Aji(4,4) = 10## : Aji(4,5) = 0.1## : Aji(4,6) = 14##

            Pji(1,1) = 0.1312## : Pji(1,2) = 0.1696## : Pji(1,3) = 0.5569##
            Pji(2,1) = 0.2329## : Pji(2,2) = 0.4135## : Pji(2,3) = 0.8307##
            Pji(3,1) = 0.2348## : Pji(3,2) = 0.1415## : Pji(3,3) = 0.3522##
            Pji(4,1) = 0.4047## : Pji(4,2) = 0.8828## : Pji(4,3) = 0.8732##

            Pji(1,4) = 0.0124## : Pji(1,5) = 0.8283## : Pji(1,6) = 0.5886##
            Pji(2,4) = 0.3736## : Pji(2,5) = 0.1004## : Pji(2,6) = 0.9991##
            Pji(3,4) = 0.2883## : Pji(3,5) = 0.3047## : Pji(3,6) = 0.6650##
            Pji(4,4) = 0.5743## : Pji(4,5) = 0.1091## : Pji(4,6) = 0.0381##
        Z = 0##
        FOR j&& = 1 TO 4
            Sum = 0##
            FOR i&& = 1 TO 6
                Xi = R(p&&,i&&,j&&)
                Sum = Sum + Aji(j&&,i&&)*(Xi-Pji(j&&,i&&))^2
            NEXT i&&
            Z = Z + Cj(j&&)*EXP(-Sum)
        NEXT j&&
        F20 = Z
END FUNCTION 'F20
'-----------
FUNCTION F21(R(),N&&,p&&,j&&) 'F21 (4-D) Shekel's Family m=5
'Global maximum = 10
        LOCAL Xi, Z, Sum, Aji(), Cj() AS EXT
        LOCAL i&&, j&&, m&&
        m&& = 5 : REDIM Aji(1 TO m&&, 1 TO 4), Cj(1 TO m&&)
            Aji(1,1) = 4## : Aji(1,2) =  4## : Aji(1,3) = 4## : Aji(1,4) =  4## : Cj(1) = 0.1##
            Aji(2,1) = 1## : Aji(2,2) =  1## : Aji(2,3) = 1## : Aji(2,4) =  1## : Cj(2) = 0.2##
            Aji(3,1) = 8## : Aji(3,2) =  8## : Aji(3,3) = 8## : Aji(3,4) =  8## : Cj(3) = 0.2##
            Aji(4,1) = 6## : Aji(4,2) =  6## : Aji(4,3) = 6## : Aji(4,4) =  6## : Cj(4) = 0.4##
            Aji(5,1) = 3## : Aji(5,2) =  7## : Aji(5,3) = 3## : Aji(5,4) =  7## : Cj(5) = 0.4##
        Z = 0##
        FOR j&& = 1 TO m&&  'NOTE:  Index j&& is used to avoid same variable name as j&&
            Sum = 0##
            FOR i&& = 1 TO 4 'Shekel's family is 4-D only
                Xi = R(p&&,i&&,j&&)
                Sum = Sum + (Xi-Aji(j&&,i&&))^2
            NEXT i&&
            Z = Z + 1##/(Sum + Cj(j&&))
        NEXT j&&
        F21 = Z
END FUNCTION 'F21
'-----------
FUNCTION F22(R(),N&&,p&&,j&&) 'F22 (4-D) Shekel's Family m=7
'Global maximum = 10
        LOCAL Xi, Z, Sum, Aji(), Cj() AS EXT
        LOCAL i&&, j&&, m&&
        m&& = 7 : REDIM Aji(1 TO m&&, 1 TO 4), Cj(1 TO m&&)
            Aji(1,1) = 4## : Aji(1,2) =  4## : Aji(1,3) = 4## : Aji(1,4) =  4## : Cj(1) = 0.1##
            Aji(2,1) = 1## : Aji(2,2) =  1## : Aji(2,3) = 1## : Aji(2,4) =  1## : Cj(2) = 0.2##
            Aji(3,1) = 8## : Aji(3,2) =  8## : Aji(3,3) = 8## : Aji(3,4) =  8## : Cj(3) = 0.2##
            Aji(4,1) = 6## : Aji(4,2) =  6## : Aji(4,3) = 6## : Aji(4,4) =  6## : Cj(4) = 0.4##
            Aji(5,1) = 3## : Aji(5,2) =  7## : Aji(5,3) = 3## : Aji(5,4) =  7## : Cj(5) = 0.4##
            Aji(6,1) = 2## : Aji(6,2) =  9## : Aji(6,3) = 2## : Aji(6,4) =  9## : Cj(6) = 0.6##
            Aji(7,1) = 5## : Aji(7,2) =  3## : Aji(7,3) = 3## : Aji(7,4) =  3## : Cj(7) = 0.3##
        Z = 0##
        FOR j&& = 1 TO m&&  'NOTE:  Index j&& is used to avoid same variable name as j&&
            Sum = 0##
            FOR i&& = 1 TO 4 'Shekel's family is 4-D only
                Xi = R(p&&,i&&,j&&)
                Sum = Sum + (Xi-Aji(j&&,i&&))^2
```

```basic
        NEXT i&&
            Z = Z + 1##/(Sum + Cj(jj&&))
    NEXT jj&&
    F22 = Z
END FUNCTION 'F22
'-----------
FUNCTION F23(R(),N0&&,p&&,j&&) 'F23 (4-D) Shekel's family m=10
'Global maximum = 10
    LOCAL Xi, Z, Sum, Aji(), Cj() AS EXT
    LOCAL i&&, jj&&, m&&
    m&& = 10 : REDIM Aji(1 TO m&&, 1 TO 4), Cj(1 TO m&&)
    Aji(1,1) = 4## : Aji(1,2) =  4## : Aji(1,3) = 4## : Aji(1,4) =  4## : Cj(1) = 0.1##
    Aji(2,1) = 1## : Aji(2,2) =  1## : Aji(2,3) = 1## : Aji(2,4) =  1## : Cj(2) = 0.2##
    Aji(3,1) = 8## : Aji(3,2) =  8## : Aji(3,3) = 8## : Aji(3,4) =  8## : Cj(3) = 0.2##
    Aji(4,1) = 6## : Aji(4,2) =  6## : Aji(4,3) = 6## : Aji(4,4) =  6## : Cj(4) = 0.4##
    Aji(5,1) = 3## : Aji(5,2) =  7## : Aji(5,3) = 3## : Aji(5,4) =  7## : Cj(5) = 0.4##
    Aji(6,1) = 2## : Aji(6,2) =  9## : Aji(6,3) = 2## : Aji(6,4) =  9## : Cj(6) = 0.6##
    Aji(7,1) = 5## : Aji(7,2) =  5## : Aji(7,3) = 3## : Aji(7,4) =  3## : Cj(7) = 0.3##
    Aji(8,1) = 8## : Aji(8,2) =  1## : Aji(8,3) = 8## : Aji(8,4) =  1## : Cj(8) = 0.7##
    Aji(9,1) = 6## : Aji(9,2) =  2## : Aji(9,3) = 6## : Aji(9,4) =  2## : Cj(9) = 0.5##
    Aji(10,1) = 7## : Aji(10,2) = 3.6## : Aji(10,3) = 7## : Aji(10,4) = 3.6## : Cj(10) = 0.5##
    Z = 0##
    FOR jj&& = 1 TO m&&    'NOTE:  Index jj&& is used to avoid same variable name as j&&
        Sum = 0##
        FOR i&& = 1 TO 4 'Shekel's family is 4-D only
            Xi = R(p&&,i&&,j&&)
            Sum = Sum + (Xi-Aji(jj&&,i&&))^2
        NEXT i&&
            Z = Z + 1##/(Sum + Cj(jj&&))
    NEXT jj&&
    F23 = Z
END FUNCTION 'F23
'=========================================== END FUNCTION DEFINITIONS ===========================================
SUB Plot2DbestProbePaths(NumPaths&&,M(),R(),Np&&,N0&&,LastIteration&&,FunctionName$)
LOCAL TrajectoryNumber&&, ProbeNumber&&, StepNumber&&, N&&, m&&, ProcID???
LOCAL MaximumFitness, MinimumFitness AS EXT
LOCAL BestProbeThisStep()
LOCAL BestFitnessThisStep(), TempFitness() AS EXT
LOCAL Annotation$, xCoord$, yCoord$, GnuPlotEXE$, PlotWithLines$
    Annotation$    = ""
    PlotWithLines$ = "YES" '"NO"
    NumPaths&& = MIN(Np&&,NumPaths&&)
    GnuPlotEXE$ = "wgnuplot.exe"
'  ---------------- Get Min/Max Fitnesses ----------------
    MaximumFitness = M(1,0) : MinimumFitness = M(1,0)  'Note:  M(p&&,j&&)
    FOR StepNumber&& = 0 TO LastIteration&&
        FOR ProbeNumber&& = 1 TO Np&&
            IF M(ProbeNumber&&,StepNumber&&) >= MaximumFitness THEN MaximumFitness = M(ProbeNumber&&,StepNumber&&)
            IF M(ProbeNumber&&,StepNumber&&) =< MinimumFitness THEN MinimumFitness = M(ProbeNumber&&,StepNumber&&)
        NEXT ProbeNumber&&
    NEXT StepNumber&&
'  ------------ Copy Fitness Array M() into TempFitness to Preserve M() ----------------
    REDIM TempFitness(1 TO Np&&, 0 TO LastIteration&&)
    FOR ProbeNumber&& = 1 TO Np&&
        FOR StepNumber&& = 0 TO LastIteration&&
            TempFitness(ProbeNumber&&,StepNumber&&) = M(ProbeNumber&&,StepNumber&&)
        NEXT StepNumber&&
    NEXT ProbeNumber&&
'  ------------ LOOP ON Paths -----------
    FOR TrajectoryNumber&& = 1 TO NumPaths&&
'          --------------- Get Trajectory Coordinate Data ----------------
            REDIM BestFitnessThisStep(0 TO LastIteration&&), BestProbeThisStep&&(0 TO LastIteration&&)
            BestFitnessThisStep(StepNumber&&) = TempFitness(1,StepNumber&&)
            FOR StepNumber&& = 0 TO LastIteration&&
                FOR ProbeNumber&& = 1 TO Np&&
                    IF TempFitness(ProbeNumber&&,StepNumber&&) >= BestFitnessThisStep(StepNumber&&) THEN
                        BestFitnessThisStep(StepNumber&&) = TempFitness(ProbeNumber&&,StepNumber&&)
                        BestProbeThisStep&&(StepNumber&&) = ProbeNumber&&
                    END IF
                NEXT ProbeNumber&&
            NEXT StepNumber&&
'   ----- Create Plot Data File -----
        N&& = FREEFILE
        SELECT CASE TrajectoryNumber&&
            CASE 1  : OPEN "p1"  FOR OUTPUT AS #N&&
            CASE 2  : OPEN "p2"  FOR OUTPUT AS #N&&
            CASE 3  : OPEN "p3"  FOR OUTPUT AS #N&&
            CASE 4  : OPEN "p4"  FOR OUTPUT AS #N&&
            CASE 5  : OPEN "p5"  FOR OUTPUT AS #N&&
            CASE 6  : OPEN "p6"  FOR OUTPUT AS #N&&
            CASE 7  : OPEN "p7"  FOR OUTPUT AS #N&&
            CASE 8  : OPEN "p8"  FOR OUTPUT AS #N&&
            CASE 9  : OPEN "p9"  FOR OUTPUT AS #N&&
            CASE 10 : OPEN "p10" FOR OUTPUT AS #N&&
```

```
        END SELECT
'  ------------ Write Plot File Data -----------

        FOR StepNumber&& = 0 TO LastIteration&&

            PRINT #N&&, USING$("######.########
######.########",R(BestProbeThisStep&&(StepNumber&&),1,StepNumber&&),R(BestProbeThisStep&&(StepNumber&&),2,StepNumber&&))

            TempFitness(BestProbeThisStep&&(StepNumber&&),StepNumber&&) = MinimumFitness 'so that same max will not be found for next trajectory

        NEXT StepNumber&&

        CLOSE #N&&

    NEXT TrajectoryNumber&&
'  ---------------------- Plot Paths ------------------------

    CALL SetUpGnuPlotIniFile(0.13##*Screenwidth&&,0.18##*Screenheight&&,0.7##*Screenwidth&&,0.7##*Screenheight&&)

    Annotation$ = ""

    N&& = FREEFILE

    OPEN "cmd2d.gp" FOR OUTPUT AS #N&&

        PRINT #N&&, "set xrange ["+REMOVE$(STR$(X!Min(1)),ANY" ")+":"+REMOVE$(STR$(X!Max(1)),ANY" ")+"]"
        PRINT #N&&, "set yrange ["+REMOVE$(STR$(X!Min(2)),ANY" ")+":"+REMOVE$(STR$(X!Max(2)),ANY" ")+"]"
        'PRINT #N&&, "set label "    + Quote$ + Annotation$ + Quote$ + " at graph " + xCoord$ + "," + yCoord$
        PRINT #N&&, "set grid xtics "  = "10"
        PRINT #N&&, "set grid ytics "  = "10"
        PRINT #N&&, "set grid mxtics"
        PRINT #N&&, "set grid mytics"
        PRINT #N&&, "show grid"
        PRINT #N&&, "set title " + Quote$ + "2D " + FunctionName$+" PATHS OF SAMPLE POINTS WITH BEST\nFITNESSES (ORDERED BY FITNESS)" + "\n" + RunID$ +
Quote$
        PRINT #N&&, "set xlabel " + Quote$ + "x1\n\n"                            + Quote$
        PRINT #N&&, "set ylabel " + Quote$ + "\nx2"                              + Quote$

        IF PlotWithLines$ = "YES" THEN

            SELECT CASE NumPaths&&

                CASE 1  : PRINT #N&&, "plot "+Quote$+"p1"+Quote$+" w 1 lw 3"
                CASE 2  : PRINT #N&&, "plot "+Quote$+"p1"+Quote$+" w 1 lw 3,"+Quote$+"p2"+Quote$+" w 1"
                CASE 3  : PRINT #N&&, "plot "+Quote$+"p1"+Quote$+" w 1 lw 3,"+Quote$+"p2"+Quote$+" w 1,"+Quote$+"p3"+Quote$+" w 1"
                CASE 4  : PRINT #N&&, "plot "+Quote$+"p1"+Quote$+" w 1 lw 3,"+Quote$+"p2"+Quote$+" w 1,"+Quote$+"p3"+Quote$+" w 1,"+Quote$+"p4"+Quote$+" w 1"
                CASE 5  : PRINT #N&&, "plot "+Quote$+"p1"+Quote$+" w 1 lw 3,"+Quote$+"p2"+Quote$+" w 1,"+Quote$+"p3"+Quote$+" w 1,"+Quote$+"p4"+Quote$+" w
1,"+Quote$+"p5"+Quote$+" w 1"
                CASE 6  : PRINT #N&&, "plot "+Quote$+"p1"+Quote$+" w 1 lw 3,"+Quote$+"p2"+Quote$+" w 1,"+Quote$+"p3"+Quote$+" w 1,"+Quote$+"p4"+Quote$+" w
1,"+Quote$+"p5"+Quote$+" w 1,"+Quote$+"p6"+Quote$+" w 1"
                CASE 7  : PRINT #N&&, "plot "+Quote$+"p1"+Quote$+" w 1 lw 3,"+Quote$+"p2"+Quote$+" w 1,"+Quote$+"p3"+Quote$+" w 1,"+Quote$+"p4"+Quote$+" w
1,"+Quote$+"p5"+Quote$+" w 1,"+Quote$+"p6"+Quote$+" w 1,"+Quote$+"p7"+Quote$+" w 1"
                CASE 8  : PRINT #N&&, "plot "+Quote$+"p1"+Quote$+" w 1 lw 3,"+Quote$+"p2"+Quote$+" w 1,"+Quote$+"p3"+Quote$+" w 1,"+Quote$+"p4"+Quote$+" w
1,"+Quote$+"p5"+Quote$+" w 1,"+Quote$+"p6"+Quote$+" w 1,"+Quote$+"p7"+Quote$+" w 1,"+Quote$+"p8"+Quote$+" w 1"
                CASE 9  : PRINT #N&&, "plot "+Quote$+"p1"+Quote$+" w 1 lw 3,"+Quote$+"p2"+Quote$+" w 1,"+Quote$+"p3"+Quote$+" w 1,"+Quote$+"p4"+Quote$+" w
1,"+Quote$+"p5"+Quote$+" w 1,"+Quote$+"p6"+Quote$+" w 1,"+Quote$+"p7"+Quote$+" w 1,"+Quote$+"p8"+Quote$+" w 1,"+Quote$+"p9"+Quote$+" w 1"
                CASE 10 : PRINT #N&&, "plot "+Quote$+"p1"+Quote$+" w 1 lw 3,"+Quote$+"p2"+Quote$+" w 1,"+Quote$+"p3"+Quote$+" w 1,"+Quote$+"p4"+Quote$+" w
1,"+Quote$+"p5"+Quote$+" w 1,"+Quote$+"p6"+Quote$+" w 1,"+Quote$+"p7"+Quote$+" w 1,"+Quote$+"p8"+Quote$+" w 1,"+Quote$+"p9"+Quote$+" w 1,"+Quote$+"p10"+Quote$+" w 1"

            END SELECT

        ELSE

            SELECT CASE NumPaths&&

                CASE 1  : PRINT #N&&, "plot "+Quote$+"p1"+Quote$+" lw 2"
                CASE 2  : PRINT #N&&, "plot "+Quote$+"p1"+Quote$+" lw 2,"+Quote$+"p2"+Quote$
                CASE 3  : PRINT #N&&, "plot "+Quote$+"p1"+Quote$+" lw 2,"+Quote$+"p2"+Quote$+" ,"+Quote$+"p3"+Quote$
                CASE 4  : PRINT #N&&, "plot "+Quote$+"p1"+Quote$+" lw 2,"+Quote$+"p2"+Quote$+" ,"+Quote$+"p3"+Quote$+" ,"+Quote$+"p4"+Quote$
                CASE 5  : PRINT #N&&, "plot "+Quote$+"p1"+Quote$+" lw 2,"+Quote$+"p2"+Quote$+" ,"+Quote$+"p3"+Quote$+" ,"+Quote$+"p4"+Quote$
+" ,"+Quote$+"p5"+Quote$
                CASE 6  : PRINT #N&&, "plot "+Quote$+"p1"+Quote$+" lw 2,"+Quote$+"p2"+Quote$+" ,"+Quote$+"p3"+Quote$+" ,"+Quote$+"p4"+Quote$
+" ,"+Quote$+"p5"+Quote$+" ,"+Quote$+"p6"+Quote$
                CASE 7  : PRINT #N&&, "plot "+Quote$+"p1"+Quote$+" lw 2,"+Quote$+"p2"+Quote$+" ,"+Quote$+"p3"+Quote$+" ,"+Quote$+"p4"+Quote$
+" ,"+Quote$+"p5"+Quote$+" ,"+Quote$+"p6"+Quote$+" ,"+Quote$+"p7"+Quote$
                CASE 8  : PRINT #N&&, "plot "+Quote$+"p1"+Quote$+" lw 2,"+Quote$+"p2"+Quote$+" ,"+Quote$+"p3"+Quote$+" ,"+Quote$+"p4"+Quote$
+" ,"+Quote$+"p5"+Quote$+" ,"+Quote$+"p6"+Quote$+" ,"+Quote$+"p7"+Quote$+" ,"+Quote$+"p8"+Quote$
                CASE 9  : PRINT #N&&, "plot "+Quote$+"p1"+Quote$+" lw 2,"+Quote$+"p2"+Quote$+" ,"+Quote$+"p3"+Quote$+" ,"+Quote$+"p4"+Quote$
+" ,"+Quote$+"p5"+Quote$+" ,"+Quote$+"p6"+Quote$+" ,"+Quote$+"p7"+Quote$+" ,"+Quote$+"p8"+Quote$+" ,"+Quote$+"p9"+Quote$
                CASE 10 : PRINT #N&&, "plot "+Quote$+"p1"+Quote$+" lw 2,"+Quote$+"p2"+Quote$+" ,"+Quote$+"p3"+Quote$+" ,"+Quote$+"p4"+Quote$
+" ,"+Quote$+"p5"+Quote$+" ,"+Quote$+"p6"+Quote$+" ,"+Quote$+"p7"+Quote$+" ,"+Quote$+"p8"+Quote$+" ,"+Quote$+"p9"+Quote$+" ,"+Quote$+"p10"+Quote$

            END SELECT

        END IF

        CLOSE #N&&

        ProcID??? = SHELL(GnuPlotEXE$+" cmd2d.gp -") : CALL Delay(0.5##)

'  ------------ CLEANUP ------------
    IF DIR$("p1") <> "" THEN KILL "p1"
    IF DIR$("p2") <> "" THEN KILL "p2"
    IF DIR$("p3") <> "" THEN KILL "p3"
    IF DIR$("p4") <> "" THEN KILL "p4"
    IF DIR$("p5") <> "" THEN KILL "p5"
    IF DIR$("p6") <> "" THEN KILL "p6"
    IF DIR$("p7") <> "" THEN KILL "p7"
    IF DIR$("p8") <> "" THEN KILL "p8"
    IF DIR$("p9") <> "" THEN KILL "p9"
    IF DIR$("p10") <> "" THEN KILL "p10"

END SUB 'Plot2DbestProbePaths()
'------------------------------------------

SUB Plot2DIndividualProbePaths(NumPaths&&,W(),R(),Np&&,Nd&&,LastIteration&&,FunctionName$)

LOCAL ProbeNumber&&, StepNumber&&, N&&, ProcID???

LOCAL Annotation$, xCoord$, yCoord$, GnuPlotEXE$, PlotWithLines$

    NumPaths&& = MIN(Np&&,NumPaths&&)

    Annotation$     = ""

    PlotWithLines$ = "YES"  '"NO"

    GnuPlotEXE$ = "wgnuplot.exe"
'  ------------- LOOP ON PROBES --------------

    FOR ProbeNumber&& = 1 TO MIN(NumPaths&&,Np&&)
'  ----- Create Plot Data File -----

        N&& = FREEFILE

        SELECT CASE ProbeNumber&&

            CASE 1  : OPEN "p1"  FOR OUTPUT AS #N&&
            CASE 2  : OPEN "p2"  FOR OUTPUT AS #N&&
            CASE 3  : OPEN "p3"  FOR OUTPUT AS #N&&
            CASE 4  : OPEN "p4"  FOR OUTPUT AS #N&&
            CASE 5  : OPEN "p5"  FOR OUTPUT AS #N&&
            CASE 6  : OPEN "p6"  FOR OUTPUT AS #N&&
            CASE 7  : OPEN "p7"  FOR OUTPUT AS #N&&
            CASE 8  : OPEN "p8"  FOR OUTPUT AS #N&&
```

```
        CASE 9  : OPEN "p9"  FOR OUTPUT AS #N&&
        CASE 10 : OPEN "p10" FOR OUTPUT AS #N&&
        CASE 11 : OPEN "p11" FOR OUTPUT AS #N&&
        CASE 12 : OPEN "p12" FOR OUTPUT AS #N&&
        CASE 13 : OPEN "p13" FOR OUTPUT AS #N&&
        CASE 14 : OPEN "p14" FOR OUTPUT AS #N&&
        CASE 15 : OPEN "p15" FOR OUTPUT AS #N&&
        CASE 16 : OPEN "p16" FOR OUTPUT AS #N&&

    END SELECT

'  ----------- Write Plot File Data -----------

    FOR StepNumber&& = 0 TO LastIteration&&

        PRINT #N&&, USING("######.######## ######.#######",R(ProbeNumber&&,1,StepNumber&&),R(ProbeNumber&&,2,StepNumber&&))

    NEXT StepNumber&&

    CLOSE #N&&

NEXT ProbeNumber&&
'  ----------------------------------------- Plot Paths -----------------------------------------
'
'usage:  CALL CreateGNUplotInitFile(PlotWindowXLC_X&&,PlotWindowLC_Y&&,PlotWindowWidth&&,PlotWindowHeight&&)

    CALL CreateGNUplotInitFile(0.17##*ScreenWidth&&,0.22##*ScreenHeight&&,0.7##*ScreenWidth&&,0.7##*ScreenHeight&&)

    Annotation$ = ""

    N&& = FREEFILE

    OPEN "cmd2d.gp" FOR OUTPUT AS #N&&

        PRINT #N&&, "set xrange ["+REMOVE$(STR$(X!Min(1)),ANY" ")+":"+REMOVE$(STR$(X!Max(1)),ANY" ")+"]"
        PRINT #N&&, "set xrange ["+REMOVE$(STR$(X!Min(2)),ANY" ")+":"+REMOVE$(STR$(X!Max(2)),ANY" ")+"]"
        PRINT #N&&, "set grid xtics " = "10"
        PRINT #N&&, "set grid ytics " = "10"
        PRINT #N&&, "set grid mxtics"
        PRINT #N&&, "show grid"
        PRINT #N&&, "set title " + Quote$ + "2D ~ FunctionName$+" INDIVIDUAL SAMPLE POINT PATHS\n(ORDERED BY SAMPLE POINT #)" + "\n" + RunID$ + Quote$
        PRINT #N&&, "set xlabel " + Quote$ + "x1\n\n" + Quote$
        PRINT #N&&, "set ylabel " + Quote$ + "\nx2" + Quote$

    IF PlotWithLines$ = "YES" THEN

        SELECT CASE NumPaths&&

            CASE 1  : PRINT #N&&, "plot "+Quote$+"p1" -Quote$+" w l lw 1"
            CASE 2  : PRINT #N&&, "plot "+Quote$+"p1" -Quote$+" lw 1 lw 1,"+Quote$+"p2"-Quote$+" w l"
            CASE 3  : PRINT #N&&, "plot "+Quote$+"p1" -Quote$+" w l lw 1,"+Quote$+"p2"-Quote$+" w l 1,"+Quote$+"p3"-Quote$+" w 1"
            CASE 4  : PRINT #N&&, "plot "+Quote$+"p1" -Quote$+" w l lw 1,"+Quote$+"p2"-Quote$+" w l 1,"+Quote$+"p3"-Quote$+" w l 1,"+Quote$+"p4"-Quote$+" w
1"
            CASE 5  : PRINT #N&&, "plot "+Quote$+"p1" -Quote$+" w l lw 1,"+Quote$+"p2"-Quote$+" w l 1,"+Quote$+"p3"-Quote$+" w l 1,"+Quote$+"p4"-Quote$+" w
1,"+Quote$+"p5"-Quote$+" w l"
            CASE 6  : PRINT #N&&, "plot "+Quote$+"p1" -Quote$+" w l lw 1,"+Quote$+"p2"-Quote$+" w l 1,"+Quote$+"p3"-Quote$+" w l 1,"+Quote$+"p4"-Quote$+" w
1,"+Quote$+"p5"-Quote$+" w 1,"+Quote$+"p6"-Quote$+" w 1"
            CASE 7  : PRINT #N&&, "plot "+Quote$+"p1" -Quote$+" w l lw 1,"+Quote$+"p2"-Quote$+" w l 1,"+Quote$+"p3"-Quote$+" w l 1,"+Quote$+"p4"-Quote$+" w
1,"+Quote$+"p5"-Quote$+" w 1,"+Quote$+"p6"-Quote$+" w 1,"
                                Quote$+"p7"-Quote$+" w 1"
            CASE 8  : PRINT #N&&, "plot "+Quote$+"p1" -Quote$+" w l lw 1,"+Quote$+"p2"-Quote$+" w l 1,"+Quote$+"p3"-Quote$+" w l 1,"+Quote$+"p4"-Quote$+" w
1,"+Quote$+"p5"-Quote$+" w 1,"+Quote$+"p6"-Quote$+" w 1,"
                                Quote$+"p7"-Quote$+" w 1,"         +Quote$+"p8"-Quote$+" w 1"
            CASE 9  : PRINT #N&&, "plot "+Quote$+"p1" -Quote$+" w l lw 1,"+Quote$+"p2"-Quote$+" w l 1,"+Quote$+"p3"-Quote$+" w l 1,"+Quote$+"p4"-Quote$+" w
1,"+Quote$+"p5"-Quote$+" w 1,"+Quote$+"p6"-Quote$+" w 1,"
                                Quote$+"p7"-Quote$+" w 1,"         +Quote$+"p8"-Quote$+" w 1,"+Quote$+"p9"-Quote$+" w 1"
            CASE 10 : PRINT #N&&, "plot "+Quote$+"p1" -Quote$+" w l lw 1,"+Quote$+"p2"-Quote$+" w l 1,"+Quote$+"p3"-Quote$+" w l 1,"+Quote$+"p4"-Quote$+" w
1,"+Quote$+"p5"-Quote$+" w 1,"+Quote$+"p6"-Quote$+" w 1,"
                                Quote$+"p7"-Quote$+" w 1,"         +Quote$+"p8"-Quote$+" w 1,"+Quote$+"p9"-Quote$+" w 1,"+Quote$+"p10"-Quote$+"
w 1"
            CASE 11 : PRINT #N&&, "plot "+Quote$+"p1" -Quote$+" w l lw 1,"+Quote$+"p2"-Quote$+" w l 1,"+Quote$+"p3"-Quote$+" w l 1,"+Quote$+"p4"-Quote$+" w
1,"+Quote$+"p5"-Quote$+" w 1,"+Quote$+"p6"-Quote$+" w 1,"
                                Quote$+"p7"-Quote$+" w 1,"         +Quote$+"p8"-Quote$+" w 1,"+Quote$+"p9"-Quote$+" w 1,"+Quote$+"p10"-Quote$+"
w 1,"+Quote$+"p11"-Quote$+" w 1"
            CASE 12 : PRINT #N&&, "plot "+Quote$+"p1" -Quote$+" w l lw 1,"+Quote$+"p2"-Quote$+" w l 1,"+Quote$+"p3"-Quote$+" w l 1,"+Quote$+"p4"-Quote$+" w
 1,"+Quote$+"p5"-Quote$+" w 1,"+Quote$+"p6"-Quote$+" w 1,"
                                Quote$+"p7"-Quote$+" w 1,"         +Quote$+"p8"-Quote$+" w 1,"+Quote$+"p9"-Quote$+" w 1,"+Quote$+"p10"-Quote$+"
w 1,"+Quote$+"p11"-Quote$+" w 1,"+Quote$+"p12"-Quote$+" w 1"
            CASE 13 : PRINT #N&&, "plot "+Quote$+"p1" -Quote$+" w l lw 1,"+Quote$+"p2"-Quote$+" w l 1,"+Quote$+"p3"-Quote$+" w l 1,"+Quote$+"p4"-Quote$+" w
 1,"+Quote$+"p5"-Quote$+" w 1,"+Quote$+"p6"-Quote$+" w 1,"
                                Quote$+"p7"-Quote$+" w 1,"         +Quote$+"p8"-Quote$+" w 1,"+Quote$+"p9"-Quote$+" w 1,"+Quote$+"p10"-Quote$+"
w 1,"+Quote$+"p11"-Quote$+" w 1,"+Quote$+"p12"-Quote$+" w 1,"
                                Quote$+"p13"-Quote$+" w 1"
            CASE 14 : PRINT #N&&, "plot "+Quote$+"p1" -Quote$+" w l lw 1,"+Quote$+"p2"-Quote$+" w l 1,"+Quote$+"p3"-Quote$+" w l 1,"+Quote$+"p4"-Quote$+" w
 1,"+Quote$+"p5"-Quote$+" w 1,"+Quote$+"p6"-Quote$+" w 1,"
                                Quote$+"p7"-Quote$+" w 1,"         +Quote$+"p8"-Quote$+" w 1,"+Quote$+"p9"-Quote$+" w 1,"+Quote$+"p10"-Quote$+"
w 1,"+Quote$+"p11"-Quote$+" w 1,"+Quote$+"p12"-Quote$+" w 1,"
                                Quote$+"p13"-Quote$+" w 1,"        +Quote$+"p14"-Quote$+" w 1"
            CASE 15 : PRINT #N&&, "plot "+Quote$+"p1" -Quote$+" w l lw 1,"+Quote$+"p2"-Quote$+" w l 1,"+Quote$+"p3"-Quote$+" w l 1,"+Quote$+"p4"-Quote$+" w
 1,"+Quote$+"p5"-Quote$+" w 1,"+Quote$+"p6"-Quote$+" w 1,"
                                Quote$+"p7"-Quote$+" w 1,"         +Quote$+"p8"-Quote$+" w 1,"+Quote$+"p9"-Quote$+" w 1,"+Quote$+"p10"-Quote$+"
w 1,"+Quote$+"p11"-Quote$+" w 1,"+Quote$+"p12"-Quote$+" w 1,"
                                Quote$+"p13"-Quote$+" w 1,"        +Quote$+"p14"-Quote$+" w 1,"+Quote$+"p15"-Quote$+" w 1"
            CASE 16 : PRINT #N&&, "plot "+Quote$+"p1" -Quote$+" w l lw 1,"+Quote$+"p2"-Quote$+" w l 1,"+Quote$+"p3"-Quote$+" w l 1,"+Quote$+"p4"-Quote$+" w
 1,"+Quote$+"p5"-Quote$+" w 1,"+Quote$+"p6"-Quote$+" w 1,"
                                Quote$+"p7"-Quote$+" w 1,"         +Quote$+"p8"-Quote$+" w 1,"+Quote$+"p9"-Quote$+" w 1,"+Quote$+"p10"-Quote$+"
w 1"
                                Quote$+"p13"-Quote$+" w 1,"        +Quote$+"p14"-Quote$+" w 1,"+Quote$+"p15"-Quote$+" w 1,"+Quote$+"p16"-Quote$+"
w 1"
            END SELECT

        ELSE

            SELECT CASE NumPaths&&

                CASE 1  : PRINT #N&&, "plot "+Quote$+"p1"-Quote$+" lw 1"
                CASE 2  : PRINT #N&&, "plot "+Quote$+"p1"-Quote$+" lw 1,"+Quote$+"p2"-Quote$
                CASE 3  : PRINT #N&&, "plot "+Quote$+"p1"-Quote$+" lw 1,"+Quote$+"p2"-Quote$+" ,"+Quote$+"p3"-Quote$
                CASE 4  : PRINT #N&&, "plot "+Quote$+"p1"-Quote$+" lw 1,"+Quote$+"p2"-Quote$+" ,"+Quote$+"p3"-Quote$+" ,"+Quote$+"p4"-Quote$
                CASE 5  : PRINT #N&&, "plot "+Quote$+"p1"-Quote$+" lw 1,"+Quote$+"p2"-Quote$+" ,"+Quote$+"p3"-Quote$+" ,"+Quote$+"p4"-Quote$+"
,"+Quote$+"p5"-Quote$+"  "+Quote$+"p6"-Quote$
                CASE 6  : PRINT #N&&, "plot "+Quote$+"p1"-Quote$+" lw 1,"+Quote$+"p2"-Quote$+" ,"+Quote$+"p3"-Quote$+" ,"+Quote$+"p4"-Quote$+"
,"+Quote$+"p5"-Quote$+" ,"+Quote$+"p6"-Quote$
                CASE 7  : PRINT #N&&, "plot "+Quote$+"p1"-Quote$+" lw 1,"+Quote$+"p2"-Quote$+" ,"+Quote$+"p3"-Quote$+" ,"+Quote$+"p4"-Quote$+"
,"+Quote$+"p5"-Quote$+" ,"+Quote$+"p6"-Quote$+" ,"
                                Quote$+"p7"-Quote$+" ,"        +Quote$+"p8"-Quote$
                CASE 8  : PRINT #N&&, "plot "+Quote$+"p1"-Quote$+" lw 1,"+Quote$+"p2"-Quote$+" ,"+Quote$+"p3"-Quote$+" ,"+Quote$+"p4"-Quote$+"
,"+Quote$+"p5"-Quote$+" ,"+Quote$+"p6"-Quote$+" ,"
                                Quote$+"p7"-Quote$+" ,"        +Quote$+"p8"-Quote$
                CASE 9  : PRINT #N&&, "plot "+Quote$+"p1"-Quote$+" lw 1,"+Quote$+"p2"-Quote$+" ,"+Quote$+"p3"-Quote$+" ,"+Quote$+"p4"-Quote$+"
,"+Quote$+"p5"-Quote$+" ,"+Quote$+"p6"-Quote$+" ,"
                                Quote$+"p7"-Quote$+" ,"        +Quote$+"p8"-Quote$+" ,"+Quote$+"p9"-Quote$+" ,"+Quote$+"p10"-Quote$
                CASE 10 : PRINT #N&&, "plot "+Quote$+"p1"-Quote$+" lw 1,"+Quote$+"p2"-Quote$+" ,"+Quote$+"p3"-Quote$+" ,"+Quote$+"p4"-Quote$+"
,"+Quote$+"p5"-Quote$+" ,"+Quote$+"p6"-Quote$+" ,"
                                Quote$+"p7"-Quote$+" ,"        +Quote$+"p8"-Quote$+" ,"+Quote$+"p9"-Quote$+" ,"+Quote$+"p10"-Quote$
,"+Quote$+"p11"-Quote$
                CASE 11 : PRINT #N&&, "plot "+Quote$+"p1"-Quote$+" lw 1,"+Quote$+"p2"-Quote$+" ,"+Quote$+"p3"-Quote$+" ,"+Quote$+"p4"-Quote$+"
,"+Quote$+"p5"-Quote$+" ,"+Quote$+"p6"-Quote$+" ,"
                                Quote$+"p7"-Quote$+" ,"        +Quote$+"p8"-Quote$+" ,"+Quote$+"p9"-Quote$+" ,"+Quote$+"p10"-Quote$
,"+Quote$+"p11"-Quote$
                CASE 12 : PRINT #N&&, "plot "+Quote$+"p1"-Quote$+" lw 1,"+Quote$+"p2"-Quote$+" ,"+Quote$+"p3"-Quote$+" ,"+Quote$+"p4"-Quote$+"
,"+Quote$+"p5"-Quote$+" ,"+Quote$+"p6"-Quote$+" ,"
                                Quote$+"p7"-Quote$+" ,"        +Quote$+"p8"-Quote$+" ,"+Quote$+"p9"-Quote$+" ,"+Quote$+"p10"-Quote$
,"+Quote$+"p11"-Quote$+" ,"+Quote$+"p12"-Quote$
```

```
                CASE 13 : PRINT #N&&, "plot "+Quote$+"p1" +Quote$+" lw 1,"+Quote$+"p2" +Quote$+" ,"+Quote$+"p3" +Quote$+" ,"+Quote$+"p4"  +Quote$+"
,"+Quote$+"p5" +Quote$+" , "+Quote$+"p6" +Quote$+" ,"_
                                           Quote$+"p7" +Quote$+" ,"+Quote$+"p8" +Quote$+" ,"+Quote$+"p9" +Quote$+" ,"+Quote$+"p10" +Quote$+"
,"+Quote$+"p11"+Quote$+" ," +Quote$+"p12"+Quote$+" ,"_
                                           Quote$+"p13"+Quote$
                CASE 14 : PRINT #N&&, "plot "+Quote$+"p1" +Quote$+" lw 1,"+Quote$+"p2" +Quote$+" ,"+Quote$+"p3" +Quote$+" ,"+Quote$+"p4"  +Quote$+"
,"+Quote$+"p5" +Quote$+" , "+Quote$+"p6" +Quote$+" ,"_
                                           Quote$+"p7" +Quote$+" ,"+Quote$+"p8" +Quote$+" ,"+Quote$+"p9" +Quote$+" ,"+Quote$+"p10" +Quote$+"
,"+Quote$+"p11"+Quote$+" ," +Quote$+"p12"+Quote$+" ,"_
                                           Quote$+"p13"+Quote$+" ," +Quote$+"p14"+Quote$
                CASE 15 : PRINT #N&&, "plot "+Quote$+"p1" +Quote$+" lw 1,"+Quote$+"p2" +Quote$+" ,"+Quote$+"p3" +Quote$+" ,"+Quote$+"p4"  +Quote$+"
,"+Quote$+"p5" +Quote$+" , "+Quote$+"p6" +Quote$+" ,"_
                                           Quote$+"p7" +Quote$+" ,"+Quote$+"p8" +Quote$+" ,"+Quote$+"p9" +Quote$+" ,"+Quote$+"p10" +Quote$+"
,"+Quote$+"p11"+Quote$+" ," +Quote$+"p12"+Quote$+" ,"_
                                           Quote$+"p13"+Quote$+" ," +Quote$+"p14" +Quote$+" ,"+Quote$+"p15"+Quote$
                CASE 16 : PRINT #N&&, "plot "+Quote$+"p1" +Quote$+" lw 1,"+Quote$+"p2" +Quote$+" ,"+Quote$+"p3" +Quote$+" ,"+Quote$+"p4"  +Quote$+"
,"+Quote$+"p5" +Quote$+" , "+Quote$+"p6" +Quote$+" ,"_
                                           Quote$+"p7" +Quote$+" ,"+Quote$+"p8" +Quote$+" ,"+Quote$+"p9" +Quote$+" ,"+Quote$+"p10" +Quote$+"
,"+Quote$+"p11"+Quote$+" ," +Quote$+"p12"+Quote$+" ,"_
                                           Quote$+"p13"+Quote$+" ," +Quote$+"p14" +Quote$+" ,"+Quote$+"p15"+Quote$+" ,"+Quote$+"p16"+Quote$
            END SELECT
        END IF

    CLOSE #N&&

    ProcID??? = SHELL(GnuPlotEXE$+" cmd2d.gp ~") : CALL Delay(0.5##)

' ----------- CLEANUP ------------
    IF DIR$("p1")  <> "" THEN KILL "p1"
    IF DIR$("p2")  <> "" THEN KILL "p2"
    IF DIR$("p3")  <> "" THEN KILL "p3"
    IF DIR$("p4")  <> "" THEN KILL "p4"
    IF DIR$("p5")  <> "" THEN KILL "p5"
    IF DIR$("p6")  <> "" THEN KILL "p6"
    IF DIR$("p7")  <> "" THEN KILL "p7"
    IF DIR$("p8")  <> "" THEN KILL "p8"
    IF DIR$("p9")  <> "" THEN KILL "p9"
    IF DIR$("p10") <> "" THEN KILL "p10"
    IF DIR$("p11") <> "" THEN KILL "p11"
    IF DIR$("p12") <> "" THEN KILL "p12"
    IF DIR$("p13") <> "" THEN KILL "p13"
    IF DIR$("p14") <> "" THEN KILL "p14"
    IF DIR$("p15") <> "" THEN KILL "p15"
    IF DIR$("p16") <> "" THEN KILL "p16"

END SUB 'Plot2DIndividualProbePaths()

'----------------------------------------------

SUB Plot3DbestProbePaths(NumPaths&&,W(),R(),Np&&,N6&&,LastIteration&&,FunctionName$) 'XYZZY

LOCAL TrajectoryNumber&&, ProbeNumber&&, StepNumber&&, N&&, m&&, ProcID???

LOCAL MaximumFitness, MinimumFitness AS EXT

LOCAL BestFitnessThisStep&&()

LOCAL BestFitnessThisStep(), TempFitness() AS EXT

LOCAL Annotation$, xCoord$, yCoord$, zCoord$, GnuPlotEXE$, PlotWithLines$

    Annotation$     = ""

    PlotWithLines$  = "NO" '"YES" '"NO"

    NumPaths&& = MIN(Np&&,NumPaths&&)

    GnuPlotEXE$ = "wgnuplot.exe"
' ------------- Get Min/Max Fitnesses ----------------
    MaximumFitness = W(1,0) : MinimumFitness = W(1,0)  'Note:  H(p&&,j&&)

    FOR StepNumber&& = 0 TO LastIteration&&

        FOR ProbeNumber&& = 1 TO Np&&

            IF W(ProbeNumber&&,StepNumber&&) >= MaximumFitness THEN MaximumFitness = W(ProbeNumber&&,StepNumber&&)

            IF W(ProbeNumber&&,StepNumber&&) <= MinimumFitness THEN MinimumFitness = W(ProbeNumber&&,StepNumber&&)

        NEXT ProbeNumber&&

    NEXT StepNumber&&

' ------------- Copy Fitness Array W() into TempFitness to Preserve W() ----------------
    REDIM TempFitness(1 TO Np&&, 0 TO LastIteration&&)

    FOR StepNumber&& = 0 TO LastIteration&&

        FOR ProbeNumber&& = 1 TO Np&&

            TempFitness(ProbeNumber&&,StepNumber&&) = W(ProbeNumber&&,StepNumber&&)

        NEXT ProbeNumber&&

    NEXT StepNumber&&

' ----------- LOOP ON Paths -----------

    FOR TrajectoryNumber&& = 1 TO NumPaths&&

' -------------- Get Trajectory Coordinate Data ----------------
        REDIM BestFitnessThisStep(0 TO LastIteration&&), BestProbeThisStep&&(0 TO LastIteration&&)

        FOR StepNumber&& = 0 TO LastIteration&&

            BestFitnessThisStep(StepNumber&&) = TempFitness(1,StepNumber&&)

            FOR ProbeNumber&& = 1 TO Np&&

                IF TempFitness(ProbeNumber&&,StepNumber&&) >= BestFitnessThisStep(StepNumber&&) THEN

                    BestFitnessThisStep(StepNumber&&) = TempFitness(ProbeNumber&&,StepNumber&&)

                    BestProbeThisStep&&(StepNumber&&)  = ProbeNumber&&

                END IF

            NEXT ProbeNumber&&

        NEXT StepNumber&&

' ----- Create Plot Data File -----

        N&& = FREEFILE

        SELECT CASE TrajectoryNumber&&

            CASE 1  : OPEN "p1"  FOR OUTPUT AS #N&&
            CASE 2  : OPEN "p2"  FOR OUTPUT AS #N&&
            CASE 3  : OPEN "p3"  FOR OUTPUT AS #N&&
            CASE 4  : OPEN "p4"  FOR OUTPUT AS #N&&
            CASE 5  : OPEN "p5"  FOR OUTPUT AS #N&&
            CASE 6  : OPEN "p6"  FOR OUTPUT AS #N&&
            CASE 7  : OPEN "p7"  FOR OUTPUT AS #N&&
            CASE 8  : OPEN "p8"  FOR OUTPUT AS #N&&
            CASE 9  : OPEN "p9"  FOR OUTPUT AS #N&&
            CASE 10 : OPEN "p10" FOR OUTPUT AS #N&&

        END SELECT
```

```
'  ------------ Write Plot File Data ------------

    FOR StepNumber&& = 0 TO LastIteration&&

        PRINT #N&&, USING("######.######## ########.######## ########.########
######.########",K(BestProbeThisStep&&(StepNumber&&),1,StepNumber&&),K(BestProbeThisStep&&(StepNumber&&),2,StepNumber&&),K(BestProbeThisStep&&(StepNumber&&),
3,StepNumber&&))+CHR$(13)

        TempFitness(BestProbeThisStep&&(StepNumber&&),StepNumber&&) = MinimumFitness 'so that same max will not be found for next trajectory

    NEXT StepNumber&&

    CLOSE #N&&

NEXT TrajectoryNumber&&
'  ------------------------- Plot Paths -------------------------
'CALL CreateGNUpLotINIFile(0.1##*ScreenWidth&,0.25##*ScreenHeight&,0.6##*ScreenWeight&,0.6##*ScreenWeight&)

Annotation$ = ""

N&& = FREEFILE

OPEN "cmd3d.gp" FOR OUTPUT AS #N&&

    PRINT #N&&, "set pm3d"
    PRINT #N&&, "show pm3d"
    PRINT #N&&, "set hidden3d"
    PRINT #N&&, "set view 45, 45, 1, 1"

    PRINT #N&&, "unset colorbox"

    PRINT #N&&, "set xrange [" + REMOVE$(STR$(X(Win(1)),ANY"" ) + "." + REMOVE$(STR$(X(XMax(1)),ANY"" ) + "]"
    PRINT #N&&, "set yrange [" + REMOVE$(STR$(X(Win(2)),ANY"" ) + "." + REMOVE$(STR$(X(XMax(2)),ANY"" ) + "]"
    PRINT #N&&, "set zrange [" + REMOVE$(STR$(X(Win(3)),ANY"" ) + "." + REMOVE$(STR$(X(XMax(3)),ANY"" ) + "]"

    PRINT #N&&, "set grid xtics ytics ztics"
    PRINT #N&&, "show grid"
    PRINT #N&&, "set title " + Quote$ + "3D " + FunctionName$ + " SAMPLE POINT PATHS" + "\n" + RunID$ + Quote$
    PRINT #N&&, "set xlabel " + Quote$ + "x1"               + Quote$
    PRINT #N&&, "set ylabel " + Quote$ + "x2"               + Quote$
    PRINT #N&&, "set zlabel " + Quote$ + "x3"               + Quote$

    IF PlotWithLines$ = "YES" THEN

        SELECT CASE NumPaths&&

            CASE 1  : PRINT #N&&, "splot "+Quote$+"p1"+Quote$+" w l lw 3"
            CASE 2  : PRINT #N&&, "splot "+Quote$+"p1"+Quote$+" w l lw 3,"+Quote$+"p2"+Quote$+" w l"
            CASE 3  : PRINT #N&&, "splot "+Quote$+"p1"+Quote$+" w l lw 3,"+Quote$+"p2"+Quote$+" w l,"+Quote$+"p3"+Quote$+" w l"
            CASE 4  : PRINT #N&&, "splot "+Quote$+"p1"+Quote$+" w l lw 3,"+Quote$+"p2"+Quote$+" w l,"+Quote$+"p3"+Quote$+" w l,"+Quote$+"p4"+Quote$+" w l"
            CASE 5  : PRINT #N&&, "splot "+Quote$+"p1"+Quote$+" w l lw 3,"+Quote$+"p2"+Quote$+" w l,"+Quote$+"p3"+Quote$+" w l,"+Quote$+"p4"+Quote$+" w
l,"+Quote$+"p5"+Quote$+" w l"
            CASE 6  : PRINT #N&&, "splot "+Quote$+"p1"+Quote$+" w l lw 3,"+Quote$+"p2"+Quote$+" w l,"+Quote$+"p3"+Quote$+" w l,"+Quote$+"p4"+Quote$+" w
l,"+Quote$+"p5"+Quote$+" w l,"+Quote$+"p6"+Quote$+" w l,"
            CASE 7  : PRINT #N&&, "splot "+Quote$+"p1"+Quote$+" w l lw 3,"+Quote$+"p2"+Quote$+" w l,"+Quote$+"p3"+Quote$+" w l,"+Quote$+"p4"+Quote$+" w
l,"+Quote$+"p5"+Quote$+" w l,"+Quote$+"p6"+Quote$+" w l,"+Quote$+"p7"+Quote$+" w l,"
            CASE 8  : PRINT #N&&, "splot "+Quote$+"p1"+Quote$+" w l lw 3,"+Quote$+"p2"+Quote$+" w l,"+Quote$+"p3"+Quote$+" w l,"+Quote$+"p4"+Quote$+" w
l,"+Quote$+"p5"+Quote$+" w l,"+Quote$+"p6"+Quote$+" w l,"+Quote$+"p7"+Quote$+" w l,"+Quote$+"p8"+Quote$+" w l"
            CASE 9  : PRINT #N&&, "splot "+Quote$+"p1"+Quote$+" w l lw 3,"+Quote$+"p2"+Quote$+" w l,"+Quote$+"p3"+Quote$+" w l,"+Quote$+"p4"+Quote$+" w
l,"+Quote$+"p5"+Quote$+" w l,"+Quote$+"p6"+Quote$+" w l,"+Quote$+"p7"+Quote$+" w l,"+Quote$+"p8"+Quote$+" w l,"+Quote$+"p9"+Quote$+" w l"
            CASE 10 : PRINT #N&&, "splot "+Quote$+"p1"+Quote$+" w l lw 3,"+Quote$+"p2"+Quote$+" w l,"+Quote$+"p3"+Quote$+" w l,"+Quote$+"p4"+Quote$+" w
l,"+Quote$+"p5"+Quote$+" w l,"+Quote$+"p6"+Quote$+" w l,"+Quote$+"p7"+Quote$+" w l,"+Quote$+"p8"+Quote$+" w l,"+Quote$+"p9"+Quote$+" w l,"+Quote$+"p10"+Quote$+" w l"

        END SELECT

    ELSE

        SELECT CASE NumPaths&&

            CASE 1  : PRINT #N&&, "splot "+Quote$+"p1"+Quote$+" lw 2"
            CASE 2  : PRINT #N&&, "splot "+Quote$+"p1"+Quote$+" lw 2,"+Quote$+"p2"+Quote$+" ."+Quote$+"p3"+Quote$
            CASE 3  : PRINT #N&&, "splot "+Quote$+"p1"+Quote$+" lw 2,"+Quote$+"p2"+Quote$+" ."+Quote$+"p3"+Quote$+" ."+Quote$+"p4"+Quote$+"
."+Quote$+"p5"+Quote$
            CASE 4  : PRINT #N&&, "splot "+Quote$+"p1"+Quote$+" lw 2,"+Quote$+"p2"+Quote$+" ."+Quote$+"p3"+Quote$+" ."+Quote$+"p4"+Quote$+"
."+Quote$+"p5"+Quote$+" ."+Quote$+"p6"+Quote$
            CASE 5  : PRINT #N&&, "splot "+Quote$+"p1"+Quote$+" lw 2,"+Quote$+"p2"+Quote$+" ."+Quote$+"p3"+Quote$+" ."+Quote$+"p4"+Quote$+"
."+Quote$+"p5"+Quote$+" ."+Quote$+"p6"+Quote$
            CASE 6  : PRINT #N&&, "splot "+Quote$+"p1"+Quote$+" lw 2,"+Quote$+"p2"+Quote$+" ."+Quote$+"p3"+Quote$+" ."+Quote$+"p4"+Quote$+"
."+Quote$+"p5"+Quote$+" ."+Quote$+"p6"+Quote$
                      Quote$+"p7"+Quote$+" ."
            CASE 7  : PRINT #N&&, "splot "+Quote$+"p1"+Quote$+" lw 2,"+Quote$+"p2"+Quote$+" ."+Quote$+"p3"+Quote$+" ."+Quote$+"p4"+Quote$+"
."+Quote$+"p5"+Quote$+" ."+Quote$+"p6"+Quote$
                      Quote$+"p7"+Quote$+" ."
            CASE 8  : PRINT #N&&, "splot "+Quote$+"p1"+Quote$+" lw 2,"+Quote$+"p2"+Quote$+" ."+Quote$+"p3"+Quote$+" ."+Quote$+"p4"+Quote$+"
."+Quote$+"p5"+Quote$+" ."+Quote$+"p6"+Quote$
                      Quote$+"p7"+Quote$+" ."    Quote$+"p8"+Quote$
            CASE 9  : PRINT #N&&, "splot "+Quote$+"p1"+Quote$+" lw 2,"+Quote$+"p2"+Quote$+" ."+Quote$+"p3"+Quote$+" ."+Quote$+"p4"+Quote$+"
."+Quote$+"p5"+Quote$+" ."+Quote$+"p6"+Quote$
                      Quote$+"p7"+Quote$+" ."    Quote$+"p8"+Quote$+" ."+Quote$+"p9"+Quote$
            CASE 10 : PRINT #N&&, "splot "+Quote$+"p1"+Quote$+" lw 2,"+Quote$+"p2"+Quote$+" ."+Quote$+"p3"+Quote$+" ."+Quote$+"p4"+Quote$+"
."+Quote$+"p5"+Quote$+" ."+Quote$+"p6"+Quote$
                      Quote$+"p7"+Quote$+" ."    Quote$+"p8"+Quote$+" ."+Quote$+"p9"+Quote$+" ."+Quote$+"p10"+Quote$

        END SELECT

    END IF

    CLOSE #N&&

    ProcID777 = SHELL(GnuPlotEXE$+" cmd3d.gp -") : CALL Delay(0.5#)
'  ------------ CLEANUP ------------
    IF DIR$("p1") <> "" THEN KILL "p1"
    IF DIR$("p2") <> "" THEN KILL "p2"
    IF DIR$("p3") <> "" THEN KILL "p3"
    IF DIR$("p4") <> "" THEN KILL "p4"
    IF DIR$("p5") <> "" THEN KILL "p5"
    IF DIR$("p6") <> "" THEN KILL "p6"
    IF DIR$("p7") <> "" THEN KILL "p7"
    IF DIR$("p8") <> "" THEN KILL "p8"
    IF DIR$("p9") <> "" THEN KILL "p9"
    IF DIR$("p10") <> "" THEN KILL "p10"

END SUB 'Plot3DbestProbePaths()
'------------------------------------
FUNCTION HasDAVGsaturated%(NavgSteps&,j&&,Np&&,Nd&&,M(),K(),DiagLength)

LOCAL A$

LOCAL k&&

LOCAL SumOfDavg, DavgStep3 AS EXT

LOCAL DavgSatTOL AS EXT

    A$ = "NO"

    DavgSatTOL = 0.0005## 'tolerance for DAVG saturation

    IF j&& < NavgSteps& + 10 THEN GOTO ExitHasDAVGsaturated 'execute at least 10 steps after averaging interval before performing this check

    DavgStep3 = DavgThisStep(j&&,Np&&,Nd&&,M(),K(),DiagLength)

    SumOfDavg = 0##

    FOR k&& = j&&-NavgSteps&+1 TO j&& 'check this step and previous (NavgSteps&-1) steps

        SumOfDavg = SumOfDavg + DavgThisStep(k&&,Np&&,Nd&&,M(),K(),DiagLength)

    NEXT k&&

    IF ABS(SumOfDavg/NavgStep$& - DavgStep3) <= DavgSatTOL THEN A$ = "YES" 'saturation if (avg value - last value) are within TOL

ExitHasDAVGsaturated:
```

```
        HasDAvGsaturated$ = A$

END FUNCTION 'HasDAvGsaturated$()
'--------------------------------

FUNCTION OscillationInDavg$(j&&,Np&&,Nd&&,M(),R(),DiagLength)

LOCAL A$

LOCAL k&&, NumSlopeChanges&&

        A$ = "NO"

        NumSlopeChanges&& = 0

        IF j&& < 15 THEN GOTO ExitOscillation 'wait at least 15 steps

        FOR k&& = j&&-10 TO j&&-1 'check previous ten steps

            IF (DavgThisStep(k&&,Np&&,Nd&&,M(),R(),DiagLength)-DavgThisStep(k&&-1,Np&&,Nd&&,M(),R(),DiagLength))* _
            - (DavgThisStep(k&&+1,Np&&,Nd&&,M(),R(),DiagLength)-DavgThisStep(k&&,Np&&,Nd&&,M(),R(),DiagLength)) < 0## THEN INCR NumSlopeChanges&&

        NEXT j&&

        IF NumSlopeChanges&& >= 3 THEN A$ = "YES"

ExitDavgOscillation:

        OscillationInDavg$ = A$

END FUNCTION 'OscillationInDavg()
'------

FUNCTION DavgThisStep(j&&,Np&&,Nd&&,M(),R(),DiagLength)

LOCAL BestFitness, TotalDistanceAllPoints, SumSQ AS EXT

LOCAL p&&, k&&, N&&, i&&, BestProbeNumber&&, BestIteration&&

'    ---------- Best Sample Point #, etc. -----------

        FOR k&& = 0 TO j&&

            BestFitness = M(1,k&&)

            FOR p&& = 1 TO Np&&

                IF M(p&&,k&&) >= BestFitness THEN

                    BestFitness = M(p&&,k&&) : BestProbeNumber&& = p&& : BestIteration&& = k&&

                END IF

            NEXT p&& 'probe #

        NEXT k&& 'time step

'    --------- Average Distance to Best Sample Point ----------

        TotalDistanceAllPoints = 0##

        FOR p&& = 1 TO Np&&

            SumSQ = 0##

            FOR i&& = 1 TO Nd&&

                SumSQ = SumSQ + (R(BestProbeNumber&&,i&&,BestIteration&&)-R(p&&,i&&,j&&))^2 'do not exclude best prove p&&=BestProbeNumber&&(j&&) from sum because it adds
zero

            NEXT i&&

            TotalDistanceAllPoints = TotalDistanceAllPoints + SQR(SumSQ)

        NEXT p&&

        DavgThisStep = TotalDistanceAllPoints/(DiagLength*(Np&&-1)) 'but exclude best prove from average

END FUNCTION 'DavgThisStep()
'-----------

SUB
PlotBestFitnessEvolution(Nd&&,Np&&,LastIteration&&,G,DeltaT,Alpha,Beta,Frep,Mbest(),PlaceInitialPoints$,InitialAcceleration$,RepositionFactor$,FunctionName$,
Gamma)

LOCAL BestFitness(), GlobalBestFitness AS EXT

LOCAL PlotAnnotation$, PlotTitle$, A$

LOCAL p&&, j&&, N&&

        REDIM BestFitness(0 TO LastIteration&&)

        CALL
GetPlotAnnotation(PlotAnnotation$,Nd&&,Np&&,LastIteration&&,G,DeltaT,Alpha,Beta,Frep,Mbest(),PlaceInitialPoints$,InitialAcceleration$,RepositionFactor$,Funct
ionName$,Gamma)

        GlobalBestFitness = -1E4200 'Mbest(1,0)

        FOR j&& = 0 TO LastIteration&&

'            BestFitness(j&&) = Mbest(1,j&&) 'orig code 03-23-2010

            BestFitness(j&&) = -1E4200 'added 03-23-2010

            FOR p&& = 1 TO Np&&

                IF Mbest(p&&,j&&) >= BestFitness(j&&)  THEN BestFitness(j&&)  = Mbest(p&&,j&&)

                IF Mbest(p&&,j&&) >= GlobalBestFitness THEN GlobalBestFitness = Mbest(p&&,j&&)

            NEXT p&& 'probe #

        NEXT j&& 'time step

        N&& = FREEFILE

        OPEN "fitness" FOR OUTPUT AS #N&&

            FOR j&& = 0 TO LastIteration&&

                PRINT #N&&, USING("##### ######.######",j&&,BestFitness(j&&))

            NEXT j&&

        CLOSE #N&&

        IF  GlobalBestFitness = 0## THEN 'format fitness number
            A$ = "0"
        ELSEIF GlobalBestFitness < 0## AND GlobalBestFitness > -1## THEN
            A$ = "-0"+REMOVE$(STR$(ROUND(ABS(GlobalBestFitness),10)),ANY"")
        ELSE
            A$ = REMOVE$(STR$(ROUND(GlobalBestFitness,10)),ANY"")
        END IF

        PlotAnnotation$ = PlotAnnotation$ + "Best Fitness = " + A$

        PlotTitle$ = "Best Fitness vs Iteration #\n" + "[" + REMOVE$(STR$(Np&&),ANY" ") + " sample points, "+ _
            REMOVE$(STR$(LastIteration&&),ANY" ")+" iterations, "+ISPDtype$+" ISPD"

        CALL CreateKuplotINIFile(0.1##*Screenwidth&,0.1##*Screenheight&,0.6##*Screenwidth&,0.6##*Screenheight&)

        CALL TwoDplot("fitness","best fitness","0.7","0.8","Iteration #\n\n.","","\nbest fitness","" _
            ,"","","",","","wgnuplot.exe"," with lines linewidth 2",PlotAnnotation$)
```

```basic
END SUB 'PlotBestFitnessEvolution()

'------
SUB
PlotAverageDistance(Nd&&,Np&&,LastIteration&&,G,DeltaT,Alpha,Beta,Frep,Hbest(),PlaceInitialPoints$,InitialAcceleration$,RepositionFactor$,FunctionName$,R(),D
iagLength$,Gamma)

LOCAL Davg(), BestFitness(), TotalDistanceAllPoints, SumSQ AS EXT

LOCAL PlotAnnotation$, PlotTitle$

LOCAL p&&, j&&, N&&, i&&, BestProbeNumber&&(), BestIteration&&()

    REDIM Davg(0 TO LastIteration&&), BestFitness(0 TO LastIteration&&), BestProbeNumber&&(0 TO LastIteration&&), BestIteration&&(0 TO LastIteration&&)

    CALL
GetPlotAnnotation(PlotAnnotation$,Nd&&,Np&&,LastIteration&&,G,DeltaT,Alpha,Beta,Frep,Hbest(),PlaceInitialPoints$,InitialAcceleration$,RepositionFactor$,Funct
ionName$,Gamma)

'   ---------- Best Sample Point #, etc. ----------

    FOR j&& = 0 TO LastIteration&&

        BestFitness(j&&) = Hbest(1,j&&)

        FOR p&& = 1 TO Np&&

            IF Hbest(p&&,j&&) >= BestFitness(j&&) THEN

                BestFitness(j&&) = Hbest(p&&,j&&) : BestProbeNumber&&(j&&) = p&& : BestIteration&&(j&&) = j&& 'only probe number is used at this time, but
other data are computed for possible future use.

            END IF

        NEXT p&& 'probe #

    NEXT j&& 'time step

    N&& = FREEFILE

'   --------- Average Distance to Best Sample Point -----------

    FOR j&& = 0 TO LastIteration&&

        TotalDistanceAllPoints = 0##

        FOR p&& = 1 TO Np&&

            SumSQ = 0##

            FOR i&& = 1 TO Nd&&

                SumSQ = SumSQ + (R(BestProbeNumber&&(j&&),i&&,j&&)-R(p&&,i&&,j&&))^2 'do not exclude p&&=BestProbeNumber&&(j&&) from sum because it adds zero

            NEXT i&&

            TotalDistanceAllPoints = TotalDistanceAllPoints + SQR(SumSQ)

        NEXT p&&

        Davg(j&&) = TotalDistanceAllPoints/(DiagLength$*Np&&-1)) 'but exclude best prove from average

    NEXT j&&

'   ----------- Create Plot Data File -----------

    OPEN "Davg" FOR OUTPUT AS #N&&

        FOR j&& = 0 TO LastIteration&&

            PRINT #N&&, USING$("###### ######.######",j&&,Davg(j&&))

        NEXT j&&

    CLOSE #N&&

    PlotTitle$ = "Average Distance of " + REMOVE$(STR$(Np&&-1),ANY" ") + " Points to Best Sample Point\nNormalized to Size of Decision Space\n" + _
                 "[" + REMOVE$(STR$(Np&&),ANY" ") + " sample points, " + REMOVE$(STR$(LastIteration&&),ANY" ") + " iterations], "+ISPDtype$+" ISPD"

    CALL CreateGnuplotINIFile(0.2##*Screenwidth&,0.2##*Screenwidth&,0.6##*Screenheight&)

    CALL TwoDplot("Davg",PlotTitle$,"0.7","0.9","Iteration #\n\n","","\n\n<b>/Ldiag","","","","0","","wgnuplot.exe"," with lines linewidth 2",PlotAnnotation$)

END SUB 'PlotAverageDistance()

'------
SUB
PlotBestProbevsIteration(Nd&&,Np&&,LastIteration&&,G,DeltaT,Alpha,Beta,Frep,Hbest(),PlaceInitialPoints$,InitialAcceleration$,RepositionFactor$,FunctionName$,
Gamma)

LOCAL BestFitness AS EXT

LOCAL PlotAnnotation$, PlotTitle$

LOCAL p&&, j&&, N&&, BestProbeNumber&&()

    REDIM BestProbeNumber&&(0 TO LastIteration&&)

    CALL
GetPlotAnnotation(PlotAnnotation$,Nd&&,Np&&,LastIteration&&,G,DeltaT,Alpha,Beta,Frep,Hbest(),PlaceInitialPoints$,InitialAcceleration$,RepositionFactor$,Funct
ionName$,Gamma)

    FOR j&& = 0 TO LastIteration&&

        BestFitness = Hbest(1,j&&)

        FOR p&& = 1 TO Np&&

            IF Hbest(p&&,j&&) >= BestFitness THEN

                BestFitness = Hbest(p&&,j&&) : BestProbeNumber&&(j&&) = p&&

            END IF

        NEXT p&& 'probe #

    NEXT j&& 'time step

    N&& = FREEFILE

    OPEN "Best Sample Point" FOR OUTPUT AS #N&&

        FOR j&& = 0 TO LastIteration&&

            PRINT #N&&, USING$("###### #####",j&&,BestProbeNumber&&(j&&))

        NEXT j&&

    CLOSE #N&&

    PlotTitle$ = "Best Sample Point # vs Iteration #\n" + "[" +REMOVE$(STR$(Np&&),ANY" ") + " sample points, " + _
                 REMOVE$(STR$(LastIteration&&),ANY" ") + " iterations], "+ISPDtype$+" ISPD"

    CALL CreateGnuplotINIFile(0.15#*Screenwidth&,0.15##*Screenwidth&,0.6##*Screenwidth&,0.6##*Screenheight&)

'USAGE: CALL
TwoDplot(PlotFileName$,PlotTitle$,xCoord$,yCoord$,XaxisLabel$,YaxisLabel$,LogXaxis$,LogYaxis$,xMin$,xMax$,yMin$,yMax$,xTics$,yTics$,GnuPlotEXE$,LineType$,Ann
otation$)

    CALL TwoDplot("Best Sample Point",PlotTitle$,"0.7","0.9","Iteration #\n\n","","\n\nBest Sample Point
#","","","","0",NoSpaces$(Np&&+1,0),"","","wgnuplot.exe"," pt 8 ps .5 lw 1",PlotAnnotation$)"pt, pointtype; ps, pointsize; lw, linewidth

END SUB 'PlotBestProbevsIteration()
```

```
'----------------------------------
SUB
GetPlotAnnotation(PlotAnnotation$,Nd&&,Np&&,LastIteration&&,G,DeltaT,Alpha,Beta,Frep,Hbest(),PlaceInitialPoints$,InitialAcceleration$,RepositionFactor$,Funct
ionname$,Gamma)

'LOCAL A$

'    A$ = "" : IF PlaceInitialPoints$ = "UNIFORM ON-AXIS" AND Nd&& > 1 THEN A$ = "("+REMOVE$(STR$(Np&&/Nd&&),ANY " ") + "/axis"

    PlotAnnotation$ = VSOversion$ + "\n" +_
                      RunID$ + "\n" +_
                      FunctionName$ + " Function" + " ("+ FormatInteger$(Nd&&) + "-D) \n" +_
                      FormatInteger$(Np&&) + " sample points \n"          + A$ + "\n" +_
                      "G = " + FormatFP$(G,2)                             + "\n" +_
                      "Alpha = "    + FormatFP$(Alpha,1)                   + "\n" +_
                      "Beta = "     + FormatFP$(Beta,1)                    + "\n" +_
                      "DelT = "     + FormatFP$(DeltaT,1)                  + "\n" +_
                      "Gamma = "    + FormatFP$(Gamma,3)                   + "\n" +_
                      "Init Points "+ PlaceInitialPoints$                 + "\n" +_
                      "Init Accel " + InitialAcceleration$                + "\n" +_
                      "Frep "       + RepositionFactor$ + "\n"

END SUB 'GetPlotAnnotation()
'-------------------------

FUNCTION FormatInteger$(m&&) : FormatInteger$ = REMOVE$(STR$(m&&),ANY " ") : END FUNCTION
'-------------------------

FUNCTION FormatFP$(X,Ndigits&&)

LOCAL A$

    IF X = O## THEN

        A$ = "0. " : GOTO ExitFormatFP

    END IF

    A$ = REMOVE$(STR$(ROUND(ABS(X),Ndigits&&)),ANY " ")

    IF ABS(X) < 1## THEN

        IF X > O## THEN

            A$ = "0" + A$

        ELSE

            A$ = "-0" + A$

        END IF

    ELSE

        IF X < O## THEN A$ = "-" + A$

    END IF

ExitFormatFP:

    FormatFP$ = A$

END FUNCTION
'-----------

SUB InitialProbeDistribution(Np&&,Nd&&,Niter&&,R(),PlaceInitialPoints$,Gamma)

LOCAL DeltaXi, DelX1, DelX2, DI AS EXT

LOCAL NumPointsPerDimension&&, p&&, i&&, k&&, NumX1points&&, NumX2points&&, x1pointNum&&, x2pointNum&&, A$

    SELECT CASE PlaceInitialPoints$

        CASE "UNIFORM ON-AXIS"

            IF Nd&& > 1 THEN

                NumPointsPerDimension&& = Np&&\Nd&& 'even #

            ELSE

                NumPointsPerDimension&& = Np&&

            END IF

            FOR i&& = 1 TO Nd&&

                FOR p&& = 1 TO Np&&

                    R(p&&,i&&) = XiWin(i&&) + Gamma*(XiMax(i&&)-XiWin(i&&))

                NEXT p&&

            NEXT i&&

            FOR i&& = 1 TO Nd&& 'place sample points probe line-by-probe line (i&& is dimension number)

                DeltaXi = (XiMax(i&&)-XiWin(i&&))/(NumPointsPerDimension&&-1)

                FOR k&& = 1 TO NumPointsPerDimension&&

                    p&& = k&& + NumPointsPerDimension&&^(i&&-1) 'probe #

                    R(p&&,i&&,0) = XiWin(i&&) + (k&&-1)*DeltaXi

                NEXT k&&

            NEXT i&&

        CASE "UNIFORM ON-DIAGONAL"

            FOR p&& = 1 TO Np&&

                FOR i&& = 1 TO Np&&

                    DeltaXi = (XiMax(i&&)-XiWin(i&&))/(Np&&-1)

                    R(p&&,i&&,0) = XiWin(i&&) + (p&&-1)*DeltaXi

                NEXT i&&

            NEXT p&&

        CASE "2D GRID"

            NumPointsPerDimension&& = SQR(Np&&) : NumX1points&& = NumPointsPerDimension&& : NumX2points&& = NumX1points&& 'broken down for possible future
use

            DelX1 = (XiMax(1)-XiWin(1))/(NumX1points&&-1)

            DelX2 = (XiMax(2)-XiWin(2))/(NumX2points&&-1)

            FOR x1pointNum&& = 1 TO NumX1points&&

                FOR x2pointNum&& = 1 TO NumX2points&&

                    p&& = NumX1points&&*(x1pointNum&&-1)+x2pointNum&& 'probe #

                    R(p&&,1,0) = XiWin(1) + DelX1*(x1pointNum&&-1) 'x1 coord
                    R(p&&,2,0) = XiWin(2) + DelX2*(x2pointNum&&-1) 'x2 coord

                NEXT x2pointNum&&
```

```
                    NEXT xIpointNum&&
            CASE "RANDOM"
                    FOR p&& = 1 TO Np&&
                        FOR i&& = 1 TO Nd&&
                            R(p&&,i&&,0) = XiWin(i&&) + RandomNum(0##,1##)*(XiMax(i&&)-XiWin(i&&))
                        NEXT i&&
                    NEXT p&&
            END SELECT
END SUB 'InitialProbeDistribution()
'------
SUB
ChangeRunParameters(NumPointsPerDimension&&,Np&&,Nd&&,Niter&&,G,Alpha,Beta,DeltaT,Frep,PlaceInitialPoints$,InitialAcceleration$,RepositionFactor$,FunctionNam
e$) 'THIS PROCEDURE NOT USED

LOCAL A$, DefaultValue$

        A$ = INPUTBOX$("# dimensions?","Change # Dimensions ("+FunctionName$+")",NoSpaces$(Nd&&,0,0)) : Nd&&    = VAL(A$) : IF Nd&& < 1 OR Nd&& > 500 THEN Nd&& =
2
        IF Nd&& = 1 THEN NumPointsPerDimension&& = 2*((NumPointsPerDimension&&+1)\2)   'require an even # sample points on each probe line to avoid overlapping at
origin (in symmetrical spaces at least...)
        IF Nd&& = 1 THEN NumPointsPerDimension&& = MAX(NumPointsPerDimension&&,3)   'at least 3 sample points on x-axis for 1-D functions
        Np&& = NumPointsPerDimension&&^Nd&&
        A$ = INPUTBOX$("# iterations?","Change # Steps ("+FunctionName$+")",NoSpaces$(Niter&&,0,0)) : Niter&&    = VAL(A$) : IF Niter&& < 3
THEN Niter&& = 50
        A$ = INPUTBOX$("Grav Const G?","Change G ("+FunctionName$+")",NoSpaces$(G,2))           : G     = VAL(A$) : IF G < -100## OR G > 100##    THEN G =
2##
        A$ = INPUTBOX$("Alpha?","Change Alpha ("+FunctionName$+")",NoSpaces$(Alpha,2))          : Alpha = VAL(A$) : IF Alpha < -50## OR Alpha > 50## THEN Alpha
= 2##
        A$ = INPUTBOX$("Beta?","Change Beta ("+FunctionName$+")",NoSpaces$(Beta,2))             : Beta  = VAL(A$) : IF Beta < -50## OR Beta > 50## THEN Beta
= 2##
        A$ = INPUTBOX$("Delta T?","Change Delta-T ("+FunctionName$+")",NoSpaces$(DeltaT,2))     : DeltaT = VAL(A$) : IF DeltaT =< 0##                   THEN DeltaT
= 1##
        A$ = INPUTBOX$("Frep [0-1]?","Change Frep ("+FunctionName$+")",NoSpaces$(Frep,3))       : Frep  = VAL(A$) : IF Frep < 0## OR Frep > 1## THEN Frep =
0.5##

'  ----------- Initial Probe Distribution -----------
        SELECT CASE PlaceInitialPoints$
            CASE "UNIFORM ON-AXIS"       : DefaultValue$ = "1"
            CASE "UNIFORM ON-DIAGONAL"   : DefaultValue$ = "2"
            CASE "2D GRID"               : DefaultValue$ = "3"
            CASE "RANDOM"                : DefaultValue$ = "4"
        END SELECT
        A$ = INPUTBOX$("Initial Points?"+CHR$(13)+"1 - UNIFORM ON-AXIS"+CHR$(13)+"2 - UNIFORM ON-DIAGONAL"+CHR$(13)+"3 - 2D GRID"+CHR$(13)+"4 - RANDOM","Initial
Probe Distribution ("+FunctionName$+")",DefaultValue$)

        IF VAL(A$) < 1 OR VAL(A$) > 4 THEN A$ = "1"

        SELECT CASE VAL(A$)
            CASE 1 : PlaceInitialPoints$ = "UNIFORM ON-AXIS"
            CASE 2 : PlaceInitialPoints$ = "UNIFORM ON-DIAGONAL"
            CASE 3 : PlaceInitialPoints$ = "2D GRID"
            CASE 4 : PlaceInitialPoints$ = "RANDOM"
        END SELECT

        IF Nd&& = 1  AND PlaceInitialPoints$ = "UNIFORM ON-DIAGONAL" THEN PlaceInitialPoints$ = "UNIFORM ON-AXIS" 'cannot do diagonal in 1-D space

        IF Nd&& <> 2 AND PlaceInitialPoints$ = "2D GRID" THEN PlaceInitialPoints$ = "UNIFORM ON-AXIS" '2D grid is available only in 2 dimensions!

'  ----------- Initial Acceleration -----------------
        SELECT CASE InitialAcceleration$
            CASE "ZERO"   : DefaultValue$ = "1"
            CASE "FIXED"  : DefaultValue$ = "2"
            CASE "RANDOM" : DefaultValue$ = "3"
        END SELECT
        A$ = INPUTBOX$("Initial Acceleration?"+CHR$(13)+"1 - ZERO"+CHR$(13)+"2 - FIXED"+CHR$(13)+"3 - RANDOM","Initial Acceleration
("+FunctionName$+")",DefaultValue$)

        IF VAL(A$) < 1 OR VAL(A$) > 3 THEN A$ = "1"

        SELECT CASE VAL(A$)
            CASE 1 : InitialAcceleration$ = "ZERO"
            CASE 2 : InitialAcceleration$ = "FIXED"
            CASE 3 : InitialAcceleration$ = "RANDOM"
        END SELECT

'  ----------- Reposition Factor --------------
        SELECT CASE RepositionFactor$
            CASE "FIXED"    : DefaultValue$ = "1"
            CASE "VARIABLE" : DefaultValue$ = "2"
            CASE "RANDOM"   : DefaultValue$ = "3"
        END SELECT
        A$ = INPUTBOX$("Reposition Factor?"+CHR$(13)+"1 - FIXED"+CHR$(13)+"2 - VARIABLE"+CHR$(13)+"3 - RANDOM","Retrieve Points
("+FunctionName$+")",DefaultValue$)

        IF VAL(A$) < 1 OR VAL(A$) > 3 THEN A$ = "1"

        SELECT CASE VAL(A$)
            CASE 1 : RepositionFactor$ = "FIXED"
            CASE 2 : RepositionFactor$ = "VARIABLE"
            CASE 3 : RepositionFactor$ = "RANDOM"
        END SELECT

END SUB 'ChangeRunParameters()
'------
'----------
FUNCTION NoSpaces$(X,NumDigits&&) :  NoSpaces$ = REMOVE$(STR$(X,NumDigits&&),ANY" ") : END FUNCTION
'----------

FUNCTION TerminateNowForSaturation(j&&,Ns&&,Np&&,Niter&&,G,DeltaT,Alpha,Beta,R(),A(),H())

LOCAL A$, i&, p&&, NumStepsForAveraging&

LOCAL BestFitness, AvgFitness, FitnessTOL AS EXT  'terminate if avg fitness does not change over NumStepsForAveraging& iterations

        FitnessTOL = 0.00001## : NumStepsForAveraging& = 10

        A$ = "NO"

        IF j&& >= NumStepsForAveraging&*10 THEN 'wait until step 10 to start checking for fitness saturation

            AvgFitness = 0##

            FOR i& = j&&-NumStepsForAveraging&+1 TO j&& 'avg fitness over current step & previous NumStepsForAveraging&-1 steps

                BestFitness = H(1,i&)

                FOR p&& = 1 TO Np&&
                    IF H(p&&,i&) >= BestFitness THEN BestFitness = H(p&&,i&)
                NEXT p&&
```

```
                AvgFitness = AvgFitness + BestFitness
            NEXT i&
            AvgFitness = AvgFitness/NumStepsForAveraging&
            IF ABS(AvgFitness-BestFitness) < FitnessTOL THEN A$ = "YES" 'compare avg fitness to best fitness at this step
        END IF
        TerminateNowForSaturation$ = A$
END FUNCTION 'TerminateNowForSaturation$()
'----------
FUNCTION MagvectorN(V(),N&&) 'returns magnitude of Nx1 column vector V

LOCAL SumSQ AS EXT

LOCAL i&&
        SumSQ = 0## : FOR i&& = 1 TO N&& : SumSQ = SumSQ + V(i&&)^2 : NEXT i&& : Magvector = SQR(SumSQ)
END FUNCTION 'Magvector()
'---
FUNCTION UnitStep(X)

LOCAL Z AS EXT
        IF X < 0## THEN
            Z = 0##
        ELSE
            Z = 1##
        END IF
        UnitStep = Z
END FUNCTION 'UnitStep()
'---
'SUB Plot1Dfunction(FunctionName$,R()) 'plots 1D function on-screen
SUB Plot1Dfunction(FunctionName$) 'plots 1D function on-screen

LOCAL NumPoints&&, i&&, N&&

LOCAL DeltaX, X, R() AS EXT
        REDIM R(1 TO 1, 1 TO 1, 0 TO 0)
        NumPoints&& = 3200I
        DeltaX = (XiMax(1)-XiMin(1))/(NumPoints&&-1)
        N&& = FREEFILE
        SELECT CASE FunctionName$
            CASE "Parrott4" 'PARROTT F4 FUNCTION
                OPEN "Parrott4" FOR OUTPUT AS #N&&
                    FOR i&& = 1 TO NumPoints&&
                        R(1,1,0) = XiMin(1) + (i&&-1)*DeltaX
                        PRINT #N&&, USING$("#.###### #.######",R(1,1,0),ParrottF4(R()),1,1,0))
                    NEXT i&&
                CLOSE #N&&
                CALL CreateGNUplotINIfile(0.2##*ScreenWidth&,0.2##*ScreenHeight&,0.6##*ScreenWidth&,0.6##*ScreenHeight&)
                CALL TwoDplot("Parrott4","Parrott F4 Function","0.7","0.7","X\n\n.","-.\n\nParrott F4(x)","","","0","1","0","1","","","wgnuplot.exe"," with lines
linewidth 2","")
            END SELECT
END SUB
'------
SUB CLEANUP 'probe coordinate plot files
        IF DIR$("P1") <> "" THEN KILL "P1"
        IF DIR$("P2") <> "" THEN KILL "P2"
        IF DIR$("P3") <> "" THEN KILL "P3"
        IF DIR$("P4") <> "" THEN KILL "P4"
        IF DIR$("P5") <> "" THEN KILL "P5"
        IF DIR$("P6") <> "" THEN KILL "P6"
        IF DIR$("P7") <> "" THEN KILL "P7"
        IF DIR$("P8") <> "" THEN KILL "P8"
        IF DIR$("P9") <> "" THEN KILL "P9"
        IF DIR$("P10") <> "" THEN KILL "P10"
        IF DIR$("P11") <> "" THEN KILL "P11"
        IF DIR$("P12") <> "" THEN KILL "P12"
        IF DIR$("P13") <> "" THEN KILL "P13"
        IF DIR$("P14") <> "" THEN KILL "P14"
        IF DIR$("P15") <> "" THEN KILL "P15"
END SUB
'------
'SUB Plot2Dfunction(FunctionName$,R())
SUB Plot2Dfunction(FunctionName$)

LOCAL A$

LOCAL NumPoints&&, i&&, k&&, N&&

LOCAL DelX1, DelX2, Z, R() AS EXT
        REDIM R(1 TO 1, 1 TO 2, 0 TO 0)
        SELECT CASE FunctionName$
            CASE "PBM_1","PBM_2","PBM_3","PBM_4","PBM_5" : NumPoints&& = 25
            CASE ELSE : NumPoints&& = 100
        END SELECT
        N&& = FREEFILE : OPEN "TwoDplot.DAT" FOR OUTPUT AS #N&&
        DelX1 = (XiMax(1)-XiMin(1))/(NumPoints&&-1) : DelX2 = (XiMax(2)-XiMin(2))/(NumPoints&&-1)
        FOR i&& = 1 TO NumPoints&&
            R(1,1,0) = XiMin(1) + (i&&-1)*DelX1 'x1 value
            FOR k&& = 1 TO NumPoints&&
                R(1,2,0) = XiMin(2) + (k&&-1)*DelX2 'x2 value
                Z = ObjectiveFunction(R(),2,1,0,FunctionName$)
                PRINT #N&&, USING$("######.###### ######.###### ######.######^^^^",R(1,1,0),R(1,2,0),Z)
```

```
            NEXT k&&

            PRINT #k&&, ""

        NEXT i&&

        CLOSE #n&&

        CALL CreateGnuplotINIFile(0.1##*Screenwidth&&,0.1##*Screenwidth&&,0.6##*Screenwidth&&,0.6##*Screenwidth&&)

        A$ = "" : IF INSTR(Functionname$,"PBH_") > 0 THEN A$ = "Coarse "

        CALL ThreeDplot2C("TwoDplot.DAT",A$+"Plot of "+Functionname$+"_function","","0.6","0.6","1.2", _
                          "x1","x2","2=f(x1,x2)","","","WGnuplot.exe","", "", "")

    END SUB
    '------
    SUB TwoDplot3curves(NumCurves&&,PlotFilename1$,PlotFilename2$,PlotFilename3$,PlotTitle$,Annotation$,xCoord$,xaxisLabel$,YaxisLabel$, _
                        Logxaxis$,Logyaxis$,xMin$,xMax$,yMin$,yMax$,xTics$,yTics$,GnuPlotExE$)

        LOCAL n&&

        LOCAL LineSize$

        LineSize$ = "2"

        n&& = FREEFILE

        OPEN "cmd2d.gp" FOR OUTPUT AS #n&&

            IF Logxaxis$ = "YES" AND Logyaxis$ = "NO"  THEN PRINT #n&&, "set logscale x"
            IF Logxaxis$ = "NO" AND Logyaxis$ = "YES"  THEN PRINT #n&&, "set logscale y"
            IF Logxaxis$ = "YES" AND Logyaxis$ = "YES" THEN PRINT #n&&, "set logscale xy"

            IF xMin$ <> "" AND xMax$ <> "" THEN  PRINT #n&&, "set xrange ["+xMin$+":"+xMax$+"]"

            IF yMin$ <> "" AND yMax$ <> "" THEN  PRINT #n&&, "set yrange ["+yMin$+":"+yMax$+"]"

            PRINT #n&&, "set label "+Quote$+Annotation$+Quote$+" at graph "+xCoord$+","+yCoord$
            PRINT #n&&, "set grid xtics"
            PRINT #n&&, "set grid ytics"
            PRINT #n&&, "set xtics "+xTics$
            PRINT #n&&, "set ytics "+yTics$
            PRINT #n&&, "set grid mxtics"
            PRINT #n&&, "set grid mytics"
            PRINT #n&&, "set title "+Quote$+PlotTitle$+Quote$
            PRINT #n&&, "set xlabel "+Quote$+xaxisLabel$+Quote$
            PRINT #n&&, "set ylabel "+Quote$+yaxisLabel$+Quote$

            SELECT CASE NumCurves&&

            CASE 1
                PRINT #n&&, "plot " + Quote$ + PlotFilename1$ + Quote$ + " with lines linewidth " + LineSize$

            CASE 2
                PRINT #n&&, "plot " + Quote$ + PlotFilename1$ + Quote$ + " with lines linewidth " + LineSize$+", " + _
                           Quote$ + PlotFilename2$ + Quote$ + " with lines linewidth " + LineSize$

            CASE 3
                PRINT #n&&, "plot " + Quote$ + PlotFilename1$ + Quote$ + " with lines linewidth " + LineSize$+"," + _
                           Quote$ + PlotFilename2$ + Quote$ + " with lines linewidth " + LineSize$+"," + _
                           Quote$ + PlotFilename3$ + Quote$ + " with lines linewidth " + LineSize$

            END SELECT

            CLOSE #n&&

            SHELL(GnuPlotExE$+" cmd2d.gp ")

            CALL Delay(0.3)

    END SUB 'TwoDplot3curves()
    '---
FUNCTION Fibonacci&&(N&&) 'RETURNS Nth FIBONACCI NUMBER

    LOCAL i&&, Fn&&, Fn1&&, Fn2&&

    LOCAL A$

        IF N&& > 91 OR N&& < 0 THEN

            MSGBOX("ERROR! Fibonacci argument"+STR$(N&&)+" > 91. Out of range or < 0...") : EXIT FUNCTION

        END IF

        SELECT CASE N&&

        CASE 0: Fn&& = 1

        CASE ELSE

            Fn&& = 0 : Fn2&& = 1 : i&& = 0

            FOR i&& = 1 TO N&&

                Fn&& = Fn1&& + Fn2&&

                Fn2&& = Fn1&&

                Fn1&& = Fn&&

            NEXT i&& 'LOOP

        END SELECT

        Fibonacci&& = Fn&&

END FUNCTION 'Fibonacci&&()
'-----------
FUNCTION RandomNum(a,b) : RandomNum = a + (b-a)*RND : END FUNCTION 'Returns random number X, a=< X < b.
'
FUNCTION GaussianDeviate(Mu,Sigma) 'returns NORMAL (Gaussian) random deviate with mean Mu and standard deviation Sigma (variance = Sigma^2)

'Refs: (1) Press, W.H., Flannery, B.P., Teukolsky, S.A., and Vetterling, W.T., "Numerical Recipes: The Art of Scientific Computing,"
'               §7.2, Cambridge University Press, Cambridge, UK, 1986.
'       (2) Shinzato, T., "Box Muller Method," 2007, http://www.sp.dis.titech.ac.jp/~shinzato/boxmuller.pdf

    LOCAL s, t, Z AS EXT

        s = RND : t = RND

        Z = Mu + Sigma*SQR(-2##*LOG(s))*COS(TwoPi*t)

        GaussianDeviate = Z

END FUNCTION 'GaussianDeviate()
'-----------
    SUB ContourPlot(PlotFilename$,PlotTitle$,Annotation$,xCoord$,yCoord$,zCoord$, _
                    xaxisLabel$,yaxisLabel$,ZaxisLabel$,zMin$,zMax$,GnuPlotExE$,A$)

        LOCAL n&&

        n&& = FREEFILE

        OPEN "cmd3d.gp" FOR OUTPUT AS #n&&

            PRINT #n&&, "show surface"
            PRINT #n&&, "set hidden3d"
            IF zMin$ <> "" AND zMax$ <> "" THEN  PRINT #n&&, "set zrange ["+zMin$+":"+zMax$+"]"
            PRINT #n&&, "set label "+Quote$+Annotation$+Quote$+" at graph "+xCoord$+","+yCoord$+","+zCoord$
            PRINT #n&&, "show label"
```

```
            PRINT #N&&, "set grid xtics ytics ztics"
            PRINT #N&&, "show grid"
            PRINT #N&&, "set title "+Quote$+PlotTitle$+Quote$
            PRINT #N&&, "set label "+Quote$+xaxisLabel$+Quote$
            PRINT #N&&, "set ylabel "+Quote$+yaxisLabel$+Quote$
            PRINT #N&&, "set zlabel "+Quote$+zaxisLabel$+Quote$
            PRINT #N&&, "splot "+Quote$+PlotFilename$+Quote$+A$   '" notitle with linespoints "A$'" notitle with lines"
          CLOSE #N&&

          SHELL(GnuPlotEXE$+" cmd3d.gp -")

        END SUB 'ContourPlot()

'...

        SUB ThreeDplot(PlotFilename$,PlotTitle$,Annotation$,xCoord$,zCoord$,_
                       xaxisLabel$,yaxisLabel$,ZaxisLabel$,zMin$,zMax$,A$)

          LOCAL N&&, ProcessID???

          N&& = FREEFILE

          OPEN "cmd3d.gp" FOR OUTPUT AS #N&&

            PRINT #N&&, "set pm3d"
            PRINT #N&&, "show pm3d"
            IF zMin$ <> "" AND zMax$ <> "" THEN  PRINT #N&&, "set zrange ["+zMin$+":"+zMax$+"]"
            PRINT #N&&, "set label "+Annotation$+Quote$ at graph "+xCoord$+","+yCoord$+","+zCoord$
            PRINT #N&&, "show label"
            PRINT #N&&, "set grid xtics ytics ztics"
            PRINT #N&&, "show grid"
            PRINT #N&&, "set title "+Quote$+PlotTitle$+Quote$
            PRINT #N&&, "set ylabel "+Quote$+yaxisLabel$+Quote$
            PRINT #N&&, "set zlabel "+Quote$+zaxisLabel$+Quote$
            PRINT #N&&, "splot "+Quote$+PlotFilename$+Quote$+A$+" notitle"' with lines"
          CLOSE #N&&

          SHELL(GnuPlotEXE$+" cmd3d.gp -") : CALL Delay(0.5##)

        END SUB 'ThreeDplot()

'...

        SUB ThreeDplot2(PlotFilename$,PlotTitle$,Annotation$,yCoord$,zCoord$,_
                        xaxisLabel$,yaxisLabel$,ZaxisLabel$,zMin$,zMax$,GnuPlotEXE$,A$,xStart$,yStart$,yStop$)

          LOCAL N&&

          N&& = FREEFILE

          OPEN "cmd3d.gp" FOR OUTPUT AS #N&&

            PRINT #N&&, "set pm3d"
            PRINT #N&&, "show pm3d"
            PRINT #N&&, "set hidden3d"
            PRINT #N&&, "set view 45, 45, 1, 1"

            IF zMin$ <> "" AND zMax$ <> "" THEN  PRINT #N&&, "set zrange ["+zMin$+":"+zMax$+"]"

            PRINT #N&&, "set xrange [" + xStart$ + ":" + xStop$ + "]"
            PRINT #N&&, "set yrange [" + yStart$ + ":" + yStop$ + "]"
            PRINT #N&&, "set label "  + Annotation$ + Quote$ at graph "+xCoord$+","+yCoord$+","+zCoord$
            PRINT #N&&, "show label"
            PRINT #N&&, "set grid xtics ytics ztics"
            PRINT #N&&, "set title " + Quote$+PlotTitle$  + Quote$
            PRINT #N&&, "set ylabel " + Quote$+yaxisLabel$ + Quote$
            PRINT #N&&, "set zlabel " + Quote$+zaxisLabel$ + Quote$
            PRINT #N&&, "splot "      + Quote$+PlotFilename$ + Quote$ + A$ + " notitle with lines"
          CLOSE #N&&

          SHELL(GnuPlotEXE$+" cmd3d.gp -")

          IF DIR$(PlotFilename$) <> "" THEN KILL PlotFilename$ 'CLEANUP

        END SUB 'ThreeDplot2()

'...

        SUB TwoDplot2Curves(PlotFilename1$,PlotFilename2$,PlotTitle$,Annotation$,xCoord$,yCoord$,XaxisLabel$,yaxisLabel$, _
                            LogXaxis$,LogYaxis$,xMin$,xMax$,yMin$,yMax$,xTics$,yTics$,GnuPlotEXE$,LineSize$)

          LOCAL N&&, ProcessID???

          N&& = FREEFILE
          OPEN "cmd2d.gp" FOR OUTPUT AS #N&&
            'print #N&&, "set output "+Quote$+"test.plt"+Quote$ 'tried this 3/11/06, didn't work...

            IF LogXaxis$ = "YES" AND LogYaxis$ = "NO"  THEN PRINT #N&&, "set logscale x"
            IF LogXaxis$ = "NO"  AND LogYaxis$ = "YES" THEN PRINT #N&&, "set logscale y"
            IF LogXaxis$ = "YES" AND LogYaxis$ = "YES" THEN PRINT #N&&, "set logscale xy"

            IF xMin$ <> "" AND xMax$ <> "" THEN PRINT #N&&, "set xrange ["+xMin$+":"+xMax$+"]"

            IF yMin$ <> "" AND yMax$ <> "" THEN PRINT #N&&, "set yrange ["+yMin$+":"+yMax$+"]"

            PRINT #N&&, "set label "+Quote$+Annotation$+Quote$ at graph "+xCoord$+","+yCoord$
            PRINT #N&&, "set grid xtics"
            PRINT #N&&, "set grid ytics"
            PRINT #N&&, "set xtics "+xTics$
            PRINT #N&&, "set ytics "+yTics$
            PRINT #N&&, "set grid mxtics"
            PRINT #N&&, "set grid mytics"
            PRINT #N&&, "set title "+Quote$+PlotTitle$+Quote$
            PRINT #N&&, "set xlabel "+Quote$+xaxisLabel$+Quote$
            PRINT #N&&, "set ylabel "+Quote$+yaxisLabel$+Quote$

            PRINT #N&&, "plot "+Quote$+PlotFilename1$+Quote$+" with lines linewidth "+REMOVE$(STR$(LineSize),ANY" ")+","+ _
                               Quote$+PlotFilename2$+Quote$+" with points pointsize 0.05" +REMOVE$(STR$(LineSize),ANY" ")

          CLOSE #N&&

          ProcessID??? = SHELL(GnuPlotEXE$+" cmd2d.gp -") : CALL Delay(0.5##)

        END SUB 'TwoDplot2Curves()

'...

        SUB Probe2Dplots(ProbePlotsFileList$,PlotTitle$,Annotation$,xCoord$,yCoord$,XaxisLabel$,yaxisLabel$, _
                         LogXaxis$,LogYaxis$,xMin$,xMax$,yMin$,yMax$,xTics$,yTics$,GnuPlotEXE$)

          LOCAL N&&, ProcessID???

          N&& = FREEFILE

          OPEN "cmd2d.gp" FOR OUTPUT AS #N&&

            IF LogXaxis$ = "YES" AND LogYaxis$ = "NO"  THEN PRINT #N&&, "set logscale x"
            IF LogXaxis$ = "NO"  AND LogYaxis$ = "YES" THEN PRINT #N&&, "set logscale y"
            IF LogXaxis$ = "YES" AND LogYaxis$ = "YES" THEN PRINT #N&&, "set logscale xy"

            IF xMin$ <> "" AND xMax$ <> "" THEN PRINT #N&&, "set xrange ["+xMin$+":"+xMax$+"]"

            IF yMin$ <> "" AND yMax$ <> "" THEN PRINT #N&&, "set yrange ["+yMin$+":"+yMax$+"]"

            PRINT #N&&, "set label "+Quote$+Annotation$+Quote$ at graph "+xCoord$+","+yCoord$
            PRINT #N&&, "set grid xtics"
            PRINT #N&&, "set grid ytics"
            PRINT #N&&, "set xtics "+xTics$
            PRINT #N&&, "set ytics "+yTics$
            PRINT #N&&, "set grid mxtics"
            PRINT #N&&, "set grid mytics"
            PRINT #N&&, "set title "+Quote$+PlotTitle$+Quote$
            PRINT #N&&, "set xlabel "+Quote$+xaxisLabel$+Quote$
```

```
        PRINT #N&&, "set ylabel "+Quote$+yaxisLabel$+Quote$

        PRINT #N&&, ProbePlotsFileList$

    CLOSE #N&&

    ProcessID??? = SHELL(GnuPlotEXE$+" cmd2d.gp -") : CALL Delay(0.5##)

END SUB 'Probe2Dplots()

'---
SUB Show2DsamplePoints(R(),Np&&,Niter&&,j&&,FunctionName$)

    LOCAL N&&, p&&

    LOCAL A$, PlotFilename$, PlotTitle$, Symbols$

    LOCAL xMin$, xMax$, yMin$, yMax$

    LOCAL s1, s2, s4 AS EXT

    PlotFilename$ = "2DsampPts("+REMOVE$(STR$(j&&),ANY" ")+")"

    IF j&& > 0 THEN 'PLOT POINTS AT THIS TIME STEP
        PlotTitle$ = "\nLOCATIONS OF "+REMOVE$(STR$(Np&&),ANY" ") + " POINTS AT ITERATION " + STR$(j&&) + " / " + REMOVE$(STR$(Niter&&),ANY" ") + "\n"
    ELSE 'PLOT INITIAL SAMPLE POINT DISTRIBUTION
        PlotTitle$ = "\nLOCATIONS OF "+REMOVE$(STR$(Np&&),ANY" ") + " INITIAL SAMPLE POINTS FOR " + FunctionName$ + " FUNCTION\n"

    END IF

    N&& = FREEFILE : OPEN PlotFilename$ FOR OUTPUT AS #N&&
        FOR p&& = 1 TO Np&& : PRINT #N&&, USING$("#####.#####    #####.####",R(p&&,1,j&&),R(p&&,2,j&&)) : NEXT p&&
    CLOSE #N&&

    s1 = 1.1## : s2 = 1.1## : s3 = 1.1## : s4 = 1.1## 'expand plots axes by 10&&

    IF XiMin(1) > 0## THEN s1 = 0.9##
    IF XiMax(1) < 0## THEN s2 = 0.9##
    IF XiMin(2) > 0## THEN s3 = 0.9##
    IF XiMax(2) < 0## THEN s4 = 0.9##

    xMin$ = REMOVE$(STR$(s1*XiMin(1),2),ANY" ")
    xMax$ = REMOVE$(STR$(s2*XiMax(1),2),ANY" ")
    yMin$ = REMOVE$(STR$(s3*XiMin(2),2),ANY" ")
    yMax$ = REMOVE$(STR$(s4*XiMax(2),2),ANY" ")

    CALL TwoDplot(PlotFilename$,PlotTitle$,"0.6","0.7","x1\n\n","\nx2","NO","NO",xMin$,xMax$,yMin$,yMax$,"5","5","wgnuplot.exe", pointsize 1 linewidth
2","")

    KILL PlotFilename$ 'erase plot data file after sample points have been displayed

END SUB 'Show2DsamplePoints()

'-------------------------------
SUB Show3DsamplePoints(R(),Np&&,N0&&,Niter&&,j&&,FunctionName$)

    LOCAL N&&, p&&, PlotWindowsLC_X&&, PlotWindowsLC_Y&&, PlotWindowWidth&&, PlotWindowHeight&&, PlotWindowOffset&&

    LOCAL A$, PlotFilename$, PlotTitle$, Symbols$, Annotation$

    LOCAL xMin$, xMax$, yMin$, yMax$, zMin$, zMax$

    LOCAL s1, s2, s3, s4, s5, s6 AS EXT

    PlotFilename$ = "Points("+REMOVE$(STR$(j&&),ANY" ")+")"

    IF j&& > 0 THEN 'PLOT POINTS AT THIS ITERATION
        PlotTitle$ = "\n\nLOCATIONS OF "+REMOVE$(STR$(Np&&),ANY" ") + " SAMPLE POINTS AT ITERATION " + STR$(j&&) + " / " + REMOVE$(STR$(Niter&&),ANY" ") +
"\n"
    ELSE 'PLOT INITIAL POINT DISTRIBUTION
        PlotTitle$ = REMOVE$(STR$(Np&&),ANY" ") + "-POINT ISPD FOR FUNCTION " + FunctionName$ +"\n\n"
    END IF

'    --------------- Point Coordinates ----------------
    N&& = FREEFILE : OPEN PlotFilename$ FOR OUTPUT AS #N&&
        PRINT #N&&, USING$("#####.####    #####.####    #####.####",R(1,1,j&&),R(1,2,j&&),R(1,3,j&&)) 'This line repeats Sample Point #1's coordinates.
It's necessary
                                                                                                       'to deal with a plotting artifact in Gnuplot!
        FOR p&& = 1 TO Np&& : PRINT #N&&, USING$("#####.####    #####.####    #####.####",R(p&&,1,j&&),R(p&&,2,j&&),R(p&&,3,j&&)) : NEXT p&&
    CLOSE #N&&

'    --------------- Principal Diagonal -----------------
    N&& = FREEFILE : OPEN "diag" FOR OUTPUT AS #N&&
        PRINT #N&&, USING$("#####.#####    #####.#####    #####.#####",XiMin(1),XiMin(2),XiMin(3))
        PRINT #N&&, _
        PRINT #N&&, USING$("#####.#####    #####.#####    #####.#####",XiMax(1),XiMax(2),XiMax(3))
    CLOSE #N&&

IF UCASE$(FunctionName$) <> "COMPRESSIONSPRING" THEN

'    --------- RE-PLOT SAMPLE POINT #1 BECAUSE OF SOME ARTIFACT THAT DROPS IT FROM POINT LINE #1 ?????? -----------------
    N&& = FREEFILE : OPEN "probe1" FOR OUTPUT AS #N&&
        PRINT #N&&, USING$("#####.####    #####.####    #####.####",R(1,1,j&&),R(1,2,j&&),R(1,3,j&&))
    CLOSE #N&&        PRINT #N&&, USING$("#####.####    #####.####    #####.####",R(1,1,j&&),R(1,2,j&&),R(1,3,j&&))

    s1 = 1.1## : s2 = s1 : s3 = s1 : s4 = s1 : s5 = s1 : s6 = s1 'expand plots axes by 10&&

    IF XiMin(1) > 0## THEN s1 = 0.9##
    IF XiMax(1) < 0## THEN s2 = 0.9##
    IF XiMin(2) > 0## THEN s3 = 0.9##
    IF XiMax(2) < 0## THEN s4 = 0.9##
    IF XiMin(3) > 0## THEN s5 = 0.9##
    IF XiMax(3) < 0## THEN s6 = 0.9##

    xMin$ = REMOVE$(STR$(s1*XiMin(1),2),ANY" ")
    xMax$ = REMOVE$(STR$(s2*XiMax(1),2),ANY" ")
    yMin$ = REMOVE$(STR$(s3*XiMin(2),2),ANY" ")
    yMax$ = REMOVE$(STR$(s4*XiMax(2),2),ANY" ")
    zMin$ = REMOVE$(STR$(s5*XiMin(3),2),ANY" ")
    zMax$ = REMOVE$(STR$(s6*XiMax(3),2),ANY" ")

END IF

'USAGE: CALL ThreeDplot(PlotFilename$,PlotTitle$,Annotation$,xCoord$,yCoord$,zCoord$,_
'                       Xaxislabel$,Yaxislabel$,Zaxislabel$,zMin$,zMax$,GnuPlotEXE$,xStart$,xStop$,yStart$,ystop$)

    PlotWindowsLC_X&& = 50 : PlotWindowsLC_Y&& = 50 : PlotWindowWidth&& = 1000 : PlotWindowHeight&& = 800
    PlotWindowOffset&& = 100

    CALL CreateGnuplotIniFile(PlotWindowsLC_X&&+PlotWindowOffset&&,PlotWindowsLC_Y&&+PlotWindowOffset&&,_
                              PlotWindowWidth&&,PlotWindowHeight&&)

    CALL ThreeDplot(PlotFilename$,PlotTitle$,Annotation$,"0.6","0.7","0.8", _
                    "x1","x2","x3",zMin$,zMax$,"wgnuplot.exe",xMin$,xMax$,yMin$,yMax$)

    KILL PlotFilename$ 'erase plot data file after sample points have been displayed

END SUB 'Show3DsamplePoints()

'--------------------------------
    SUB ThreeDplot(PlotFilename$,PlotTitle$,Annotation$,xCoord$,yCoord$,zCoord$,_
                   Xaxislabel$,Yaxislabel$,Zaxislabel$,zMin$,zMax$,GnuPlotEXE$,xStart$,xStop$,yStart$,ystop$)

        LOCAL N&&, ProcID???

        N&& = FREEFILE

        OPEN "cmd3d.gp" FOR OUTPUT AS #N&&

            PRINT #N&&, "set pm3d"
            PRINT #N&&, "show pm3d"
            PRINT #N&&, "set hidden3d"
```

```
'               PRINT #%&&, "set view 45, 45, 1, 1"

               IF zMin$ <> "" AND zMax$ <> "" THEN  PRINT #%&&, "set zrange ["+zMin$+":"+zMax$+"]"

               PRINT #%&&, "set xrange [" + xstart$ + ":" + xstop$ + "]"
               PRINT #%&&, "set yrange [" + ystart$ + ":" + ystop$ + "]"

               PRINT #%&&, "set label "  + Quote$ + Annotation$ + Quote$ + " at graph "+xCoord$+","+yCoord$+","+zCoord$
               PRINT #%&&, "show label"
               PRINT #%&&, "set grid xtics ytics ztics"
               PRINT #%&&, "show grid"
               PRINT #%&&, "set title "  + Quote$+PlotTitle$    + Quote$
               PRINT #%&&, "set xlabel " + Quote$+XaxisLabel$   + Quote$
               PRINT #%&&, "set ylabel " + Quote$+YaxisLabel$   + Quote$
               PRINT #%&&, "set zlabel " + Quote$+ZaxisLabel$   + Quote$
'               PRINT #%&&, "unset colorbox"
               print #%&&, "set style fill"
'
'               PRINT #%&&, "splot "      + Quote$+PlotFileName$ + Quote$ + " notitle lw 1 pt 8," _
'                                        + Quote$ + "diag"      + Quote$ + " notitle w l," _
'                                        + Quote$ + "probeline1" + Quote$ + " notitle w l," _
'                                        + Quote$ + "probeline2" + Quote$ + " notitle w l" _
'               PRINT #%&&, "splot "      + Quote$+PlotFileName$ + Quote$ + " notitle lw 1 pt 8," _
                                         + Quote$ + "diag"      + Quote$ + " notitle w l," _
                                         + Quote$ + "probe1"    + Quote$ + " notitle w l!"

            CLOSE #%&&

            ProcID???? = SHELL(GnuPlotEXE$+" cmd3d.gp -")

            CALL Delay(0.5##)

         END SUB 'ThreeDplot3()

'----
         SUB TwoDplot(PlotFileName$,PlotTitle$,xCoord$,yCoord$,xaxisLabel$,YaxisLabel$, _
                     LogxaxisS,LogyaxisS,xMin$,xMax$,yMin$,yMax$,xTics$,yTics$,GnuPlotEXE$,LineType$,Annotation$)

            LOCAL %&&, ProcessID???

            %&& = FREEFILE

            OPEN "cmd2d.gp" FOR OUTPUT AS #%&&

               IF LogxaxisS = "YES" AND LogyaxisS = "NO"  THEN PRINT #%&&, "set logscale x"
               IF LogxaxisS = "NO"  AND LogyaxisS = "YES" THEN PRINT #%&&, "set logscale y"
               IF LogxaxisS = "YES" AND LogyaxisS = "YES" THEN PRINT #%&&, "set logscale xy"

               IF xMin$ <> "" AND xMax$ <> "" THEN  PRINT #%&&, "set xrange ["+xMin$+":"+xMax$+"]"
               IF yMin$ <> "" AND yMax$ <> "" THEN  PRINT #%&&, "set yrange ["+yMin$+":"+yMax$+"]"

               PRINT #%&&, "set label "  + Quote$ + Annotation$ + Quote$ + " at graph " + xCoord$ + "," + yCoord$
               PRINT #%&&, "set grid xtics " + xTics$
               PRINT #%&&, "set grid ytics " + yTics$
               PRINT #%&&, "set grid mxtics"
               PRINT #%&&, "set grid mytics"
               PRINT #%&&, "show grid"
               PRINT #%&&, "set title "  + Quote$+PlotTitle$+Quote$
               PRINT #%&&, "set xlabel " + Quote$+XaxisLabel$+Quote$
               PRINT #%&&, "set ylabel " + Quote$+YaxisLabel$+Quote$

               PRINT #%&&, "plot "+Quote$+PlotFileName$+Quote$+" notitle"+LineType$

            CLOSE #%&&

            ProcessID??? = SHELL(GnuPlotEXE$+" cmd2d.gp -") : CALL Delay(0.5##)

         END SUB 'TwoDplot()

'-----
         SUB CreateGnuplotINIfile(PlotWindowULC_X&&,PlotWindowULC_Y&&,PlotwindowWidth&&,PlotwindowHeight&&)

            LOCAL %&&, WinPath$, A$, B$, WindowsDirectory$

            WinPath$ = UCASE$(ENVIRON$("Path"))'DIR$("C:\WINDOWS",23)

            DO
               B$ = A$

               A$ = EXTRACT$(WinPath$,";")

               WinPath$ = REMOVE$(WinPath$,A$+";")

               IF RIGHT$(A$,7) = "WINDOWS" OR A$ = B$ THEN EXIT LOOP

               IF RIGHT$(A$,5) = "WINNT"  OR A$ = B$ THEN EXIT LOOP

            LOOP

            WindowsDirectory$ = A$

            %&& = FREEFILE

'        ---------- WGNUPLOT.INPUT FILE ----------
'         OPEN WindowsDirectory$+"\wgnuplot.ini" FOR OUTPUT AS #%&&

               PRINT #%&&,"[WGNUPLOT]"
               PRINT #%&&,"TextOrigin=0 0"
               PRINT #%&&,"TextSize=640 150"
               PRINT #%&&,"TextFont=Terminal,9"
               PRINT #%&&,"GraphOrigin="+REMOVE$(STR$(PlotWindowULC_X&&),ANY" ")+" "+REMOVE$(STR$(PlotWindowULC_Y&&),ANY" ")
               PRINT #%&&,"GraphSize="  +REMOVE$(STR$(PlotwindowWidth&&),ANY" ")+" "+REMOVE$(STR$(PlotwindowHeight&&),ANY" ")
               PRINT #%&&,"GraphFont=Arial,10"
               PRINT #%&&,"GraphColor=1"
               PRINT #%&&,"GraphToTop=1"
               PRINT #%&&,"GraphBackground=255 255 255"
               PRINT #%&&,"Border=0 0 0 0 0"
               PRINT #%&&,"Axis=192 192 192 2 2"
               PRINT #%&&,"Line1=0 0 255 0 0"
               PRINT #%&&,"Line2=0 255 0 0 1"
               PRINT #%&&,"Line3=255 0 0 0 2"
               PRINT #%&&,"Line4=255 0 255 0 3"
               PRINT #%&&,"Line5=0 128 0 0 4"

            CLOSE #%&&

         END SUB 'CreateGnuplotINIfile()

'------
         SUB Delay(NumSecs)

            LOCAL StartTime, StopTime AS EXT

            StartTime = TIMER

            DO UNTIL (StopTime-StartTime) >= NumSecs

               StopTime = TIMER

            LOOP

         END SUB 'Delay()

'------
SUB MathematicaConstants
   EulerConst   = 0.577215664901532860606512##
   PI           = 3.14159265358979323846264##
   Pi2          = Pi/2##
   Pi4          = Pi/4##
   TwoPi        = 2##*PI
   FourPi       = 4##*Pi
   FivePi       = 5##*PI
```

```basic
    e              = 2.718281828459045235360287##
    Root2          = 1.414213562373095048884##
END SUB

'------

SUB AlphabetAndDigits
    Alphabet$      = "ABCDEFGHIJKLMNOPQRSTUVWXYZabcdefghijklmnopqrstuvwxyz"
    Digits$        = "0123456789"
END SUB

'------

SUB SpecialSymbols
    Quote$         = CHR$(34) 'Quotation mark "
    SpecialCharacters$ = "`(),#:;/_
END SUB

'------

SUB EMconstants
    Mu0  = 4E-7##*Pi                   'hy/meter
    Eps0 = 8.8544##*1E-12 'fd/meter
    c    = 2.998E8##       'velocity of light, 1##/SQR(Mu0*Eps0) 'meters/sec
    eta0 = SQR(Mu0/Eps0)  'impedance of free space, ohms
END SUB

'------

SUB ConversionFactors
    Rad2Deg        = 180##/Pi
    Deg2Rad        = 1##/Rad2Deg
    Feet2Meters    = 0.3048##
    Meters2Feet    = 1##/Feet2Meters
    Inches2Meters  = 0.0254##
    Meters2Inches  = 1##/Inches2Meters
    Miles2Meters   = 1609.344##
    Meters2Miles   = 1##/Miles2Meters
    Naut Mi2Meters = 1852##
    Meters2NautMi  = 1##/NautMi2Meters
END SUB

'------

SUB ShowConstants 'puts up msgbox showing all constants

    LOCAL A$

    A$ = _
    "Mathematical Constants:"+CHR$(13)+_
    "Euler const="+STR$(EulerConst)+CHR$(13)+_
    "Pi="+STR$(Pi)+CHR$(13)+_
    "Pi/2="+STR$(Pi/2)+CHR$(13)+_
    "Pi/4="+STR$(Pi/4)+CHR$(13)+_
    "2Pi="+STR$(TwoPi)+CHR$(13)+_
    "4Pi="+STR$(FourPi)+CHR$(13)+_
    "e="+STR$(e)+CHR$(13)+_
    "Sqr2="+STR$(Root2)+CHR$(13)+CHR$(13)+_
    "Alphabet, Digits & Special Characters: "+CHR$(13)+_
    "Alphabet: "+Alphabet$+CHR$(13)+_
    "Digits: "+Digits$+CHR$(13)+_
    "Spec chars="+SpecialCharacters$+CHR$(13)+CHR$(13)+_
    "E&M Constants:"+CHR$(13)+_
    "Mu0="+STR$(Mu0)+CHR$(13)+_
    "Eps0="+STR$(Eps0)+CHR$(13)+_
    "c="+STR$(c)+CHR$(13)+_
    "eta0="+STR$(eta0)+CHR$(13)+CHR$(13)+_
    "Conversion Factors: "+CHR$(13)+_
    "Rad2Deg="+STR$(Rad2Deg)+CHR$(13)+_
    "Deg2Rad="+STR$(Deg2Rad)+CHR$(13)+_
    "Ft2Meters="+STR$(Feet2Meters)+CHR$(13)+_
    "Meters2Ft="+STR$(Meters2Feet)+CHR$(13)+_
    "Inches2Meters="+STR$(Inches2Meters)+CHR$(13)+_
    "Meters2Inches="+STR$(Meters2Inches)+CHR$(13)+_
    "Miles2Meters="+STR$(Miles2Meters)+CHR$(13)+_
    "Meters2Miles="+STR$(Meters2Miles)+CHR$(13)+_
    "NautMi2Meters="+STR$(NautMi2Meters)+CHR$(13)+_
    "Meters2NautMi="+STR$(Meters2NautMi)+CHR$(13)+CHR$(13)

    MSGBOX(A$)

END SUB

'------

SUB DisplayMatrix(Np&&,Nd&&,Niter&&,R())

    LOCAL p&&, i&&, j&&, A$

        A$ = "Position vector Matrix R()"+CHR$(13)

        FOR p&& = 1 TO Np&&

            FOR i&& = 1 TO Nd&&
                FOR j&& = 0 TO Niter&&

                    A$ = A$ + "R("+STR$(p&&)+", "+STR$(i&&)+", "+STR$(j&&)+") = "+STR$(R(p&&,i&&,j&&)) + CHR$(13)

                NEXT j&&
            NEXT i&&
            MSGBOX(A$)

        NEXT p&&
END SUB

'------

SUB DisplayMatrixThisIteration(Np&&,Nd&&,p&&,R(),Gamma)

    LOCAL p&&, i&&, A$, B$

        A$ = "Position vector Matrix R() at step "+STR$(j&&)+", Gamma ="+STR$(Gamma)+":"+CHR$(13)+CHR$(13)

        FOR p&& = 1 TO Np&&

            A$ = A$ + "Probe#"+REMOVE$(STR$(p&&),ANY" ")+": "

            B$ = ""

            FOR i&& = 1 TO Nd&&
                B$ = B$ + "  " + USING$("####.##",R(p&&,i&&,j&&))
            NEXT i&&
            A$ = A$ + B$ + CHR$(13)

        NEXT p&&
        MSGBOX(A$)

END SUB

'------

SUB DisplayAmatrix(Np&&,Nd&&,Niter&&,A())

    LOCAL p&&, i&&, j&&, A$

        A$ = "Acceleration Vector Matrix A()"+CHR$(13)

        FOR p&& = 1 TO Np&&
```

```
            FOR i&& = 1 TO Nd&&
                FOR j&& = 0 TO Niter&&
                    A$ = A$ + "A("+STR$(p&&)+", "+STR$(i&&)+", "+STR$(j&&+" ") ="+STR$(A(p&&,i&&,j&&)) + CHR$(13)
                NEXT j&&
            NEXT i&&
        NEXT p&&
        MSGBOX(A$)
    END SUB
    '------
    SUB DisplayAmatrixThisIteration(Np&&,Nd&&,j&&,A())
    LOCAL p&&, i&&, A$
        A$ = "Acceleration matrix A() at step "+STR$(j&&)+":"+CHR$(13)
        FOR p&& = 1 TO Np&&
            FOR i&& = 1 TO Nd&&
                A$ = A$ + "A("+STR$(p&&)+", "+STR$(i&&)+", "+STR$(j&&+" ") ="+STR$(A(p&&,i&&,j&&)) + CHR$(13)
            NEXT i&&
        NEXT p&&
        MSGBOX(A$)
    END SUB
    '------
    SUB DisplayHmatrix(Np&&,Niter&&,H())
    LOCAL p&&, j&&, A$
        A$ = "Fitness Matrix H()"+CHR$(13)
        FOR p&& = 1 TO Np&&
            FOR j&& = 0 TO Niter&&
                A$ = A$ + "H("+STR$(p&&)+", "+STR$(j&&+" ") ="+STR$(H(p&&,j&&)) + CHR$(13)
            NEXT j&&
        NEXT p&&
        MSGBOX(A$)
    END SUB
    '------
    SUB DisplayHbestMatrix(Np&&,Niter&&,Hbest())
    LOCAL p&&, j&&, A$
        A$ = "Np= "+STR$(Np&&)+"  nt="STR$(Niter&&)+CHR$(13)+"Fitness Matrix Hbest()"+CHR$(13)
        FOR p&& = 1 TO Np&&
            FOR j&& = 0 TO Niter&&
                A$ = A$ + "Hbest("+STR$(p&&)+", "+STR$(j&&+" ") ="+STR$(Hbest(p&&,j&&)) + CHR$(13)
            NEXT j&&
        NEXT p&&
        MSGBOX(A$)
    END SUB
    '------
    SUB DisplayHmatrixThisIteration(Np&&,j&&,H())
    LOCAL p&&, A$
        A$ = "Fitness matrix H() at step "+STR$(j&&)+":"+CHR$(13)
        FOR p&& = 1 TO Np&&
            A$ = A$ + "H("+STR$(p&&)+", "+STR$(j&&+" ") ="+STR$(H(p&&,j&&)) + CHR$(13)
        NEXT p&&
        MSGBOX(A$)
    END SUB
    '------
    SUB DisplayXiMinMax(Nd&&,XiMin(),XiMax())
    LOCAL i&&, A$
        A$ = ""
        FOR i&& = 1 TO Nd&&
            A$ = A$ + "xiMin("+STR$(i&&)+" ) = "+STR$(XiMin(i&&))+"  xiMax("+STR$(i&&)+" ) = "+STR$(XiMax(i&&)) + CHR$(13)
        NEXT i&&
        MSGBOX(A$)
    END SUB
    '------
    SUB DisplayRunParameters2(FunctionName$,Nd&&,Np&&,Niter&&,G,DeltaT,Alpha,Beta,Frep,PlaceInitialPoints$,InitialAcceleration$,RepositionFactor$)
    LOCAL A$
        A$ = "Function = "+ FunctionName$+CHR$(13)+_
            "Nd = "+STR$(Nd&&)+CHR$(13)+_
            "Np = "+STR$(Np&&)+CHR$(13)+_
            "Nt = "+STR$(Niter&&)+CHR$(13)+_
            "G  = "+STR$(G)+CHR$(13)+_
            "DeltaT = "+STR$(DeltaT)+CHR$(13)+_
            "Alpha  = "+STR$(Alpha)+CHR$(13)+_
            "Beta   = "+STR$(Beta)+CHR$(13)+_
            "Frep   = "+STR$(Frep)+CHR$(13)+_
            "Init Points: "+PlaceInitialPoints$+CHR$(13)+_
            "Init Accel: "+InitialAcceleration$+CHR$(13)+_
            "Retrive Method: "+RepositionFactor$+CHR$(13)
        MSGBOX(A$)
    END SUB
    '------
    SUB
    Tabulate1DprobeCoordinates(MaxiDsamplePointsPlotted&&,Nd&&,Np&&,LastIteration&&,G,DeltaT,Alpha,Beta,Frep,R(),H(),PlaceInitialPoints$,InitialAcceleration$,Rep
    ositionFactor$,FunctionName$,Gamma)

    LOCAL N&&, ProbeNum&&, FileHeader$, A$, B$, C$, D$, E$, F$, H$, Stepnum&, FieldNumber&&  'kludgy, yes, but it accomplishes its purpose...
```

```
              CALL
GetPlotAnnotation(FileHeader$,N&&&,Np&&,LastIteration&&,G,DeltaT,Alpha,Beta,Frep,M(),PlaceInitialPoints$,InitialAcceleration$,RepositionFactor$,FunctionName$
,Gamma)

            REPLACE "\n" WITH ", " IN FileHeader$

            FileHeader$ = LEFT$(FileHeader$,LEN(FileHeader$)-2)

            FileHeader$ = "PROBE COORDINATES" + CHR$(13) +_
                          "-----------------" + CHR$(13) + FileHeader$

            N&& = FREEFILE : OPEN "ProbeCoordinates.DAT" FOR OUTPUT AS #N&&

            A$ = "    Step #    " : B$ = "   ------   " : C$ = ""

            FOR ProbeNum&& = 1 TO Np&& 'create out data file header

                SELECT CASE ProbeNum&&
                    CASE    1 TO   9 : E$ = ""   : F$ = "       "          : H$ = "        "
                    CASE   10 TO  99 : E$ = "-"  : F$ = "      "           : H$ = "      "
                    CASE  100 TO 999 : E$ = "--" : F$ = "     "            : H$ = "     "
                END SELECT

                A$ = A$ + "P" + NoSpaces$(ProbeNum&&,0,0) + F$ 'note: adding zero to ProbeNum&& necessary to convert to floating point...

                B$ = B$ + E$ + "--" + H$

                C$ = C$ + "######.###   "

                C$ = C$ + "##.######"

            NEXT ProbeNum&&

            PRINT #N&&, FileHeader$ + CHR$(13) : PRINT #N&&, A$ | PRINT #N&&, B$

            FOR StepNum& = 0 TO LastIteration&&

                D$ = USING$("#####    ",StepNum&)

                FOR ProbeNum&& = 1 TO Np&& : D$ = D$ + USING$(C$,R(ProbeNum&&,1,StepNum&)) : NEXT ProbeNum&&

                PRINT #N&&, D$

            NEXT StepNum&

        CLOSE #N&&

END SUB 'TabulateIdProbeCoordinates()
'------

SUB
PlotIdProbePositions(MaxIdSamplePointsPlotted&&,N&&&,Np&&,LastIteration&&,G,DeltaT,Alpha,Beta,Frep,R(),M(),PlaceInitialPoints$,InitialAcceleration$,Repositio
nFactor$,FunctionName$,Gamma)
        'plots on-screen 1d motion probe positions vs time step if Np =< 15

LOCAL ProcessID???, N&&, n1&&, n2&&, n3&&, n4&&, n5&&, n6&&, n7&&, n8&&, n9&&, n10&&, n11&&, n12&&, n13&&, n14&&, n15&&, ProbeNum&&, StepNum&, A$

LOCAL PlotAnnotation$

'MSGBOX("1: Np="+STR$(Np&&)+"   Max ID Pts = "+STR$(MaxIdSamplePointsPlotted&&))

    IF Np&& > MaxIdSamplePointsPlotted&& THEN EXIT SUB

'MSGBOX("2: Np="+STR$(Np&&)+"   Max ID Pts = "+STR$(MaxIdSamplePointsPlotted&&))

    CALL CLEANUP 'delete old "Px" plot files, if any

    DO 'create output data files, probe-by-probe
        INCR ProbeNum&& : n1&& = FREEFILE : OPEN "P"+REMOVE$(STR$(ProbeNum&&),ANY" ") FOR OUTPUT AS #n1&&  : IF ProbeNum&& = Np&& THEN EXIT LOOP
        INCR ProbeNum&& : n2&& = FREEFILE : OPEN "P"+REMOVE$(STR$(ProbeNum&&),ANY" ") FOR OUTPUT AS #n2&&  : IF ProbeNum&& = Np&& THEN EXIT LOOP
        INCR ProbeNum&& : n3&& = FREEFILE : OPEN "P"+REMOVE$(STR$(ProbeNum&&),ANY" ") FOR OUTPUT AS #n3&&  : IF ProbeNum&& = Np&& THEN EXIT LOOP
        INCR ProbeNum&& : n4&& = FREEFILE : OPEN "P"+REMOVE$(STR$(ProbeNum&&),ANY" ") FOR OUTPUT AS #n4&&  : IF ProbeNum&& = Np&& THEN EXIT LOOP
        INCR ProbeNum&& : n5&& = FREEFILE : OPEN "P"+REMOVE$(STR$(ProbeNum&&),ANY" ") FOR OUTPUT AS #n5&&  : IF ProbeNum&& = Np&& THEN EXIT LOOP
        INCR ProbeNum&& : n6&& = FREEFILE : OPEN "P"+REMOVE$(STR$(ProbeNum&&),ANY" ") FOR OUTPUT AS #n6&&  : IF ProbeNum&& = Np&& THEN EXIT LOOP
        INCR ProbeNum&& : n7&& = FREEFILE : OPEN "P"+REMOVE$(STR$(ProbeNum&&),ANY" ") FOR OUTPUT AS #n7&&  : IF ProbeNum&& = Np&& THEN EXIT LOOP
        INCR ProbeNum&& : n8&& = FREEFILE : OPEN "P"+REMOVE$(STR$(ProbeNum&&),ANY" ") FOR OUTPUT AS #n8&&  : IF ProbeNum&& = Np&& THEN EXIT LOOP
        INCR ProbeNum&& : n9&& = FREEFILE : OPEN "P"+REMOVE$(STR$(ProbeNum&&),ANY" ") FOR OUTPUT AS #n9&&  : IF ProbeNum&& = Np&& THEN EXIT LOOP
        INCR ProbeNum&& : n10&& = FREEFILE : OPEN "P"+REMOVE$(STR$(ProbeNum&&),ANY" ") FOR OUTPUT AS #n10&& : IF ProbeNum&& = Np&& THEN EXIT LOOP
        INCR ProbeNum&& : n11&& = FREEFILE : OPEN "P"+REMOVE$(STR$(ProbeNum&&),ANY" ") FOR OUTPUT AS #n11&& : IF ProbeNum&& = Np&& THEN EXIT LOOP
        INCR ProbeNum&& : n12&& = FREEFILE : OPEN "P"+REMOVE$(STR$(ProbeNum&&),ANY" ") FOR OUTPUT AS #n12&& : IF ProbeNum&& = Np&& THEN EXIT LOOP
        INCR ProbeNum&& : n13&& = FREEFILE : OPEN "P"+REMOVE$(STR$(ProbeNum&&),ANY" ") FOR OUTPUT AS #n13&& : IF ProbeNum&& = Np&& THEN EXIT LOOP
        INCR ProbeNum&& : n14&& = FREEFILE : OPEN "P"+REMOVE$(STR$(ProbeNum&&),ANY" ") FOR OUTPUT AS #n14&& : IF ProbeNum&& = Np&& THEN EXIT LOOP
        INCR ProbeNum&& : n15&& = FREEFILE : OPEN "P"+REMOVE$(STR$(ProbeNum&&),ANY" ") FOR OUTPUT AS #n15&& : IF ProbeNum&& = Np&& THEN EXIT LOOP
    LOOP

    ProbeNum&& = 0

    DO 'output probe positions as a function of time step
        INCR ProbeNum&& : FOR StepNum& = 0 TO LastIteration&& : PRINT #n1&&,  USING$("#####  #####.#######",StepNum&,R(ProbeNum&&,1,StepNum&)) : NEXT
StepNum& : IF ProbeNum&& = Np&& THEN EXIT LOOP
        INCR ProbeNum&& : FOR StepNum& = 0 TO LastIteration&& : PRINT #n2&&,  USING$("#####  #####.#######",StepNum&,R(ProbeNum&&,1,StepNum&)) : NEXT
StepNum& : IF ProbeNum&& = Np&& THEN EXIT LOOP
        INCR ProbeNum&& : FOR StepNum& = 0 TO LastIteration&& : PRINT #n3&&,  USING$("#####  #####.#######",StepNum&,R(ProbeNum&&,1,StepNum&)) : NEXT
StepNum& : IF ProbeNum&& = Np&& THEN EXIT LOOP
        INCR ProbeNum&& : FOR StepNum& = 0 TO LastIteration&& : PRINT #n4&&,  USING$("#####  #####.#######",StepNum&,R(ProbeNum&&,1,StepNum&)) : NEXT
StepNum& : IF ProbeNum&& = Np&& THEN EXIT LOOP
        INCR ProbeNum&& : FOR StepNum& = 0 TO LastIteration&& : PRINT #n5&&,  USING$("#####  #####.#######",StepNum&,R(ProbeNum&&,1,StepNum&)) : NEXT
StepNum& : IF ProbeNum&& = Np&& THEN EXIT LOOP
        INCR ProbeNum&& : FOR StepNum& = 0 TO LastIteration&& : PRINT #n6&&,  USING$("#####  #####.#######",StepNum&,R(ProbeNum&&,1,StepNum&)) : NEXT
StepNum& : IF ProbeNum&& = Np&& THEN EXIT LOOP
        INCR ProbeNum&& : FOR StepNum& = 0 TO LastIteration&& : PRINT #n7&&,  USING$("#####  #####.#######",StepNum&,R(ProbeNum&&,1,StepNum&)) : NEXT
StepNum& : IF ProbeNum&& = Np&& THEN EXIT LOOP
        INCR ProbeNum&& : FOR StepNum& = 0 TO LastIteration&& : PRINT #n8&&,  USING$("#####  #####.#######",StepNum&,R(ProbeNum&&,1,StepNum&)) : NEXT
StepNum& : IF ProbeNum&& = Np&& THEN EXIT LOOP
        INCR ProbeNum&& : FOR StepNum& = 0 TO LastIteration&& : PRINT #n9&&,  USING$("#####  #####.#######",StepNum&,R(ProbeNum&&,1,StepNum&)) : NEXT
StepNum& : IF ProbeNum&& = Np&& THEN EXIT LOOP
        INCR ProbeNum&& : FOR StepNum& = 0 TO LastIteration&& : PRINT #n10&&, USING$("#####  #####.#######",StepNum&,R(ProbeNum&&,1,StepNum&)) : NEXT
StepNum& : IF ProbeNum&& = Np&& THEN EXIT LOOP
        INCR ProbeNum&& : FOR StepNum& = 0 TO LastIteration&& : PRINT #n11&&, USING$("#####  #####.#######",StepNum&,R(ProbeNum&&,1,StepNum&)) : NEXT
StepNum& : IF ProbeNum&& = Np&& THEN EXIT LOOP
        INCR ProbeNum&& : FOR StepNum& = 0 TO LastIteration&& : PRINT #n12&&, USING$("#####  #####.#######",StepNum&,R(ProbeNum&&,1,StepNum&)) : NEXT
StepNum& : IF ProbeNum&& = Np&& THEN EXIT LOOP
        INCR ProbeNum&& : FOR StepNum& = 0 TO LastIteration&& : PRINT #n13&&, USING$("#####  #####.#######",StepNum&,R(ProbeNum&&,1,StepNum&)) : NEXT
StepNum& : IF ProbeNum&& = Np&& THEN EXIT LOOP
        INCR ProbeNum&& : FOR StepNum& = 0 TO LastIteration&& : PRINT #n14&&, USING$("#####  #####.#######",StepNum&,R(ProbeNum&&,1,StepNum&)) : NEXT
StepNum& : IF ProbeNum&& = Np&& THEN EXIT LOOP
        INCR ProbeNum&& : FOR StepNum& = 0 TO LastIteration&& : PRINT #n15&&, USING$("#####  #####.#######",StepNum&,R(ProbeNum&&,1,StepNum&)) : NEXT
StepNum& : IF ProbeNum&& = Np&& THEN EXIT LOOP
    LOOP

    ProbeNum&& = 0

    DO 'close output data files
        INCR ProbeNum&& : CLOSE #n1&&  : IF ProbeNum&& = Np&& THEN EXIT LOOP
        INCR ProbeNum&& : CLOSE #n2&&  : IF ProbeNum&& = Np&& THEN EXIT LOOP
        INCR ProbeNum&& : CLOSE #n3&&  : IF ProbeNum&& = Np&& THEN EXIT LOOP
        INCR ProbeNum&& : CLOSE #n4&&  : IF ProbeNum&& = Np&& THEN EXIT LOOP
        INCR ProbeNum&& : CLOSE #n5&&  : IF ProbeNum&& = Np&& THEN EXIT LOOP
        INCR ProbeNum&& : CLOSE #n6&&  : IF ProbeNum&& = Np&& THEN EXIT LOOP
        INCR ProbeNum&& : CLOSE #n7&&  : IF ProbeNum&& = Np&& THEN EXIT LOOP
        INCR ProbeNum&& : CLOSE #n8&&  : IF ProbeNum&& = Np&& THEN EXIT LOOP
        INCR ProbeNum&& : CLOSE #n9&&  : IF ProbeNum&& = Np&& THEN EXIT LOOP
        INCR ProbeNum&& : CLOSE #n10&& : IF ProbeNum&& = Np&& THEN EXIT LOOP
        INCR ProbeNum&& : CLOSE #n11&& : IF ProbeNum&& = Np&& THEN EXIT LOOP
        INCR ProbeNum&& : CLOSE #n12&& : IF ProbeNum&& = Np&& THEN EXIT LOOP
        INCR ProbeNum&& : CLOSE #n13&& : IF ProbeNum&& = Np&& THEN EXIT LOOP
        INCR ProbeNum&& : CLOSE #n14&& : IF ProbeNum&& = Np&& THEN EXIT LOOP
        INCR ProbeNum&& : CLOSE #n15&& : IF ProbeNum&& = Np&& THEN EXIT LOOP
    LOOP

    ProbeNum&& = 0 : A$ = ""

    DO 'create file string for plot command file
        INCR ProbeNum&& : A$ = A$ + Quote$ + "P"+REMOVE$(STR$(ProbeNum&&),ANY" ") + Quote$ + " w l lw 2, " : IF ProbeNum&& = Np&& THEN EXIT LOOP
```

```
         INCR ProbeNum&& : A$ = A$ + Quote$ + "P"+REMOVE$(STR$(ProbeNum&&),ANY" ") + Quote$ + " w 1 lw 2 : IF ProbeNum&& >= Np&& THEN EXIT LOOP
         INCR ProbeNum&& : A$ = A$ + Quote$ + "P"+REMOVE$(STR$(ProbeNum&&),ANY" ") + Quote$ + " w 1 lw 2 : IF ProbeNum&& >= Np&& THEN EXIT LOOP
         INCR ProbeNum&& : A$ = A$ + Quote$ + "P"+REMOVE$(STR$(ProbeNum&&),ANY" ") + Quote$ + " w 1 lw 2 : IF ProbeNum&& >= Np&& THEN EXIT LOOP
         INCR ProbeNum&& : A$ = A$ + Quote$ + "P"+REMOVE$(STR$(ProbeNum&&),ANY" ") + Quote$ + " w 1 lw 2 : IF ProbeNum&& >= Np&& THEN EXIT LOOP
         INCR ProbeNum&& : A$ = A$ + Quote$ + "P"+REMOVE$(STR$(ProbeNum&&),ANY" ") + Quote$ + " w 1 lw 2 : IF ProbeNum&& >= Np&& THEN EXIT LOOP
         INCR ProbeNum&& : A$ = A$ + Quote$ + "P"+REMOVE$(STR$(ProbeNum&&),ANY" ") + Quote$ + " w 1 lw 2 : IF ProbeNum&& >= Np&& THEN EXIT LOOP
         INCR ProbeNum&& : A$ = A$ + Quote$ + "P"+REMOVE$(STR$(ProbeNum&&),ANY" ") + Quote$ + " w 1 lw 2 : IF ProbeNum&& >= Np&& THEN EXIT LOOP
         INCR ProbeNum&& : A$ = A$ + Quote$ + "P"+REMOVE$(STR$(ProbeNum&&),ANY" ") + Quote$ + " w 1 lw 2 : IF ProbeNum&& >= Np&& THEN EXIT LOOP
         INCR ProbeNum&& : A$ = A$ + Quote$ + "P"+REMOVE$(STR$(ProbeNum&&),ANY" ") + Quote$ + " w 1 lw 2 : IF ProbeNum&& >= Np&& THEN EXIT LOOP
         INCR ProbeNum&& : A$ = A$ + Quote$ + "P"+REMOVE$(STR$(ProbeNum&&),ANY" ") + Quote$ + " w 1 lw 2 : IF ProbeNum&& >= Np&& THEN EXIT LOOP
         INCR ProbeNum&& : A$ = A$ + Quote$ + "P"+REMOVE$(STR$(ProbeNum&&),ANY" ") + Quote$ + " w 1 lw 2 : IF ProbeNum&& >= Np&& THEN EXIT LOOP
         INCR ProbeNum&& : A$ = A$ + Quote$ + "P"+REMOVE$(STR$(ProbeNum&&),ANY" ") + Quote$ + " w 1 lw 2 : IF ProbeNum&& >= Np&& THEN EXIT LOOP
      LOOP

      A$ = LEFT$(A$,LEN(A$)-2)

      CALL
GetPlotAnnotation(PlotAnnotation$,Nd$&&,Np&&,LastIteration&&,G,DeltaT,Alpha,Beta,Frep,H(),PlaceInitialPoints$,InitialAcceleration$,RepositionFactor$,FunctionN
ame$,Gamma)

      N$&& = FREEFILE

      OPEN "cmd2d.gp" FOR OUTPUT AS #N$&&

         PRINT #N$&&, "set label "       + Quote$ + PlotAnnotation$ + Quote$ + " at graph 0.5,0.95"
         PRINT #N$&&, "set grid xtics"
         PRINT #N$&&, "set grid ytics"
         PRINT #N$&&, "set title "  + Quote$ + "Evolution of " + FunctionName$ + " Probe Positions" + "\n" + RunID$ + Quote$
         PRINT #N$&&, "set xlabel " + Quote$ + "Iteration #" + Quote$
         PRINT #N$&&, "set ylabel " + Quote$ + "Probe Coordinate" + Quote$
         PRINT #N$&&, "plot "       + A$

      CLOSE #N$&&

      CALL CreateGNUplotIniFile(0.2##*ScreenWidth&&,0.2##*ScreenHeight&&,0.6##*ScreenWidth&&,0.6##*ScreenHeight&&) 'USAGE: CALL
CreateGNUplotIniFile(PlotWindowLocX_&&,PlotWindowLocY_&&,PlotWindowWidth&&,PlotWindowHeight&&)

      ProcessID??? = SHELL("wgnuplot.exe"+" cmd2d.gp -") : CALL Delay(S##) 'before SUB Cleanup is called

END SUB 'PlotlDprobePositions()

'------------------------------------

SUB
DisplayRunParameters(FunctionName$,Nd$&&,Np&&,Niter&&,G,DeltaT,Alpha,Beta,Frep,R(),A(),H(),PlaceInitialPoints$,InitialAcceleration$,RepositionFactor$,RunVS0$,
ShrinkDS$,CheckForEarlyTermination$)

   LOCAL A$, B$, Yn&

      B$ = "" : IF PlaceInitialPoints$ = "UNIFORM ON-AXIS" AND Nd$&& > 1 THEN B$ = "  ["+REMOVE$(STR$(Np&&/Nd$&&),ANY" ") + "/axis]"

      RunVS0$ = "NO"

   A$ = "RUN CFO WITH THE " + CHR$(13) +_
        "FOLLOWING PARAMETERS?"                                     + CHR$(13) + CHR$(13) +_
        "Function         " + FunctionName$                         + "  (" + REMOVE$(STR$(Nd$&&),ANY" ") + "-D)" + CHR$(13) +_
        "# Iterations     " + REMOVE$(STR$(Niter&&),ANY" ")         + CHR$(13) +_
        "Grav Const G =   " + REMOVE$(STR$(G,2),ANY" ")             + CHR$(13) +_
        "Delta-T =        " + REMOVE$(STR$(DeltaT,3),ANY" ")        + CHR$(13) +_
        "Exp Alpha =      " + REMOVE$(STR$(Alpha,3),ANY" ")         + CHR$(13) +_
        "Exp Beta =       " + REMOVE$(STR$(Beta,3),ANY" ")          + CHR$(13) +_
        "Frep =           " + REMOVE$(STR$(Frep,4),ANY" ")          + " (*RepositionFactor$ = " + RepositionFactor$ + ")" + CHR$(13) +_
        "Initial Points:  " + PlaceInitialPoints$                   + CHR$(13) +_
        "Initial Accel:   " + InitialAcceleration$                  + CHR$(13) +_
        "Check for Early Termination? " + CheckForEarlyTermination$ + CHR$(13) +_
        "Shrink Decision Space?       " + ShrinkDS$ + CHR$(13) +CHR$(13)

'     lResult& = MSGBOX(txt$ [, [style&], title$])

      Yn& = MSGBOX(A$,%MB_YESNO,"CONFIRM RUN")

      IF Yn& = %IDYES THEN RunVS0$ = "YES"

   END SUB

'------

SUB StatusWindow(FunctionName$,StatusWindowHandle???)

   GRAPHIC WINDOW "Run Progress, "+FunctionName$,0.08##*ScreenWidth&&,0.08##*ScreenHeight&&,0.25##*ScreenWidth&&,0.17##*ScreenHeight&& TO StatusWindowHandle???

   GRAPHIC ATTACH StatusWindowHandle???,0,REDRAW

   GRAPHIC FONT "Lucida Console",8,0 'Courier New',8,0 'Fixed width fonts

   GRAPHIC SET PIXEL (35,15) : GRAPHIC PRINT " Initializing...      " : GRAPHIC REDRAW

END SUB

'------

SUB GetTestFunctionNumber(FunctionName$)

   LOCAL hOlg AS DWORD

   LOCAL N&&, n&&

   LOCAL FrameWidth1&, FrameWidth2&, FrameHeight&, BoxWidth&, BoxHeight&

   BoxWidth& = 276 : BoxHeight& = 325 : FrameWidth1& = 82 : FrameWidth2& = 90 : FrameHeight& = BoxHeight&-25

   DIALOG NEW 0, "CENTRAL FORCE OPTIMIZATION TEST FUNCTIONS",,, BoxWidth&, BoxHeight&,%WS_CAPTION OR %WS_SYSMENU, 0 TO hOlg

'------------------------------

   CONTROL ADD FRAME,  hOlg, %IDC_FRAME1,  "Test Functions",        5, 2, FrameWidth1&, FrameHeight&
   CONTROL ADD FRAME,  hOlg, %IDC_FRAME2,  "GSO Test Functions",  95, 2, FrameWidth2&, 255

   CONTROL ADD OPTION, hOlg, %IDC_Function_Number1,  "Parrott F4",    10, 14, 70, 10,%WS_GROUP OR %WS_TABSTOP
   CONTROL ADD OPTION, hOlg, %IDC_Function_Number2,  "SGO",           10, 24, 70, 10
   CONTROL ADD OPTION, hOlg, %IDC_Function_Number3,  "Goldstein-Price",10, 34, 70, 10
   CONTROL ADD OPTION, hOlg, %IDC_Function_Number4,  "Step",          10, 44, 70, 10
   CONTROL ADD OPTION, hOlg, %IDC_Function_Number5,  "Schwefel 2.26", 10, 54, 70, 10
   CONTROL ADD OPTION, hOlg, %IDC_Function_Number6,  "Colville",      10, 64, 70, 10
   CONTROL ADD OPTION, hOlg, %IDC_Function_Number7,  "Griewank",      10, 74, 70, 10

   CONTROL ADD OPTION, hOlg, %IDC_Function_Number31, "PBM #1",        10, 84, 70, 10
   CONTROL ADD OPTION, hOlg, %IDC_Function_Number32, "PBM #2",        10, 94, 70, 10
   CONTROL ADD OPTION, hOlg, %IDC_Function_Number33, "PBM #3",        10,104, 70, 10
   CONTROL ADD OPTION, hOlg, %IDC_Function_Number34, "PBM #4",        10,114, 70, 10
   CONTROL ADD OPTION, hOlg, %IDC_Function_Number35, "PBM #5",        10,124, 70, 10
   CONTROL ADD OPTION, hOlg, %IDC_Function_Number36, "Himmelblau",    10,134, 70, 10
   CONTROL ADD OPTION, hOlg, %IDC_Function_Number37, "Rosenbrock",    10,144, 70, 10
   CONTROL ADD OPTION, hOlg, %IDC_Function_Number38, "Sphere",        10,154, 70, 10
   CONTROL ADD OPTION, hOlg, %IDC_Function_Number39, "HimmelblauHLO", 10,164, 70, 10
   CONTROL ADD OPTION, hOlg, %IDC_Function_Number40, "Tripod",        10,174, 70, 10
   CONTROL ADD OPTION, hOlg, %IDC_Function_Number41, "Rosenbrock F6", 10,184, 70, 10
   CONTROL ADD OPTION, hOlg, %IDC_Function_Number42, "Comp Spring",   10,194, 70, 10
   CONTROL ADD OPTION, hOlg, %IDC_Function_Number43, "Gear Train",    10,204, 70, 10
   CONTROL ADD OPTION, hOlg, %IDC_Function_Number44, "Loaded Monopole",10,214, 70, 10
   CONTROL ADD OPTION, hOlg, %IDC_Function_Number45, "Ackley",        10,224, 70, 10
   CONTROL ADD OPTION, hOlg, %IDC_Function_Number46, "Cosine Mixture",10,234, 70, 10
   CONTROL ADD OPTION, hOlg, %IDC_Function_Number47, "Exponential",   10,244, 70, 10
   CONTROL ADD OPTION, hOlg, %IDC_Function_Number48, "Rastrigin",     10,254, 70, 10
   CONTROL ADD OPTION, hOlg, %IDC_Function_Number49, "Reserved",      10,264, 70, 10
   CONTROL ADD OPTION, hOlg, %IDC_Function_Number50, "Reserved",      10,274, 70, 10

'------------ Test Functions from GSO Paper -------------
   CONTROL ADD OPTION, hOlg, %IDC_Function_Number8,  "F1",          : 120, 14, 40, 10
   CONTROL ADD OPTION, hOlg, %IDC_Function_Number9,  "F2",          : 120, 24, 40, 10
   CONTROL ADD OPTION, hOlg, %IDC_Function_Number10, "F3",          : 120, 34, 40, 10
   CONTROL ADD OPTION, hOlg, %IDC_Function_Number11, "F4",          : 120, 44, 40, 10
   CONTROL ADD OPTION, hOlg, %IDC_Function_Number12, "F5",          : 120, 54, 40, 10
   CONTROL ADD OPTION, hOlg, %IDC_Function_Number13, "F6",          : 120, 64, 40, 10
```

```
CONTROL ADD OPTION, hDlg, %IDC_Function_Number14, "F7(Random!)" , 120, 74, 47, 10
CONTROL ADD OPTION, hDlg, %IDC_Function_Number15, "F8"          , 120, 84, 40, 10
CONTROL ADD OPTION, hDlg, %IDC_Function_Number16, "F9"          , 120, 94, 40, 10
CONTROL ADD OPTION, hDlg, %IDC_Function_Number17, "F10"         , 120, 104, 40, 10
CONTROL ADD OPTION, hDlg, %IDC_Function_Number18, "F11"         , 120, 114, 40, 10
CONTROL ADD OPTION, hDlg, %IDC_Function_Number19, "F12"         , 120, 124, 40, 10
CONTROL ADD OPTION, hDlg, %IDC_Function_Number20, "F13"         , 120, 134, 40, 10
CONTROL ADD OPTION, hDlg, %IDC_Function_Number21, "F14"         , 120, 144, 40, 10
CONTROL ADD OPTION, hDlg, %IDC_Function_Number22, "F15"         , 120, 154, 40, 10
CONTROL ADD OPTION, hDlg, %IDC_Function_Number23, "F16"         , 120, 164, 40, 10
CONTROL ADD OPTION, hDlg, %IDC_Function_Number24, "F17"         , 120, 174, 40, 10
CONTROL ADD OPTION, hDlg, %IDC_Function_Number25, "F18"         , 120, 184, 40, 10
CONTROL ADD OPTION, hDlg, %IDC_Function_Number26, "F19"         , 120, 194, 40, 10
CONTROL ADD OPTION, hDlg, %IDC_Function_Number27, "F20"         , 120, 204, 40, 10
CONTROL ADD OPTION, hDlg, %IDC_Function_Number28, "F21"         , 120, 214, 40, 10
CONTROL ADD OPTION, hDlg, %IDC_Function_Number29, "F22"         , 120, 224, 40, 10
CONTROL ADD OPTION, hDlg, %IDC_Function_Number30, "F23"         , 120, 234, 40, 10

CONTROL SET OPTION  hDlg, %IDC_Function_Number1, %IDC_Function_Number1, %IDC_Function_Number3 'default to Parrott F4

'----------------------------------------------------------------------

CONTROL ADD BUTTON, hDlg, %IDOK, "&Ok", 200, 0.45##*BoxHeight&, 50, 14

'----------------------------------------------------------------------

DIALOG SHOW MODAL hDlg CALL DlgProc

CALL Delay(0.5##)

IF FunctionNumber&& < 1 OR FunctionNumber&& > 48 THEN

    FunctionNumber&& = 1 : MSGBOX("Error in function number...")

END IF

SELECT CASE FunctionNumber&&

    CASE 1  : FunctionName$ = "ParrottF4"
    CASE 2  : FunctionName$ = "SGO"
    CASE 3  : FunctionName$ = "GP"
    CASE 4  : FunctionName$ = "STEP"
    CASE 5  : FunctionName$ = "SCHWEFEL_226"
    CASE 6  : FunctionName$ = "COLVILLE"
    CASE 7  : FunctionName$ = "GRIEWANK"
    CASE 8  : FunctionName$ = "F1"
    CASE 9  : FunctionName$ = "F2"
    CASE 10 : FunctionName$ = "F3"
    CASE 11 : FunctionName$ = "F4"
    CASE 12 : FunctionName$ = "F5"
    CASE 13 : FunctionName$ = "F6"
    CASE 14 : FunctionName$ = "F7"
    CASE 15 : FunctionName$ = "F8"
    CASE 16 : FunctionName$ = "F9"
    CASE 17 : FunctionName$ = "F10"
    CASE 18 : FunctionName$ = "F11"
    CASE 19 : FunctionName$ = "F12"
    CASE 20 : FunctionName$ = "F13"
    CASE 21 : FunctionName$ = "F14"
    CASE 22 : FunctionName$ = "F15"
    CASE 23 : FunctionName$ = "F16"
    CASE 24 : FunctionName$ = "F17"
    CASE 25 : FunctionName$ = "F18"
    CASE 26 : FunctionName$ = "F19"
    CASE 27 : FunctionName$ = "F20"
    CASE 28 : FunctionName$ = "F21"
    CASE 29 : FunctionName$ = "F22"
    CASE 30 : FunctionName$ = "F23"
    CASE 31 : FunctionName$ = "PBM_1"
    CASE 32 : FunctionName$ = "PBM_2"
    CASE 33 : FunctionName$ = "PBM_3"
    CASE 34 : FunctionName$ = "PBM_4"
    CASE 35 : FunctionName$ = "PBM_5"
    CASE 36 : FunctionName$ = "HIMMELBLAU"
    CASE 37 : FunctionName$ = "ROSENBROCK"
    CASE 38 : FunctionName$ = "SPHERE"
    CASE 39 : FunctionName$ = "HIMMELBLAUNLO"
    CASE 40 : FunctionName$ = "TRIPOD"
    CASE 41 : FunctionName$ = "ROSENBROCKF6"
    CASE 42 : FunctionName$ = "COMPRESSIONSPRING"
    CASE 43 : FunctionName$ = "GEARTRAIN"
    CASE 44 : FunctionName$ = "LO_HONO"
    CASE 45 : FunctionName$ = "ACKLEY"
    CASE 46 : FunctionName$ = "COSINEMIX"
    CASE 47 : FunctionName$ = "EXPONENTIAL"
    CASE 48 : FunctionName$ = "RASTRIGIN"
END SELECT

END SUB

'-----------

CALLBACK FUNCTION DlgProc() AS LONG
'------------------------------------------------------
' Callback procedure for the main dialog
'------------------------------------------------------
LOCAL c, lRes AS LONG, sText AS STRING

SELECT CASE AS LONG CBMSG

CASE %WM_INITDIALOG' &&WM_INITDIALOG is sent right before the dialog is shown.

CASE %WM_COMMAND            ' <- a control is calling

    SELECT CASE AS LONG CBCTL   ' <- look at control's id

    CASE %IDOK                  ' <- Ok button or enter key was pressed

        IF CBCTLMSG = %BN_CLICKED THEN
            '----------------------------------
            ' Loop through the Function_number controls
            ' to see which one is selected
            '----------------------------------
            FOR c = %IDC_Function_Number1 TO %IDC_Function_Number50

                CONTROL GET CHECK CBHNDL, c TO lRes

                IF lRes THEN EXIT FOR

            NEXT 'c holds the id for selected test function.

            FunctionNumber&& = c-120

            DIALOG END CBHNDL

        END IF

    END SELECT

END SELECT

END FUNCTION

'------------------------- PBM ANTENNA BENCHMARK FUNCTIONS ----------------------------

'Reference for benchmarks PBM_1 through PBM_5:
'Pantoja, M.F., Bretones, A. R., Martin, R. G., "Benchmark Antenna Problems for Evolutionary
'Optimization Algorithms, IEEE Trans. Antennas & Propagation, vol. 55, no. 4, April 2007,
'pp. 1111-1121

FUNCTION PBM_1(R()),NG&&,p&&,j&&) 'PBM Benchmark #1: Max O for variable-Length CF Dipole

    LOCAL Z, Lengthwaves, ThetaRadians AS EXT

    LOCAL N&&, Nsegs&&, FeedSegNum&&
```

```
          LOCAL NumSegs$, FeedSeg$, HalfLength$, Radius$, ThetaDeg$, Lyne$, GainDB$

'msgbox("entered PBM_1")

          Lengthwaves  = R(p&&,1,j&&)

          ThetaRadians = R(p&&,2,j&&)

          ThetaDeg$ = REMOVE$(STR$(ROUND(ThetaRadians*Rad2Deg,2)),ANY" ")

          IF TALLY(ThetaDeg$,".") = 0 THEN ThetaDeg$ = ThetaDeg$&"."

          Nsegs&& = 2*(INT(100*Lengthwaves\2)+1 '100 segs per wavelength, must be an odd #, VOLTAGE SOURCE

          FeedSegNum&& = Nsegs&&\2 + 1 'center segment number, VOLTAGE SOURCE

          NumSegs$    = REMOVE$(STR$(Nsegs&&),ANY" ")

          FeedSeg$    = REMOVE$(STR$(FeedSegNum&&),ANY" ")

          HalfLength$ = REMOVE$(STR$(ROUND(Lengthwaves/2##,6)),ANY" ")

          IF TALLY(HalfLength$,".") = 0 THEN HalfLength$ = HalfLength$&"."

          Radius$     = "0.00001" 'REMOVE$(STR$(ROUND(Lengthwaves/1000##,6)),ANY" ")

          N&& = FREEFILE

          OPEN "PBM1.NEC" FOR OUTPUT AS #N&&

                  PRINT #N&&,"CM File: PBM1.NEC"
                  PRINT #N&&,"CM Run ID "+DATE$+" "+TIME$
                  PRINT #N&&,"CM N="+STR$(N&&)+", p="STR$(p&&)+", j="+STR$(j&&)
                  PRINT #N&&,"CM R(p,1,j)="+STR$(R(p&&,1,j&&))+", R(p,2,j)="+STR$(R(p&&,2,j&&))
                  PRINT #N&&,"CE"
                  PRINT #N&&,"GW 1,"+NumSegs$+",0.,0.,-"+HalfLength$+",0.,0.,"+HalfLength$+","+Radius$
                  PRINT #N&&,"GE"
                  PRINT #N&&,"EX 0,1,"+FeedSeg$+",0,1.,0." 'VOLTAGE SOURCE
                  PRINT #N&&,"FR 0,1,0,0,299.79564,0."
                  PRINT #N&&,"RP 0,1,1,1001,"+ThetaDeg$+",0.,0.,1000." 'gain at 1000 wavelengths range
                  PRINT #N&&,"XQ"
                  PRINT #N&&,"EN"

          CLOSE #N&&
'              - - ANGLES - -    - - - POWER GAINS -      - - - - POLARIZATION - -    - - - E(THETA)- - -   - - - E(PHI) - - -
' THETA     PHI     VERT.   HOR.   TOTAL    AXIAL   TILT  SENSE    MAGNITUDE  PHASE    MAGNITUDE  PHASE
' DEGREES  DEGREES    DB     DB      DB      RATIO   DEG.            VOLTS/M  DEGREES    VOLTS/M  DEGREES
' 90.00     0.00   1.91 -999.99   1.91   0.00000   0.00 LINEAR    1.295046-04   5.37  0.000000+00   -5.24
'123456789s123456789s123456789s123456789s123456789s123456789s123456789s123456789s123456789s123456789s123456789s123456789s
'       10        20        30        40        50        60        70        80        90       100       110       120

          SHELL "n41_2k1.exe",0

          N&& = FREEFILE

          OPEN "PBM1.OUT" FOR INPUT AS #N&&

                  WHILE NOT EOF(N&&)

                      LINE INPUT #N&&, Lyne$

                          IF INSTR(Lyne$,"DEGREES  DEGREES") > 0 THEN EXIT LOOP

                  WEND 'position at next data line

                  LINE INPUT #N&&, Lyne$

          CLOSE #N&&

          GainDB$ = REMOVE$(MID$(Lyne$,37,8),ANY" ")

          PBM_1 = 10^(VAL(GainDB$)/10##) 'Directivity

END FUNCTION 'PBM_1()
'----------------------

FUNCTION PBM_2(R(),N&&&,p&&,j&&) 'PBM Benchmark #2: Max D for Variable-Separation Array of CF Dipoles

          LOCAL Z, DipoleSeparationwaves, ThetaRadians AS EXT

          LOCAL N&&, i&&

          LOCAL NumSegs$, FeedSeg$, Radius$, ThetaDeg$, Lyne$, GainDB$, Xcoord$, Wirenum$

          DipoleSeparationwaves = R(p&&,1,j&&)

          ThetaRadians = R(p&&,2,j&&)

          ThetaDeg$ = REMOVE$(STR$(ROUND(ThetaRadians*Rad2Deg,2)),ANY" ")

          IF TALLY(ThetaDeg$,".") = 0 THEN ThetaDeg$ = ThetaDeg$&"."

          NumSegs$ = "49"

          FeedSeg$ = "25"

          Radius$  = "0.00001"

          N&& = FREEFILE

          OPEN "PBM2.NEC" FOR OUTPUT AS #N&&

                  PRINT #N&&,"CM File: PBM2.NEC"
                  PRINT #N&&,"CM Run ID "+DATE$+" "+TIME$
                  PRINT #N&&,"CM N="+STR$(N&&)+", p="STR$(p&&)+", j="+STR$(j&&)
                  PRINT #N&&,"CM R(p,1,j)="+STR$(R(p&&,1,j&&))+", R(p,2,j)="+STR$(R(p&&,2,j&&))
                  PRINT #N&&,"CE"

                  FOR i&& = -9 TO 9 STEP 2
                      Wirenum$ = REMOVE$(STR$((i&&+11)\2),ANY" ")
                      Xcoord$ = REMOVE$(STR$(i&&*DipoleSeparationwaves/2##),ANY" ")
                      PRINT #N&&,"GW "+Wirenum$+","+NumSegs$+","+Xcoord$+",0.,-0.25,"+Xcoord$+",0.,0.25,"+Radius$
                  NEXT i&&

                  PRINT #N&&,"GE"

                  FOR i&& = 1 TO 10
                      PRINT #N&&,"EX 0,"+REMOVE$(STR$(i&&),ANY" ")+","+FeedSeg$+",0,1.,0." 'VOLTAGE SOURCE
                  NEXT i&&
                  PRINT #N&&,"FR 0,1,0,0,299.79564,0."
                  PRINT #N&&,"RP 0,1,1,1001,"+ThetaDeg$+",90.,0.,0.,1000." 'gain at 1000 wavelengths range
                  PRINT #N&&,"XQ"
                  PRINT #N&&,"EN"

          CLOSE #N&&
'              - - ANGLES - -    - - - POWER GAINS -      - - - - POLARIZATION - -    - - - E(THETA)- - -   - - - E(PHI) - - -
' THETA     PHI     VERT.   HOR.   TOTAL    AXIAL   TILT  SENSE    MAGNITUDE  PHASE    MAGNITUDE  PHASE
' DEGREES  DEGREES    DB     DB      DB      RATIO   DEG.            VOLTS/M  DEGREES    VOLTS/M  DEGREES
' 90.00     0.00   1.91 -999.99   1.91   0.00000   0.00 LINEAR    1.295046-04   5.37  0.000000+00   -5.24
'123456789s123456789s123456789s123456789s123456789s123456789s123456789s123456789s123456789s123456789s123456789s123456789s
'       10        20        30        40        50        60        70        80        90       100       110       120

          SHELL "n41_2k1.exe",0

          N&& = FREEFILE

          OPEN "PBM2.OUT" FOR INPUT AS #N&&

                  WHILE NOT EOF(N&&)

                      LINE INPUT #N&&, Lyne$

                          IF INSTR(Lyne$,"DEGREES  DEGREES") > 0 THEN EXIT LOOP

                  WEND 'position at next data line
```

```
        LINE INPUT #N&&, Lyne$

    CLOSE #N&&

    GainDB$ = REMOVE$(MID$(Lyne$,37,8),ANY" ")

    IF AddNoiseToPBM2$ = "YES" THEN

        Z = 10^(VAL(GainDB$)/10##) + GaussianDeviate(0##,0.4472##) 'Directivity with Gaussian noise (zero mean, 0.2 variance)

    ELSE

        Z = 10^(VAL(GainDB$)/10##) 'Directivity without noise

    END IF

    PBM_2 = Z

END FUNCTION 'PBM_2()

'----

FUNCTION PBM_3(R(),NG&&,p&&,j&&) 'PBM Benchmark #3: Max D for Circular Dipole Array

    LOCAL Beta, ThetaRadians, Alpha, Rev, ImV AS EXT

    LOCAL N&&, i&&

    LOCAL NumSegs$, FeedSeg$, Radius$, ThetaDeg$, Lyne$, GainDB$, Xcoord$, Ycoord$, WireNum$, ReEX$, ImEX$

    Beta       = R(p&&,1,j&&)

    ThetaRadians = R(p&&,2,j&&)

    ThetaDeg$ = REMOVE$(STR$(ROUND(ThetaRadians*Rad2Deg,2)),ANY" ")

    IF TALLY(ThetaDeg$,".") = 0 THEN ThetaDeg$ = ThetaDeg$+"."

    NumSegs$ = "49"

    FeedSeg$ = "25"

    Radius$  = "0.00001"

    N&& = FREEFILE

    OPEN "PBM3.NEC" FOR OUTPUT AS #N&&

        PRINT #N&&,"CM File: PBM3.NEC"
        PRINT #N&&,"CM Run ID "+DATE$+" "+TIME$
        PRINT #N&&,"CM Nd="+STR$(NG&&)+", p="STR$(p&&)+", j="+STR$(j&&)
        PRINT #N&&,"CM R(p,1,j)="+STR$(R(p&&,1,j&&))+", R(p,2,j)="+STR$(R(p&&,2,j&&))
        PRINT #N&&,"CE

        FOR i&& = 1 TO 8

            WireNum$ = REMOVE$(STR$(i&&),ANY" ")

            SELECT CASE i&&
                CASE 1 : Xcoord$ = "1"        : Ycoord$ = "0"
                CASE 2 : Xcoord$ = "0.70711"  : Ycoord$ = "0.70711"
                CASE 3 : Xcoord$ = "0"        : Ycoord$ = "1"
                CASE 4 : Xcoord$ = "-0.70711" : Ycoord$ = "0.70711"
                CASE 5 : Xcoord$ = "-1"       : Ycoord$ = "0"
                CASE 6 : Xcoord$ = "-0.70711" : Ycoord$ = "-0.70711"
                CASE 7 : Xcoord$ = "0"        : Ycoord$ = "-1"
                CASE 8 : Xcoord$ = "0.70711"  : Ycoord$ = "-0.70711"
            END SELECT

            PRINT #N&&,"GW "+WireNum$+","+NumSegs$+","+Xcoord$+","+Ycoord$+",-0.25,"+Xcoord$+","+Ycoord$+",0.25,"+Radius$

        NEXT i&&

        PRINT #N&&,"GE"

        FOR i&& = 1 TO 8

            Alpha = -COS(TwoPi*Beta*(i&&-1))

            ReV = COS(Alpha)
            ImV = SIN(Alpha)

            ReEX$ = REMOVE$(STR$(ROUND(ReV,6)),ANY" ")
            ImEX$ = REMOVE$(STR$(ROUND(ImV,6)),ANY" ")

            IF TALLY(ReEX$,".") = 0 THEN ReEX$ = ReEX$+"."
            IF TALLY(ImEX$,".") = 0 THEN ImEX$ = ImEX$+"."

            PRINT #N&&,"EX "+WireNum$+","+FeedSeg$+",0,"+ReEX$+","+ImEX$ 'VOLTAGE SOURCE

        NEXT i&&

        PRINT #N&&,"FR 0,1,0,0,299.79564,0."
        PRINT #N&&,"RP 0,1,1,1001,"+ThetaDeg$+",0.,0.,0.,1000." 'gain at 1000 wavelengths range
        PRINT #N&&,"XQ"
        PRINT #N&&,"EN"

    CLOSE #N&&

'       - - ANGLES - -       - - POWER GAINS -       - - POLARIZATION -       - - E(THETA)        - - E(PHI) - -
'        THETA     PHI       VERT.  HOR.   TOTAL      AXIAL   TILT   SENSE    MAGNITUDE   PHASE    MAGNITUDE   PHASE
'       DEGREES   DEGREES      DB     DB      DB       RATIO   DEG.            VOLTS/M   DEGREES    VOLTS/M   DEGREES
'        90.00     0.00      3.91  -999.99  3.91     0.00000   0.00 LINEAR   1.291046-04   5.37  0.00000E+00  -1.24
'123456789x123456789x123456789x123456789x123456789x123456789x123456789x123456789x123456789x123456789x123456789x
'        10        20        30        40        50        60        70        80        90       100       110       120

    SHELL "H41_2K1.exe",0

    N&& = FREEFILE

    OPEN "PBM3.OUT" FOR INPUT AS #N&&

        WHILE NOT EOF(N&&)

            LINE INPUT #N&&, Lyne$

            IF INSTR(Lyne$,"DEGREES  DEGREES") > 0 THEN EXIT LOOP

        WEND 'position at next data line

        LINE INPUT #N&&, Lyne$

    CLOSE #N&&

    GainDB$ = REMOVE$(MID$(Lyne$,37,8),ANY" ")

    PBM_3 = 10^(VAL(GainDB$)/10##) 'Directivity

END FUNCTION 'PBM_3()

'----

FUNCTION PBM_4(R(),NG&&,p&&,j&&) 'PBM Benchmark #4: Max D for Vee Dipole

    LOCAL TotalLengthWaves, AlphaRadians, ArmLength, XLength, ZLength, Lfeed AS EXT

    LOCAL N&&, i&&, Nsegs&&, Feed2coord$

    LOCAL NumSegs$, Lyne$, GainDB$, Xcoord$, Zcoord$

    TotalLengthWaves = 2##*R(p&&,1,j&&)

    AlphaRadians = R(p&&,2,j&&)

    Lfeed        = 0.01##

    Feed2coord$ = REMOVE$(STR$(Lfeed),ANY" ")

    ArmLength = (TotalLengthWaves-2##*Lfeed)/2##

    XLength = ROUND(ArmLength*COS(AlphaRadians),6)
```

```
Xcoord$ = REMOVE$(STR$(Xlength),ANY " ") : IF TALLY(Xcoord$,".") = 0 THEN Xcoord$ = Xcoord$+"."

Zlength = ROUND(ArmLength*SIN(AlphaRadians),6)

Zcoord$ = REMOVE$(STR$(Zlength+Lfeed),ANY " ") : IF TALLY(Zcoord$,".") = 0 THEN Zcoord$ = Zcoord$+"."

Nsegs&& = 2*(INT(TotalLengthWaves*100)\2) 'even number, total # segs

NumSegs$ = REMOVE$(STR$(Nsegs&&\2),ANY " ") '# segs per arm

N&& = FREEFILE

OPEN "PBM4.NEC" FOR OUTPUT AS #N&&

    PRINT #N&&,"CM FILE: PBM4.NEC"
    PRINT #N&&,"CM Run ID "+DATE$+" "+TIME$
    PRINT #N&&,"CM Nd="+STR$(Nd&&)+", p="STR$(p&&)+", j="+STR$(j&&)
    PRINT #N&&,"CM R(p,1,j)="+STR$(R(p&&,1,j&&))+", R(p,2,j)="+STR$(R(p&&,2,j&&))
    PRINT #N&&,"CE"

    PRINT #N&&,"GW 1,5,0.,0.,-"+Feedzcoord$+",0.,0.,-"+Feedzcoord$+",0.00001" 'feed wire, 1 segment, 0.01 wvln

    PRINT #N&&,"GW 2,"+NumSegs$+",0.,0.,-"+Feedzcoord$+","+Xcoord$+",0.,-"+Zcoord$+",0.00001" 'upper arm

    PRINT #N&&,"GW 3,"+NumSegs$+",0.,0.,-"+Feedzcoord$+","+Xcoord$+",0.,-"+Zcoord$+",0.00001" 'lower arm

    PRINT #N&&,"GE"

    PRINT #N&&,"EX 0,1,3,0,1.,0." 'VOLTAGE SOURCE

    PRINT #N&&,"FR 0,1,0,0,299.79564,0."
    PRINT #N&&,"RP 0,1,1,1001,90.,0.,0.,0.,1000." 'ENDFIRE gain at 1000 wavelengths range
    PRINT #N&&,"XQ"
    PRINT #N&&,"EN"

CLOSE #N&&

' - - ANGLES - -      - - POWER GAINS - -          - - POLARIZATION - - -    - - - E(THETA) - - -    - - E(PHI) - -
'   THETA     PHI     VERT.   HOR.    TOTAL     AXIAL    TILT   SENSE    MAGNITUDE    PHASE     MAGNITUDE    PHASE
' DEGREES  DEGREES     DB      DB       DB       RATIO    DEG.           VOLTS/M    DEGREES     VOLTS/M    DEGREES
'  90.00     0.00    1.91  -999.99    1.91    0.00000    0.00  LINEAR   1.29504E-04     5.37   0.00000E+00    -5.24
'123456789i123456789e123456789s123456789t123456789i123456789e123456789s123456789t123456789i123456789e123456789s
'    10       20      30      40       50       60       70      80       90          100         110        120

SHELL "n41_2k1.exe",0

N&& = FREEFILE

OPEN "PBM4.OUT" FOR INPUT AS #N&&

    WHILE NOT EOF(N&&)
        LINE INPUT #N&&, Lyne$
        IF INSTR(Lyne$,"DEGREES  DEGREES") > 0 THEN EXIT LOOP
    WEND 'position at next data line
    LINE INPUT #N&&, Lyne$

CLOSE #N&&

Gain0B$ = REMOVE$(MID$(Lyne$,37,8),ANY " ")
PBM_4 = 10^(VAL(Gain0B$)/10##) 'Directivity
END FUNCTION 'PBM_4()

'----

FUNCTION PBM_5(R(),Nd&&,p&&,j&&) 'PBM Benchmark #5: N-element collinear array (Nd=N-1)

    LOCAL TotalLengthWaves, Di(), Ystart, Y1, Y2, SumDi AS EXT

    LOCAL N&&, i&&, q&&

    LOCAL Lyne$, Gain0B$

    REDIM Di(1 TO Nd&&)

    FOR i&& = 1 TO Nd&&

        Di(i&&) = R(p&&,i&&,j&&) 'dipole separation, wavelengths

    NEXT i&&

    TotalLengthwaves = 0##

    FOR i&& = 1 TO Nd&&

        TotalLengthwaves = TotalLengthwaves + Di(i&&)

    NEXT i&&

    TotalLengthwaves = TotalLengthwaves + 0.5## 'add half-wavelength of 1 meter at 299.8 MHz

    Ystart = -TotalLengthwaves/2##

    N&& = FREEFILE

    OPEN PBM5.NEC FOR OUTPUT AS #N&&

        PRINT #N&&,"CM FILE: PBM5.NEC"
        PRINT #N&&,"CM Run ID "+DATE$+" "+TIME$
        PRINT #N&&,"CM Nd="+STR$(Nd&&)+", p="STR$(p&&)+", j="+STR$(j&&)
        PRINT #N&&,"CM R(p,1,j)="+STR$(R(p&&,1,j&&))+", R(p,2,j)="+STR$(R(p&&,2,j&&))
        PRINT #N&&,"CE"

        FOR i&& = 1 TO Nd&&+1

            SumDi = 0##

            FOR q&& = 1 TO i&&-1

                SumDi = SumDi + Di(q&&)

            NEXT q&&

            Y1 = ROUND(Ystart + SumDi,6)

            Y2 = ROUND(Y1+0.5##,6) 'add one-half wavelength for other end of dipole

            PRINT #N&&,"GW "+REMOVE$(STR$(i&&),ANY " ")+",49,0.,"+REMOVE$(STR$(Y1),ANY " ")+",0.,0.,"+REMOVE$(STR$(Y2),ANY " ")+",0.,0.00001"

        NEXT i&&

        PRINT #N&&,"GE"

        FOR i&& = 1 TO Nd&&+1
            PRINT #N&&,"EX 0,"+REMOVE$(STR$(i&&),ANY " ")+",25,0,1.,0." 'VOLTAGE SOURCES
        NEXT i&&

        PRINT #N&&,"FR 0,1,0,0,299.79564,0."
        PRINT #N&&,"RP 0,1,1,1001,90.,0.,0.,0.,1000." 'gain at 1000 wavelengths range
        PRINT #N&&,"XQ"
        PRINT #N&&,"EN"

' - - ANGLES - -      - - POWER GAINS - -          - - POLARIZATION - - -    - - - E(THETA) - - -    - - E(PHI) - -
'   THETA     PHI     VERT.   HOR.    TOTAL     AXIAL    TILT   SENSE    MAGNITUDE    PHASE     MAGNITUDE    PHASE
' DEGREES  DEGREES     DB      DB       DB       RATIO    DEG.           VOLTS/M    DEGREES     VOLTS/M    DEGREES
'  90.00     0.00    1.91  -999.99    1.91    0.00000    0.00  LINEAR   1.29504E-04     5.37   0.00000E+00    -5.24
'123456789i123456789e123456789s123456789t123456789i123456789e123456789s123456789t123456789i123456789e123456789s
'    10       20      30      40       50       60       70      80       90          100         110        120

    SHELL "n41_2k1.exe",0
```

```
        N&& = FREEFILE

        OPEN "PBM5.OUT" FOR INPUT AS #N&&

            WHILE NOT EOF(N&&)
                LINE INPUT #N&&, Lyne$
                IF INSTR(Lyne$,"DEGREES  DEGREES") > 0 THEN EXIT LOOP
            WEND 'position at next data line
            LINE INPUT #N&&, Lyne$

        CLOSE #N&&

        GainDB$ = REMOVE$(MID$(Lyne$,37,8),ANY" ")
        PBM_5 = 10^(VAL(GainDB$)/10##) 'Directivity

END FUNCTION 'PBM_5()

'----------------------
SUB CopyBestNECInputFile(NECFile$)
LOCAL N&&, M&&, Lyne$
        N&& = FREEFILE : OPEN NECFile$ FOR INPUT AS #N&&
        M&& = FREEFILE : OPEN "Best"+NECFile$ FOR OUTPUT AS #M&&
        WHILE NOT EOF(N&&) : LINE INPUT #N&&, Lyne$ : PRINT #M&&, Lyne$ : WEND
        CLOSE #N&& : CLOSE #M&&
END SUB

'------
FUNCTION LD_MONO(R(),N&&&,p&&,j&&) 'DIPOLE-LOADED MONOPOLE (Altshuler)

        LOCAL N&&, I&&, Nsegs&&, NumFreqs&&, NumThetaAngles&&, NumPhiAngles&&, NumPatternPoints&&, WireNum&&

        LOCAL Lyne$, FN1$, FileStatus$, FileID$, OutFileID$, Fit1$

        LOCAL a_avln, SegLength, X1, X2, Z1, Z2, Z3, Z4 AS EXT

        LOCAL Fitness, FcMHz, DeltaFreqMHz, AGT AS EXT

        LOCAL TotPowerGainDBI, AvgGainDBI, VarGainDBI, MinGainDBI, MaxGainDBI, Etheta(), Ephi(), Etotal() AS SINGLE

        LOCAL xEnd1, yEnd1, zEnd1, xEnd2, yEnd2, zEnd2 AS EXT

        a_avln = 0.0005## 'wire radius WAVELENGTHS at 299.8 MHz (1 mm diam)

        NumFreqs&& = 1 : FcMHz = 299.8## : DeltaFreqMHz = 0##

        NumThetaAngles&& = 19 : NumPhiAngles&& = 8 : NumPatternPoints&& = NumThetaAngles&&*NumPhiAngles&&

        Fit1$ = "1/Sum(Gtot(Th,Ph)^2)" ; G=Pwr Gain(dBi); Th,Ph=angles"

'    PROCESS SET PRIORITY %NORMAL_PRIORITY_CLASS 'NORMAL PRIORITY TO AVOID PROBLEMS WRITING/READING NEC FILES
'    PROCESS SET PRIORITY %REALTIME_PRIORITY_CLASS 'PREEMPTS ALL OTHER PROCESSES -> CAN CAUSE PROBLEMS...

        N&& = FREEFILE

        OPEN "MONO.NEC" FOR OUTPUT AS #N&&

            FileID$ = REMOVE$(DATE$+TIME$,ANY Alphabet$+" -:/")

            PRINT #N&&,"CM File: MONO.NEC"
            PRINT #N&&,"CM DIPOLE-LOADED MONOPOLE (Altshuler, 1997)"
            PRINT #N&&,"CM Fc ="+STR$(FcMHz)+" MHz"
            IF DeltaFreqMHz <> 0## THEN PRINT #N&&,"CM Delta Freq = +/-"+REMOVE$(STR$(DeltaFreqMHz),ANY " ")+" MHz"
            PRINT #N&&,"CM Run ID: "+RunID$
            PRINT #N&&,"CM Fitness: " + Fit1$
            PRINT #N&&,"CM File ID "+FileID$
            PRINT #N&&,"CM Nds="+STR$(N&&&)+", p="STR$(p&&)+", j="+STR$(j&&)
            PRINT #N&&,"CE"

'DIPOLE-LOADED MONOPOLE GEOMETRY DS (SEE FIG. 1 IN REF [1])
'========================================================
'            XiMin(1) = 0.05## : XiMax(1) = 0.50## 'X1, wavelengths
'            XiMin(2) = 0.05## : XiMax(2) = 0.50## 'X2, wavelengths
'            XiMin(3) = 0.03## : XiMax(3) = 0.15## 'Z1, wavelengths
'            XiMin(4) = 0.01## : XiMax(4) = 0.10## 'Z2, wavelengths
'            XiMin(5) = 0.01## : XiMax(5) = 0.10## 'Z3, wavelengths
'            XiMin(6) = 0.01## : XiMax(6) = 0.10## 'Z4, wavelengths

'REFS:
'[1] Altshuler, E. E. and Linden, D. S., "Wire-Antenna Designs using Genetic Algorithms,"
'IEEE Ant. Prop. Mag., vol. 39, No. 2, April 1997, pp. 33-43.
'[2] Altshuler, E. E., "A Monopole Antenna Loaded with a Modified Folded Dipole,"
'IEEE Trans. Ant. Prop., vol. 41, No. 7, July 1993, pp. 871-876.
'========================================================
            X1 = ROUND(R(p&&,1,j&&),4)
            X2 = ROUND(R(p&&,2,j&&),4)
            Z1 = ROUND(R(p&&,3,j&&),4)
            Z2 = ROUND(R(p&&,4,j&&),4)
            Z3 = ROUND(R(p&&,5,j&&),4)
            Z4 = ROUND(R(p&&,6,j&&),4)

            WireNum&& = 1 : xEnd1 = 0## : yEnd1 = 0## : zEnd1 = 0##    : xEnd2 = 0## : yEnd2 = 0## : zEnd2 = Z1
            CALL PrintGWcard(N&&,WireNum&&,xEnd1,yEnd1,zEnd1,xEnd2,yEnd2,zEnd2,a_avln)

            WireNum&& = 2 : xEnd1 = 0## : yEnd1 = 0## : zEnd1 = Z1    : xEnd2 = X1 : yEnd2 = 0## : zEnd2 = Z1
            CALL PrintGWcard(N&&,WireNum&&,xEnd1,yEnd1,zEnd1,xEnd2,yEnd2,zEnd2,a_avln)

            WireNum&& = 3 : xEnd1 = X1 : yEnd1 = 0## : zEnd1 = Z1    : xEnd2 = X1 : yEnd2 = 0## : zEnd2 = Z2+Z1
            CALL PrintGWcard(N&&,WireNum&&,xEnd1,yEnd1,zEnd1,xEnd2,yEnd2,zEnd2,a_avln)

            WireNum&& = 4 : xEnd1 = X1 : yEnd1 = 0## : zEnd1 = Z2+Z1    : xEnd2 = 0## : yEnd2 = 0## : zEnd2 = Z2+Z1
            CALL PrintGWcard(N&&,WireNum&&,xEnd1,yEnd1,zEnd1,xEnd2,yEnd2,zEnd2,a_avln)

            WireNum&& = 5 : xEnd1 = 0## : yEnd1 = 0## : zEnd1 = Z2+Z1    : xEnd2 = -X2 : yEnd2 = 0## : zEnd2 = Z2+Z1
            CALL PrintGWcard(N&&,WireNum&&,xEnd1,yEnd1,zEnd1,xEnd2,yEnd2,zEnd2,a_avln)

            WireNum&& = 6 : xEnd1 = -X2 : yEnd1 = 0## : zEnd1 = Z2+Z1    : xEnd2 = -X2 : yEnd2 = 0## : zEnd2 = Z3+Z2+Z1
            CALL PrintGWcard(N&&,WireNum&&,xEnd1,yEnd1,zEnd1,xEnd2,yEnd2,zEnd2,a_avln)

            WireNum&& = 7 : xEnd1 = -X2 : yEnd1 = 0## : zEnd1 = Z3+Z2+Z1 : xEnd2 = -X2 : yEnd2 = 0## : zEnd2 = Z4+Z3+Z2+Z1
            CALL PrintGWcard(N&&,WireNum&&,xEnd1,yEnd1,zEnd1,xEnd2,yEnd2,zEnd2,a_avln)

            PRINT #N&&, "GE1" 'wires ending on gnd plane are connected to it

            PRINT #N&&, "GN1" 'PEC ground plane

            PRINT #N&&, "FR 0,"+Int2STR$(NumFreqs&&)+",0,0,"+FP2STR$(FcMHz-DeltaFreqMHz)+","+FP2STR$(DeltaFreqMHz) 'frequency card

            PRINT #N&&, "EX 0,1,1,1,1.,0." 'excite base of wire #1 with 1+j0 volts

            PRINT #N&&, "RP0,"+Int2STR$(NumThetaAngles&&)+","+Int2STR$(NumPhiAngles&&)+",1001,0.,0.,"+
                       FP2STR$(90##/(NumThetaAngles&&-1))+","+FP2STR$(360##/NumPhiAngles&&)+",100000." 'hemispheric gain, cuts 5 deg in Theta, 45 deg in
Phi

            PRINT #N&&, "EN"

        CLOSE #N&&

        SHELL "N4l2D_lkSegs_lkLD_lkIS.EXE MONO.NEC MONO.OUT",0

        IF DIR$("MONO.OUT") = "" THEN MSGBOX("MONO.OUT not found!")

        FileStatus$ = "NOK" : AGT = 0##
        N&& = FREEFILE
        OPEN "MONO.OUT" FOR INPUT AS #N&&
            IF LOF(N&&) = 0 THEN MSGBOX("MONO.OUT is empty!")
```

```
        WHILE NOT EOF(N&&)
            LINE INPUT #N&&, Lyne$
            IF INSTR(Lyne$,"AVERAGE POWER GAIN=") > 0 THEN AGT = VAL(MID$(Lyne$,23,12))
            IF INSTR(Lyne$,"File ID") > 0 THEN
                OutFileID$ = REMOVE$(Lyne$,ANY AlphaBet$+" ")
                IF FileID$ <> OutFileID$ THEN
                    MSGBOX("LOADED MONOPOLE file ID's don't match!") : EXIT LOOP
                END IF
            END IF
            IF INSTR(Lyne$,"RUN TIME =") > 0 THEN
                FileStatus$ = "OK" : EXIT LOOP
            END IF
        WEND
    CLOSE #N&&

    Fitness = -98765## 'default value

    IF FileStatus$ = "OK" THEN

        REDIM TotPowerGainDBI(1 TO NumThetaAngles&&*NumPhiAngles&&),Etheta(1 TO NumThetaAngles&&*NumPhiAngles&&),_
              EPhi(1 TO NumThetaAngles&&*NumPhiAngles&&),Etotal(1 TO NumThetaAngles&&*NumPhiAngles&&)
        N&& = FREEFILE
        OPEN "MONO_OUT" FOR INPUT AS #N&&
            Lyne$ = ""
            DO UNTIL INSTR(Lyne$,"- - - RADIATION PATTERNS - - -") > 0
                LINE INPUT #N&&, Lyne$
            LOOP

            FOR k&& = 1 TO 7 : LINE INPUT #N&&, Lyne$ : NEXT k&& 'skip 7 lines

            FOR k&& = 1 TO NumPatternPoints&& 'read power gains
                LINE INPUT #N&&, Lyne$
                TotPowerGainDBI(k&&) = VAL(MID$(Lyne$,38,7))
                Etheta(k&&)          = VAL(MID$(Lyne$,74,14))*1E6 'field MAGNITUDES in uv/m
                EPhi(k&&)            = VAL(MID$(Lyne$,98,14))*1E6
                Etotal(k&&)         = SQR(Etheta(k&&)^2+EPhi(k&&)^2)
            NEXT k&&
        CLOSE #N&&

        AvgGainDBI = 0## : AvgEtotal = 0## 'get min/max gains & compute average power gain & average total field
        MinGainDBI = 1E4200 : MaxGainDBI = -1E4200
        FOR k&& = 1 TO NumPatternPoints&&
            IF TotPowerGainDBI(k&&) =< MinGainDBI THEN MinGainDBI = TotPowerGainDBI(k&&)
            IF TotPowerGainDBI(k&&) => MaxGainDBI THEN MaxGainDBI = TotPowerGainDBI(k&&)
            AvgGainDBI = AvgGainDBI + TotPowerGainDBI(k&&)
            AvgEtotal = AvgEtotal + Etotal(k&&)
        NEXT k&&
        AvgGainDBI = AvgGainDBI/NumPatternPoints&&
        AvgEtotal = AvgEtotal/NumPatternPoints&&

'       VarGainDBI = 0## 'compute gain variance
'       FOR k&& = 1 TO NumPatternPoints&& : VarGainDBI = VarGainDBI + TotPowerGainDBI(k&&)^2 : NEXT k&&
'       VarGainDBI = VarGainDBI/NumPatternPoints&& - AvgGainDBI^2

        Fitness = 0##
        FOR k&& = 1 TO NumPatternPoints&&
            Fitness = Fitness + ABS(TotPowerGainDBI(k&&)-AvgGainDBI)
        NEXT k&&
        Fitness = 1##/Fitness

    END IF 'FileStatus$ = "OK"

    N&& = FREEFILE
    OPEN "MONO.NEC" FOR APPEND AS #N&&
        PRINT #N&&,"" : PRINT #N&&,""
        PRINT #N&&,"DIPOLE-LOADED MONOPOLE (Altshuler, 1997)"
        PRINT #N&&,"Band center frequency, Fc = 299.8 MHz"
        IF DeltaFreqMHz <> 0## THEN PRINT #N&&,"CH Delta Freq = +/-"+REMOVE$(STR$(DeltaFreqMHz),ANY " ")+" MHz"
        PRINT #N&&,"Run ID: "+RunID$
        PRINT #N&&,"Fit Fnc: " + Fit1$
        PRINT #N&&,"Fitness      ="+STR$(Fitness)
        PRINT #N&&,"AGT          ="+STR$(AGT)
        PRINT #N&&,"Min/Max Gains ="+STR$(MinGainDBI)+"/"+STR$(MaxGainDBI)+" dBi"
        PRINT #N&&,"Delta Gain   ="+STR$(MaxGainDBI-MinGainDBI)+" dBi"
        PRINT #N&&,""
    CLOSE #N&&

    N&& = FREEFILE
    OPEN "MONO.OUT" FOR APPEND AS #N&&
        PRINT #N&&,"" : PRINT #N&&,""
        PRINT #N&&,"DIPOLE-LOADED MONOPOLE (Altshuler, 1997)"
        PRINT #N&&,"Band center frequency, Fc = 299.8 MHz"
        IF DeltaFreqMHz <> 0## THEN PRINT #N&&,"CH Delta Freq = +/-"+REMOVE$(STR$(DeltaFreqMHz),ANY " ")+" MHz"
        PRINT #N&&,"Run ID: "+RunID$
        PRINT #N&&,"Fit Fnc: " + Fit1$
        PRINT #N&&,"Fitness      ="+STR$(Fitness)
        PRINT #N&&,"AGT          ="+STR$(AGT)
        PRINT #N&&,"Min/Max Gains ="+STR$(MinGainDBI)+"/"+STR$(MaxGainDBI)+" dBi"
        PRINT #N&&,"Delta Gain   ="+STR$(MaxGainDBI-MinGainDBI)+" dBi"
        PRINT #N&&,""
    CLOSE #N&&

    'NEC OUTPUT FILE FORMAT
    '==========================
    '     - - ANGLES - -   - - - - POWER GAINS - - - -     - - - POLARIZATION - - -   - - - E(THETA) - - -   - - - E(PHI) - -
    '    THETA   PHI     VERT.   HOR.    TOTAL    AXIAL   TILT   SENSE   MAGNITUDE  PHASE     MAGNITUDE   PHASE
    '   DEGREES  DEGREES   DB     DB      DB      RATIO   DEG.           VOLTS/M   DEGREES     VOLTS/M   DEGREES
    '    90.00   0.00   3.91 -999.99   3.91   0.00000    0.00  LINEAR   1.295D4E-04    5.37     0.00000E+00    -5.24
    '123456789x123456789x123456789x123456789x123456789x123456789x123456789x123456789x123456789x123456789x123456789x123456789x
    '     10       20       30       40       50       60       70       80       90      100      110      120
    '  AVERAGE POWER GAIN= 2.04127E+00       SOLID ANGLE USED IN AVERAGING= ( 1.7500)*PI STERADIANS.

    LD_MONO = ROUND(Fitness,5)

END FUNCTION 'LD_MONO()

'--------------------

SUB PrintGWcard(N&&,WireNum&&,xEnd1,yEnd1,zEnd1,xEnd2,yEnd2,zEnd2,a_wvln)

    LOCAL WireLength AS EXT

    LOCAL Nsegs&&

    WireLength = SQR((xEnd2-xEnd1)^2+(yEnd2-yEnd1)^2+(zEnd2-zEnd1)^2)
    Nsegs&& = MAX(WireLength/0.01##,1) 'seg length of 0.01 wvln, all dims in METERS @ 299.8 MHz because Lambda = 1.000 m @ 299.8 MHz
    PRINT #N&&,"GW"+Int2STR$(WireNum&&)+","+Int2STR$(Nsegs&&)+","+FP2STR$(xEnd1)+","+FP2STR$(yEnd1)+","+FP2STR$(zEnd1)+_
                                                         ","+FP2STR$(xEnd2)+","+FP2STR$(yEnd2)+","+FP2STR$(zEnd2)+","+FP2STR$(a_wvln)

END SUB 'PrintGWcard()

'--------------------

SUB GetNECdata(NECoutputFile$,NumFreqs&&,NumRadPatAngles&&,Zo,FrequencyMHZ(),RadEfficiencyPCT(),MaxGainDBI(),WinGainDBI(),RinOhms(),XinOhms(),VSWR(),ForwardGain
DBI(),RearGainDBI(),FileStatus$,FileID$)

    LOCAL N&&, idx&&, AngleNum&&

    LOCAL Lyne$, Dum$

    LOCAL GmaxDBI, GminDBI, FwdGainDBI, RearGain AS EXT

    REDIM FrequencyMHZ(1 TO NumFreqs&&),RadEfficiencyPCT(1 TO NumFreqs&&),MaxGainDBI(1 TO NumFreqs&&),WinGainDBI(1 TO NumFreqs&&),_
          RinOhms(1 TO NumFreqs&&),XinOhms(1 TO NumFreqs&&),VSWR(1 TO NumFreqs&&),ForwardGainDBI(1 TO NumFreqs&&), RearGainDBI(1 TO NumFreqs&&)

    FileStatus$ = "NOK"

    OPEN NECoutputFile$ FOR INPUT AS #N&&

        WHILE NOT EOF(N&&)

            LINE INPUT #N&&, Lyne$

            IF INSTR(Lyne$,"RUN TIME") > 0 THEN FileStatus$ = "OK"

        WEND

    CLOSE #N&&
```

```
        IF FileStatus$ <> "OK" THEN EXIT SUB

        OPEN NECoutputFile$ FOR INPUT AS #n&&

            idx&& = 1

            LINE INPUT #n&&, Lyne$

            WHILE NOT EOF(n&&)

                LINE INPUT #n&&, Lyne$

                IF INSTR(Lyne$,"File ID") > 0 THEN 'CHECK THAT NEC OUTPUT FILE WAS COMPUTED FROM CURRENT NEC INPUT FILE BY COMPARING FILE ID'S (TO AVOID
CACHE/BUFFER PROBLEMS CREATED BY OS)
                    IF FileID$ <> REMOVE$(Lyne$,ANY Alphabet$+" ") THEN MSGBOX("WARNING! NEC I/O File ID's Don't Match!"+CHR$(13)+Lyne$   =
"+Lyne$+CHR$(13)+"FileID$ = "+FileID$)
                END IF

                IF INSTR(Lyne$,"FREQUENCY=") > 0 THEN
                    Lyne$ = REMOVE$(Lyne$,"MHZ") : Lyne$ = REMOVE$(Lyne$,"FREQUENCY= ") : FrequencyMHZ(idx&&) = VAL(Lyne$)
'MSGBOX("idx="+STR$(idx&&)+"   F="+STR$(FrequencyMHZ(idx&&)))

                IF INSTR(Lyne$,"INPUT PARAMETERS") > 0 THEN
                    LINE INPUT #n&&, Dum$ : LINE INPUT #n&&, Dum$ : LINE INPUT #n&&, Dum$ 'skip three lines
                    LINE INPUT #n&&, Lyne$ 'input next line with impedance data
                    XinOhms(idx&&) = VAL(MID$(Lyne$,61,12)) : XinOhms(idx&&) = VAL(MID$(Lyne$,73,12)) : VSWR(idx&&) =
StandingwaveRatio(Zo,XinOhms(idx&&),XinOhms(idx&&))
                IF INSTR(Lyne$,"EFFICIENCY") > 0 THEN RadEfficiencyPCT(idx&&) = VAL(MID$(Lyne$,ANY Alphabet$+" ="))
                IF INSTR(Lyne$,"E(THETA)") > 0 THEN
                    LINE INPUT #n&&, Dum$ : LINE INPUT #n&&, Dum$ 'skip two lines
                    GmaxDBI = -9999## : GminDBI = -GmaxDBI
'               ------------ FORWARD PATTERN -----------
                    FOR AngleNum&& = 1 TO NumRadPattAngles&&
                        LINE INPUT #n&&, Lyne$ 'input next NumRadPattAngles&& lines with pattern data
                        IF VAL(MID$(Lyne$,38,7)) >= GmaxDBI THEN GmaxDBI = VAL(MID$(Lyne$,38,7)) 'get max gain
                        IF (VAL(MID$(Lyne$,38,7)) =< GminDBI AND VAL(MID$(Lyne$,38,7)) >= -999.99##) THEN GminDBI = VAL(MID$(Lyne$,38,7)) 'get min gain
                        IF AngleNum&& = NumRadPattAngles&& THEN FwdGainDBI = VAL(MID$(Lyne$,38,7)) 'forward gain
                    NEXT AngleNum&&
                    MaxGainDBI(idx&&) = GmaxDBI
                    MinGainDBI(idx&&) = GminDBI
                    ForwardGainDBI(idx&&) = FwdGainDBI

'               ------------ REAR PATTERN -----------
                    FOR AngleNum&& = 1 TO NumRadPattAngles&&
                        LINE INPUT #n&&, Lyne$ 'input next NumRadPattAngles&& lines with pattern data
                        IF AngleNum&& = NumRadPattAngles&& THEN RearGain = VAL(MID$(Lyne$,38,7)) 'rear gain
                    NEXT AngleNum&&
                    RearGainDBI(idx&&) = RearGain

                INCR idx&&
            END IF
'msgbox("idx="+STR$(idx&&))

        WEND

        CLOSE #n&&
'  TAG   SEG.     VOLTAGE (VOLTS)         CURRENT (AMPS)       IMPEDANCE (OHMS)      ADMITTANCE (MHOS)       POWER
'  NO.   NO.     REAL        IMAG.        REAL       IMAG.     REAL      IMAG.       REAL       IMAG.       (WATTS)
'   1    1  1.00000E+00 0.00000E+00 7.17910E-06 8.91191E-04 8.99811E+00-1.11951E+03 7.17910E-06 8.91191E-04 3.58955E-06
'123456789x123456789x123456789x123456789x123456789x123456789x123456789x123456789x123456789x123456789x123456789x123456789x
'      10          20          30          40          50         60         70         80          90         100        110        120        130

' - - ANGLES - -       - - - POWER GAINS -        - - - - - POLARIZATION - - - -     - - - - E(THETA) - - -      - - - - E(PHI) - - -
'  THETA    PHI       VERT.   HOR.   TOTAL        AXIAL   TILT  SENSE    MAGNITUDE   PHASE       MAGNITUDE   PHASE
'  DEGREES DEGREES     DB      DB     DB          RATIO   DEG.           VOLTS/M     DEGREES     VOLTS/M     DEGREES
'   0.00   0.00   -999.99 -999.99 -999.99    0.00000     0.00            0.00000E+00 -240.17     0.00000E+00 -240.17
'  10.00   0.00    -18.97 -999.99  -18.97    0.00000     0.00  LINEAR    2.69380E-08  -61.44     0.00000E+00 -240.17
'123456789x123456789x123456789x123456789x123456789x123456789x123456789x123456789x123456789x123456789x123456789x123456789x
'      10          20          30          40          50         60         70         80          90         100        110        120        130

        END SUB 'GetNECdata()
'-----------------------

FUNCTION StandingwaveRatio(Zo,ReZ,ImZ)

LOCAL ReRho, ImRho, MagRho, SWR AS EXT

        SWR = 9999##

        CALL ComplexDivide(ReZ-Zo,ImZ,ReZ+Zo,ImZ,ReRho,ImRho)

        MagRho = SQR(ReRho*ReRho+ImRho*ImRho)   'reflection coefficient

        IF MagRho <> 1## THEN SWR=(1##+MagRho)/(1##-MagRho)

        StandingwaveRatio = SWR

END FUNCTION 'StandingwaveRatio()
'-----------------------

SUB ComplexMultiply(ReA,ImA,ReB,ImB,ReC,ImC)
'Returns real and imaginary parts of product C=A*B
        ReC = ReA*ReB-ImA*ImB
        ImC = ImA*ReB+ReA*ImB
END SUB
'-----------

SUB ComplexDivide(ReA,ImA,ReB,ImB,ReC,ImC)
'Returns real and imaginary parts of quotient C=A/B
LOCAL Denom AS EXT
        Denom = ReB*ReB+ImB*ImB
        ReC = (ReA*ReB+ImA*ImB)/Denom
        ImC = (ImA*ReB-ReA*ImB)/Denom
END SUB
'-----------

'**************************************** END PROGRAM 'VSO_07-02-2013' ****************************************
```